\documentclass[a4paper,11pt, titlepage]{book}

\usepackage{amsmath}
\usepackage{graphicx}
\usepackage{amsfonts}
\usepackage{amssymb}
\usepackage{epsfig}
\usepackage{color}
\usepackage{psfrag}

\newcommand{\EQ}{\begin{equation}}
\newcommand{\EN}{\end{equation}}

\newcommand{\capitolo}[1]{%
   \renewcommand{\theequation}{\thesection.\arabic{equation}} \chapter{#1}}

\newcounter{sezioneapp}[chapter]
\renewcommand{\thesezioneapp}{\Alph{sezioneapp}}   
\newcommand{\sezioneapp}{\newpage\secdef\appendi\sappendi}%
\newcommand{\appendi}[2][?]{\refstepcounter{sezioneapp}%
\addcontentsline{toc}{section}{\protect\numberline{\thechapter.\thesezioneapp}#2}
   \markright {\MakeUppercase{\thesezioneapp. \  #2}}
   \setcounter{equation}{0}
   \renewcommand{\theequation}{\thechapter.\thesezioneapp.\arabic{equation}}
{\noindent\Large\bfseries Appendix \thechapter.\thesezioneapp.~ #2}%
\\[2.5ex]\protect{\noindent}}%
\newcommand{\sappendi}[1]{}

\begin{document}

\begin{titlepage}

\begin{figure}[t]
\begin{center}
\vspace{-1cm}
\includegraphics[height=2.5cm]{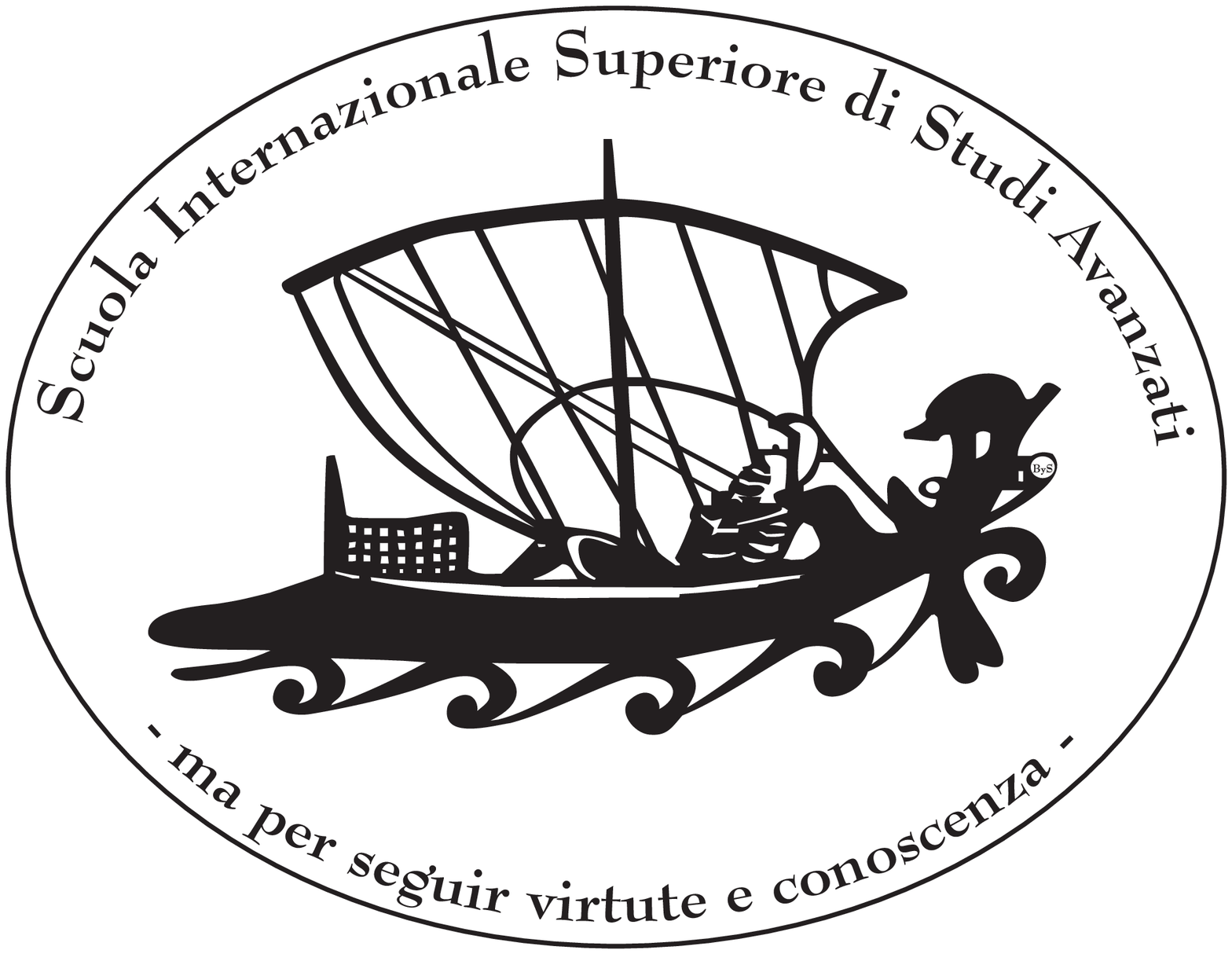}
\vspace{4cm}
\end{center}
\end{figure}

\bigskip
\bigskip
\bigskip
\huge
\begin{center}
\textbf{Semiclassical methods in 2D QFT:}\\
\textbf{spectra and finite--size effects}


\end{center}

\vspace{3cm}

\large
\begin{center}
Thesis submitted for the degree of \\
``Doctor Philosophi\ae''
\end{center}

\vspace{4cm}

\begin{tabular}{l l}
Candidate:  \hspace{7cm} & Supervisor: \\
Valentina Riva & Prof. Giuseppe Mussardo \vspace{0.5cm}\\
& External referee:\\
& Prof. Germ\'an Sierra
\end{tabular}

\vspace{3cm}

\begin{center}
October 2004
\end{center}

\vspace{-24.8cm}
\begin{center}
{\Large \bf SISSA \hspace{4cm} ISAS } \\
\vspace{1.2cm}
{\small SCUOLA INTERNAZIONALE SUPERIORE DI STUDI AVANZATI  \\
\vspace{0.1cm}
INTERNATIONAL  SCHOOL FOR ADVANCED STUDIES}\\
\end{center}

\vspace{1cm}

\end{titlepage}

\begin{center}
\large \bfseries Abstract
\end{center}
\normalsize

In this thesis, we describe some recent results obtained in the
analysis of two--dimensional quantum field theories by means of
semiclassical techniques. These achievements represent a natural
development of the non--perturbative studies performed in the past
years for conformally invariant and integrable theories, which
have lead to analytical predictions for several measurable
quantities in the universality classes of statistical systems.
Here we propose a semiclassical method to control analytically the
spectrum and the finite--size effects in both integrable and
non--integrable theories. The techniques used are appropriate
generalization of the ones introduced in seminal works during the
Seventies by Dashen, Hasllacher and Neveu and by Goldstone and
Jackiw. Their approaches, which do not require integrability and
therefore can be applied to a large class of system, are best
suited to deal with those quantum field theories characterized by
a non--linear interaction potential with different degenerate
minima. In fact, these systems display kink excitations which
generally have a large mass in the small coupling regime. Under
these circumstances, although the results obtained are based on a
small coupling assumption, they are nevertheless non-perturbative,
since the kink backgrounds around which the semiclassical
expansion is performed are non-perturbative too.

\tableofcontents

\pagestyle{headings}

\chapter*{Introduction} \setcounter{equation}{0}
\addcontentsline{toc}{chapter}{Introduction}

\pagestyle{plain}

Non--perturbative methods in quantum field theory (QFT) play a
central r\^ole in theoretical physics, with applications in many
areas, from string theory to condensed matter. In the last years,
a considerable progress has been registered in the study of
two--dimensional systems. For these models, in fact, exact results
have been obtained in the particular situations when the systems
are conformally invariant or integrable.

Conformal invariance does not merely represent a restriction on
the solvable problems. In fact, it provides a fruitful link
between QFT and statistical mechanics, realized through the ideas
of scaling and renormalization, which explain the basic features
of phase transitions and are naturally encoded in the QFT
framework. In fact, physical systems at a second order phase
transition (called critical) are characterized by the divergence
of the correlation length, which makes unimportant the fine
details of their microscopic structure and leads to an
organization of the critical behaviours in universality classes.
As a consequence, the order parameter fluctuations can be
described with the language of conformal field theories (CFT). The
corresponding dynamics can be exactly solved in two-dimensions,
since in this case conformal invariance permits to determine
universal amplitudes and to extract as well the entire spectrum of
the theory.

Furthermore, dynamics in the vicinity of a critical point can be
described by perturbed CFT, obtained by adding to the action
operators which break the conformal symmetry and introduce a mass
scale in the system, inducing a renormalization group flow.
Suitable choices of the perturbing operator make the off-critical
massive field theory integrable, with consequent elasticity and
factorization of the scattering. In two dimensions, this fact
together with a simplified kinematics leads to the exact
computation of the scattering amplitudes, which encode in their
analytical structure the complete information about the spectrum.
Moreover, the knowledge of the $S$--matrix permits to implement
the so--called form factors approach, which makes the analysis of
off--shell correlators possible.

Within this program, during the the last years measurable
universal quantities for many statistical models have been
computed. The studied systems include, for instance, the Ising
model in a magnetic field, the tricritical Ising model, the
$q$-state Potts model, percolation and self-avoiding walks. In all
other cases, the understanding of two dimensional QFT has been
reached up to now either by conformal perturbation theory or
numerical methods.

A complete understanding of QFT also requires the control of
finite--size effects, both for practical and theoretical reasons.
In fact, the finite--volume energy spectrum contains a lot of
information about the properties of the theory in infinite volume.
On the one hand, this permits to control the systematic error
induced by the finiteness of the samples in the extrapolation
procedure of numerical simulations. On the other hand, scattering
data can be extracted from the numerical analysis through this
correspondence. Moreover, defining a theory in finite volume one
can explicitly follow the renormalization group flow between its
ultraviolet (UV) conformal limit and the infrared (IR) massive
behaviour, since these two limits correspond, respectively, to
volumes much smaller or larger than the correlation length.
Finally, QFT in finite volume are intimately related to another
important subject, i.e. QFT at finite temperature.

At present, finite--size effects have been studied
non--perturbatively only in the above mentioned cases of conformal
or integrable QFT. At criticality, in fact, powerful analytical
techniques are available to extract the entire spectrum of the
transfer matrix. Furthermore, in integrable massive theories, the
knowledge of the exact $S$--matrix permits to study the system at
finite temperature, with the so--called Thermodynamic Bethe
Ansatz. This technique provides integral equations, mostly solved
numerically, for the ground state energy of the theory in finite
volume. In all other cases, the control of finite--size effects
has been reached either by perturbative or numerical methods.

\vspace{0.5cm}

A natural development of the above mentioned studies of
two--dimensional QFT consists in looking for some techniques to
control analytically non--integrable systems and finite--size
effects, both for theoretical reasons and their application to
several condensed--matter or statistical systems. My original
contribution to this research stream has been presented in the
papers [1-4] and will be described in this thesis. The basic tool
used is an appropriate generalization and extension of
semiclassical methods, which have proved to be efficient in
analysing non-perturbative effects in QFT since their introduction
in the seminal works \cite{DHN,GJ}. The semiclassical approach
does not require integrability, therefore it can be applied on a
large class of systems. At the same time, it permits to face
problems which are not fully understood even in the integrable
cases, such as the analytic study of QFT in finite volume. In
particular, it has led to new non--perturbative results on form
factors at a finite volume \cite{finvolff}, spectra of
non--integrable models \cite{dsgmrs} and energy levels of QFT on
finite geometries \cite{SGscaling,SGstrip}.

The semiclassical method is best suited to deal with those quantum
field theories characterized by a non--linear interaction
potential with different degenerate minima. In fact, these systems
display kink excitations, associated to static classical
backgrounds which interpolate between the degenerate vacua and
generally have a large mass in the small coupling regime. Under
these circumstances, although the results obtained are based on a
small coupling assumption, they are nevertheless non-perturbative,
since the kink backgrounds around which the semiclassical
expansion is performed are non-perturbative too. In any case, the
restriction on the variety of examinable theories imposed by the
above requirements is rather mild, since non--linearity is the
main feature of a wealth of relevant physical problems.

In the study of non--integrable spectra and finite--size effects,
we have used two basic results. The first is the well--known
semiclassical quantization technique, introduced by Dashen,
Hasslacher and Neveu (DHN) in \cite{DHN}. It consists in solving a
Schr\"odinger--like problem, associated to the \lq\lq stability
equation" for the small quantum fluctuations around the classical
backgrounds, and in building the energy levels of the system in
terms of the classical energy and the stability frequencies. The
second tool is a covariant refinement of a result due to Goldstone
and Jackiw \cite{GJ}. They have shown that the classical kink
backgrounds can be interpreted at quantum level as the Fourier
transform of the form factors between kink states. In
\cite{finvolff}, we have overcome the drawbacks of the original
result, which was expressed non-covariantly in terms of the
particles space momentum, using the so-called rapidity variable,
which is Lorentz invariant. This refinement opened the way to the
use of semiclassical form factors in the analysis of particle
spectra and correlation functions, because it permits to go in the
crossed channel and write the form factor between the vacuum and a
kink-antikink state. Since the analytical structure of these form
factors encodes the information about the masses of all the
kink-antikink bound states, in this way it is possible to explore
non-perturbatively the complete spectrum of non--integrable
theories, a purpose for which there are no other known analytical
techniques. In \cite{finvolff}, a prediction about the spectrum of
the $\phi^{4}$ field theory in the broken $\mathbb{Z}_{2}$
symmetry phase was made, in agreement with previous approximate
results. In the following work \cite{dsgmrs}, a detailed analysis
of the spectrum was performed in the double Sine-Gordon model,
which turns out to be a relevant description of many concrete
physical systems and displays appealing features such as false
vacuum decay and phase transition phenomena.

The study of finite--size effects can be tackled after having
identified suitable classical solutions to describe the kinks in
finite volume. With this respect, the key result of
\cite{finvolff} is the construction of such backgrounds for the
Sine-Gordon and the broken $\phi^4$ theories on a cylindrical
geometry with antiperiodic boundary conditions. The form factors
are then expressed as a Fourier series expansion of the classical
solutions, and they are used to obtain an estimate of the spectral
representation of correlation functions in finite volume. This
result adds to few others, which however strictly rely on specific
integrable structures of the considered models, while it can be in
principle extended to any theory displaying topologically
non-trivial backgrounds. The complete DHN quantization has been
then performed in \cite{SGscaling} for static classical solutions
on a finite geometry. In particular, the example of the
Sine--Gordon model with periodic boundary conditions has been
explicitly treated, reconstructing the scaling function and the
excited energy levels in terms of the size of the system. Finally,
the same theory, defined this time on a strip with Dirichlet
boundary conditions, has been studied in \cite{SGstrip}. The
semiclassical achievements provide an explicit analytical control
on the interpolation between the massive field theory and its UV
limit, and can be compared with the numerical results obtained  in
the sine--Gordon model with different techniques. However, their
application is more general, and permits to estimate the scaling
functions also in non--integrable theories.

The thesis is organized as follows. In Chapter\,\ref{chapint} we
review the non--perturbative results available for conformal and
integrable QFT in two dimensions, with particular emphasis on
their applications to statistical mechanics. In
Chapter\,\ref{chapSM} we introduce the semiclassical techniques
established in the past for QFT in infinite volume, and we test
the efficiency of the semiclassical approximation by comparing the
results obtained in the sine--Gordon model with the exact ones
provided by the integrability of the theory.
Chapter\,\ref{chapnonint} is devoted to the analysis of the
spectra of excitations in the non--integrable systems defined by
the $\phi^4$ field theory in the broken $\mathbb{Z}_{2}$ symmetry
phase and by the Double Sine--Gordon model. Finally, in
Chapter\,\ref{chapfinitesize} we present the study of finite--size
effects. In particular, we describe the form factors in the
sine--Gordon and broken $\phi^4$ theories defined on a twisted
cylinder, and the energy levels in the sine--Gordon model defined
both on a periodic cylinder and on a strip with Dirichlet boundary
conditions.

\capitolo{Quantum field theories in 2D and statistical physics}
\label{chapint}\setcounter{equation}{0}

\pagestyle{headings}

One of the most fruitful applications that QFT has found in recent
years is the analysis of the universality classes of
two--dimensional statistical models near their critical points,
which correspond to second order phase transitions. Critical
systems, in fact, fall into universality classes which can be
classified by conformally invariant field theories (CFT). The
corresponding dynamics can be exactly solved in two-dimensions
through a systematic computation of correlators, allowed by
conformal invariance. Furthermore, dynamics in the vicinity of a
critical point can be described by perturbed CFT, obtained by
adding to the action operators which break the conformal symmetry
and introduce a mass scale in the system. Suitable choices of the
perturbing operator make the off-critical massive field theory
integrable, with consequent elasticity and factorization of the
scattering. In two dimensions, this fact together with the
simplified kinematics leads to the possibility of computing
exactly the scattering amplitudes, which encode in their
analytical structure the complete information about the spectrum.
Moreover, the knowledge of the $S$--matrix permits to implement
the so--called form factors approach, which makes the analysis of
off--shell correlators possible. Within this program, during the
the last years measurable universal quantities for many
statistical models have computed. The studied systems include, for
instance, the Ising model in a magnetic field, the tricritical
Ising model, the $q$-state Potts model, percolation and
self-avoiding walks.

In this Chapter, we will introduce the basic concepts related to
conformal invariance and integrability. Section\,\ref{secCFT}
presents a brief overview of conformal field theories, with few
examples which will be referred to in the following Chapters. In
Section\,\ref{secintegrable} we discuss the main features of
massive scattering in integrable quantum field theories,
exploiting the powerful consequences of the simplified
two--dimensional kinematics. Finally, Section\,\ref{secLG}
describes a useful effective Lagrangian description of CFT.

\section{Critical systems and conformal field theories}\label{secCFT}
\setcounter{equation}{0}

A statistical mechanical system is said to be critical when its
correlation length $\xi$, defined as the typical distance over
which the order parameters are statistically correlated, increases
infinitely (\,$\xi\rightarrow\infty$). Correspondingly, length
scales lose their relevance, and scale invariance emerges. This is
peculiar of the continuous (\,or second order) phase transitions,
which consist in a sudden change of the macroscopic properties of
the system as some parameters (\,e.g. the temperature) are varied,
without finite jumps in the energy (\,characteristic of first
order transitions). A remarkable property of these systems is that
the fine details of their microscopic structure become
unimportant, and the various possible critical behaviours are
organized in universality classes, which depend only on the space
dimensionality and on the underlying symmetry. This allows a
description of the order parameter fluctuations in the language of
a field theory, which is invariant under the global scale
transformations
\begin{equation*}
x^{\mu}\rightarrow x'\,^{\mu}=\lambda x^{\mu} \,,
\end{equation*}
provided that the fields transform as
\begin{equation*}
\Phi(x)\rightarrow \Phi'(x')=\lambda^{-\Delta}\Phi(x) \,,
\end{equation*}
where $\Delta$ is called the scaling dimension of the field
$\Phi$.

The use of conformal invariance to describe statistical mechanical
systems at criticality is motivated by a theorem, due to Polyakov,
which states that local field theories which are scaling invariant
are also conformally invariant \cite{pol}. Therefore, every
universality class of critical behaviour can be identified with a
conformal field theory (CFT), i.e. a quantum field theory that is
invariant under conformal symmetry. This way of studying critical
systems started with a pioneering paper by Belavin, Polyakov and
Zamolodchikov \cite{bpz}, and is systematically presented in many
review articles and text books (\,see for instance
\cite{ginsp,zamrev,dif}). We will now concisely summarize the main
properties of CFT.

An infinitesimal coordinate transformation $x^{\mu}\rightarrow
x^{\mu}+\xi^{\mu}(x)$ is called conformal if it leaves the metric
tensor $g_{\mu\nu}$ invariant up to a local scale factor, i.e.
\begin{equation*}\label{metr}
g_{\mu\nu}(x)\rightarrow \varrho(x)g_{\mu\nu}(x).
\end{equation*}
These transformations, which include rotations, translations and
dilatations, preserve the angle between two vectors and satisfy
the condition
\begin{equation}\label{csi}
\partial_{\mu}\xi_{\nu}+\partial_{\nu}\xi_{\mu}=\frac{2}{d}\,\eta_{\mu\nu}(\partial\cdot\xi),
\end{equation}
where $d$ is the dimension of space-time. In two dimensions, since
the conformal group enlarges to an infinite set of
transformations, it is possible to solve exactly the dynamics of a
critical system, assuming conformal invariance and a
short-distance operator product expansion (OPE) for the
fluctuating fields. In fact, if we describe euclidean
two-dimensional space-time with complex coordinates
$$
z=x^{0}+ix^{1}\;,\qquad  \qquad \bar{z}=x^{0}-ix^{1}\;,
$$
eq.(\ref{csi}) specializes to the Cauchy-Riemann equations for
holomorphic functions. Therefore the solutions are holomorphic or
anti-holomorphic transformations, $z\rightarrow f(z)$ and
$\bar{z}\rightarrow \bar{f}(\bar{z})$, such that
$\partial_{\bar{z}}f=\partial_z\bar{f}=0$. These functions admit
the Laurent expansion
\begin{equation*}\label{lau}
f(z)=\sum_{n=-\infty}^{\infty}a_{n}\,z^{n+1}\;, \qquad \qquad
\bar{f}(\bar{z})=\sum_{n=-\infty}^{\infty}a'_{n}\,\bar{z}^{n+1}\;,
\end{equation*}
which has an infinite number of parameters. In this way, the
conformal group enlarges to an infinite set of transformations.

Defining now a two--dimensional quantum field theory invariant
under conformal transformation, we can associate to each field an
holomorphic and an antiholomorphic comformal dimensions $h$ and
$\bar{h}$, defined in terms of the scaling dimension $\Delta$ and
of the spin $s$ as
\begin{equation}\label{h}
h=\frac{1}{2}(\Delta+s)\;,\qquad\qquad
\bar{h}=\frac{1}{2}(\Delta-s)\;.
\end{equation}
A field is called primary if it transforms under a local conformal
transformation $z\rightarrow w=f(z)$ as
\begin{equation}
\phi'(w,\bar{w})=\left(\frac{dw}{dz}\right)^{-h}\left(\frac{d\bar{w}}{d\bar{z}}\right)^{-\bar{h}}\phi(z,\bar{z})\;.
\end{equation}
Conformal invariance fixes the form of the correlators of two and
three primary fields up to a multiplicative constant:
\begin{equation}
\langle\phi_{1}(z_{1},\bar{z}_{1})\,\phi_{2}(z_{2},\bar{z}_{2})\rangle\,=\,\left\{
\begin{array}{ll}
\frac{C_{12}}{z_{12}^{2h}\,\bar{z}_{12}^{2\bar{h}}} &
\quad  \textrm{if}\;\;h_{1}=h_{2}=h \;\;\textrm{and}\;\;\bar{h}_{1}=\bar{h}_{2}=\bar{h}\medskip \\
0 & \quad \textrm{otherwise}\,\,\, \label{2pt}
\end{array}
\right. \,\, ,
\end{equation}
\begin{equation}\label{3pt}
\langle\phi_{1}(z_{1},\bar{z}_{1})\,\phi_{2}(z_{2},\bar{z}_{2})\,\phi_{3}(z_{3},\bar{z}_{3})\rangle\,=\,
C_{123}\frac{1}{z_{12}^{h_{1}+h_{2}-h_{3}}\,z_{23}^{h_{2}+h_{3}-h_{1}}\,z_{13}^{h_{3}+h_{1}-h_{2}}}
\end{equation}
\begin{displaymath}
\qquad\qquad\qquad\qquad\times\frac{1}{\bar{z}_{12}^{\bar{h}_{1}+\bar{h}_{2}-\bar{h}_{3}}\,
\bar{z}_{23}^{\bar{h}_{2}+\bar{h}_{3}-\bar{h}_{1}}\,\bar{z}_{13}^{\bar{h}_{3}+\bar{h}_{1}-\bar{h}_{2}}}\,,
\end{displaymath}
where $z_{ij}=z_{i}-z_{j}$ and
$\bar{z}_{ij}=\bar{z}_{i}-\bar{z}_{j}$.

It is typical of correlation functions to have singularities when
the positions of two or more fields coincide. The operator product
expansion (OPE) is the representation of a product of operators
(at positions $x$ and $y$) by a sum of terms involving single
operators multiplied by functions of $x$ and $y$, possibly
diverging as $x\rightarrow y$. This expansion has a weak sense,
being valid within correlation functions, and leads to the
construction of an algebra of scaling fields defined by
\begin{equation}\label{OPE}
\phi_{i}(x)\phi_{j}(y)=\sum_{k}\hat{C}_{ij}^{k}(x,y)\phi_{k}(y),
\end{equation}
where $\hat{C}_{ij}^{k}(x,y)$ are the structure constants.
Translation and scaling invariance forces these functions to have
the following form:
\begin{equation*}
\hat{C}_{ij}^{k}(x,y)=\frac{C_{ij}^{j}}{|x-y|^{\Delta_{i}+\Delta_{j}-\Delta_{k}}},
\end{equation*}
where $C_{ij}^{j}$ are exactly the undetermined multiplicative
constants of the tree-point correlators (\ref{3pt}).

A particularly important operator is the stress--energy tensor
$T^{\mu\nu}$, which expresses the variation of the action under a
transformation of coordinates $x^{\mu}\rightarrow
x^{\mu}+\xi^{\mu}(x)$:
\begin{equation*}\label{action}
\delta S=-\frac{1}{2\pi}\int d^{2}x
\,T^{\mu\nu}(x)\partial_{\mu}\xi_{\nu}\;.
\end{equation*}
Conformal invariance is equivalent to the vanishing of $\delta S$
under the condition (\ref{csi}), and it is guaranteed by the
tracelessness of the stress-energy tensor. Together with
translation and rotation invariance
($\partial_{\mu}T^{\mu\nu}=0$), the condition $T^{\mu}_{\mu}=0$ is
expressed in complex coordinates as
\begin{equation*}\label{T}
\partial_{\bar{z}} T=0 \;\qquad \textrm{and}\qquad \partial_z\bar{T}=0,
\end{equation*}
where $T(z)=T_{11}-T_{22}+2iT_{12}$ and
$\bar{T}(\bar{z})=T_{11}-T_{22}-2iT_{12}$. Therefore the
stress-energy tensor splits into a holomorphic and an
antiholomorphic part. In two dimensions, it is possible to deduce
the following OPE for the stress-energy tensor and a primary field
of dimension $(h,\bar{h})$:
\begin{eqnarray}\label{Tprim}
T(z)\,\phi(w,\bar{w})=\frac{h}{(z-w)^{2}}\,\phi(w,\bar{w})+\frac{1}{z-w}\,\partial_{w}\phi(w,\bar{w})\:\textrm{+
regular terms },\\
\bar{T}(\bar{z})\,\phi(w,\bar{w})=
\frac{\bar{h}}{(\bar{z}-\bar{w})^{2}}\,\phi(w,\bar{w})+\frac{1}{\bar{z}-\bar{w}}\,\partial_{\bar{w}}\phi(w,\bar{w})
\:\textrm{+ regular terms }.\nonumber
\end{eqnarray}
Furthermore, it is possible to show that the OPE of the
stress-energy tensor with itself has the form:
\begin{equation}\label{TT}
T(z)T(w)=\frac{c/2}{(z-w)^{4}}+\frac{2}{(z-w)^{2}}\,T(w)+\frac{1}{z-w}\,\partial
T(w)\:\textrm{+ regular terms },
\end{equation}
where the constant $c$, called central charge, depends on the
specific model. A similar expression holds for the antiholomorphic
component. The holomorphic and antiholomorphic components of the
stress-energy tensor can be expanded in Laurent series
respectively on modes $L_{n}$ and $\bar{L}_{n}$, which are the
quantum generators of the local conformal transformations
\begin{equation}\label{Ln}
T(z)=\sum_{n=-\infty}^{\infty}\frac{L_{n}}{z^{n+2}}\;,\qquad\qquad
\bar{T}(\bar{z})=\sum_{n=-\infty}^{\infty}\frac{\bar{L}_{n}}{\bar{z}^{n+2}}\;,
\end{equation}
and obey the Virasoro algebra
\begin{eqnarray}\label{vir}
&& [L_{n},L_{m}]=(n-m)\,L_{n+m}+\frac{c}{12}\,n(n^{2}-1)\,\delta_{n+m,0}\;,\nonumber\\
&& [\bar{L}_{n},\bar{L}_{m}]=(n-m)\,\bar{L}_{n+m}+\frac{c}{12}\,n(n^{2}-1)\,\delta_{n+m,0}\;,\\
&& [L_{n},\bar{L}_{m}]=0\;.\nonumber
\end{eqnarray}

In virtue of the decomposition of (\ref{vir}) in the direct sum of
two algebras, one in the holomorphic and the other in the
antiholomorphic sector, the general properties of CFT have the
same form in the two sectors, and from now on we will only
restrict to the holomorphic part.

Comparing definition (\ref{Ln}) with the OPE (\ref{Tprim}), we can
deduce the action of some generators on a primary field:
\begin{eqnarray}\label{Lprim}
\left(L_{0}\phi\right)(z)&=&h\,\phi(z)\nonumber\\
\left(L_{-1}\phi\right)(z)&=&\partial\phi(z)\\
\left(L_{n}\phi\right)(z)&=&0\quad\textrm{if}\quad n\geq
1\nonumber
\end{eqnarray}
The relation $[L_{0},L_{n}]=-nL_{n}$ leads to the interpretation
of generators $L_{n}$ with $n>0$ as destruction operators and with
$n<0$ as creation operators. Hence primary fields define highest
weight representations of the Virasoro algebra, being annihilated
by all destruction operators. The action of creation operators on
these fields is encoded in the regular part of the OPE
(\ref{Tprim}), and defines the so-called descendant fields
\begin{equation*}\label{desc}
\phi^{(n_{1},n_{2},...,n_{k})}=\left(L_{-n_{1}}L_{-n_{2}}...L_{-n_{k}}\right)\phi
\,,
\end{equation*}
which are again eigenvectors of $L_{0}$:
\begin{equation*}\label{hdesc}
L_{0}\left[\phi^{(n_{1},n_{2},...,n_{k})}\right]=\left(h+\sum_{i=1}^{k}n_{i}\right)\phi^{(n_{1},n_{2},...,n_{k})}\,.
\end{equation*}
The number $N=\sum_{i=1}^{k}n_{i}$ is called level of the
descendant. As an example, the stress-energy tensor is a level two
descendant of the identity $(T=L_{-2}\mathbb{I})$. The set
$[\phi]$ constituted by all the descendant fields of a primary
operator $\phi$ is called conformal family. It is possible to show
that every correlation function involving descendant fields can be
computed by applying a linear differential operator to the
correlation function of the corresponding primary fields.

The Hilbert space of states of a CFT is built by acting on the
vacuum with the operators evaluated at $z=0$. Therefore, the
primary states are given by
$$
|\,h\,\rangle\,\equiv\,\phi(0)|\,0\,\rangle\;,
$$
and the descendent states can be obtained from them as
$L_{-n_{1}}L_{-n_{2}}...L_{-n_{k}}|\,h\,\rangle$.

\begin{figure}[h]
\psfrag{x}{$x$}\psfrag{t}{$\tau$}\psfrag{R}{$R$}\psfrag{T}{\hspace{-0.6cm}$z\to
w(z)$}
\hspace{2.5cm}\psfig{figure=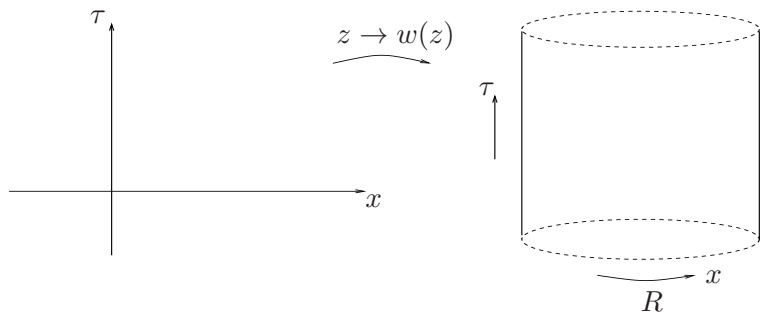,height=4cm,width=10cm}
\caption{Conformal map (\ref{cyl}) from plane to cylinder.}
\label{figplanecyl}
\end{figure}

In concluding this section, it is worth mentioning how the central
charge $c$ has the physical meaning of measuring the response of
the system to the introduction of a macroscopic scale
\cite{finitesize,cardy,casimir2}. In fact, the complex plane can
be conformally mapped to an infinite cylinder of circumference $R$
by the transformation (see Fig.\,\ref{figplanecyl})
\begin{equation}\label{cyl}
z\rightarrow w(z)=\frac{R}{2\pi}\ln z\;.
\end{equation}
Implementing the above transformation on the stress--energy tensor
one gets
\begin{equation*}
T_{cyl.}(w)=\left(\frac{2\pi}{R}\right)^{2}\left[T_{pl.}(z)z^{2}-\frac{c}{24}\right].
\end{equation*}
If we assume that the vacuum energy density $\langle
T_{pl.}\rangle$ vanishes on the plane, we see that it is non--zero
on the cylinder:
\begin{equation*}\label{cas}
\langle T_{cyl.}\rangle=-\frac{c\pi^{2}}{6R^{2}}.
\end{equation*}
The central charge is then proportional to the Casimir energy,
which naturally goes to zero as the macroscopic scale $R$ goes to
infinity. In particular, the hamiltonian and the momentum are
expressed on the cylinder in terms of the Virasoro generators as
\begin{equation}\label{hamcyl}
H=\frac{2\pi}{R}\left(L_{0}+\bar{L}_{0}-\frac{c}{12}\right) \qquad
\qquad P=\frac{2\pi i }{R}\left(L_{0}-\bar{L}_{0}\right)
\end{equation}

\vspace{0.5cm}

\subsection{Examples}

We will now describe few examples of CFT, which will be further
discussed in the thesis. We limit ourselves to the statement of
the main results, whose proofs can be found in the literature.

\subsubsection{Gaussian CFT}

An interesting CFT, only apparently trivial, is given by the
gaussian action which describes a free bosonic field
\begin{equation}\label{gaussian}
{\cal A}_{\textrm{G}} \,=\,\frac{1}{2}\,g \int d^2 x\;\,
\partial_{\mu} \phi \,\partial^{\mu} \phi \;,
\end{equation}
where $g$ is a normalization parameter. The two--point
function of the field $\phi$ is expressed as
$$
\langle\,\phi(z)\phi(w)\,\rangle\,=\,-\frac{1}{4\pi
g}\,\log(z-w)+\text{const.}\;.
$$
Using Wick's theorem, one can easily show that the OPE between the
energy--momentum tensor $ T(z)\equiv-2\pi
g\,:\partial\phi\,\partial\phi :\,$ and the operator
$\partial\phi$ is given by
$$
T(z)\partial\phi(w)\,=\,
\frac{\partial\phi(w)}{(z-w)^2}\,+\,\frac{\partial^2\phi(w)}{(z-w)}\,+\,\text{regular
terms}\;,
$$
from which it follows that $\partial\phi$ is a primary field with
conformal dimensions $h=\bar{h}=1$. Furthermore, the OPE of $T$
with itself
$$
T(z)T(w)\,=\, \frac{1/2}{(z-w)^4}\,+\,\frac{2
T(w)}{(z-w)^2}\,+\,\frac{\partial T(w)}{(z-w)}\,+\,\text{regular
terms}\;
$$
shows that the theory is characterized by a central charge $c=1$.
Besides $\partial\phi$, other primary operators can be built by
normal ordering the exponentials of $\phi$:
$$
V_{\alpha}(z,\bar{z}) \,=\,:\, e^{i \alpha\phi(z,\bar{z})} \,:\;.
$$
These fields, called vertex operators, have conformal dimensions
$h=\bar{h}=\frac{\alpha^2}{8\pi g}$.

A generalization of the above theory is obtained compactifying the
bosonic field, i.e. making it an angular variable on a circle of
radius ${\cal R}$. This can be done by defining the theory on a
cylinder of circumference $R$ with boundary conditions
\begin{equation}\label{compact}
\phi(x + R,t) \,=\,\phi(x,t) + 2\pi n\,{\cal R} \;.
\end{equation}
The index $n$, called winding number, labels the various sectors
of this CFT, together with an index $s$ related to the discrete
eigenvalues $\frac{2\pi s}{{\cal R}}$ of the momentum operator on
the circle. In each sector, the states $|\, s,n \rangle$ with
lowest anomalous dimension are created by the vertex operators
\begin{equation}\label{vertex}
 V_{s,n}(z,\bar{z}) \,=\,:\, \exp\left[i \alpha_{s,n}^+ \varphi(z)
+ i \alpha_{s,n}^- \bar{\varphi}(\bar{z})\right] \,:\;,
\end{equation}
i.e.
$$
\mid s,n\rangle \,=\,V_{s,n}(0,0) \,\mid \textrm{vac}\rangle
\,\,\,,
$$
where
\begin{eqnarray*}
\alpha_{s,n}^{\pm} & = &\frac{s}{{\cal R}}\pm 2\pi g\,n {\cal R}
\,\,\,\,; \\
\phi(x,t) & = & \varphi(z) + \bar{\varphi}(\bar{z}) \,\,\,\,\,.
\nonumber
\end{eqnarray*}
Their conformal dimensions are given by
\begin{equation}\label{dimvertex}
h_{s,n}  \,=\, 2\pi g\left(\frac{s}{4\pi g\,{\cal
R}}+\frac{1}{2}\,n {\cal R}\right)^{2} \;,\qquad \bar{h}_{s,n}
\,=\, 2\pi g\left(\frac{s}{4\pi g\,{\cal R}}-\frac{1}{2}\,n {\cal
R}\right)^{2} \,\,\,.
\end{equation}

\subsubsection{Minimal models}

Virasoro minimal models are particular CFT characterized by a
finite set of conformal families, in virtue of a truncation of the
operator algebra. These theories can be labelled as ${\cal
M}(p,p')$ with two integers $p$ and $p'$, in terms of which the
central charge and the conformal dimensions of primary fields are
expressed as
\begin{eqnarray}
&& c=1-6\frac{(p-p')^{2}}{p\, p'}\;,\\
&& h_{r,s}=\frac{(p\, r-p'\,s)^{2}-(p-p')^{2}}{4 p \,p'}\;,
\end{eqnarray}
with
\begin{equation*}
1\leq r< p'\qquad\textrm{and}\qquad 1\leq s< p \; .
\end{equation*}
The conformal dimensions are organized in a rectangle in the
$(r,s)$ plane, called Kac table. The number of distinct fields is
$(p-1)(p'-1)/2$, since there is a symmetry $h_{r,s}=h_{p'-r,p-s}$
which makes half of the Kac rectangle redundant. It can be shown
that minimal models are unitary only if $|p-p'|=1$, and in this
case they are usually labelled with $p'=m$ and $p=m+1$.

\begin{figure}[h]
\psfrag{1}{\small$1$}\psfrag{2}{\small$2$}\psfrag{3}{\small$3$}\psfrag{a}{\small$1$}
\psfrag{b}{\small$2$}\psfrag{r}{$r$}
\psfrag{s}{$s$}\psfrag{I}{$\mathbb{I}$}\psfrag{sg}{$\sigma$}\psfrag{e}{$\varepsilon$}
\hspace{5cm}\psfig{figure=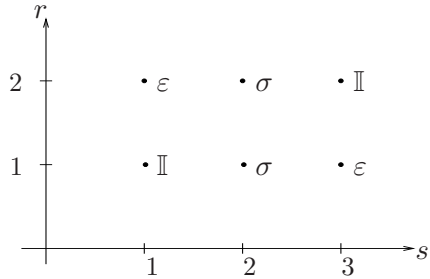,height=3.5cm,width=5.5cm}
\caption{Kac table of the minimal model ${\cal
M}(4,3)$}\label{kacising}
\end{figure}

The simplest unitary minimal model is ${\cal M}(4,3)$, which has
central charge $c=\frac{1}{2}$ and the Kac table shown in
Fig.\,\ref{kacising}.

This field theory is in the same universality class as the lattice
Ising model \cite{bpz}, defined by the usual configuration energy
\begin{equation*}\label{ising}
E[\sigma]=-J\sum_{\langle
i,j\rangle}\sigma_{i}\sigma_{j}-h\sum_{i}\sigma_{i} \;, \qquad
\sigma_{i}\in\{-1,1\} \;.
\end{equation*}
Besides the identity operator $\phi_{1,1}=\mathbb{I}$, the theory
contains the operator $\phi_{1,2}=\sigma$ with conformal dimension
$h_{\sigma}=\frac{1}{16}$, which is the continuum version of the
lattice spin $\sigma_{i}$, and $\phi_{1,3}=\varepsilon$, with
$h_{\varepsilon}=\frac{1}{2}$, which corresponds to the
interaction energy $\sigma_{i}\sigma_{i+1}$. The algebra defined
by the OPE (\ref{OPE}) can be schematically represented in this
model by the fusion rules
\begin{eqnarray*}\label{fusis}
&&\sigma\times\sigma\,\sim\,\mathbb{I}+\varepsilon\;,\\
&&\sigma\times\varepsilon\,\sim\,\sigma\;,\\
&&\varepsilon\times\varepsilon\,\sim\,\mathbb{I}\;.
\end{eqnarray*}
This notation means, for instance, that the OPE of $\sigma$ with
$\sigma$ (or fields belonging to their families) may contain terms
belonging only to the conformal families of $\mathbb{I}$ and
$\varepsilon$.

\begin{figure}[h]
\psfrag{1}{\small$1$}\psfrag{2}{\small$2$}\psfrag{3}{\small$3$}\psfrag{4}{\small$4$}\psfrag{a}{\small$1$}
\psfrag{b}{\small$2$}\psfrag{c}{\small$3$}\psfrag{r}{$r$}
\psfrag{s}{$s$}\psfrag{I}{$\mathbb{I}$}\psfrag{sg}{$\sigma$}\psfrag{sgp}{$\sigma'$}
\psfrag{e}{$\varepsilon$}\psfrag{ep}{$\varepsilon'$}\psfrag{es}{$\varepsilon''$}
\hspace{5cm}\psfig{figure=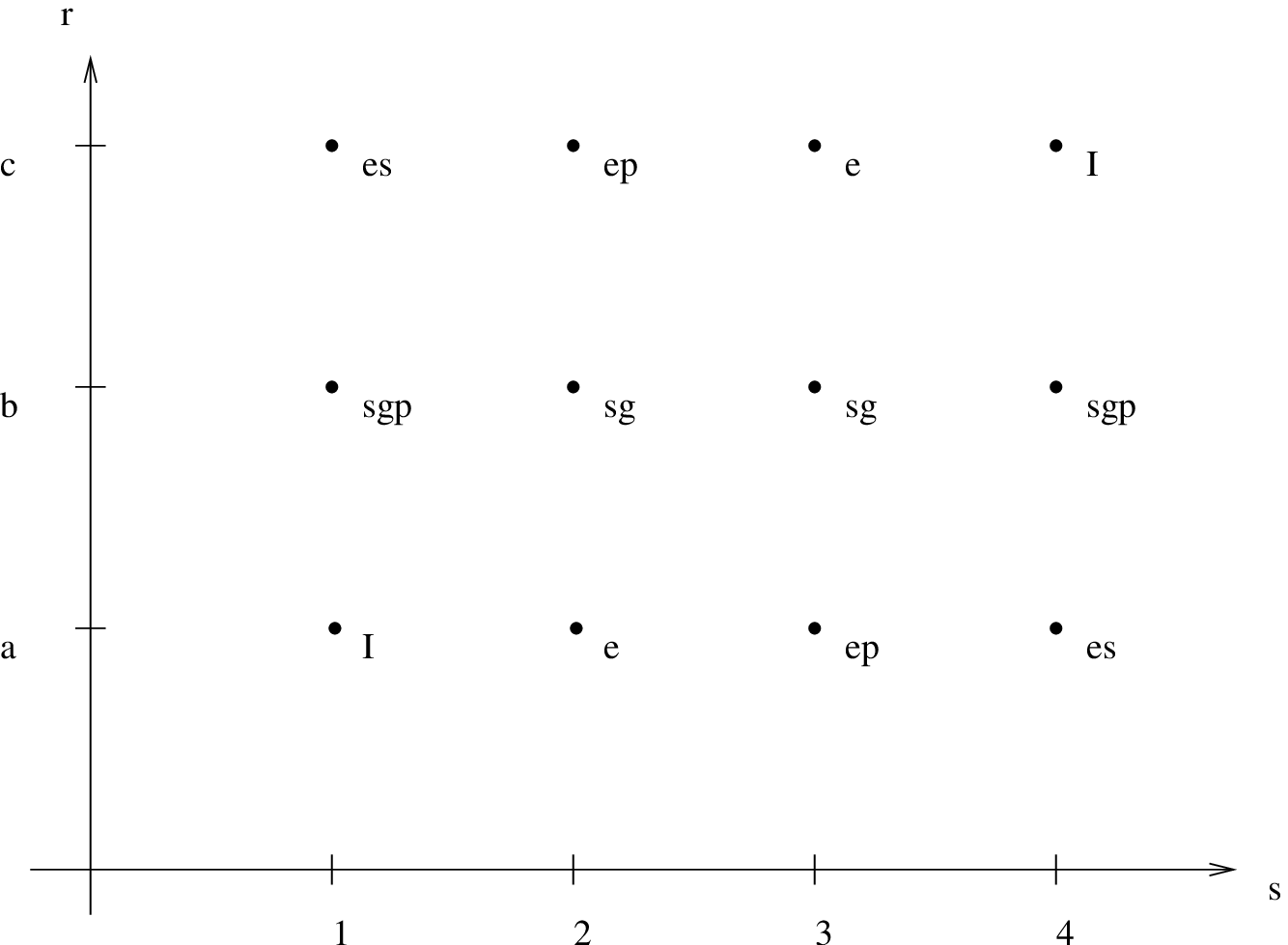,height=4cm,width=6cm}
\caption{Kac table of the minimal model ${\cal
M}(5,4)$}\label{kacTIM}
\end{figure}

The next unitary model, ${\cal M}(5,4)$, displays a richer
structure. The Kac table of this model, which has central charge
$c=\frac{7}{10}$, is shown in Fig.\,\ref{kacTIM}.

\begin{figure}[h]
\psfrag{J}{$J$}\psfrag{m}{$\mu^{-1}$}\psfrag{I}{$II$}\psfrag{II}{$I$}\psfrag{(Jc,mc)}{\hspace{-0.8cm}$(J_c,\mu_c^{-1})$}
\psfrag{(JI,0)}{$(J_I,0)$}
\hspace{4cm}\psfig{figure=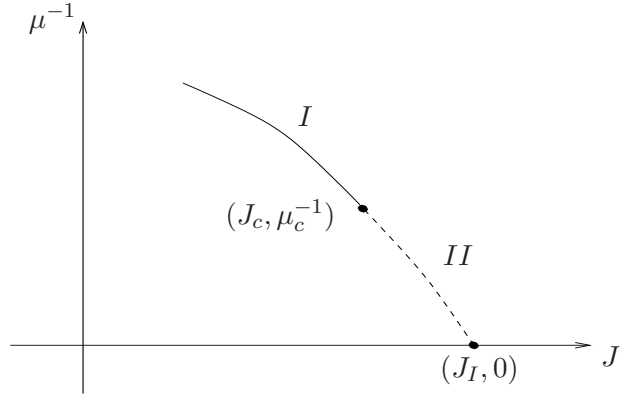,height=5cm,width=8cm}
\caption{Phase diagram of the TIM.} \label{TIMphasediagr}
\end{figure}

It was recognized in \cite{M54} that the lattice model associated
with this conformal field theory is the dilute Ising model at its
tricritical fixed point (TIM), defined by
\begin{equation*}\label{TIM}
E[\sigma,t]=-J\sum_{\langle
i,j\rangle}\sigma_{i}\sigma_{j}t_{i}t_{j}-\mu\sum_{i}(t_{i}-1) \;,
\qquad \sigma_{i}\in\{-1,1\},\,t_{i}\in\{0,1\} \;,
\end{equation*}
where $\mu$ is the chemical potential and $t_{i}$ is the vacancy
variable. The corresponding phase diagrams is drawn in Figure
\ref{TIMphasediagr}, where I and II denote respectively a first
and second order phase transition, and the point $(J_{I},0)$
represents the Ising model, with all lattice's site occupied.

The field $\phi_{1,2}=\varepsilon$, with
$h_{\varepsilon}=\frac{1}{10}$, corresponds to the energy density,
while $\phi_{1,3}=\varepsilon'$, with
$h_{\varepsilon'}=\frac{6}{10}$, is the vacancy (or subleading
energy) operator. The leading and subleading magnetization fields
are respectively $\phi_{2,2}=\sigma$ and $\phi_{2,1}=\sigma'$,
with $h_{\sigma}=\frac{3}{80}$ and $h_{\sigma'}=\frac{7}{16}$. The
remaining field $\phi_{1,4}=\varepsilon''$ has conformal dimension
$h_{\varepsilon''}=\frac{3}{2}$. Dividing the operators in even
and odd with respect to the $\mathbb{Z}_{2}$ symmetry of the model
under $\sigma_i\to -\sigma_i$, we can list the fusion rules in the
following way:

\begin{center}
\begin{tabular}{|l l l l l|} \hline
$\textrm{even}\times \textrm{even}$ & \hspace{0.5cm} & $\textrm{even} \times \textrm{odd}$ & \hspace{0.5cm} &
$\textrm{odd} \times \textrm{odd}$ \\
\hline $\varepsilon \times \varepsilon=\mathbb{I}+\varepsilon'$ &
& $\varepsilon \times \sigma=\sigma+\sigma'$ &
   & $\sigma\times \sigma= \mathbb{I}+\varepsilon+\varepsilon'+\varepsilon''$\\
$\varepsilon \times \varepsilon'=\varepsilon+\varepsilon''$  & &
$\varepsilon \times \sigma'=\sigma$ &
     & $\sigma\times \sigma'= \varepsilon+\varepsilon'$ \\
$\varepsilon \times \varepsilon''=\varepsilon'$ & & $\varepsilon'
\times \sigma=\sigma+\sigma'$ &
        & $\sigma' \times \sigma'= \mathbb{I}+\varepsilon''$\\
$\varepsilon' \times \varepsilon'=\mathbb{I}+\varepsilon'$ &
& $\varepsilon' \times \sigma'=\sigma$ &                     &\\
$\varepsilon' \times \varepsilon''=\varepsilon$  &                 & $\varepsilon'' \times \sigma=\sigma$ &          &\\
$\varepsilon'' \times \varepsilon''=\mathbb{I}$  &               & $\varepsilon'' \times \sigma'=\sigma'$ &        &\\
\hline
\end{tabular}
\end{center}

\vspace{0.5cm}

\section{Integrable quantum field theories}\label{secintegrable}
\setcounter{equation}{0}

The scaling region in the vicinity of second order phase
transitions can be described by a given CFT perturbed by its
relevant operators $\Phi_i$ (characterized by an anomalous
dimension $\Delta_i < 2$), with the action
\begin{equation}\label{Rgflow}
{\cal A}\, = \, {\cal A}_{CFT} \,+\,\sum\limits_i \lambda_i\int
d^{2}x\,\Phi_i(x)\;,
\end{equation}
where the couplings have mass dimension $\lambda_i\sim
[m]^{\,2-\Delta_i}\;$. The relevant operators, being
superrenormalizable with respect to UV divergencies, do not affect
the behaviour of the system at short distances, but they change it
at large scales. Any combination of the relevant fields defines a
Renormalization Group (RG) trajectory which starts from the given
CFT and can reach another critical point (defined by a different
CFT) or a non--critical fixed point, corresponding to a massive
QFT. From now on, we will only consider this second case. It was
shown in \cite{zamCFTpert} that, depending on the choice of the
perturbing operator, the off-critical massive field theory can be
integrable, with consequent elasticity and factorization of the
scattering. Obviously, this kind of theories covers only a class
of the statistical systems of interest, however this class
includes relevant physical problems. For instance, integrable QFT
correspond to the Ising model with thermal or magnetic
perturbation separately, described respectively by
$$
{\cal A}_{{\cal M}(4,3)} \,+\,\lambda_{\varepsilon}\int
d^{2}x\,\phi_{1,3}(x) \qquad\qquad \text{and}\qquad\qquad  {\cal
A}_{{\cal M}(4,3)} \,+\, \lambda_{\sigma}\int
d^{2}x\,\phi_{1,2}(x)\;,
$$
or to the tricritical Ising model perturbed by the leading energy
operator
$$
{\cal A}_{{\cal M}(5,4)} \,+\,\lambda_{\varepsilon}\int
d^{2}x\,\phi_{1,2}(x)\;.
$$

We will now present a brief overview of integrable massive quantum
field theories, underlying their basic features (for an exhaustive
review, see \cite{gmrep}). This will also give us the opportunity
of introducing the most important kinematical quantities used in
the following.

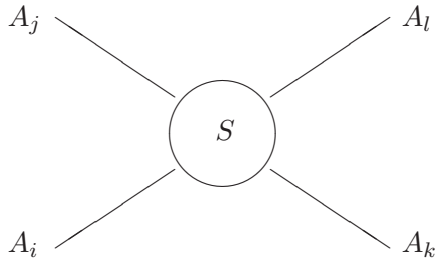
\begin{figure}[h]
\setlength{\unitlength}{0.0125in}
\hspace{0.3cm}\begin{picture}(40,90)(60,420) \put(292,467){$S$}
\put(295,470){\circle{55}} \put(315,485){\line(3,2){50}}
\put(315,455){\line(3,-2){50}} \put(275,485){\line(-3,2){50}}
\put(275,455){\line(-3,-2){50}}
\put(205,420){$A_{i}$}\put(205,515){$A_{j}$}\put(370,515){$A_{l}$}\put(370,420){$A_{k}$}
\end{picture}
\caption{Two-particle $S$-matrix}\label{figS}
 \end{figure}

An integrable QFT is characterized by the presence of an infinite
set of conserved charges, which make the corresponding scattering
theory purely elastic and factorized \cite{zams}. This implies
that an arbitrary $n$-particle collision process can be described
by the product of $n(n-1)/2$ elastic pair collisions. Hence the
determination of the complete $S$ matrix reduces to that of the
two-particle amplitudes, which are defined as
\begin{equation}\label{2partS}
|A_{i}\left(p_{1}\right)A_{j}\left(p_{2}\right)\rangle_{in}\,=\,
S_{ij}^{kl}\:|A_{k}\left(p_{3}\right)A_{l}\left(p_{4}\right)\rangle_{out}\,,
\end{equation}
where $A_{i}\left(p_{1}\right)$ and $A_{j}\left(p_{2}\right)$
denote the incoming particles (with 2-momenta $p_{1}^{\mu}$ and
$p_{2}^{\mu}$), and $A_{k}\left(p_{3}\right)$ and
$A_{l}\left(p_{4}\right)$ the outgoing states (see
Fig.\,\ref{figS}).

\vspace{1cm}

\vspace{0.5cm}

Lorentz invariance fixes the two body $S$-matrix to be a function
of the Mandelstam variables $s=\left(p_{1}+p_{2}\right)^{2}$,
$t=\left(p_{1}-p_{3}\right)^{2}$ and
$u=\left(p_{1}-p_{4}\right)^{2}$, which satisfy the relation
$s+t+u=\sum_{i=1}^{4}m_{i}^{2}$. Since in (1+1) dimensions and for
elastic scattering only one of these variables is independent, it
is convenient to introduce a parameterization of the momenta in
terms of the so-called rapidity variable $\theta$:
\begin{equation}\label{rapidity}
p_{i}^{0}=m_{i}\cosh\theta_{i}\,,\qquad\qquad
p_{i}^{1}=m_{i}\sinh\theta_{i}\,,
\end{equation}
which corresponds to the following expression for the Mandelstam
variable $s$:
\begin{equation}\label{smand}
s\,=\,\left(p_{1}+p_{2}\right)_{\mu}\left(p_{1}+p_{2}\right)^{\mu}=m_{i}^{2}+m_{j}^{2}+2m_{i}m_{j}\cosh\theta_{ij}\,,
\end{equation}
with $\theta_{ij}=\theta_{i}-\theta_{j}$. The functions
$S_{ij}^{kl}$ will then depend only on the rapidity difference of
the involved particles:
\begin{equation*}
|A_{i}\left(\theta_{1}\right)A_{j}\left(\theta_{2}\right)\rangle_{in}\,=\,S_{ij}^{kl}\left(\theta_{12}\right)\,|
A_{k}\left(\theta_{2}\right)A_{l}\left(\theta_{1}\right)\rangle_{out}\,.
\end{equation*}

\begin{figure}[h]
\psfrag{s}{$s$}\psfrag{t}{\small$\theta$}\psfrag{(mi+mj)}{\footnotesize$\hspace{-0.2cm}\left(m_i+m_j\right)^2$}
\psfrag{(mi-mj)}{\footnotesize$\hspace{-0.5cm}\left(m_i-m_j\right)^2$}
\psfrag{ipi}{\small$i\pi$}\psfrag{-ipi}{\small$-i\pi$}\psfrag{physicalstrip}{physical
strip}\psfrag{A}{\footnotesize$\hspace{0.8cm}R_+$}\psfrag{B}{\footnotesize$\hspace{0.8cm}R_-$}
\psfrag{C}{\footnotesize$\hspace{-0.7cm}L_+$}\psfrag{D}{\footnotesize$\hspace{-0.7cm}L_-$}
\psfrag{a}{\footnotesize$R_+$}\psfrag{b}{\footnotesize$R_-$}
\psfrag{c}{\footnotesize$L_+$}\psfrag{d}{\footnotesize$L_-$}
\hspace{1.5cm}\psfig{figure=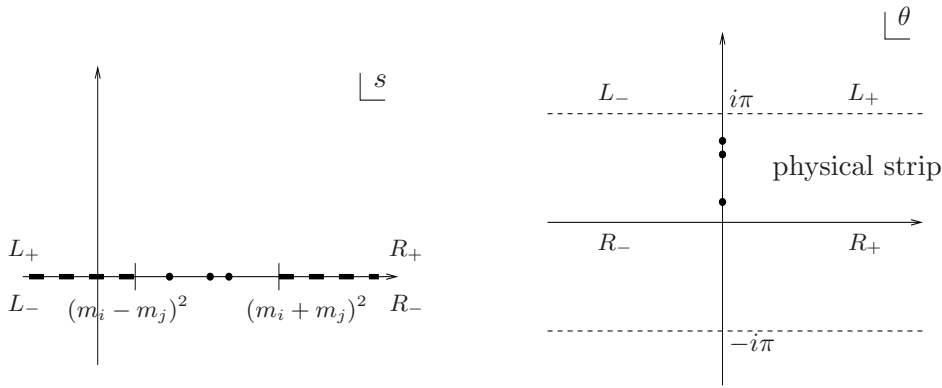,height=5cm,width=12cm}
\caption{Analytic structure of the elastic $S$--matrix in the
variables $s$ and $\theta$.} \label{figtheta}
\end{figure}

Elasticity of the scattering processes implies a drastic
simplification in the analytic structure of the $S$--matrix, which
can be extended to be an analytic function in the complex
$s$--plane \cite{bookS}. In fact, contrary to the generic case,
where many branch cuts are present, in the elastic case the
two--particle $S$-matrix only displays two square root branching
points at the two--particle thresholds
$\left(m_{i}-m_{j}\right)^{2}$ and $\left(m_{i}+m_{j}\right)^{2}$,
and is real valued on the interval of the real axis between them.
From (\ref{smand}) it follows that the functions
$S_{ij}^{kl}\left(\theta\right)$ are meromorphic in $\theta$, and
real at $\textrm{Re}(\theta)=0$. The two cuts in the $s$ variable,
in fact, are unfolded by the transformation (\ref{smand}): for
instance, the upper side of the cut along
$[\left(m_{i}+m_{j}\right)^{2},\infty]$ is mapped into the
positive semiaxis $0<\theta<\infty$, while the lower side is
mapped into the negative semiaxis $-\infty<\theta<0$. The physical
sheet of the $s$--plane goes into the strip $0\leq
\textrm{Im}(\theta)\leq\pi$, while the second Riemann sheet is
mapped into $-\pi\leq \textrm{Im}(\theta)\leq 0$. The structure in
the $\theta$ plane repeats then with periodicity $2\pi i$. See
Fig.\,\ref{figtheta} for a representation of the analytic
structure of the $S$--matrix in the two variables $s$ and
$\theta$.

The two-particle $S$-matrices satisfy the usual requirements of
unitarity, expressed as
\begin{equation}\label{unitbulk}
\sum_{n,m}S_{ij}^{nm}\left(\theta\right)S_{nm}^{kl}\left(-\theta\right)=\delta_{i}^{k}\delta_{j}^{l}\,,
\end{equation}
and crossing symmetry, given by
\begin{equation}\label{crossbulk}
S_{ik}^{lj}\left(\theta\right)=S_{ij}^{kl}\left(i\pi-\theta\right)\,,
\end{equation}
since the analytic continuation $s\to t$ from the $s$-channel to
the $t$-channel corresponds to the change of variable
$\theta\rightarrow i\pi-\theta$. Furthermore, the amplitudes are
restricted by the star-triangle (or Yang-Baxter) equations
\begin{equation}\label{YBbulk}
S_{i_{1}i_{2}}^{k_{1}k_{2}}\left(\theta_{12}\right)S_{k_{1}k_{3}}^{j_{1}j_{3}}\left(\theta_{13}\right)
S_{k_{2}i_{3}}^{j_{2}k_{3}}\left(\theta_{23}\right)=S_{i_{1}i_{3}}^{k_{1}k_{3}}\left(\theta_{13}\right)
S_{k_{1}k_{2}}^{j_{1}j_{2}}\left(\theta_{12}\right)S_{i_{2}k_{3}}^{k_{2}j_{3}}\left(\theta_{23}\right)\,,
\end{equation}
where a sum over the intermediate indices is understood. The
system of equations (\ref{unitbulk}), (\ref{crossbulk}) and
(\ref{YBbulk}) is in many cases sufficient to solve the kinematics
of the problem, determining a consistent solution for the
two-particle $S$-matrix, up to a so-called CDD ambiguity, which
consists in multiplying the solution by factors that alone satisfy
the same equations.

\begin{figure}[h]
\psfrag{Ai}{$A_i$}\psfrag{Aj}{$A_j$}\psfrag{Ak}{$A_k$}\psfrag{Al}{$A_l$}\psfrag{An}{$A_n$}
\psfrag{Gijn}{$\hspace{-0.2cm}\Gamma_{ij}^{n}$}\psfrag{Gnkl}{$\Gamma_{n}^{kl}$}
\hspace{5cm}\psfig{figure=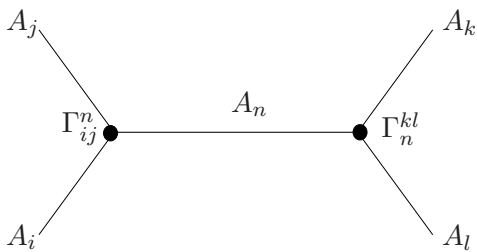,height=3cm,width=6cm}
\caption{First-order pole in the $S$--matrix} \label{figpoleS}
\end{figure}

Since the bound states of a theory correspond to singularities of
the $S$-matrix, its analytic structure encodes the spectrum of the
system. Stable bound states are usually associated to simple poles
in the $s$ variable located in the real interval
$\left(m_{i}-m_{j}\right)^{2}<s<\left(m_{i}+m_{j}\right)^{2}$.
These are mapped by (\ref{smand}) onto the imaginary
$\theta$--axis of the physical strip (see Fig.\,\ref{figtheta}).
If a two--particle amplitude with initial states $A_{i}$ and
$A_{j}$ has a simple pole in the $s$-channel at
$\theta=iu_{ij}^{n}$ ($A_{n}$ is the associated intermediate bound
state), in the vicinity of this singularity we have
\begin{equation}\label{resS}
S_{ij}^{kl}(\theta)\sim\frac{iR_{ij}^{n}}{\left(\theta-iu_{ij}^{n}\right)}\;,\qquad\text{with}\qquad
R_{ij}^{n}\,=\,\Gamma_{ij}^{n}\Gamma_{n}^{kl}
\end{equation}
where $\Gamma_{ij}^{n}$ are the on-shell three--particle coupling
constants of the underlying quantum field theory (see
Fig.\,\ref{figpoleS}). Remembering that the corresponding
singularity in the $s$ variable is of the form
$\left(s-m_{n}^{2}\right)^{-1}$ and using relation (\ref{smand}),
we get the following expression for the mass of the bound state:
\begin{equation}\label{boundstatesmasses}
m_{n}^{2}\,=\,m_{i}^{2}+m_{j}^{2}+2\,m_{i}\,m_{j}\,\cos
u_{ij}^{n}.
\end{equation}

\vspace{0.5cm}

The dynamics of the system can be determined by implementing the
so--called \lq\lq bootstrap principle", which consists in
identifying the bound states with some of the particles appearing
as asymptotic states. This leads to further equations which permit
to fix the CDD ambiguities mentioned above and to identify the
particle content of the theory.

Finally, it is worth recalling that unstable bound states
(resonances) are associated to $s$--variable poles in the second
Riemann sheet at $s=\left(m_k-i\frac{\Gamma_k}{2}\right)^2$, where
$\Gamma_k$ is the inverse life--time of the particle. These
correspond to poles in $\theta$ located in the strip $-\pi\leq
\textrm{Im}(\theta)\leq 0$ at positions
$\theta=-iu_{ij}^{k}+\alpha_{ij}^k$ satisfying
\begin{eqnarray}
&& m_k^2-\frac{\Gamma_k^2}{4}\,=\,m_{i}^{2}+m_{j}^{2}+2\,
m_{i}\,m_{j}\,\cos
u_{ij}^{k}\;\cosh\alpha_{ij}^k\;,\nonumber\\
&& m_k\,\Gamma_k\,=\,2\,m_i \,m_j \,\sin
u_{ij}^{k}\;\sinh\alpha_{ij}^k\;.\label{formres}
\end{eqnarray}

\vspace{0.5cm}

\begin{figure}[h]
\psfrag{a1}{$a_1$}\psfrag{a2}{$a_2$}\psfrag{a3}{$a_3$}\psfrag{an}{$a_n$}\psfrag{phi}{$\hspace{0.1cm}\phi$}
\hspace{5cm}\psfig{figure=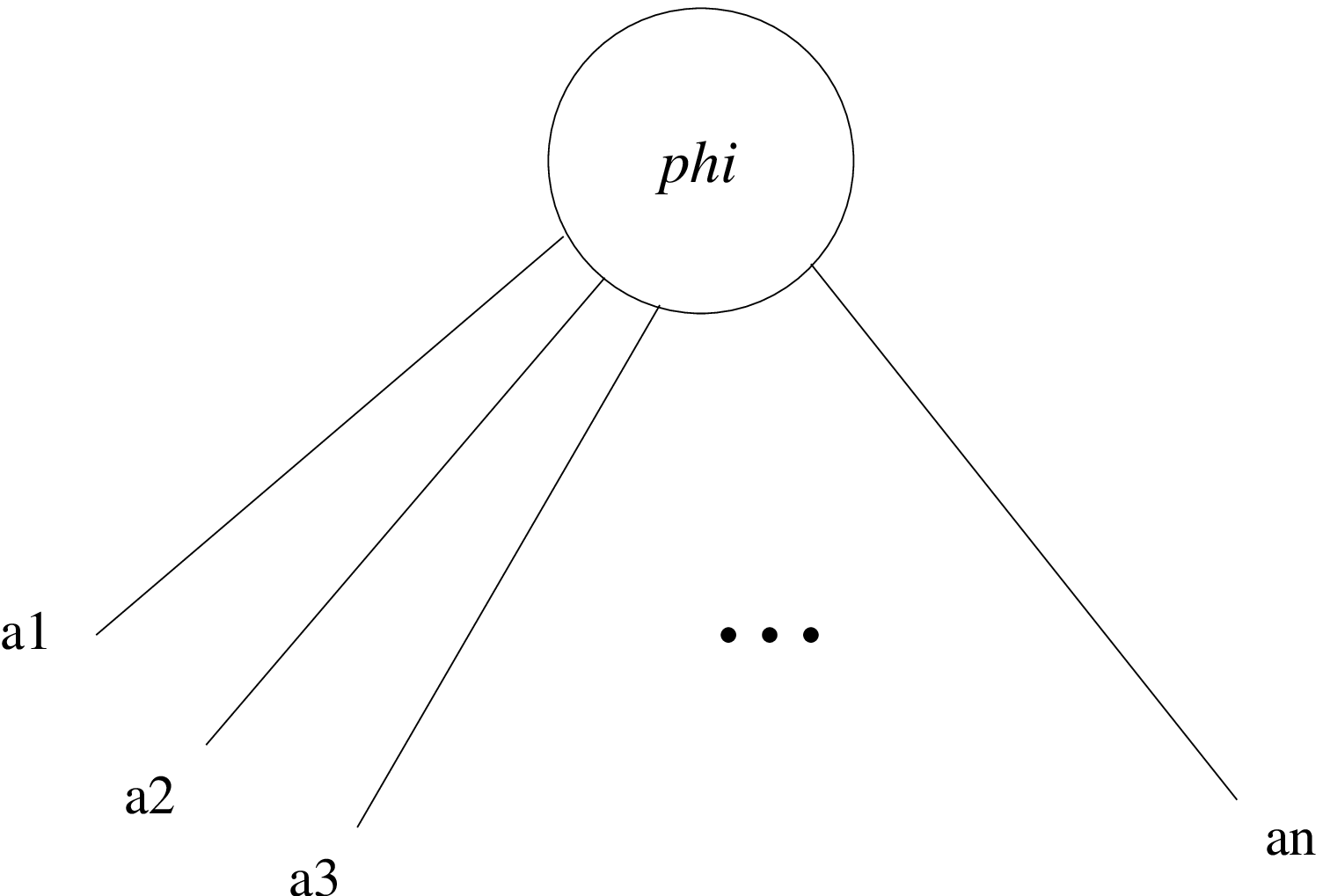,height=3.5cm,width=6cm}
\caption{Form factor (\ref{npartff})} \label{fignpartff}
\end{figure}

The knowledge of the exact $S$-matrix further permits to extract
non--perturbative information on the correlation functions of the
theory. In fact, the spectral function $\rho^{(\Phi)}$ associated
to a given operator $\Phi$, defined as
\begin{equation*}
\langle\,0\,|\,\Phi(x)\Phi(0)|\,0\rangle\,\equiv\int
\frac{d^{2}p}{(2\pi)^{2}}\,\rho^{(\Phi)}(p^{2})\,e^{i p\cdot x}\;,
\end{equation*}
can be expanded as a sum over complete sets of particle states
\begin{eqnarray}\nonumber
\rho^{(\Phi)}(p^{2})=2\pi\sum\limits_{n}\frac{1}{n!}\int &
d\Omega_{1}...d\Omega_{n}&\delta(p^{0}-p^{0}_{1}...-p^{0}_{n})\,
\delta(p^{1}-p^{1}_{1}...-p^{1}_{n})\,\times\,\\
  && \times\, \left|\,F^{\Phi}_{a_1\ldots
a_n}(\theta_1,\ldots,\theta_n) \,\right|^2\;, \label{spectral}
\end{eqnarray}
where $\,d\Omega\equiv\frac{d p}{2\pi\,2
E}=\frac{d\theta}{4\pi}\,$ and
\begin{equation}\label{npartff}
F^\Phi_{a_1\ldots a_n}(\theta_1,\ldots,\theta_n)=\langle
\,0\,|\,\Phi(0)\,|\, A_{a_1}(\theta_1)\ldots A_{a_n}(\theta_n)
\,\rangle\;.
\end{equation}
The matrix elements (\ref{npartff}), called \lq\lq form factors"
and pictorially depicted in Fig.\,\ref{fignpartff}, are subject to
the Watson equations, which relate them to the scattering
amplitudes. In the case of integrable theories these equations
take a simplified form \cite{KW,YZff,smirnovbook}, which permit
the determination of form factors once the $S$-matrix is known.
Furthermore, it has been shown in a series of works (see, for
instance \cite{fastconv}) that the spectral representation has a
fast convergent behaviour, therefore accurate estimates of
correlators and other related physical quantities can be obtained
by just using few exact terms in the series (\ref{spectral}),
having consequently a great simplification of the problem.

\begin{figure}[h]
\psfrag{phi}{$\phi$}\psfrag{a}{$a$}\psfrag{b}{$b$}\psfrag{c}{$c$}\psfrag{Gabc}{$\Gamma_{ab}^{c}$}
\hspace{4cm}\psfig{figure=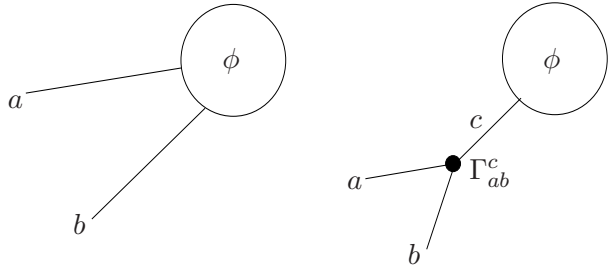,height=3.5cm,width=8cm}
\caption{Dynamical pole (\ref{ffdynpole}) in the two--particle
form factor $F_{ab}^{\Phi}(\theta)$} \label{figdynpoles}
\end{figure}

Here we limit ourselves to the description of the main equations
ruling the simplest non--trivial form factors, i.e. the
two--particle ones, that we will use and discuss in the following.
The discontinuity of the matrix elements across the unitarity cut
leads to the following relation with the two--particle scattering
amplitudes:
\begin{equation}\label{ffunitarity}
F_{ab}^{\Phi}(\theta) \,=\, S_{ab}^{cd}(\theta)
F_{cd}^{\Phi}(-\theta)\,\,,
\end{equation}
where $\theta = \theta_1 - \theta_2$. The crossing symmetry of the
form factor is expressed as
\begin{equation}\label{ffcrossing}
F_{a\bar{a}}^{\Phi}(\theta+2i\pi) \,= \,e^{-2i\pi\gamma_{\Phi,a}}
F_{\bar{a}a}^{\Phi}(-\theta)\,\,,
\end{equation}
where the phase factor $e^{-2i\pi \gamma_{\Phi,a}}$ is inserted to
take into account a possible semi-locality of the operator which
interpolates the particle $a$ (i.e. any operator $\varphi_a$ such
that $\langle 0|\varphi_a|a\rangle\neq 0$) with respect to the
operator $\Phi(x)$. When $\gamma_{\Phi,a} =0$, there is no
crossing symmetric counterpart to the unitarity cut but when
$\gamma_{\Phi,a}\neq 0$, there is instead a non-locality
discontinuity in the plane of the Mandelstam variable $s$ defined
in (\ref{smand}), with $s=0$ as branch point. In the rapidity
parameterization there is however no cut because the different
Riemann sheets of the $s$-plane are mapped onto different sections
of the $\theta$-plane; the branch point $s=0$ is mapped onto the
points $\theta=\pm i\pi$ which become therefore the locations of
simple annihilation poles. The residues at these poles are given
by
\begin{equation}\label{ffannih}
{\mbox Res}_{\theta=\pm i\pi}F_{ab}^{\Phi}(\theta)=i
\delta_{\bar{a}b}(1-e^{\mp 2i\pi\gamma_{\Phi,a}})\langle
0|\Phi|0\rangle\,\,.
\end{equation}
Finally, the two--particle form factors inherit the $s$--channel
bound state poles of the $S$--matrix, and the corresponding
residua are given by
\begin{equation}\label{ffdynpole}
{\mbox Res}_{\theta=iu_{ab}^{c}}F_{ab}^{\Phi}(\theta)=i
\Gamma_{ab}^{c}F_{c}^{\Phi}\,\,.
\end{equation}
where the couplings $\Gamma_{ab}^{c}$ coincide with the ones in
(\ref{resS}). This relation, pictorially represented in
Fig.\,\ref{figdynpoles}, is the property of form factors that we
will mostly exploit in the following.

\vspace{0.5cm}

\section{Landau--Ginzburg theory}\label{secLG}\setcounter{equation}{0}

As we have seen in Sect.\,\ref{secCFT}, statistical models at the
critical point can be described by conformal field theories (CFT),
and many systems of physical relevance have been identified with
the Virasoro minimal models. The operators of these theories can
be organized in a finite number of families, and this simplifies
the dynamics allowing in principle for a complete solution. The
only disadvantage of this kind of theories is that they have no
Lagrangian formulation, therefore they cannot be studied by a
path--integral approach and the underlying physics is not always
transparent. However, for a class of minimal theories, the unitary
ones ${\cal M}(m+1,m)$, there is a simple effective Lagrangian
description, suggested in \cite{LG}, which is realized by a
self--interacting field $\phi$ subjected to a power--like
potential. The field $\phi$ stands for the order parameter of the
statistical system, and the potential $V(\phi)$, whose extrema
correspond to the various critical phases of the system, is
usually chosen to be invariant under the reflection $\phi\to
-\phi$. For a potential of degree $2(m-1)$, this ensures the
existence of $m-1$ minima separated by $m-2$ maxima. Several
critical phases of the system can coexist if the corresponding
extrema coincide. The most critical potential is therefore a
monomial of the form
$$
V_m(\phi)\,=\,\phi^{\,2(m-1)}\;.
$$
Starting from the field $\phi$, one can construct composite fields
$:\phi^{\,k}:$ by normal ordering its powers. These have been
shown in \cite{LG} to display the same fusion properties as the
operators present in the minimal model ${\cal M}(m+1,m)$,
supporting in this way the correspondence. In particular, the
field $\phi$ is always associated to the primary operator
$\phi_{2,2}$.

One of the nicest features of the Landau--Ginzburg description is
that it provides a very intuitive picture of the perturbation of
CFT away from the critical points, since this simply corresponds
to adding opportune powers of $\phi$ to $V_m$.

\begin{figure}[ht]\hspace{1.5cm}
\begin{tabular}{p{4cm}p{4cm}p{4cm}}
\psfrag{phi}{$\phi$}\psfrag{V}{\hspace{-2cm}$(a)$\hspace{1.5cm}$V(\phi)$}
\psfig{figure=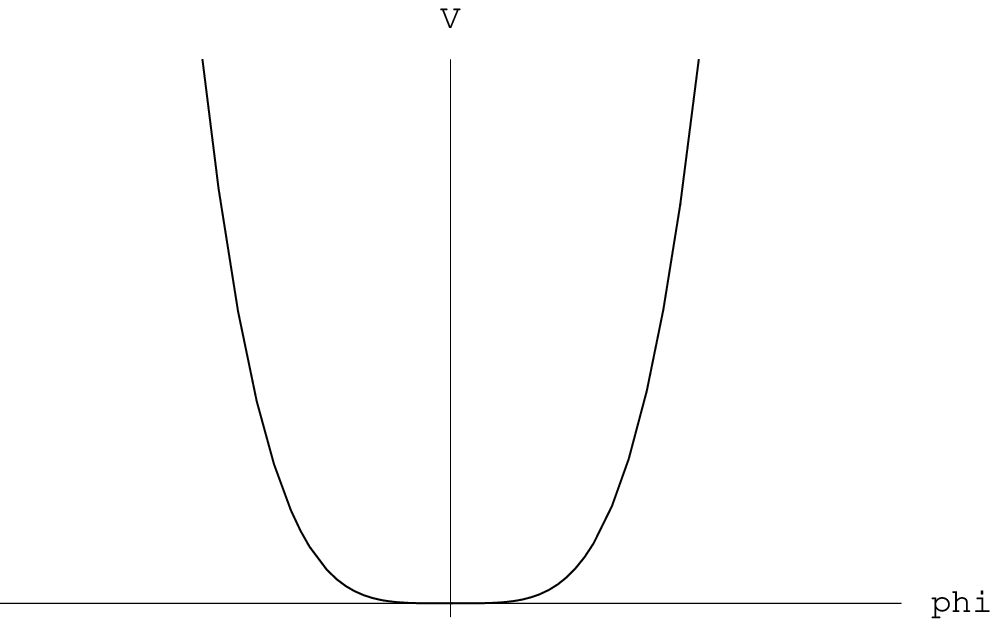,height=3cm,width=3.5cm} &
\psfrag{phi}{$\phi$}\psfrag{V}{\hspace{-2cm}$(b)$\hspace{1.5cm}$V(\phi)$}
\psfig{figure=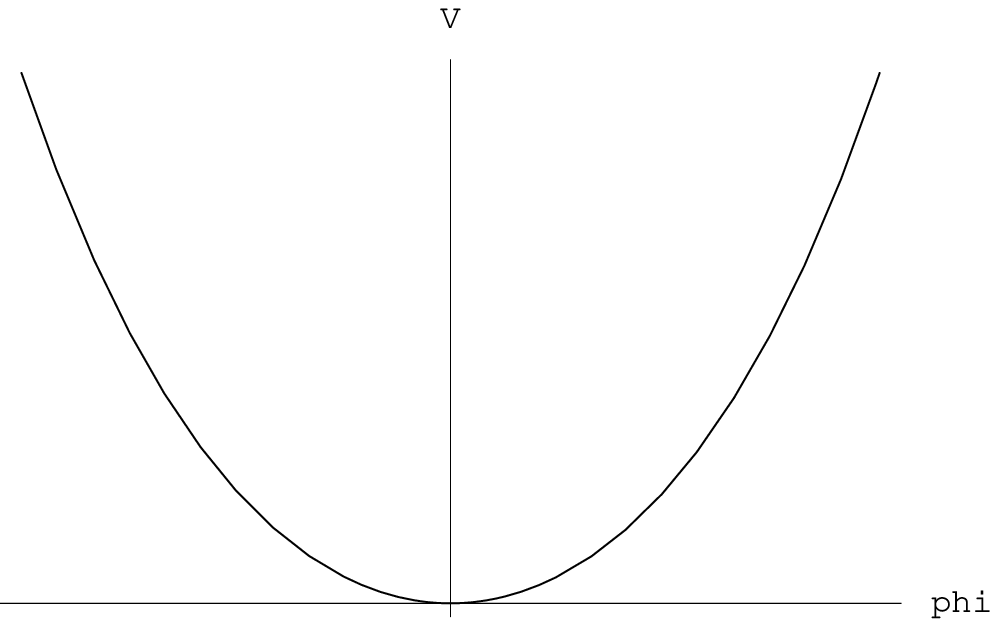,height=3cm,width=3.5cm} &
\psfrag{phi}{$\phi$}\psfrag{V}{\hspace{-2cm}$(c)$\hspace{1.5cm}$V(\phi)$}
\psfig{figure=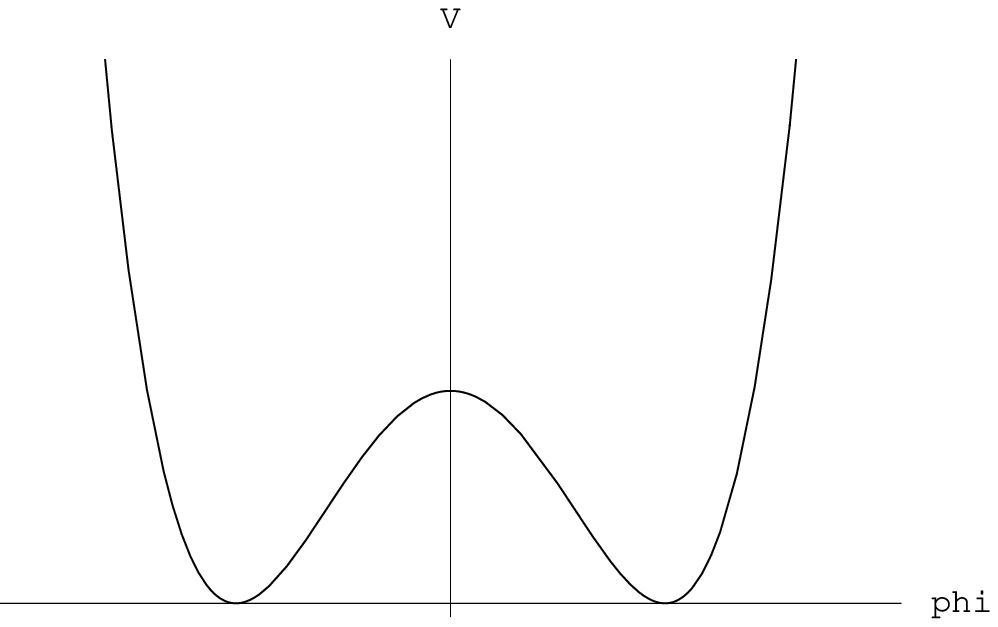,height=3cm,width=3.5cm}
\end{tabular}
\caption{Landau--Ginzburg potential for the Ising model $(a)$ at
the critical point, $(b)$ at high temperature, $(c)$ at low
temperature.}\label{figLGphi4}
\end{figure}

For instance, the universality class of the Ising model is
described by $m=3$, the spin operator corresponds to
$\sigma\sim\phi$ and the energy operator to
$\varepsilon\sim:\phi^{\,2}:$. Therefore the thermal perturbation
of the critical point is described by the Landau--Ginzburg theory
$$
V(\phi)\,=\,A\,\phi^{\,4}+ B\,\phi^{\,2}+C\;,
$$
where the sign of $B$ refers to high or low temperature,
respectively (see Fig.\,\ref{figLGphi4}).

\vspace{1cm}

\vspace{0.5cm}

\begin{figure}[ht]\hspace{1.5cm}
\begin{tabular}{p{4cm}p{4cm}p{4cm}}
\psfrag{phi}{$\phi$}\psfrag{V}{\hspace{-2cm}$(a)$\hspace{1.5cm}$V(\phi)$}
\psfig{figure=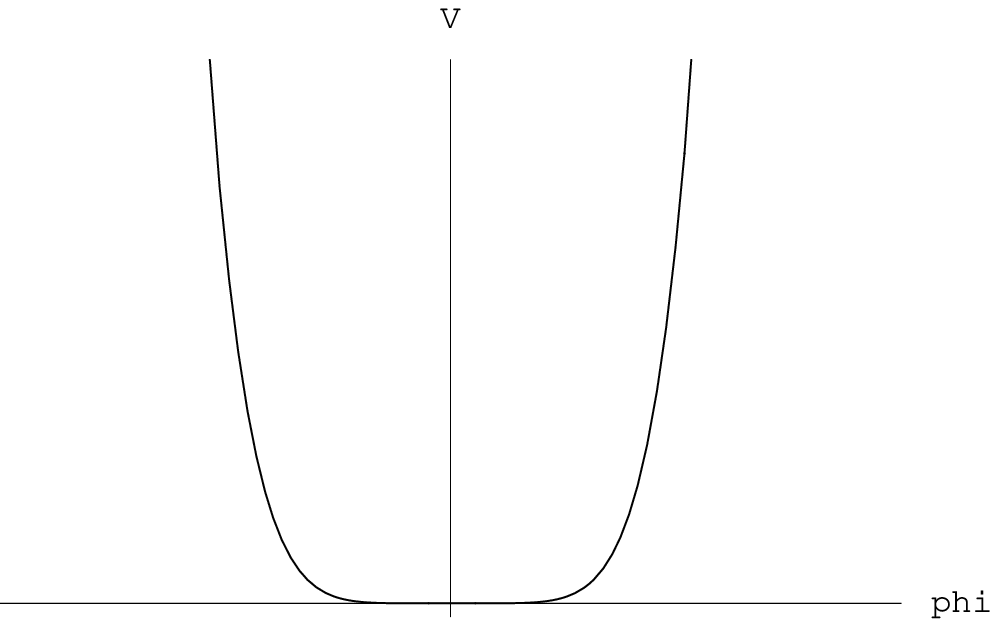,height=3cm,width=3.5cm} &
\psfrag{phi}{$\phi$}\psfrag{V}{\hspace{-2cm}$(b)$\hspace{1.5cm}$V(\phi)$}
\psfig{figure=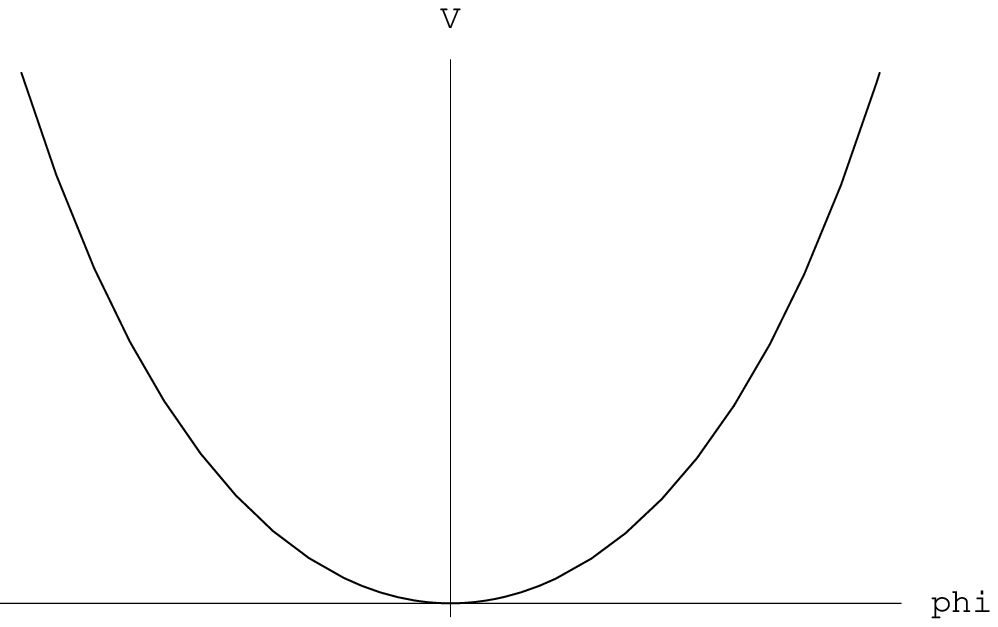,height=3cm,width=3.5cm} &
\psfrag{phi}{$\phi$}\psfrag{V}{\hspace{-2cm}$(c)$\hspace{1.5cm}$V(\phi)$}
\psfig{figure=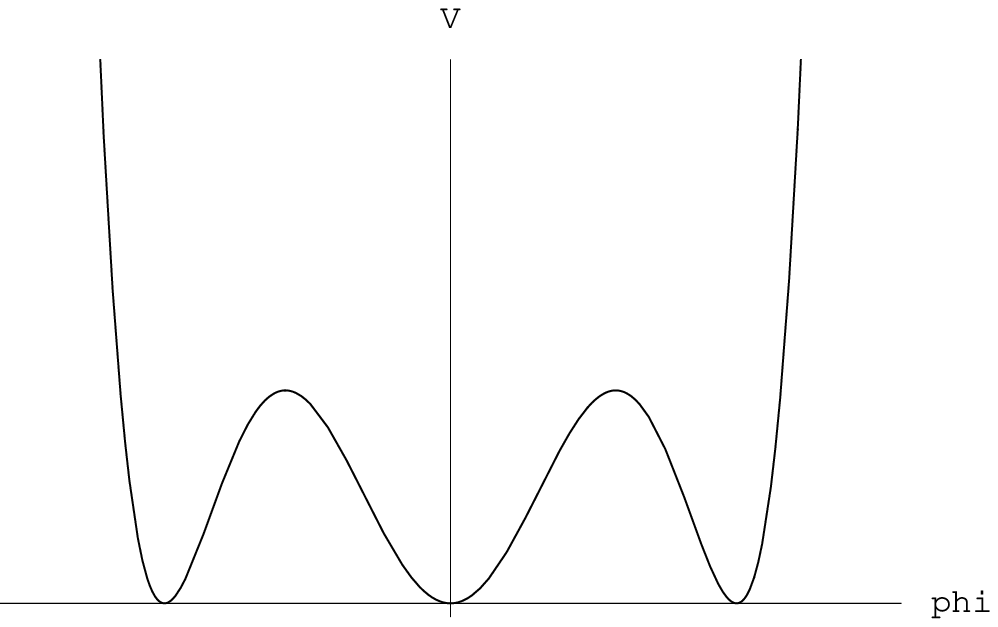,height=3cm,width=3.5cm}
\end{tabular}
\caption{Landau--Ginzburg potential for the tricritical Ising
model $(a)$ at the tricritical point, $(b)$ with positive leading
and subleading energy perturbations, $(c)$ with positive leading
and negative subleading energy perturbations.}\label{figLGphi6}
\end{figure}

Another very interesting example is given by the case $m=4$, which
corresponds to the tricritical Ising model. The operator content
of the theory is in this case identified as
$$
\begin{cases}
\sigma\sim\phi\\
\varepsilon\sim:\phi^{\,2}:\\
\sigma'\sim:\phi^{\,3}:\\
\varepsilon'\sim:\phi^{\,4}:\\
\varepsilon''\sim:\phi^{\,6}:\\
\end{cases}
$$
Hence, for instance, the perturbation of the tricritical point by
leading and subleading energy densities is described as
$$
V(\phi)\,=\,A\,\phi^{\,6}+ B\,\phi^{\,4}+C\,\phi^{\,2}+D\;,
$$
and some of the resulting potentials are shown in
Fig.\,\ref{figLGphi6}.

\vspace{1cm}

\vspace{0.5cm}

In concluding this Section it is worth to remark how the above
mentioned correspondence between Lagrangians and minimal models is
based on euristic arguments which are not rigorously proven.
Furthermore, quantum corrections can be very strong, hence a
quantitative analysis of statistical system is generally not
possible starting from the Lagrangian formulation. However, the
qualitative picture offered by the Landau--Ginzburg theory, being
essentially based on the symmetries of the model, remains valid
also at quantum level. In particular, this description nicely
illustrates the underlying physics in cases when the potential
displays degenerate minima, since the classical solutions
interpolating between the minima have a direct meaning of quantum
particles, as we are going to show in the next Chapter.

\capitolo{Semiclassical Quantization}
\label{chapSM}\setcounter{equation}{0}

In this Chapter we will describe the two basic technical tools
used in the following to investigate non--integrable spectra and
finite--size effects. The first is represented by the
semiclassical quantization technique introduced for relativistic
field theories in a series of papers by Dashen, Hasslacher and
Neveu (DHN) \cite{DHN} by using an appropriate generalization of
the WKB approximation in quantum mechanics. The second is a result
due to Goldstone and Jackiw \cite{GJ}, which relates the form
factors of the basic field between kink states to the Fourier
transform of the classical solution describing the kink.

Here we only focus on the description of these two results, that
we will fully exploit in the following. However, it should be
mentioned that the semiclassical study of QFT was developed during
the Seventies by few independent groups and with complementary
techniques, such as the path--integral approach, the inclusion of
fermions and the analysis of higher order corrections in the
semiclassical expansion. It is worth mentioning, for instance, the
work of Callan and Gross \cite{callgross}, Gervais, Jevicki and
Sakita \cite{pathint}, Faddeev and Korepin \cite{faddkor}. A
complete review of these beautiful achievements can be found in
\cite{raj}.

We will restrict to the study of two--dimensional theories, in
virtue of their simplified kinematics that allows for a powerful
applications of the semiclassical techniques. However, it is well
known that the semiclassical methods are naturally formulated for
QFT in any dimension $(d+1)$, and have provided, for instance,
relevant insight in the study of four--dimensional gauge theories.
In fact, the main feature that makes interesting their use is the
presence of non--linear interaction, which gives rise to
topologically non--trivial classical solutions. Classical
non--linear equations have been always recognized to play a
crucial role in the description of physical phenomena, in virtue
of their intriguing complex features (see for instance
\cite{whitham,scott}). The expectation of finding interesting
properties also in the associated quantum field theories, first
acquired by Skyrme \cite{skyrme}, opened the way to the
non--perturbative semiclassical studies that we are going to
describe.

The Chapter is organized as follows. In Section\,\ref{mainidea} we
introduce the DHN quantization technique for static
non--perturbative backgrounds. In Section\,\ref{secsemFF} we
describe the Goldstone and Jackiw's result and its relativistic
formulation, proposed in \cite{finvolff}, which is the basic tool
used in the following to study the spectrum of non--integrable
QFT. Finally, in Section\,\ref{SGinfinite} we apply the above
techniques to the sine--Gordon model, where, in virtue of the
integrability of the theory, exact result to be compared with the
semiclassical ones are available.

\section{DHN method}\label{mainidea}
\setcounter{equation}{0}

The semiclassical quantization of a field theory defined by a
Lagrangian
\begin{equation}\label{generallagr}
{\cal L}\,=\,\frac{1}{2}\left(\partial_\mu
\phi\right)\left(\partial^\mu \phi\right)-V(\phi)
\end{equation}
is based on the identification of a classical background
$\phi_{cl}(x,t)$ which satisfies the Euler--Lagrange equation of
motion
\begin{equation} \label{eom}
\partial_{\mu}\partial^{\mu}\phi_{cl}+V'(\phi_{cl})=0\;.
\end{equation}
The procedure is particularly simple and interesting if one
considers finite--energy static classical solutions $\phi_{cl}(x)$
in 1+1 dimensions, usually called \lq\lq kinks" or \lq\lq
solitons". Their presence is one of the main features of a large
class of 1+1--dimensional field theories defined by a non--linear
interaction $V(\phi)$ displaying discrete degenerate minima
$\phi_{i}$, which are constant solutions of the equation of motion
and are called \lq\lq vacua". The (anti)kinks interpolate between
two next neighbouring minima of the potential, and consequently
they carry topological charges $Q=\pm 1$, where
\begin{equation}
Q\,\equiv\,\frac{1}{\phi_{i+
1}-\phi_{i}}\,\int\limits_{-\infty}^{\infty}dx\,\frac{d\phi(x,t)}{dx}\;.
\end{equation}
The conservation of this quantity, which can be easily deduced
from the associated conserved current
$$
j^{\mu}=\epsilon^{\mu\nu}\partial_{\nu}\,\phi\;,\qquad\qquad
Q=\frac{1}{\phi_{i+ 1}-\phi_{i}}\,
\int\limits_{-\infty}^{\infty}dx\,j^0\;,
$$
will play a crucial role in the following.

Being static solutions of the equation of motion, i.e. time
independent in their rest frame, the kinks can be simply obtained
by integrating the first order differential equation related to
(\ref{eom})
\begin{equation}
\frac{1}{2}\left(\frac{\partial \phi_{cl}}{\partial x}\right)^{2}
\,=\, V(\phi_{cl}) + A \,\,\,, \label{firstorder}
\end{equation}
further imposing that $\phi_{cl}(x)$ reaches two different minima
of the potential at $x \rightarrow \pm \infty$. These boundary
conditions, which describe the infinite volume case, require the
vanishing of the integration constant $A$. As we will see in the
following, the kink solutions in a finite volume correspond
instead to a non--zero value of $A$, related to the size of the
system.

\begin{figure}[ht]
\begin{tabular}{p{5cm}p{5cm}p{5cm}}
\psfrag{phi}{$\phi$}\psfrag{V}{$V(\phi)$}\psfrag{b}{$\frac{m}{\sqrt{\lambda}}$}
\psfrag{a}{$\hspace{-0.3cm}-\frac{m}{\sqrt{\lambda}}$}
\psfig{figure=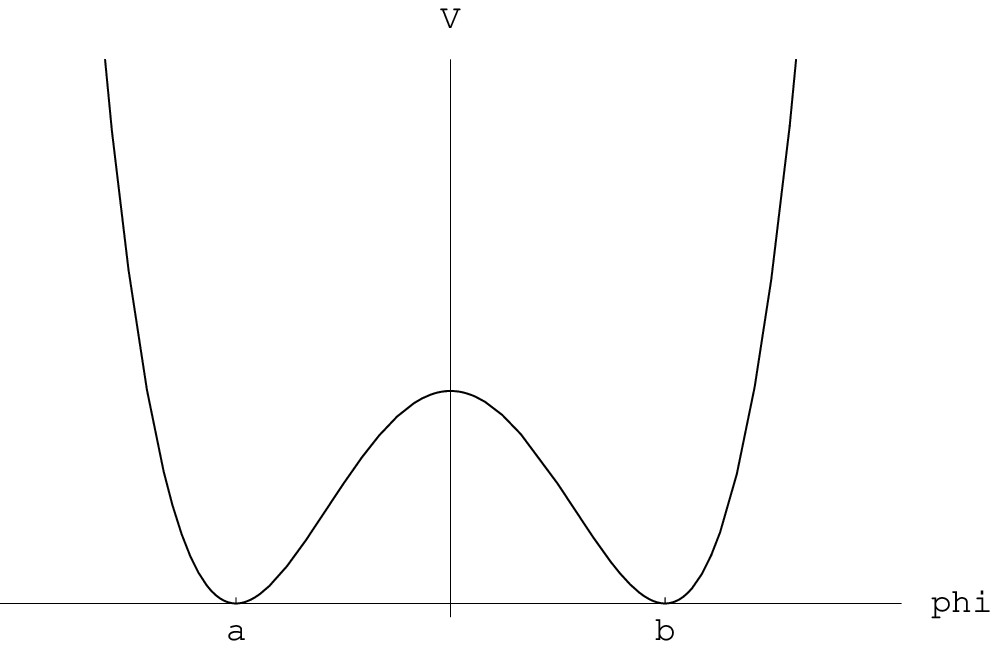,height=4cm,width=4.5cm} &
\psfrag{x}{$x$}\psfrag{phi}{$\phi_{cl}(x)$}\psfrag{c}{$\hspace{-0.4cm}\frac{m}{\sqrt{\lambda}}$}
\psfrag{d}{$\hspace{0.3cm}-\frac{m}{\sqrt{\lambda}}$}
\psfig{figure=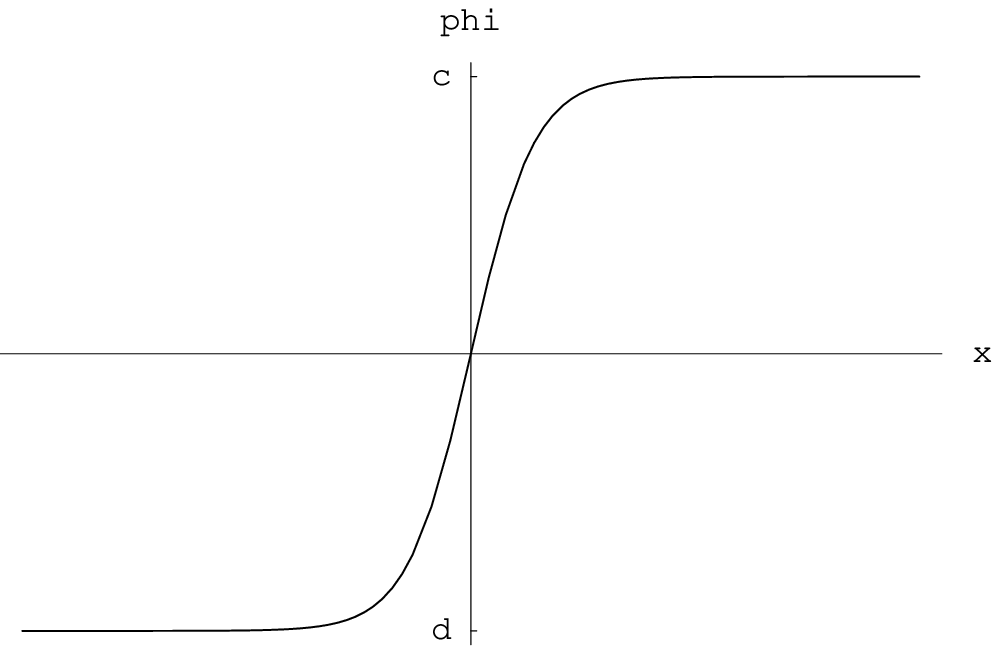,height=4cm,width=4.5cm} &
\psfrag{x}{$x$}\psfrag{ecl}{$\varepsilon_{cl}(x)$}
\psfig{figure=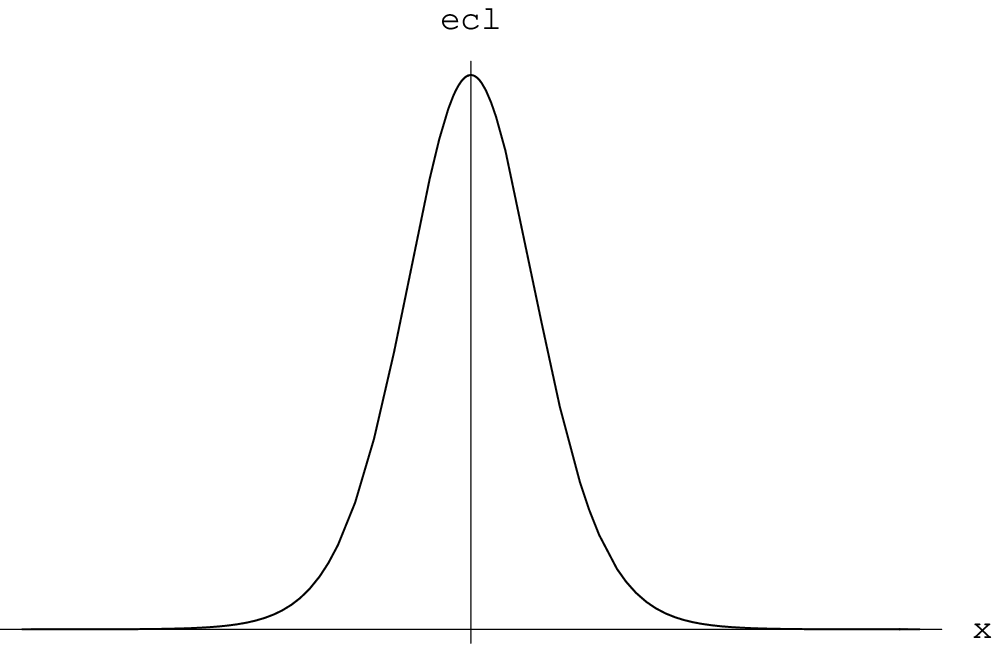,height=4cm,width=4.5cm}
\end{tabular}
\caption{Potential $V(\phi)$, kink $\phi_{cl}(x)$ and energy
density $\varepsilon_{cl}(x)$.}\label{figphi4}
\end{figure}

For definiteness in the illustration of the method, we will focus
on the example of the $\phi^{4}$ theory in the broken
$\mathbb{Z}_2$ symmetry phase, defined by the potential
\begin{equation}\label{phi4pot}
V(\phi) = \frac{\lambda}{4}\,\phi^{4}-\frac{m^{2}}{2}\,\phi^{2} +
\frac{m^{4}}{4\lambda}\;.
\end{equation}
This theory displays two degenerate minima at
$\phi_{\pm}=\pm\frac{m}{\sqrt{\lambda}}$, and a static (anti)kink
interpolating between them
\begin{equation}\label{phi4kinkinf}
\phi_{cl}(x) \, = \,(\pm) \,\frac{m}{\sqrt{\lambda}}\;
\tanh\frac{m }{\sqrt{2}}(x-x_0)\;,
\end{equation}
where the arbitrariness of the center of mass position $x_0$ is
due to the translational invariance of the theory. The
corresponding classical energy is obtained by integrating the
energy density $\varepsilon_{cl}(x)\equiv
\frac{1}{2}\left(\frac{d\phi_{cl}}{dx}\right)^2+V(\phi_{cl})\;$:
\begin{equation}\label{phi4kinkinfecl}
{\cal
E}_{cl}\equiv\int\limits_{-\infty}^{\infty}dx\;\varepsilon_{cl}(x)=\frac{2\sqrt{2}}{3}\frac{m^{3}}{\lambda}\;.
\end{equation}
Fig.\,\ref{figphi4} shows the potential, the classical kink and
its energy density.

\vspace{0.5cm}

\vspace{0.5cm}

The kinks exhibit certain particle properties. For instance, they
are localized and topologically stable objects, since the
conservation of topological charge forbids their decay into mesons
with $Q = 0$. Moreover, in integrable theories their scattering is
dispersionless and, in the collision processes, they preserve
their form simply passing through each other. All the above
properties are an indication that the kinks can survive the
quantization, giving rise to the quantum states in one--particle
sector of the corresponding QFT. As it is well known, one cannot
apply directly to them the standard perturbative methods of
quantization around the free field theory since the kinks are
entirely non--perturbative solutions of the interacting theory.
Their classical mass, for instance, is usually inversely
proportional to the coupling constant (see (\ref{phi4kinkinfecl})
in the broken $\phi^4$ example). In infinite volume, an effective
method for the semiclassical quantization of such solutions (as
well as of the vacua ones) has been developed in a series of
papers by Dashen, Hasslacher and Neveu (DHN) \cite{DHN} by using
an appropriate generalization of the WKB approximation in quantum
mechanics.

The DHN method consists in initially splitting the field
$\phi(x,t)$ in terms of the static classical solution (which can
be either one of the vacua or the kink configuration) and its
quantum fluctuations, i.e.
\begin{equation*}
\phi(x,t) \,=\,\phi_{cl}(x) + \eta(x,t) \,\,\, \,\,\, , \,\,\,
\,\,\, \eta(x,t) \,=\,\sum_{k} e^{i \omega_k t} \,\eta_{k}(x)
\,\,\,,
\end{equation*}
and in further expanding the action of the theory in powers of
$\eta$, obtaining for instance, in the example (\ref{phi4pot}),
the expansion in $\lambda$
\begin{equation}\label{lagrexp}
{\cal S}(\phi)=\int dx\,dt \,{\cal L}(\phi_{cl})+\int
dx\,dt\;\frac{1}{2}\,\eta(x,t)\left(\frac{d^{2}}{dt^{2}}-\frac{d^{2}}{dx^{2}}-m^{2}+3\lambda\phi_{cl}^{2}\right)\eta(x,t)
+\lambda\int
dx\,dt\,\left(\phi_{cl}\,\eta^{3}+\frac{1}{4}\eta^{4}\right)\;.
\end{equation}
The semiclassical approximation is realized by keeping only the
quadratic terms, and as a result of this procedure, $\eta_{k}(x)$
satisfies the so called \lq\lq stability equation"
\begin{equation}\label{stability}
\left[-\frac{d^2}{d x^2} + V''(\phi_{cl}) \right] \, \eta_{k}(x)
\,=\, \omega_k^2 \,\eta_k(x) \,\,\,,
\end{equation}
together with certain boundary conditions. The semiclassical
energy levels in each sector are then built in terms of the energy
of the corresponding classical solution and the eigenvalues
$\omega_i$ of the Scr\"odinger--like equation (\ref{stability}),
i.e.
\begin{equation}
E_{\{n_i\}} \,=\,{\cal E}_{cl} + \hbar \,\sum_{k}\left(n_k +
\frac{1}{2}\right) \, \omega_k + O(\hbar^2) \,\,\,, \label{tower}
\end{equation}
where $n_k$ are non--negative integers. In particular the ground
state energy in each sector is obtained by choosing all $n_k = 0$
and it is therefore given by\footnote{From now on we will fix
$\hbar=1$, since the semiclassical expansion in $\hbar$ is
equivalent to the expansion in the interaction coupling
$\lambda$.}
\begin{equation}
E_{0} \,=\,{\cal E}_{cl} + \frac{\hbar}{2} \,\sum_{k} \omega_k +
O(\hbar^2) \,\,\,. \label{e0}
\end{equation}

The semiclassical quantization technique was applied in \cite{DHN}
to the kink background (\ref{phi4kinkinf}), in order to compute
the first quantum corrections to its mass, given at leading order
by the classical energy. The stability equation (\ref{stability})
can be cast in the hypergeometric form in the variable
$z=\frac{1}{2}(1+\tanh \frac{mx}{\sqrt{2}})$, and the solution is
\begin{equation*}
\eta(x)\,=\,z^{\sqrt{1-\frac{\omega^{2}}{2
m^{2}}}}(1-z)^{-\sqrt{1-\frac{\omega^{2}}{2 m^{2}}}}\,
F\left(3,-2,1+2\sqrt{1-\frac{\omega^{2}}{2 m^{2}}};\;z\right)\;.
\end{equation*}
The corresponding spectrum is given by the two discrete
eigenvalues
\begin{equation}\label{omega0phi4}
\omega_{0}^{2}=0\;,\qquad\text{with}\qquad
\eta_0(x)=\frac{1}{\cosh^2\frac{m x}{\sqrt{2}}}\;,
\end{equation}
and
\begin{equation}\label{omega1phi4}
\omega_{1}^{2}=\frac{3}{2}\,m^{2}\;,\qquad\text{with}\qquad
\eta_1(x)=\frac{\sinh\frac{m x}{\sqrt{2}}}{\cosh^2\frac{m
x}{\sqrt{2}}}\;,
\end{equation}
plus the continuous part, labelled by $q\in\mathbb{R}$,
\begin{equation}\label{omegaqphi4}
\omega_{q}^{2}=m^{2}\left(2+\frac{1}{2}\,q^{2}\right)\;,\qquad\text{with}\qquad
\eta_q(x)=e^{i q m x/\sqrt{2}}\left(3\tanh^2\frac{m
x}{\sqrt{2}}-1-q^2-3 i q \tanh\frac{m x}{\sqrt{2}}\right)\;.
\end{equation}
The presence of the zero mode $\omega_0$ is due to the arbitrary
position of the center of mass $x_0$ in (\ref{phi4kinkinf}), while
$\omega_1$ and $\omega_q$ represent, respectively, an internal
excitation of the kink particle and the scattering of the kink
with mesons\footnote{The mesons represent the excitations over the
vacua, i.e. the constant backgrounds
$\phi_{\pm}=\pm\frac{m}{\sqrt{\lambda}}$, therefore their square
mass is given by $V''(\phi_{\pm})=2\,m^2$.} of mass $\sqrt{2}\,m$
and momentum $m q/\sqrt{2}$. This interpretation will find a clear
explanation in Sect.\,\ref{secsemFF}.

\begin{figure}[h]
\psfrag{k}{$k$}\psfrag{l}{$\frac{\lambda}{4}$}
\hspace{4.5cm}\psfig{figure=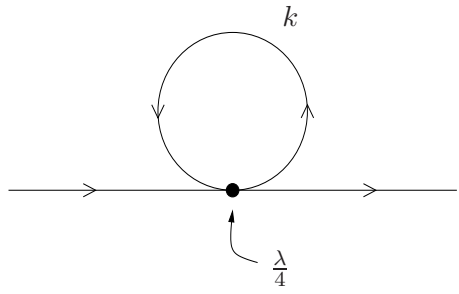,height=3.5cm,width=6cm}
\caption{Tadpole diagram giving rise to $\delta m^{2}$.}
\label{figtadpole}
\end{figure}

The semiclassical correction to the kink mass can be now computed
as the difference between the ground state energy in the kink
sector and the one of the vacuum sector, plus a mass counterterm
due to normal ordering (see Fig.\,\ref{figtadpole}):
\begin{equation*}
M={\cal
E}_{cl}+\frac{1}{2}\,m\sqrt{\frac{3}{2}}+\frac{1}{2}\sum_{n}\left[m\sqrt{2+\frac{1}{2}\,q_{n}^{2}}-\sqrt{k_{n}^{2}
+2m^{2}}\right]-\frac{1}{2}\,\delta
m^{2}\int\limits_{-\infty}^{\infty}dx\left[\phi^{2}_{cl}(x)-\frac{m^{2}}{\lambda}\right]\;,
\end{equation*}
with
\begin{equation*}
\delta m^{2}=\frac{3\lambda}{4\pi}\int\limits_{-\infty}^{\infty}
\frac{d k}{\sqrt{k^{2}+2m^{2}}}\;.
\end{equation*}

The discrete values $q_{n}$ and $k_{n}$ are obtained putting the
system in a big finite volume of size $R$ with periodic boundary
conditions:
\begin{equation*}
2 n\pi= k_{n}R = q_{n}\frac{m R}{\sqrt{2}}+\delta(q_{n})\;,
\end{equation*}
where the phase shift $\delta(q)$ is extracted from $\eta_q(x)$ in
(\ref{omegaqphi4}) as
$$
\eta_q
(x)\;{\mathrel{\mathop{\kern0pt\longrightarrow}\limits_{x\to
\pm\infty
}}}\;e^{i\left[q\,\frac{mx}{\sqrt{2}}\pm\frac{1}{2}\,\delta(q)\right]}\;,\qquad\qquad
\delta(q)=-2\arctan\left(\frac{3 q}{2-q^2}\right)\;.
$$
Sending $R\rightarrow\infty$ and computing the integrals we
finally have
\begin{equation}
M\,=\,\frac{2\sqrt{2}}{3}\,\frac{m^{3}}{\lambda}+m\left(\frac{1}{6}\,\sqrt{\frac{3}{2}}-\frac{3}{\pi\sqrt{2}}\right)\;.
\end{equation}

In concluding this section, it is worth mentioning that the
construction of the complete Hilbert space requires to consider
time--dependent multi--kink solutions with finite energy. Their
semiclassical quantization can be performed with an appropriate
modification of the DHN method \cite{DHN} to include the
time--dependence. Moreover, the semiclassical computations can be
extended at higher order in the small--coupling expansion, keeping
cubic (and higher) powers of $\eta$ in (\ref{lagrexp}). This
procedure is highly non--trivial, especially due to the presence
of the zero mode $\omega_0=0$, which causes some divergencies at
higher order in $\lambda$. However, effective methods for solving
problem have been developed and applied in
\cite{callgross,pathint}, in the context of a path integral
technique.

\section{Classical solutions and form factors}\label{secsemFF}
\setcounter{equation}{0}

\subsection{Goldstone and Jackiw's result}\label{secGJ}

A direct relation between the kink states and the corresponding
classical solutions has been established by Goldstone and Jackiw
\cite{GJ}, who have shown that the matrix element of the field
$\phi$ between kink states is given, at leading order in the
semiclassical limit, by the Fourier transform of the kink
background. A detailed explanation of this result can be found in
\cite{raj}, and a similar achievement in a non--relativistic
context is presented in \cite{Fateev}.

The technique to derive this result relies on the following
assumptions, which consolidate the basic ideas already exposed in
Sect.\,\ref{mainidea}, and that we illustrate here in the case of
the broken $\phi^{4}$ theory (\ref{phi4pot}):

\begin{enumerate}

\item The Hilbert space of the theory contains, in addition to the
\lq vacuum sector' (i.e. the vacuum and multi--meson states), the
so--called kink sector. This is spanned by the states
$|\,p\,\rangle$ and $|\,p^{*}\rangle$, representing the kink
particle of momentum $p\;$ in its ground state and excited state,
respectively, and the states $|\,p,k_{1},...,k_{m}\,\rangle$ and
$|\,p^{*},k_{1},...,k_{m}\,\rangle$, representing the scattering
states of the kink particle and $m$ mesons.

\item The kink sector is orthogonal to the vacuum sector, hence
the kink is stable against decay into mesons. In fact, although
the states in the kink sector have higher energy than the vacuum
sector ones, they are prevented from decaying by the conservation
of the topological charge, which holds also in the quantum theory.

\item The mass of the quantum kink behaves as
$M\sim\frac{1}{\lambda}$ in the weak coupling limit (see
(\ref{phi4kinkinfecl})).

\item The matrix elements of $\phi(x,t)$ between the above states
behave in the weak coupling limit as
\begin{eqnarray*}
&&\langle p'\,|\,\phi\,|\,p\,\rangle \,\sim \,
O\left(1/\sqrt{\lambda}\right)\\
&&\langle p'\,|\,\phi\,|\,p,k\,\rangle \;\text{and}\;\langle
p'\,|\,\phi\,|\,p^{*}\rangle\,\sim \,
O\left(1\right)\\
&&\langle
p',k'_{1},...,k'_{l}\,|\,\phi\,|\,p,k_{1},...,k_{m}\,\rangle
\,\sim \, O\left(\lambda^{(l+m-1)/2}\right)
\end{eqnarray*}
This assumption, which will find confirmation \textit{a
posteriori}, relies on the fact that the kink classical background
itself is of order $1/\sqrt{\lambda}$, and that the emission or
absorption of every additional meson and the internal excitation
of the kink carry a factor $\sqrt{\lambda}$. This can be
intuitively understood by noticing that, in the expansion
(\ref{lagrexp}) of the interaction $V(\phi)$, the leading
perturbative term is of order $\lambda\phi_{cl}$, i.e. of order
$\sqrt{\lambda}$.

\end{enumerate}

Let's now define the matrix element of the basic field $\phi(x,t)$
between in and out one--kink states, with momenta $p_{1}$ and
$p_{2}$, as the Fourier transform of a function $\hat{f}(a)$, to
be determined:
\begin{equation}
\langle\,p_{2}\,|\,\phi(0)\,|\,p_{1}\,\rangle\;=\,\int
da\;e^{i(p_{1}-p_{2})a}\;\hat{f}(a)\;.
\end{equation}
Next we consider the Heisenberg equation of motion for the quantum
field $\phi(x,t)$
\begin{equation}
\left(\partial_{t}^{2}-\partial_{x}^{2}\right)\phi(x,t)\,=\,m^{2}\,\phi(x,t)\,-\,\lambda\,\phi^{3}(x,t)\;,
\end{equation}
we take the matrix elements of both sides\footnote{Lorentz
invariance imposes the relation
$\langle\,p_{2}\,|\,\phi(x,t)\,|\,p_{1}\,\rangle\,=\,
e^{-i(p_{1}-p_{2})_{\mu}x^{\mu}}\langle\,p_{2}\,|\,\phi(0)\,|\,p_{1}\,\rangle$}
\begin{equation*}
\left[-(p_{1}-p_{2})_{\mu}(p_{1}-p_{2})^{\mu}\right]
e^{-i(p_{1}-p_{2})_{\mu}x^{\mu}}\langle\,p_{2}\,|\,\phi(0)\,|\,p_{1}\,\rangle
\,=\,
\end{equation*}
\begin{equation}\label{heis}
=\,e^{-i(p_{1}-p_{2})_{\mu}x^{\mu}}\,\left\{\,m^{2}\,\langle\,p_{2}\,|\,\phi(0)\,|\,p_{1}\,\rangle
\,-\,\lambda\,\langle\,p_{2}\,|\,\phi^{3}(0)\,|\,p_{1}\,\rangle\,\right\}\;,
\end{equation}
and we equate them to leading order $1/\sqrt{\lambda}$. Assumption
3 implies that the kink momentum is very small compared to its
mass, therefore in the left hand side of (\ref{heis}) we can
neglect the energy difference
$$
\left(E_{1}-E_{2}\right)^{2}=\left(\frac{p_{1}^{2}-p_{2}^{2}}{2M}+...\right)^{2}=O(\lambda^{2})\;.
$$
Hence the left hand side gives, at leading order,
$$
\int
da\;e^{i(p_{1}-p_{2})a}\;\left(-\frac{d^{2}}{da^{2}}\,\hat{f}(a)\right)\;.
$$
In the last term of the right hand side of (\ref{heis}), the cubic
power of $\phi$ can be expanded over a complete set of states in
the kink sector; in virtue of assumption 4, the leading term is
obtained when the intermediate states are one--kink states:
\begin{equation}\label{intermediate}
-\lambda\sum_{p,q}
\langle\,p_{2}\,|\,\phi(0)\,|\,p\,\rangle\langle\,p\,|\,\phi(0)\,|\,q\,\rangle\langle
\,q\,|\,\phi(0)\,|\,p_{1}\,\rangle\,=\,-\lambda \int
da\;e^{i(p_{1}-p_{2})a}\;\,[\,\hat{f}(a)\,]^{3}\;.
\end{equation}
Hence, at leading order in $\lambda$, the function $\hat{f}(a)$
obeys the same differential equation satisfied by the kink
solution, i.e.
\begin{equation}
\frac{d^{2}}{da^{2}}\hat{f}(a)=\lambda[\,\hat{f}(a)\,]^{3}-m^{2}\hat{f}(a)\;.
\end{equation}
This means that we can take $\hat{f}(a)  = \phi_{cl}(a)$,
adjusting its boundary conditions by an appropriate choice for the
value of the constant $A$ in eq.\,(\ref{firstorder}).

Therefore, we finally obtain
\begin{equation}\label{finalGJ}
\langle\,p_{2}\,|\,\phi(0)\,|\,p_{1}\,\rangle\;=\,\int
da\;e^{i(p_{1}-p_{2})a}\;\phi_{cl}(a)\;+\;\text{higher order
terms}\;.
\end{equation}
Along the same lines, it is easy to prove that the form factor of
an operator expressible as a function of $\phi$ is given by the
Fourier transform of the same function of $\phi_{cl}$. For
instance, the form factor of the energy density operator
$\varepsilon$ can be computed performing the Fourier transform of
$\varepsilon_{cl}(x) =
\frac{1}{2}\left(\frac{d\phi_{cl}}{dx}\right)^{2} +
V\left(\phi_{cl}\right)$.

Eq.\,(\ref{finalGJ}) is the crucial achievement that opens the way
to our study of non--integrable theories. However, it is worth to
mention a further result obtained in \cite{GJ}, which confirms the
interpretation of the stability frequencies $\omega_1$ and
$\omega_q$ described in the previous Section. Let's consider the
form factor of $\phi$ between two states containing one kink and
one kink plus one meson, respectively, and define a function
$\hat{f}_k(a)$ as
\begin{equation}
\langle\,p_{2},k\,|\,\phi(0)\,|\,p_{1}\,\rangle\;=\,\int
da\;e^{i(p_{1}-p_{2}-k)a}\;\hat{f}_{k}(a)\;.
\end{equation}
Calling $\omega_k$ the energy of the meson with momentum $k$, and
equating again at leading order the two sides of the Heisenberg
equation of motion evaluated between the above asymptotic states,
we obtain the equation
\begin{equation}
\left[-\frac{d^2}{d a^2} + V''(\phi_{cl}) \right] \, \hat{f}_k(a)
\,=\, \omega_k^2 \,\hat{f}_k(a) \,\,\,,
\end{equation}
and consequently we can choose $\hat{f}_k(a)=\eta_k
(a)/\sqrt{2\omega_k}$, where $\eta_k (a)$ is one of the
eigenfunctions of (\ref{stability}) and the constant
$1/\sqrt{2\omega_k}$ is chosen to realize a familiar boson
normalization. This completely justifies the identification of the
eigenvalues of (\ref{stability}) with the mesons energies.
Similarly, if we define the form factor of $\phi$ between a kink
in the ground state and one in the internal excited state as
\begin{equation}\label{ffexc}
\langle\,p^*_{2}\,|\,\phi(0)\,|\,p_{1}\,\rangle\;=\,\int
da\;e^{i(p_{1}-p_{2})a}\;\hat{f}^*(a)\;,
\end{equation}
and we call $\omega_1$ the energy of the internal excitation, we
obtain
\begin{equation}\label{fhatexc}
\left[-\frac{d^2}{d a^2} + V''(\phi_{cl}) \right] \,
\hat{f}^{*}(a) \,=\, \omega_1^2 \,\hat{f}^{*}(a) \,\,\,,
\end{equation}
which leads to $\hat{f}^*(a)=\eta_1 (a)/\sqrt{2\omega_1}$.

\subsection{Relativistic formulation of Goldstone and Jackiw's
result}\label{sectGJimprov}

The remarkable Goldstone and Jackiw's result has, unfortunately,
two serious drawbacks: expressing the form factor as a function of
the difference of momenta, Lorentz covariance is lost and
moreover, the antisymmetry under the interchange of momenta makes
problematic any attempt to go in the crossed channel and obtain
the matrix element between the vacuum and a kink--antikink state.

In order to overcome these problems, in \cite{finvolff} we have
refined the method proposed in \cite{GJ}. This was done by using,
instead of the space--momenta of the kinks, their rapidity
variable $\theta$, defined in (\ref{rapidity}). The approximation
of large kink mass used by Goldstone and Jackiw can be realized
considering the rapidity as very small. For example, in the
$\phi^{4}$ theory (\ref{phi4pot}), where the kink mass $M$ is of
order $1/\lambda$, we work under the hypothesis that $\theta$ is
of order $\lambda$. In this way we get consistently
\begin{equation}\label{rapidityapprox}
E\equiv M \cosh\theta\simeq M\;,\qquad p\equiv M \sinh\theta\simeq
M\,\theta \ll M\;.
\end{equation}
It is easy to see that the proof of (\ref{finalGJ}) outlined in
Sect.\,\ref{secGJ} still holds, if we define the form factor
between kink states as the Fourier transform with respect to the
rapidity difference:
\begin{equation}\label{ffinf}
\langle \,p_{1}|\,\phi(0)\,|p_{2}\,\rangle\equiv f(\theta)\equiv
M\int da\,e^{i\,M\theta a}\phi_{cl}(a)\;,
\end{equation}
with the inverse Fourier transform defined as
\begin{equation}
\phi_{cl}(a)\equiv \int \frac{d\theta}{2\pi}\,e^{-i\,M\theta
a}f(\theta)\;.
\end{equation}
In fact, the analysis of the r.h.s. of (\ref{heis}) remains
unchanged, while the one of the l.h.s. is even improved, since we
can exploit the approximation (\ref{rapidityapprox}) directly on
the covariant expression
\begin{equation*}
-(p_{1}-p_{2})_{\mu}(p_{1}-p_{2})^{\mu} \,=\,-2
M^{2}\left(1-\cosh\theta\right)\simeq M^{2}\theta^{2}\;.
\end{equation*}
Our formalism only displays a small inessential difference with
respect to \cite{GJ}, in the orders in $\lambda$ of the various
form factors (see assumption 4 in Sect.\,\ref{secGJ}). In fact,
the $M$ factor in front of (\ref{ffinf}), which is a natural
consequence of considering the rapidity as the basic variable,
increases by $1/\lambda$ the order of all form factors, without
however altering any conclusion.

With the above considerations, it is now possible to express the
crossed channel form factor through the variable transformation
$\theta\rightarrow i\pi-\theta$. We then have
\begin{equation}\label{f2}
F_{2}(\theta) \,\equiv\,
\langle\,0|\,\phi(0)|\,p_{1},\bar{p}_{2}\,\rangle =
f(i\pi-\theta)\;.
\end{equation}

\begin{figure}[h]
\psfrag{phi}{$\hspace{0.15cm}\phi$}\psfrag{K}{$K$}\psfrag{Kb}{$\bar{K}$}
\psfrag{b}{$b$}\psfrag{Gabc}{$\Gamma_{K\bar{K}}^{b}$}\psfrag{(a)}{$(a)$}\psfrag{(b)}{$(b)$}\psfrag{(c)}{$(c)$}
\hspace{1cm}\psfig{figure=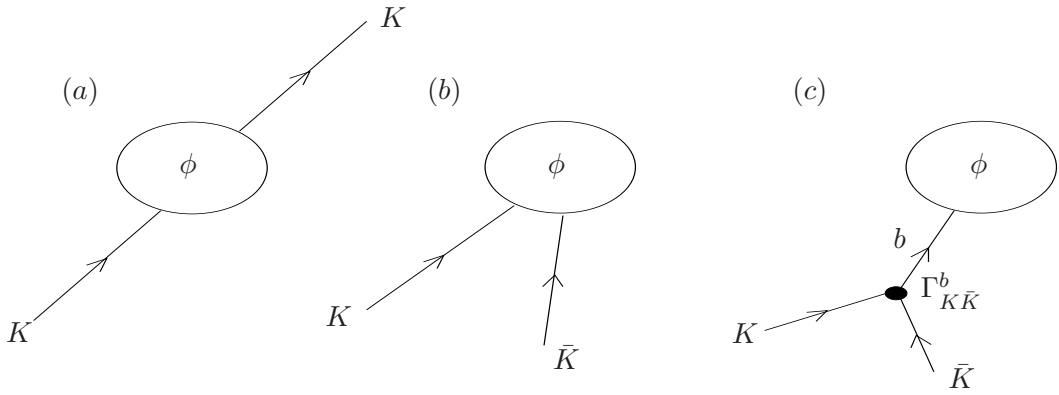,height=5cm,width=14cm}
\caption{Pictorial representation of $(a)$ the form factor
$f(\theta)$, $(b)$ the crossed channel form factor $F_2(\theta)$,
$(c)$ the form factor $F_2(\theta^*)$ at the dynamical pole
$\theta^{*}$.} \label{figGJref}
\end{figure}

The analysis of this quantity allows us, in particular, to get
information about the spectrum of the theory. As we have shown in
(\ref{ffdynpole}), in fact, its dynamical poles located at
$\theta^{*}=i(\pi-u)$, with $0<u<\pi$, correspond to
kink--antikink bound states with masses given by
(\ref{boundstatesmasses}), which in this case specializes to
\begin{equation}
m_{(b)} \,= \, 2 M \sin\frac{u}{2}\;.
\end{equation}
(see Fig.\,\ref{figGJref}). It is worth noticing that this
procedure for extracting the semiclassical bound states masses is
remarkably simpler than the standard DHN method of quantizing the
corresponding classical backgrounds, because in general these
solutions depend also on time and have a much more complicated
structure than the kink ones. Moreover, in non--integrable
theories these backgrounds could even not exist as exact solutions
of the field equations: this happens for example in the $\phi^{4}$
theory, where the DHN quantization has been performed on some
approximate backgrounds \cite{DHN}.

Once the matrix elements (\ref{f2}) are known, one can estimate
the leading behaviour in $\lambda$ of the spectral function
$\rho^{(\phi)}$ in a regime of the momenta dictated by our
assumption of small kink rapidity. The delta functions in
(\ref{spectral}) make meaningful the use of our form factors,
derived in the small $\theta$ approximation, only if we consider a
regime in which $p^{0}\simeq M$ and $p^{1}\ll M$ and, from now on,
we will always understand this restriction. The leading
$O(1/\lambda)$ contribution to the spectral function, denoted in
the following by $\hat\rho^{\,(\phi)}(p^2)$, is given by the
trivial vacuum term plus the kink-antikink contribution:
\begin{equation}\label{rhoinf}
\hat\rho^{\,(\phi)}(p^{2}) =
2\pi\delta(p^{0})\delta(p^{1})|\,\langle\,0|\,\phi(0)|\,0\,\rangle\,|\,^{2}
+ \frac{\pi}{4}\,\frac{\delta\left(\frac{p^{0}}{M}-2\right)}{
M^{2} }\int \frac{d\theta_{1}}{2\pi}
\left|F_{2}\left(2\theta_{1}+i\pi-\frac{p^{1}}{M}\right)\right|^{2}\;,
\end{equation}
where the range of integration of the above quantity is of order
$p^{1}/M$ (note that, being $p^{1}/M\ll 1$, the integral can be
roughly estimated by evaluating $|F_{2}|^{2}$ at $\theta_{1}=0$:
this is what we will do in the following).

\section{The Sine--Gordon model in infinite volume}\label{secSGinfvol}
\setcounter{equation}{0}\label{SGinfinite}

The Sine--Gordon model, defined by the potential
\begin{equation}\label{SGpot}
V(\phi) = \frac{m^{2}}{\beta^{2}}\left(1-\cos\beta\phi\right)\;,
\end{equation}
gives us the chance of testing the validity of the semiclassical
approximation. In fact, the integrability of this model has led to
the exact computation of its spectrum and form factors
\cite{zams,KW,YZff,smirnovbook}, which can be evaluated in the
small-$\beta$ limit to verify the semiclassical results.

This model has been object of intensive research efforts, since it
plays a relevant role in several physical contexts, either as a
classical non--linear system or as a quantum field theory. For
instance, its non--linear equation of motion appears in the study
of propagation of dislocations in crystals, self--induced
transparency effects in non--linear optics and spin wave
propagation in liquid $^3$He (see \cite{scott} for a review). As a
quantum field theory, it plays a crucial role in the description
of quantum spin chains, in virtue of the bosonization procedure,
which permits to express fermionic fields as exponentials of
bosonic ones. In particular, an exact correspondence has been
established in \cite{SGMT} between the sine--Gordon and the
Thirring model, which describes massive fermions interacting
through a four--fermion coupling.

The exact solution of the model shows that the spectrum of
fundamental excitations is constituted by a soliton, an
antisoliton and a given number of neutral particles, called \lq\lq
breathers" and obtained as soliton--antisoliton bound states
\cite{zams}. The number of breathers depend on $\beta$ as the
integer part of $\frac{8\pi}{\gamma}$, where
\begin{equation} \label{dressing}
\gamma \,=\,\frac{\beta^2}{1 - \beta^2/8\pi}
\end{equation}
is the renormalized coupling constant. In particular,
$\beta^2=4\pi$ represent a free point, above which no bound states
exist and the theory is in a repulsive regime, while the regime of
small $\beta$, that we will explore semiclassically, is deeply
attractive.

The potential (\ref{SGpot}) displays infinite degenerate minima,
and the soliton solutions of the equations of motion are given by
configurations which interpolate between two neighbouring vacua.
Focusing on the two minima at $\phi = 0$ and $\phi =
\frac{2\pi}{\beta}$, these are connected by the classical soliton
\begin{equation}
\phi_{cl}(x)=\frac{4}{\beta} \arctan\left(e^{ m (x -x_0)
}\right)\; ,\label{SGkinkinfvol}
\end{equation}
which is solution of eq.\,(\ref{firstorder}) with $A = 0$ and has
classical mass $M_{cl} = 8\frac{m}{\beta^{2}}$. The semiclassical
quantization of this background has been performed in \cite{DHN}:
eq. (\ref{stability}) can be cast in the hypergeometric form by
using the variable $z=\frac{1}{2}(1+\tanh mx)$, and its solution
is expressed as
\begin{equation*}
\eta(x)\,=\,\frac{1}{\beta}\,z^{\frac{1}{2}\sqrt{1-\frac{\omega^{2}}{m^{2}}}}\,
(1-z)^{-\frac{1}{2}\sqrt{1-\frac{\omega^{2}}{m^{2}}}}\,
F\left(2,-1,1+\sqrt{1-\frac{\omega^{2}}{m^{2}}},z\right)\;.
\end{equation*}
The corresponding spectrum is given by the discrete zero mode
$$
\omega_{0}^{2} = 0\;,\qquad \text{with}\qquad
\eta_0(x)=\frac{1}{2\,\cosh mx}\;,
$$
and by the continuous part
$$
\omega_{q}^{2} = m^{2}(1+q^{2})\;,\qquad \text{with}\qquad
\eta_q(x)=e^{imqx}\;\frac{iq-\tanh mx}{1+iq}\;.
$$
For this integrable theory we can easily verify the interpretation
of the continuous spectrum as representing the scattering states
of the kink with neutral particles. In fact, the semiclassical
phase shift $\delta(q)$, defined as
\begin{equation}\label{phaseshiftSG}
\eta_q
(x)\;{\mathrel{\mathop{\kern0pt\longrightarrow}\limits_{x\to
\pm\infty }}}\;e^{i\left[q m
x\pm\frac{1}{2}\,\delta(q)\right]}\;,\qquad\text{with}\qquad
\delta(q)=2\arctan\frac{1}{q}=-i\log\frac{q+i}{q-i}\;,
\end{equation} correctly reproduces the small-$\beta$ limit of
the exact one
$$
\delta_{\text{exact}}(\theta)=-i\log S_{k b}(\theta)\;,
$$
where the kink--breather scattering amplitude \cite{zams}
$$
S_{k
b}\,=\,\frac{\sinh\theta+i\cos\frac{\gamma}{16}}{\sinh\theta-i\cos\frac{\gamma}{16}}
$$
is evaluated in the kink rest frame, in which the breather
momentum can be expressed as $mq=m\sinh\theta$.

Similarly to the broken $\phi^4$ case discussed in
Sect.\,\ref{mainidea}, the semiclassical correction to the kink
mass is given by
\begin{equation}
M - M_{cl}\,=\,
\frac{1}{2}\sum_{n}\left[m\sqrt{1+q_{n}^{2}} -
\sqrt{k_{n}^{2}+m^{2}}\right]- \frac{\delta \mu^{2}}{\beta^{2}}
\int\limits_{-\infty}^{\infty}dx\left[1-\cos\beta\phi_{cl}(x)\right]\;,
\label{subinfvol}
\end{equation}
with
\begin{equation}
\delta \mu^{2} \,=\, -\frac{m^{2}\beta^{2}}{8\pi}
\int\limits_{-\infty}^{\infty} \frac{d k}{\sqrt{k^{2}+m^{2}}}\;.
\end{equation}
Determining the discrete values $q_{n}$ and $k_{n}$ in a large
finite volume of size $R$ with periodic boundary conditions and
sending $R\rightarrow\infty$, one finally obtains the
semiclassical quantum correction to the mass of the kink
\begin{equation}
M\,=\,\frac{8m}{\beta^{2}}-\frac{m}{\pi}\;. \label{piccoloh}
\end{equation}
As we have already mentioned, it is known that the coupling
constant renormalises as $\beta^2 \rightarrow \gamma$, and the
exact quantum mass of the soliton is given by $M = \frac{8
m}{\gamma}$, which coincides with the above expression
(\ref{piccoloh}). The equality of the semiclassical and the exact
results for the masses is a remarkable property of the SG model.

In \cite{finvolff} we have derived and discussed the semiclassical
form factor (\ref{ffinf}) relative to the kink background
(\ref{SGkinkinfvol}). Since the first quantum corrections to
$M_{cl}=8\frac{m}{\beta^{2}}$ are of higher order in $\beta^{2}$,
we consistently approximate the kink mass $M$ with this value and
we assume that the (anti)soliton rapidity is of order $\beta^{2}$.
The semi--classical form factor (\ref{ffinf}) is explicitly given
by
\begin{equation}
f(\theta)=\frac{2\pi
i}{\beta}\,\frac{1}{\theta\,\cosh\left[\frac{4\pi}{\beta^{2}}\,
\theta\right]}+\frac{2\pi^{2}}{\beta^{3}}
\delta\left(\frac{\theta}{\beta^{2}}\right)\;.
\end{equation}
In order to compare the corresponding $F_2(\theta)$ given by
(\ref{f2}) with the semiclassical limit of the exact one, the only
thing to take into account is that, in the definition of the exact
form factor of the fundamental field $\phi(x)$, the asymptotic
two--particle state is actually given by the antisymmetric
combination of soliton and antisoliton. Since at our level the
form factor between antisoliton states is simply
\begin{equation*}
<\bar{p}_{1}|\,\phi(0)|\,\bar{p}_{2}>=\frac{4
M}{\beta}\int\limits_{-\infty}^{\infty}da\, e^{i (M \theta) \,
a}\,\arctan\left(e^{-m a }\right) = f(-\theta)\;,
\end{equation*}
we finally obtain
\begin{equation}\label{f2SG}
F_{2}(\theta)=\frac{4\pi i}{\beta}\,\frac{1}{(i\pi-\theta)\,
\cosh\left[\frac{4\pi}{\beta^{2}}\,(i\pi-\theta)\right]}\;.
\end{equation}
In the regime $i\pi-\theta\simeq O(\beta^{2})$, this function
indeed agrees with the exact result \cite{KW}
\begin{equation}
F_{2}^{exact}(\theta)=\frac{2\pi i
}{\beta}\;\frac{1}{\sinh\frac{i\pi-\theta}{2}}\;\frac{\cosh\frac{i\pi-\theta}{2}}{
\cosh\left[\frac{1}{2\xi}\,(i\pi-\theta)\right]}\;G(i\pi-\theta)\;,
\end{equation}
where $\xi\,=\,\gamma/8\pi$ and
\begin{equation} \label{G}
G(\theta)\,=\,\text{exp}\left[\int\limits_{0}^{\infty}\frac{dt}{t}\;
\frac{\sinh\frac{t}{2}(1-\xi)}{\sinh\frac{t}{2}\,\xi\;\;\cosh\frac{t}{2}}\;
\frac{\sin^{2}\frac{\theta \,t }{2\pi}}{\sinh t}\right] \,\,\,.
\end{equation}
Furthermore, we can also check that the dynamical poles of
(\ref{f2SG}), located at
\begin{equation*}
\theta_{n}=i\pi\left[1-\frac{\beta^{2}}{8\pi}(2n+1)\right]\,,\quad\qquad
-\frac{1}{2}<n<-\frac{1}{2}+\frac{4\pi}{\beta^{2}} \,,
\end{equation*}
consistently reproduce the odd part of the well-known
semiclassical breathers spectrum \cite{DHN}
\begin{equation}
m_{b}^{(2n+1)} = 2M\sin\left[\frac{\beta^{2}}{16}(2n+1)\right] =
(2n+1)\,m\left[1-\frac{(2n+1)^{2}}{3\times 8^{3}
}\beta^{4}+...\right]
\end{equation}
Since in the vacuum sector $\langle\,0\,|\,\phi\,|\,0\,\rangle=0$,
in this model the $1/\beta^{2}$ leading contribution to the
spectral function takes the form:
\begin{equation}
\hat\rho(p^{2}) \,=
\,4\pi^{3}\,\delta\left(\frac{p^{0}}{M}-2\right)
\frac{1}{\beta^{2}(p^{1})^{2}\,\cosh^{2}
\left[\frac{\pi}{2}\,\frac{p^{1}}{m}\right]}\;.
\end{equation}

\begin{figure}[ht]
\begin{tabular}{p{8cm}p{8cm}}
\psfrag{Fs}{$F\hspace{3cm} \beta=0.1$}\psfrag{t}{$\theta$}
\psfig{figure=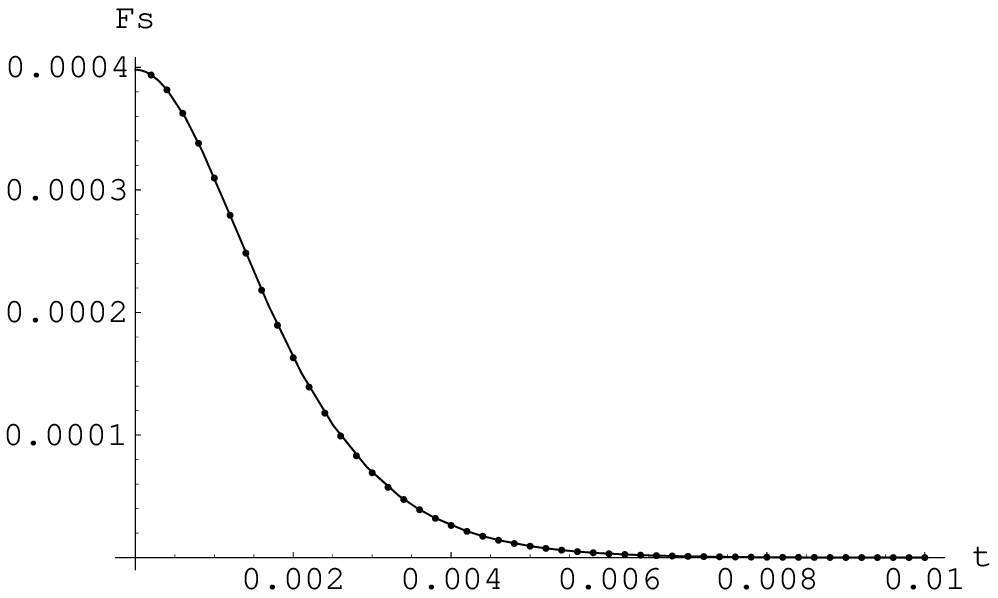,height=5cm,width=7cm}&
\psfrag{Fs}{$F\hspace{3cm} \beta=0.5$}\psfrag{t}{$\theta$}
\psfig{figure=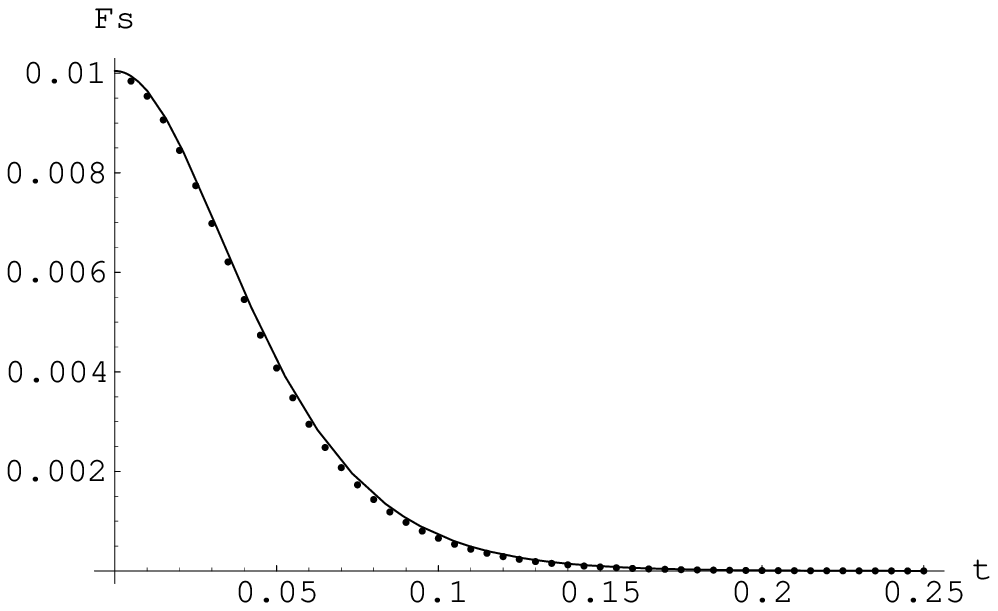,height=5cm,width=7cm}
\end{tabular}
\caption{Comparison between the exact function $F$ given by
(\ref{Fex}) (continuous line) and its semiclassical approximation
(\ref{Fsem}) (dotted line), at $\beta=0.1$ and
$\beta=0.5$.}\label{figcheckquant}
\end{figure}

\begin{figure}[ht]
\begin{tabular}{p{8cm}p{8cm}}
\psfrag{Fs}{$F\hspace{3cm} \beta=1\qquad(a)$}\psfrag{t}{$\theta$}
\psfig{figure=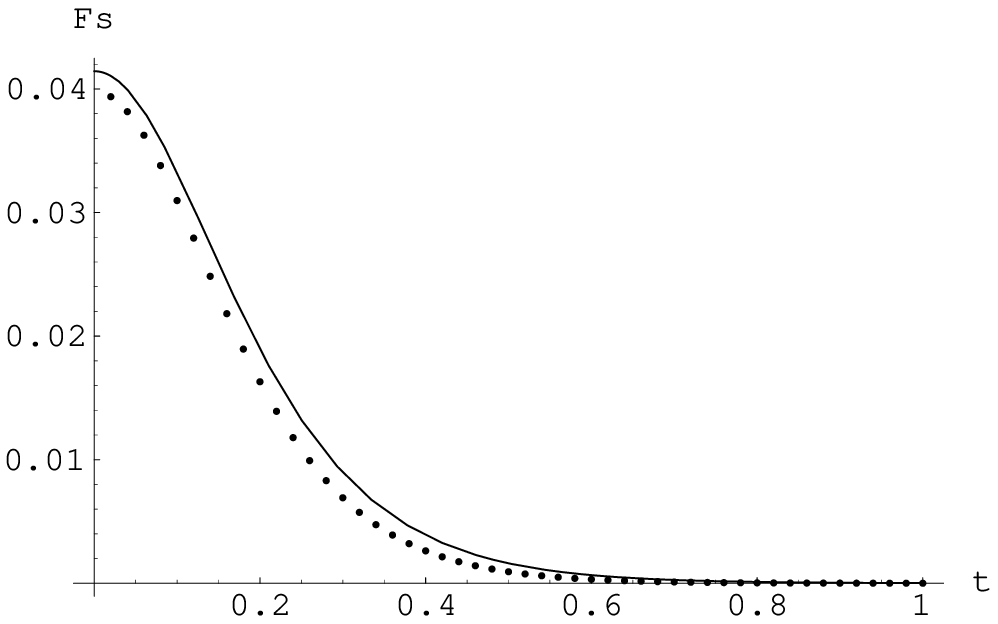,height=5cm,width=7cm}&
\psfrag{Fs}{$F\hspace{3cm} \beta=1\qquad(b)$}\psfrag{t}{$\theta$}
\psfig{figure=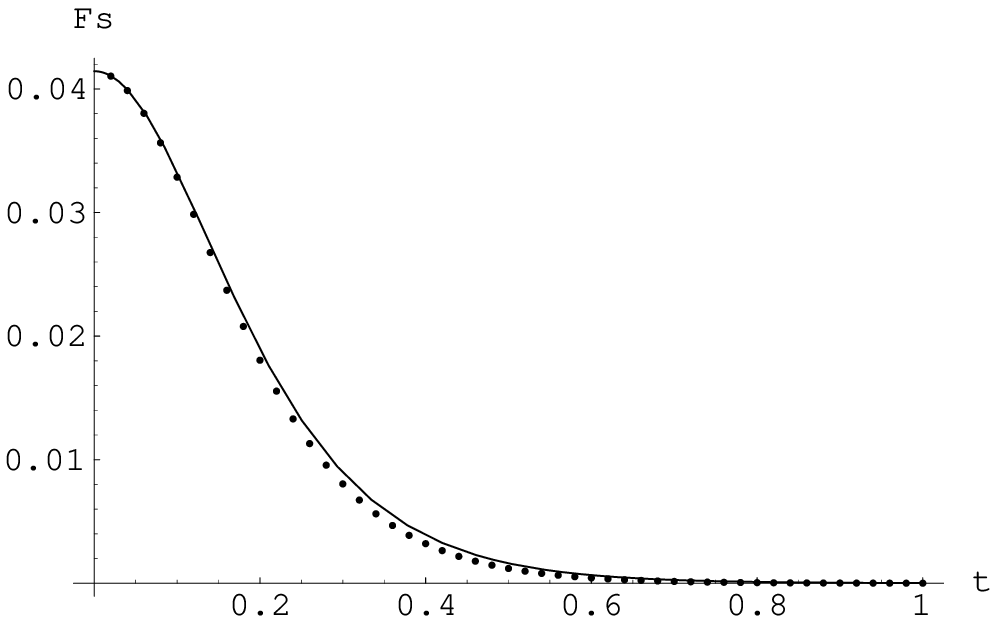,height=5cm,width=7cm}
\end{tabular}
\caption{Comparison, at $\beta=1$, between (a) the exact function
$g$ given by (\ref{Fex}) (continuous line) and its semiclassical
approximation (\ref{Fsem}) (dotted line), (b) the exact function
$g$ given by (\ref{Fex}) (continuous line) and its semiclassical
approximation (\ref{Fsem}) with the substitution
$\beta^{2}\to\gamma$ (dotted line).}\label{figcheckquantdress}
\end{figure}

Furthermore, in \cite{SGscaling} we have presented a more
quantitative comparison which permits to conclude that formula
(\ref{ffinf}), though proven under the semiclassical assumption of
small coupling and small rapidities, remarkably extends its
validity to finite values of the coupling and to a large range of
the rapidities. Consider, for instance, the form factor of the
energy operator (up to a normalization $N$)
$g(\theta)\,=\,N\langle\,\theta_{2}\,
|\,\epsilon(0)\,|\,\theta_{1}\,\rangle$, whose semiclassical and
exact expressions are given, respectively, by
\begin{eqnarray}\label{Fsem}
g_{\text{semicl.}}(\theta)&=&\hspace{3mm}\frac{\theta}{2}
\hspace{7mm}\frac{1}{\sinh \left[\frac{4\pi}{\beta^{2}}
\,\theta\right]}
\\\label{Fex}
g_{\text{exact}}(\theta)&=&\sinh\frac{\theta}{2}\hspace{6mm}
\frac{1}{\sinh\frac{\theta} {2\xi}} \hspace{5mm} G(\theta)
\end{eqnarray}
with $G(\theta)$ defined in (\ref{G}). Fig.\,\ref{figcheckquant}
shows how, for small values of the coupling, the agreement between
the two functions is very precise for the whole range of the
rapidity. Furthermore, the discrepancy between exact and
semiclassical formulas at larger values of $\beta$ can be simply
cured, in our example, by substituting the bare coupling $\beta^2$
with its renormalized expression $\gamma$ into the semiclassical
result (\ref{Fsem}), as shown in Fig.\,\ref{figcheckquantdress}.
Hence we can conclude that the monodromy factor (\ref{G}), which
is the relevant quantity missing in our approximation, does not
play a significant role in the quantitative evaluation of the form
factor even for certain finite values of the coupling\footnote{ It
is easy to understand the reason of this conclusion in the above
example: at small values of $\theta$ we have $G(\theta) \simeq 1$,
whereas for $\theta \rightarrow \infty$, when $G(\theta)$ may
contribute, the whole form factor goes anyway to zero. Similar
conclusion can be reached for all other form factors which vanish
at $\theta \rightarrow \infty$.}.

\newpage

\vspace{0.5cm}

\vspace{1cm}

\capitolo{Non-integrable quantum field theories}\label{chapnonint}
\setcounter{equation}{0}

As a natural development of the studies on integrable quantum
field theories, there has been recently an increasing interest in
studying the properties of non--integrable quantum field theories
in $(1+1)$ dimensions, both for theoretical reasons and their
application to several condensed--matter or statistical systems.
However, contrary to the integrable models, many features of these
quantum field theories are still poorly understood: in most of the
cases, in fact, their analysis is only qualitative and even some
of their basic data, such as the mass spectrum, are often not
easily available. Although one could always rely on numerical
methods to shed some light on their properties, it is obviously
important to develop and apply some theoretical tools to control
them analytically. In this respect, there has been recently some
progress, thanks to two different and complementary approaches.

The first approach, called the Form Factor Perturbation Theory
(FFPT) \cite{DMS}, is best suited to deal with those
non--integrable theories close to the integrable ones. The second
approach, given by the semiclassical method introduced in Chapter
\ref{chapSM}, is on the other hand best suited to deal with those
quantum field theories having kink excitations of large mass in
their semiclassical limit. Although this method is restricted to
work in a semiclassical regime, it permits however to analyse
non--integrable theories in the whole coupling--constants space,
even far from the integrable points.

In this Chapter, after illustrating the FFPT technique
(Sect.\,\ref{secFFPT}), we describe the original results obtained
with the semiclassical method for the $\phi^4$ field theory in the
broken symmetry phase (Sect.\,\ref{secphi4}) and for the Double
Sine--Gordon model (Sect.\,\ref{secDSG}). The Appendices are
devoted to some technical results obtained in the study of the
Double Sine--Gordon theory.

\section{Form Factor Perturbation Theory}\label{secFFPT}
\setcounter{equation}{0}

The Form Factor Perturbation Theory technique (FFPT), introduced
in \cite{DMS}, permits to obtain quantitative predictions on the
mass spectrum, scattering amplitudes and other physical quantities
in non--integrable theories which can be seen as a perturbation of
integrable ones. As any other perturbation scheme, it works finely
as far as the non--integrable theory is an adiabatic deformation
of the original integrable model, i.e. when the two theories are
isospectral. This happens when the field which breaks the
integrability is {\em local} with respect to the operator which
creates the particles. If, on the contrary, the field which moves
the theory away from integrability is {\em non--local} with
respect to the particles, the resulting non--integrable model
generally displays confinement phenomena and, in this case, some
caution has to be taken in interpreting these perturbative
results.

The method can be applied when the action ${\cal A}$ of the theory
in exam is represented by a deformation of an integrable one
${\cal A}_{0}$ through a given operator $\Psi$:
\begin{equation}
{\cal A}\, = \, {\cal A}_{0} + g\int d^{2}x\,\Psi(x)\;.
\end{equation}
One of the first consequences of moving away from integrability is
a change in the spectrum of the theory: the first order
corrections to the mass of the particle $a$ belonging to the
spectrum of the unperturbed theory is in fact given by
\begin{equation}
\label{FFPTmass} \delta m_{a}^{2} \,= \, 2 g
F_{a\bar{a}}^{\Psi}(i\pi) + O(g^{2})\;,
\end{equation}
where the particle--antiparticle form factor of the operator
$\Psi(x)$, defined by the matrix element
$$
F_{a\bar{a}}^{\Psi}(\theta_1 - \theta_2) \,=\, \langle 0 \mid
\Psi(0) \mid a(\theta_1) \bar{a}(\theta_2) \rangle \,\,\,,
$$
is introduced as function of the difference of the rapidity
variables, defined in (\ref{rapidity}). The mass correction
(\ref{FFPTmass}) may be finite or divergent, depending on the
locality properties of the operator $\Psi(x)$ with respect to the
particle $a$. The situation was clarified in \cite{dm} and it is
worth recalling the main conclusion of that analysis.

As we have seen in $(\ref{ffcrossing}),\,(\ref{ffannih})$, if the
operator which interpolate particle $a$ is non--local with respect
to the perturbing operator $\Psi$, the form factor
$F_{a\bar{a}}^{\Psi}(\theta)$ displays an annihilation pole at
$\theta=i\pi$, which is exactly the value at which the form factor
is evaluated in (\ref{FFPTmass}). Therefore, in this case the mass
correction received by particle $a$ under the perturbation $\Psi$
is formally infinite, unless the residue in $(\ref{ffannih})$
vanishes.

Let's consider as an example the case in which the unperturbed
theory is the sine--Gordon model (\ref{SGpot}) with frequency
$\beta$, and the perturbing operator is the exponential
$\Psi_{\alpha} = e^{i \alpha \varphi}$.  This vertex operator is
semi--local with respect to the soliton particle $s$, with the
index $\gamma_{\alpha,s}$ entering (\ref{ffcrossing}) given by
$\gamma_{\alpha,s} = \alpha/\beta$, whereas it is local with
respect to the breather particles $b$ (i.e.
$\gamma_{\alpha,b}=0$). This implies that formula (\ref{FFPTmass})
can be safely applied to compute the first order correction to the
mass of the breathers, whereas a divergence may appear in an
analogous computation of the mass correction of the solitons. This
divergence has to be seen as the mathematical signal that the
solitons of the original integrable model no longer survive as
asymptotic particles of the perturbed theory, i.e. they are
confined.

\section{Broken $\phi^{4}$ theory }\label{secphi4}
\setcounter{equation}{0}

In this Section, we will describe the original results obtained in
\cite{finvolff} and \cite{dsgmrs} about the spectrum of the
$\phi^4$ field theory in the $\mathbb{Z}_2$ broken symmetry phase.
This non--integrable theory, defined by the potential
(\ref{phi4pot})
$$
V(\phi) = \frac{\lambda}{4}\,\phi^{4}-\frac{m^{2}}{2}\,\phi^{2} +
\frac{m^{4}}{4\lambda}\;,
$$
is invariably referred to as a paradigm for a wealth of physical
phenomena. Just to mention a relevant example, in the
Landau--Ginzburg language discussed in Sect.\,\ref{secLG}, it
describes the universality class of the Ising model at low
temperature. In spite of this deep interest, however, its
non--perturbative features are still poorly understood.

\subsection{Semiclassical spectrum}

The main properties of the potential (\ref{phi4pot}) and its kink
background (\ref{phi4kinkinf}) have been already discussed in
Sect.\,\ref{mainidea}, where the semiclassical quantization of the
kink has been described. Here we directly move to the computation
of the kink--antikink form factors, in order to extract from their
analytical structure information about the complete spectrum of
the theory.

With our formulation in terms of the rapidity, we can write an
unambiguous\footnote{This has to be contrasted with the tentative
covariant extrapolation discussed in \cite{GJ}.} Lorentz covariant
expression for the form factor (\ref{ffinf})
\begin{eqnarray}
\langle\,p_{2}|\,\phi(0)|\,p_{1}\,\rangle & = & M\frac{
m}{\sqrt{\lambda}}\int\limits_{-\infty}^{\infty}da\,e^{iM\theta
a}\,\tanh\left(\frac{m a}{\sqrt{2}}\right) = \nonumber \\
&& = \frac{4}{3}i\pi \left(\frac{m}{\sqrt{\lambda}}\right)^{3}
\frac{1}{\sinh\left(\frac{2}{3}\pi\frac{m^{2}}{\lambda}\,\theta\right)}\;,
\label{phi4ffinfvol}
\end{eqnarray}
where the kink mass is approximated at leading order by the
classical energy $M=\frac{2\sqrt{2}}{3}\frac{m^{3}}{\lambda}$.

It is of great interest to analyse in this case the dynamical
poles of $F_{2}(\theta)$ for extracting information about the
spectrum. They are located at
\begin{equation}
\theta_{n}=i\pi\left[1-\frac{3}{2\pi}\,\frac{\lambda}{m^{2}}\,n\right]\,,
\quad\qquad 0<n<\frac{2\pi}{3}\frac{m^{2}}{\lambda} \,,
\end{equation}
and the corresponding bound states masses are given by
\begin{equation}\label{phi4boundst}
m_{b}^{(n)} = 2M\sin\left[\frac{3}{4}\,
\frac{\lambda}{m^{2}}\,n\right] =
n\,\sqrt{2}\,m\left[1-\frac{3}{32}\,
\frac{\lambda^{2}}{m^{4}}\,n^{2}+...\right]\,.
\end{equation}
Note that the leading term is consistently given by multiples of
$\sqrt{2}m$, which is the known mass of the elementary boson of
this theory\footnote{The elementary bosons represent the
excitations over the vacua, i.e. the constant backgrounds
$\phi_{\pm}=\pm\frac{m}{\sqrt{\lambda}}$, therefore their square
mass is given by $V''(\phi_{\pm})=2\,m^2$.}. Contrary to the
Sine-Gordon model, we now have all integer multiples of this mass
and not only the odd ones: this is because we are in the broken
phase of the theory, where the invariance under $\phi\rightarrow
-\phi$ is lost. Furthermore, this spectrum exactly coincides with
the one derived in \cite{DHN} by building approximate classical
solutions to represent the "breathers".

Another important information can be extracted from the residue of
$F_{2}(\theta)$ on the pole corresponding to the lightest bound
state $b^{(1)}$. This quantity, indeed, has to be proportional to
the one-particle form factor
$\langle\,0|\,\phi|\,b^{(1)}\,\rangle$ through the semiclassical
3-particle on-shell coupling of kink, antikink and elementary
boson $\Gamma_{k\bar{k}b}$, as shown in (\ref{ffdynpole}). Since
in our normalization the one-particle form factor takes the
constant value $1/\sqrt{2}$, at leading order in the coupling we
get
\begin{equation}\label{gphi4}
\Gamma_{k\,\bar{k}\,b}\,=\,2\sqrt{2}\,\frac{m}{\sqrt{\lambda}}\;,
\end{equation}
a quantity so far unknown in the non--integrable $\phi^4$ theory.

Finally, the $1/\lambda$ leading contribution of the spectral
function is given in this case by
\begin{equation}
\hat\rho(p^{2}) =
\frac{2\pi}{\lambda}\,\delta\left(p^{0}/m\right)\delta\left(p^{1}/m\right)
+\frac{\pi^{3}}{2\lambda}\,\delta\left(\frac{p^{0}}{M}-2\right)\frac{1}{
\sinh^{2}\left[\frac{\pi}{\sqrt{2}}\,\frac{p^{1}}{m}\right]}\;.
\end{equation}

\subsection{Resonances}

The appearance of resonances in the classical kink--antikink
scattering has been studied for this theory with numerical
techniques in \cite{campbellphi4}. In this work, the key
ingredient for the presence of resonances was identified in the
presence of the so--called \lq\lq shape mode", which is the
discrete eigenvalue $\omega_1$ of the small oscillations around
the kink background, given in (\ref{omega1phi4}), corresponding to
the internal excitation of the kink. We have shown in
\cite{dsgmrs} how this mechanism can be analytically explained in
our formalism. In fact, Goldstone and Jackiw's result on the form
factor of the field $\phi$ between asymptotic states containing a
simple and an excited kink, described by formulas (\ref{ffexc})
and (\ref{fhatexc}), can be also refined in terms of the rapidity
variable. In our case, where the fluctuation $\eta_1$ is given in
(\ref{omega1phi4}), the covariant form factor, analytically
continued in the crossed channel, is expressed as
\begin{equation}
\langle\, 0|\,\phi(0)\,|\,\bar{p}_{2}\,p^{*}_{1}\,\rangle \, =
\,-i\,\frac{M\,\pi}{6^{1/4}\,m^{5/2}}\,
\frac{M\,(i\pi-\theta)}{\cosh\left[\frac{\pi}{\sqrt{2}\,m}\,
M\,(i\pi-\theta)\right]} \;.
\end{equation}
The dynamical poles of this object correspond to bound states with
masses
\begin{equation}
\label{phi4excboundst} \left(m_{b^{*}}^{(n)}\right)^{2} \, = \, 4
M (M + \omega_{1}) \sin^{2}\left[\frac{3}{8}\,
\frac{\lambda}{m^{2}}\,(2n+1)\right] + \omega_{1}^{2} \;.
\end{equation}
The states with
\begin{equation}
\label{resnumber} \frac{8}{3}\,\frac{m^{2}}{\lambda}\,
\arcsin\sqrt{\frac{4M^{2} - \omega_{1}^{2}}{4M ( M + \omega_{1})}}
< 2n+1 < \frac{4}{3}\,\frac{m^{2}}{\lambda}\,\pi
\end{equation}
have masses in the range
\begin{equation}
2M <m_{b^{*}}^{(n)} < 2M + \omega_{1}\;,
\end{equation}
and, therefore, they can be seen as resonances in the
kink-antikink scattering.

Since the numerical analysis done in \cite{campbellphi4} is
independent of the coupling constant\footnote{Classically, in
fact, one can always rescale the field and eliminate the coupling
constant $\lambda$.}, a quantitative comparison with our
semiclassical result is rather difficult, due to the dependence on
$\lambda$ of (\ref{resnumber}). However, the presence of many
resonance states seen at classical level is qualitatively
confirmed to persist also in the quantum field theory at small
$\lambda$, i.e. in its semiclassical regime, according to
(\ref{resnumber}).

\section{Double Sine--Gordon model}\label{secDSG}
\setcounter{equation}{0}

An interesting non--integrable model where both FFPT and
Semiclassical Method can be used is the so--called Double
Sine--Gordon Model (DSG). It is defined by the potential
\begin{equation}\label{potDSG}
V(\varphi) \,= \, -\frac{\mu}{\beta^{2}}\,\cos\beta\,\varphi -
\frac{\lambda}{\beta^{2}} \,\cos\left(\frac{\beta}{2}\,
\varphi+\delta\right) + C\;,
\end{equation}
where $C$ is a constant that has be chosen such that to have a
vanishing potential energy of the vacuum state. The classical
dynamics of this model has been extensively studied in the past by
means of both analytical and numerical techniques (see
\cite{campbellDSG} for a complete list of the results), while its
thermodynamics has been studied in \cite{condat} by using the
transfer integral method \cite{transferintegral}, and with the
path integral technique (see for instance \cite{tognetti}).

With $\lambda$ or $\mu$ equal to zero, the DSG reduces to the
ordinary integrable Sine--Gordon (SG) model with frequency $\beta$
or $\beta/2$ respectively. Hence the DSG model with a small value
of one of the couplings can be regarded as a deformation of the
corresponding SG model and studied, therefore, by means of the
FFPT \cite{dm}. On the other hand, for $\beta \rightarrow 0$,
irrespectively of the value of the coupling constants $\lambda$
and $\mu$, the DSG model reduces to its semiclassical limit.
Despite the non--integrable nature of the DSG model, its classical
kink solutions are -- remarkably enough -- explicitly known
\cite{campbellDSG,condat} and therefore the Semiclassical Method
can be successfully applied to recover the (semi--classical)
spectrum of the theory. As we will see in the following, the two
approaches turn out to be complementary in certain regions of the
coupling constants, i.e. both are needed in order to get the whole
mass spectrum of the theory, whereas in other regions they provide
the same picture about the spectrum of the excitations.

Apart from the theoretical interest in testing the efficiency of
the two methods on this specific model where both are applicable,
the study of the DSG is particularly important since this model
plays a relevant role in several physical contexts, either as a
classical non--linear system or as a quantum field theory. At the
classical level, its non--linear equation of motion can be used in
fact to study ultra--short optical pulses in resonant degenerate
medium or texture dynamics in He$^3$ (see, for instance,
\cite{bullough} and references therein). As a quantum field
theory, depending on the values of the parameters $\lambda,\mu,
\beta,\delta$ in its Lagrangian, it displays a variety of physical
effects, such as the decay of a false vacuum or the occurrence of
a phase transition, the confinement of the kinks or the presence
of resonances due to unstable bound states of excited
kink--antikink states. Moreover, it finds interesting applications
in the study of several systems, such as the massive Schwinger
field theory or the Ashkin--Teller model \cite{dm}, as well as in
the analysis of the O(3) non--linear sigma model with $\theta$
term \cite{cm}, i.e. the quantum field theory relevant for
understanding the dynamics of quantum spin chains
\cite{haldane,afflecklecture}. The DSG model also matters in the
investigation of other interesting condensed matter phenomena,
such as the soliton confinement of spin--Peierls antiferromagnets
\cite{affleck}, the dynamics of the spin chains in a staggered
external field or the electron interaction in a staggered
potential \cite{fabr}.

Motivated by the above combined theoretical and physical
interests, we have performed in \cite{dsgmrs} a thorough study of
the spectrum of the DSG model. In some regions of the coupling
constants, the obtained original semiclassical results have
revealed to be necessary to complete the spectrum analysis, while
in other regions they have been compared with the analogous ones
produced by FFPT. Here we present the main result of the
semiclassical analysis of the spectrum of the DSG model and its
comparison with the results coming from FFPT
(Sect.\,\ref{secsemanal}). Furthermore, the phenomenon of false
vacuum decay will be also discussed (Sect.\,\ref{secfalse}).
Finally, this Chapter's appendices are devoted to some technical
results obtained in this study. In particular, Appendix
\ref{chapnonint}.\ref{kmasscorr} presents the computation of the
kink mass corrections by using the FFPT, Appendix
\ref{chapnonint}.\ref{secFF} lists the relevant expressions of the
semiclassical form factors in DSG model, Appendix
\ref{chapnonint}.\ref{breathers} discusses the analysis of neutral
states in comparison with the Sine--Gordon model, and Appendix
\ref{chapnonint}.\ref{secDShG} describes the basic results in a
closely related model, i.e. the Double Sinh--Gordon model.

\subsection{Semiclassical spectrum}\label{secsemanal}

We will study the theory defined by (\ref{potDSG}) in a regime of
small $\beta$, where the semiclassical results are expected to
give a valuable approximation of the spectrum\footnote{By applying
the stability conditions found in \cite{dm} to this model, they
reduce to the condition $\beta^{2} < 8\pi$. Hence, for these
values of $\beta$ and, in particular in the semiclassical limit
$\beta \to 0$, the potential (\ref{potDSG}) is stable under
renormalization and no countertems have to be added.}. At the
quantum level, the different Renormalization Group trajectories
originating from the gaussian fixed point described by the kinetic
term $\frac{1}{2} (\partial_{\mu} \varphi)^2$ of the lagrangian
are labelled by the dimensionless scaling variable $\eta = \lambda
\mu^{-(8 \pi - \beta^2/4) / (8\pi - \beta^2)}$ which simply
reduces to the ratio $\eta = \frac{\lambda}{\mu}$ in the
semiclassical limit. When $\lambda$ or $\mu$ are equal to zero,
the DSG model coincides with an ordinary Sine-Gordon model with
coupling $\beta$ or $\beta/2$, and mass scale $\sqrt{\mu}$ or
$\sqrt{\lambda/4}$, respectively.

Since for general values of the couplings the potential
(\ref{potDSG}) presents a $\frac{4\pi}{\beta}$-periodicity, it was
noticed in \cite{hungDSG} that one has an adiabatic perturbation
of an integrable model only if the $\lambda=0$ theory is regarded
as a two--folded Sine-Gordon model. This theory is a modification
of the standard Sine-Gordon model, where the period of the field
$\phi$ is defined to be $\frac{4\pi}{\beta}$, instead of
$\frac{2\pi}{\beta}$ \cite{foldedSG}. As a consequence of this new
periodicity assignment, such a theory has two different degenerate
vacua $|k\,\rangle$, with $k=0,1$ and
$|k+2\,\rangle\equiv|k\,\rangle$, which are defined by
$\langle\,k|\,\phi\,|\,k\,\rangle=\frac{2\pi}{\beta} \,k$. Hence
it has two different kinks, related to the classical backgrounds
by the formula
\begin{equation}\label{twofoldedkinks}
K_{k,k+1}^{cl}(x)\,=\,\frac{2k\pi}{\beta} +
\frac{4}{\beta}\arctan\,e^{\,m\,x}\,\qquad k=0,1\;,
\end{equation}
and two corresponding antikinks, related to the classical
solutions by the expression
\begin{eqnarray}
K_{k+1,k}^{cl}(x)&=&\frac{2k\pi}{\beta} +
\frac{4}{\beta}\arctan\,e^{\,-m\,x}\nonumber\\
&=&\frac{2(k+1)\pi}{\beta} -
\frac{4}{\beta}\arctan\,e^{\,m\,x}\,\qquad k=0,1\;.
\end{eqnarray}
Finally, in the spectrum there are also two sets of kink-antikink
bound states $b_{n}^{(l)}$, with $l=0,1$ and $n=1,...,
\left[\frac{8\pi}{\xi}\right]$.

The flow between the two limiting Sine-Gordon models (with
frequency $\beta$ or $\beta/2$, respectively) displays a variety
of different qualitative features, including confinement and phase
transition phenomena, depending on the signs of $\lambda$ and
$\mu$, and on the value of the relative phase $\delta$. However,
it was observed in \cite{dm} that the only values of $\delta$
which lead to inequivalent theories are those given by
$|\delta|\leq\frac{\pi}{2}$. Furthermore, in virtue of the
relations
$$
\begin{array}{rcl}
V_{\delta}\left(\phi + \frac{\pi}{\beta},\lambda,\mu\right) & = &
V_{\delta+\pi/2}\left(\phi,\lambda,-\mu\right)\;,\\
V_{\delta}(-\phi,\lambda,\mu) & = &
V_{-\delta}(\phi,\lambda,\mu)\;,
\end{array}
$$
we can describe all the inequivalent possibilities keeping
$\mu$ positive and the relative phase in the range $0 \leq \delta
\leq \frac{\pi}{2}$. The sign of the coupling $\lambda$, instead,
simply corresponds to a shift or a reflection of the potential,
without changing its qualitative features. As we are going to show
in the following, the case $\delta = \frac{\pi}{2}$ displays
peculiar features, while a common description is possible for any
other value of $\delta$ in the range $0 \leq \delta <
\frac{\pi}{2}$.

It is worth mentioning that the possibility of writing exact
classical solutions for all the different kinds of topological
objects in this model finds a deep explanation in the relation
between the trigonometric potential (\ref{potDSG}) and power-like
potentials. In fact, defining
\begin{equation}
\varphi\,=\,\frac{n\pi}{\beta}\pm \frac{4}{\beta}\arctan
Y\;,\qquad n=0,1,2,3\;,
\end{equation}
one can easily see that the first order equation which determines
the kink solution
\begin{equation*}
\frac{1}{2}\left(\frac{d\varphi}{dx}\right)^{2} \,=\,
-\frac{\mu}{\beta^{2}}\,\cos\beta\,\varphi -
\frac{\lambda}{\beta^{2}}\,\cos\left(\frac{\beta}{2}\, \varphi +
\delta\right) + C
\end{equation*}
is mapped into the equation for $Y$
\begin{equation}
\frac{1}{2}\left(\frac{dY}{dx}\right)^{2}\,=\,U(Y)\;,
\end{equation}
where $U(Y)$ describes various kinds of algebraic potentials,
depending on the values of $n$, $\delta$ and $C$. The $\delta=0$
case was analysed in \cite{relationphi4} and its classical
solutions are very simple because $U(Y)$ only contains quartic and
quadratic powers of $Y$. It is easy to see that a similar
situation also occurs in the $\delta = \frac{\pi}{2}$ case; for
instance, choosing $n=1$ and $C=-\frac{1}{\beta^{2}}\left(\mu +
\frac{\lambda^{2}}{8\mu}\right)$, one obtains the quartic
potential \EQ U(Y)\,=\, \frac{(4\mu + \lambda)^{2}}{128\mu}
\left(\frac{4\mu - \lambda}{4\mu + \lambda}\;Y^{2}-1\right)^{2}\;,
\EN which has the well known classical background \EQ
Y(x)\,=\,\sqrt{\frac{4\mu + \lambda}{4\mu - \lambda}}\,
\tanh\left(\sqrt{\mu-\frac{\lambda^{2}}{16\,\mu}}\;\frac{x}{2}\right)
\,\,\,. \EN For generic $\delta$, instead, also cubic and linear
powers of $Y$ appear, making more complicated the analysis of the
classical solutions.

\subsubsection{$\delta=0$ case}

It is convenient to start our discussion with the case $\delta=0$.
This case, in fact, displays those topological features which are
common to all other models with $0 <\delta < \frac{\pi}{2}$, but
it admits a simpler technical analysis, due to the fact that
parity invariance survives the deformation of the original SG
model. As we will see explicitly, in this case the results of the
FFPT and the Semiclassical Method are complementary, since they
describe different kinds of excitations present in the theory.

\vspace{0.5cm}

\psfrag{phiopi}{$\frac{\phi}{\pi}$}

\begin{figure}[ht]
\begin{tabular}{p{8cm}p{8cm}}
\psfrag{V}{$V(\mu=1,\,\lambda=1/2)$}
\psfig{figure=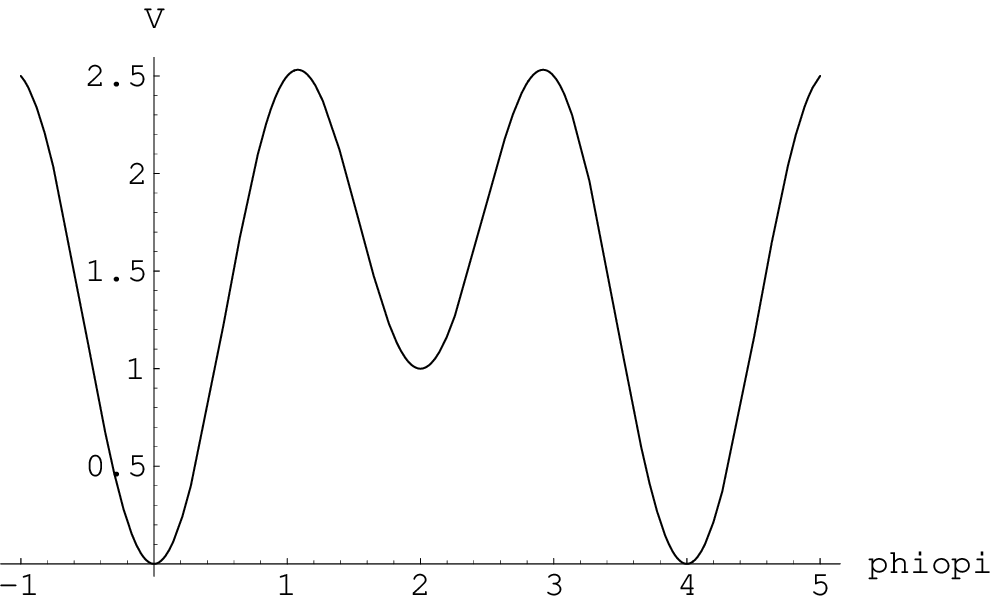,height=5cm,width=7cm} &
\psfrag{V}{$V(\mu=1,\,\lambda=4)$}
\psfig{figure=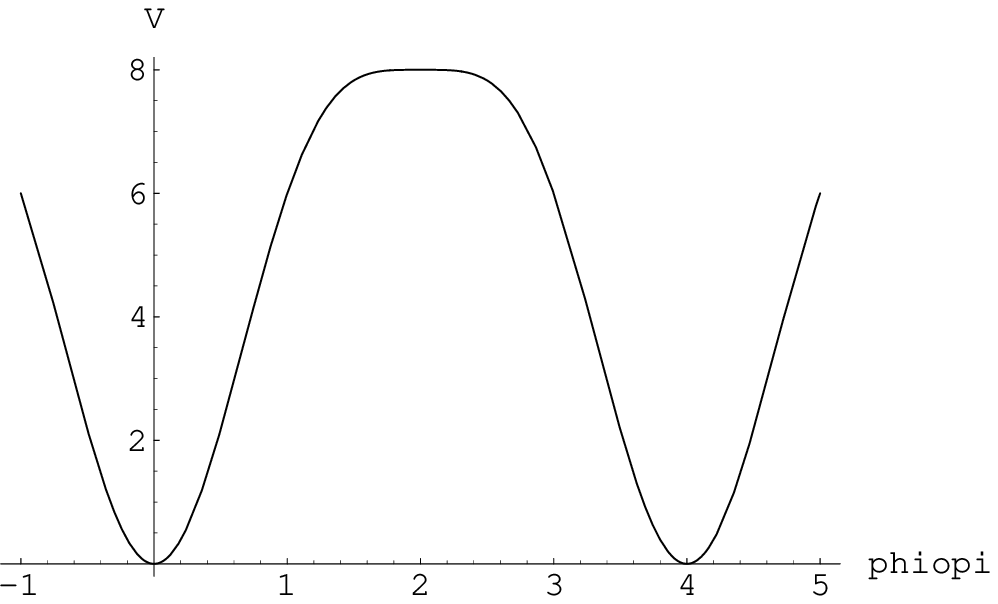,height=5cm,width=7cm}
\end{tabular}
\caption{DSG potential in the case $\delta=0$.}\label{figdelta0}
\end{figure}

Fig.\ref{figdelta0}  shows the shape of this DSG potential in the
two different regimes, i.e. (i) $0<\lambda<4 \mu$ and (ii)
$\lambda>4\mu$. The absolute minimum persists in the position
$0\;(\text{mod}\, \frac{4\pi}{\beta})$ for any values of the
couplings, while the other minimum at
$\frac{2\pi}{\beta}\;(\text{mod}\,\frac{4\pi}{\beta})$ becomes
relative and disappears at the point $\lambda=4\mu$. The breaking
of the degeneration between the two initial vacua in the
two--folded SG causes the confinement of the original SG solitons,
as it can be explicitly checked by applying the FFPT. The linearly
rising potential, responsible for the confinement of the SG
solitons, gives rise then to a discrete spectrum of bound states
whose mass is beyond $2 M_{SG}$, where $M_{SG}$ is the mass of the
SG solitons \cite{dm,affleck}.

The disappearing of the initial solitons represents, of course, a
drastic change in the topological features of the spectrum. At the
same time, however, a stable new static kink solution appears for
$\lambda\neq 0$, interpolating between the new vacua at $0$ and
$\frac{4\pi}{\beta}$. The existence of this new topological
solution is at the origin of the complementarity between the FFPT
and the Semiclassical Method. By the first technique, in fact, one
can follow adiabatically the deformation of the SG breathers
masses: these are neutral objects that persist in the theory
although the confinement of the original kinks has taken place. It
is of course impossible to see these particle states by using the
Semiclassical Method, since the corresponding solitons, which
originate these breathers as their bound states, have disappeared.
Semiclassical Method can instead estimate the masses of other
neutral particles, i.e. those which appear as bound states of the
new stable kink present in the deformed theory.

\begin{figure}[h]
\psfrag{x}{$x$}\psfrag{2 pi}{$2\pi$}\psfrag{4
pi}{$4\pi$}\psfrag{phiK(x)}{$\phi_{K}(x)$} \psfrag{- R}{$-R$}
\psfrag{R}{$R$}
\hspace{4cm}\psfig{figure=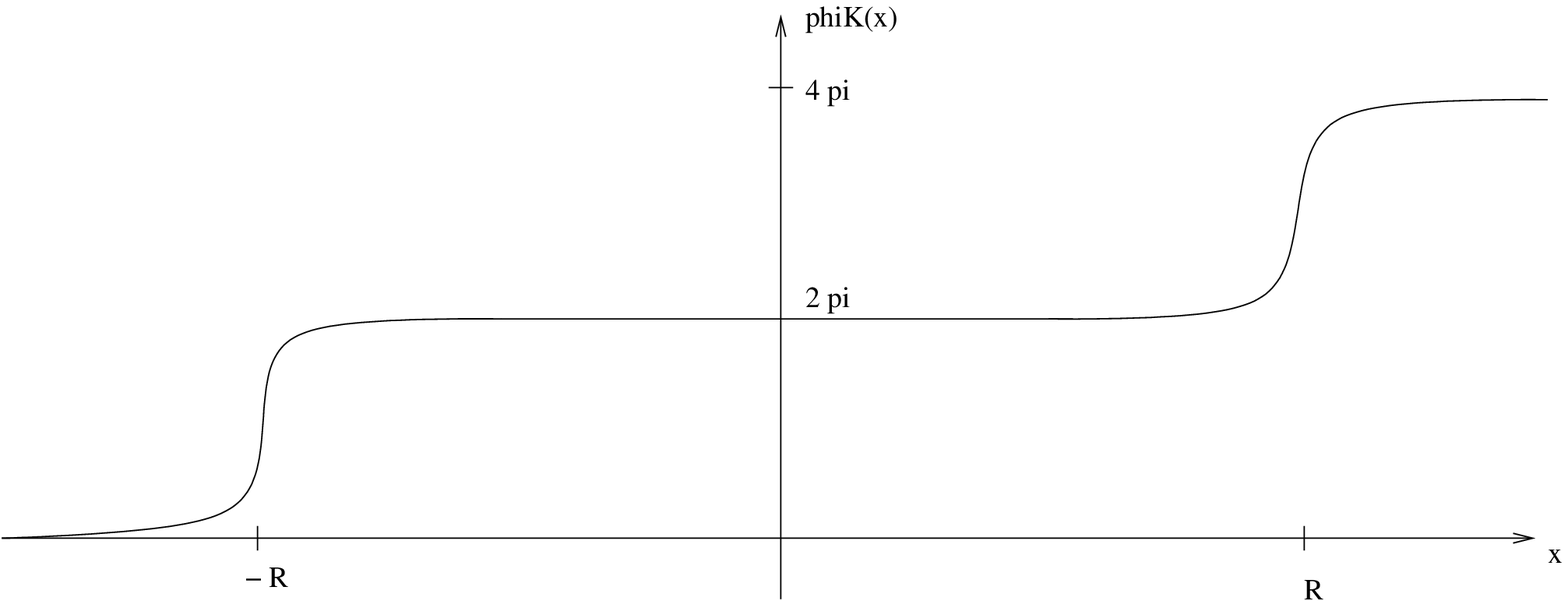,height=5cm,width=7cm}
\caption{Kink solution (\ref{doublekink})} \label{figdoublekink}
\end{figure}

This new kink solution, interpolating between $0$ and
$\frac{4\pi}{\beta}$, is given explicitly by
\begin{equation}\label{doublekink}
\varphi_{K}(x) = \frac{2\pi}{\beta}+
\frac{4}{\beta}\arctan\left[\sqrt{
\frac{\lambda}{\lambda+4\mu}}\,\sinh\left(m\,x\right)\right]\;,
\end{equation}
where
\begin{equation}\label{curvaturedeltazero}
m^{2}=\mu+\frac{\lambda}{4}ù
\end{equation}
is the curvature of the absolute minimum. Interestingly enough
\cite{campbellDSG}, this background admits an equivalent
expression in terms of the superposition of two solitons of the
unperturbed Sine-Gordon model, centered at the fixed points $\pm
R$
\begin{equation*}
\varphi_{K}(x)=\varphi_{\text{SG}}(x+R)+\varphi_{\text{SG}}(x-R)\;,
\end{equation*}
where $\varphi_{\text{SG}}(x) =
\frac{4}{\beta}\arctan\left[e^{m\,x}\right]$ are the usual
Sine-Gordon solitons with the deformed mass parameter
(\ref{curvaturedeltazero}) whereas their distance $2R$ is
expressed in terms of the couplings by
\[
R \, =\,
\frac{1}{m}\,\text{arccosh}\sqrt{\frac{4\mu}{\lambda}+1}\,\,\,.
\]
By looking at Fig.\ref{figdoublekink}, it is clear that this
background, in the small $\lambda$ limit, describes the two
confined solitons of SG, which become free in the $\lambda=0$
point, i.e. where $R\to\infty$.

The classical energy of this kink is given by
\begin{equation}\label{doublekinkmass}
M_{K} \,= \, \frac{16\,m}{\beta^{2}}\,
\left\{1+\frac{\lambda}{\sqrt{4\mu(\lambda+4\mu)}}\,
\text{arctanh} \sqrt{\frac{4\mu}{\lambda+4\mu}}\right\}\;,
\end{equation}
and in the $\lambda\to 0$ limit it tends to twice the classical
energy of the Sine-Gordon soliton, i.e.
\begin{equation*}
M_{K}\;{\mathrel{\mathop{\kern0pt\longrightarrow}\limits_{\lambda\to
0 }}}\;\frac{16\sqrt{\mu}}{\beta^{2}}\;,
\end{equation*}
therefore confirming the above picture. In the $\mu\to 0$ limit,
the asymptotic value of the above expression is instead the mass
of the soliton in the Sine-Gordon model with coupling $\beta/2$.
The expansion for small $\mu$
\begin{equation}
\label{doublekinkfirst}
M_{K}\;{\mathrel{\mathop{\kern0pt\longrightarrow} \limits_{\mu\to
0 }}}\; \frac{8\sqrt{\lambda/4}}{(\beta/2)^{2}} +
\frac{\mu}{\beta^{2}}\,\frac{32}{3\sqrt{\lambda}} + O(\mu^{2})\;,
\end{equation}
gives the first order correction which is in agreement with the
result of the FFPT in the semiclassical limit (see Appendix
\ref{chapnonint}.\ref{kmasscorr}).

The bound states created by the kink (\ref{doublekink}) and its
antikink can be obtained by looking at the poles of the
semiclassical form factors of the fields $\varphi(x)$ and
$\varepsilon(x)$, reported in Appendix
\ref{chapnonint}.\ref{secFF}, and their mass are given
by\footnote{Due to parity invariance, the dynamical poles of the
form factor of $\varphi$ between kink states only give the bound
states with $n$ odd. The even states can be obtained from the form
factor of the energy operator $\epsilon(x)$.}
\begin{equation}
\label{doubleboundstates} m_{(K)}^{(n)} \, = \,
2M_{K}\sin\left(n\,\frac{m}{2M_{K}}\right) \,\,\,\,\,\,\, ,
\,\,\,\,\,\,\, 0 < n < \pi\frac{M_{K}}{m}\;.
\end{equation}
For small $\mu$ we easily recognize the perturbation of the
standard breathers in Sine-Gordon with $\beta/2$:
\begin{equation}\label{breathersmu}
m_{(K)}^{(n)}\;{\mathrel{\mathop{\kern0pt\longrightarrow}
\limits_{\mu\to 0}}}\;
\frac{64}{\beta^{2}}\sqrt{\frac{\lambda}{4}}
\sin\left(n\,\frac{\beta^{2}}{64}\right) +
\frac{2}{3}\,\frac{\mu}{\sqrt{\lambda}}
\left[\frac{32}{\beta^{2}}\sin\left(n\,
\frac{\beta^{2}}{64}\right)+n\,\cos\left(n\,
\frac{\beta^{2}}{64}\right) \right]+O(\mu^{2})\;,
\end{equation}
while the expansion of the bound states masses for small $\lambda$
\begin{eqnarray*}
m_{(K)}^{(n)}\;&{\mathrel{\mathop{\kern0pt\longrightarrow}
\limits_{\lambda\to 0 }}}\;&
\frac{32\sqrt{\mu}}{\beta^{2}}\sin\left(n\,
\frac{\beta^{2}}{32}\right)
+ \\
&& +\frac{1}{8}\,\frac{\lambda}{\sqrt{\mu}}
\left[\left(1-\ln\frac{\lambda}{16\mu}\right)
\frac{32}{\beta^{2}}\sin\left(n\,\frac{\beta^{2}}{32}\right) +
n\,\ln\frac{\lambda}{16\mu}\,\cos\left(n\,
\frac{\beta^{2}}{32}\right) \right]+O(\lambda^{2})\;
\end{eqnarray*}
deserves further comments: in fact, although the above masses have
well-defined asymptotic values, they do not correspond however to
any state of the unperturbed SG theory. The reason is that the
classical background (\ref{doublekink}) in the $\lambda\to 0$
limit does not describe any longer a localized single particle.
This implies that its Fourier transform cannot be interpreted as
the two-particle form factor and, consequently, its poles cannot
be associated to any bound states.

A technical signal of the disappearing of the above mentioned
bound states in the $\lambda\to 0$ limit can be found by computing
the three particle coupling among the kink, the antikink and the
lightest bound state, given by (\ref{ffdynpole}). At leading order
in $\beta$ we get
\begin{equation*}
\Gamma_{K\bar{K}\,b}=\frac{16\sqrt{2}}{\beta}\;\frac{m}{\sqrt{\lambda}}\;.
\end{equation*}
The divergence of the coupling as $\lambda\to 0$ indicates that
the considered scattering processes cannot be seen anymore as a
bound state creation, i.e. the corresponding bound state
disappears from the theory. A general discussion of the same
qualitative phenomenon for the ordinary Sine-Gordon model can be
found in \cite{imcoupl}, where the disappearing from the theory of
a heavy breather at specific values of $\beta$ is explicitly
related to the divergence or to the imaginary nature of the three
particle coupling among this breather and two lightest ones.

\begin{figure}[h]
\hspace{4cm}\psfig{figure=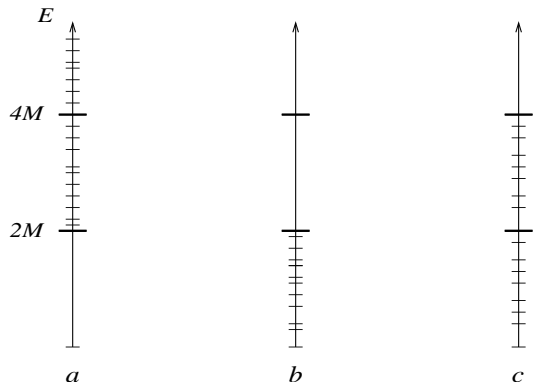,height=5cm,width=7cm}
\normalsize\caption{Neutral states coming from: a) solitons
confinement, b) deformations of SG breathers, c) bound states of
the kink (\ref{doublekink})} \label{figneutral}
\end{figure}

Summarizing, in this model we have three kinds of neutral objects,
i.e. meson particles. The first kind $(a)$ is given by the bound
states originating from the confinement potential of the original
solitons. These discrete states have masses above the threshold
$2M_{SG}$, where $M_{SG}$ is the mass of the SG solitons, and
merge in the continuum spectrum of the non-confined solitons in
the $\lambda\to 0$ limit \cite{dm,affleck}. The second kind $(b)$
is represented by the deformations of SG breathers, that can be
followed by means of the FFPT and have masses, for small
$\lambda$, in the range $[0,2M_{SG}]$. Finally, the third kind
$(c)$ is given by the bound states (\ref{doubleboundstates}) of
the stable kink of the DSG theory and they have masses in the
range $[0,4M_{SG}]$. All these mass spectra are drawn in Fig.
\ref{figneutral}.

Obviously, due to the non--integrable nature of this quantum field
theory not all these particles belong to the stable part of its
spectrum. Apart from a selection rule coming from the conservation
of parity, decay processes are expected to be simply controlled by
phase--space considerations, i.e. a heavier particle with mass
$M_h$ will decay in lighter particles of masses $m_i$ satisfying
the condition
\begin{equation*}
M_h \geq \sum_i m_i \,\,\,.
\end{equation*}
Hence, to determine the stable particles of the theory, one has
initially to identify the lightest mesons of odd and even parity
with mass $m^{\star}_-$ and $m^{\star}_+$ ($m^{\star}_- < m
^{\star}_+$), respectively. Then, the stable particles of even
parity are those with mass $m$ below the threshold $2 m^{\star}_-$
whereas the stable particles of odd parity are those with mass $m
< m^{\star}_- + m^{\star}_+$. For instance, in the $\mu\to 0$
limit we know that the only stable mesons are those given by the
particles $(c)$, as confirmed by the expansion
(\ref{breathersmu}). Hence, in this limit no one of the other
neutral particles is present as asymptotic states. For the mesons
of type $(a)$, this can be easily understood since they are all
above the threshold dictated by the lightest neutral particle. The
situation is more subtle, instead, for the states $(b)$. However,
their absence in the theory with $\mu \to 0$ clearly indicates
that at some particular value of $\lambda$ even the lightest of
these objects acquires a mass above the threshold
$2m_{(K)}^{(1)}$, with $m_{(K)}^{(1)}$ given by
(\ref{doubleboundstates}). Analogous analysis can be done for
other values of the couplings so that the general conclusion is
that most of the above neutral states are nothing else but
resonances of the DSG model. As it is well known, unstable
particles should have a non--zero imaginary part in their masses,
as we have discussed in (\ref{formres}), but expression
(\ref{doubleboundstates}) is real. This is due to the fact that,
at semiclassical level, the fingerprint of instability is the
imaginary nature of some of the frequencies $\omega_i$,
eigenvalues of the stability equation (\ref{stability}), which are
not considered in the leading order expression
(\ref{doubleboundstates}) for the bound states masses. Hence,
although at the leading order we are missing the imaginary
contributions to the masses, we always have to keep in mind that
they come from some of the $\omega_{i}$.

Another kind of resonances has been numerically observed in the
classical scattering of kink and antikinks of the type
(\ref{doublekink}). The corresponding study, performed in
\cite{campbellDSG}, relies on the ideas developed in
\cite{campbellphi4} for the broken $\phi^4$ theory, and it is
essentially based on the presence of the \lq\lq shape mode". Along
the lines of the discussion presented in Sect.\,\ref{secphi4},
this mechanism can be easily interpreted in our formalism.
Unfortunately, in the case of DSG we were not able to solve
analytically the stability equation around the kink background
(\ref{doublekink}). However, the conclusion that this kind of
resonances is described by the bound states of a normal and an
excited kink remains unchanged.

\begin{figure}[h]
\psfrag{x}{$x$}\psfrag{2 pi}{$2\pi$}\psfrag{4
pi}{$4\pi$}\psfrag{phiB(x)}{$\phi_{B}(x)$} \psfrag{R}{$R$}
\psfrag{- R}{$-R$}
\hspace{4cm}\psfig{figure=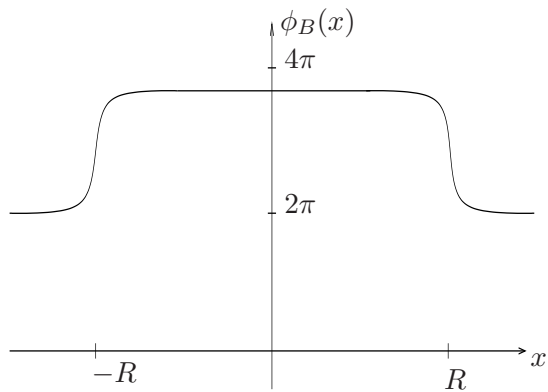,height=5cm,width=7cm}
\caption{Bounce-like solution (\ref{bounce})}\label{figbounce}
\end{figure}

In addition to the above scenario of kink states and bound state
thereof, in the region $\lambda < 4 \mu$ there is another
non-trivial static solution of the theory, defined over the false
vacuum placed at $\varphi = \frac{2\pi}{\beta}$. It interpolates
between the two values $\frac{2\pi}{\beta}$ and
$\frac{4\pi}{\beta} - \frac{2}{\beta}\arccos(1-\lambda/2\mu)$, and
then it comes back (see Fig.\,\ref{figbounce}). Its explicit
expression is given by
\begin{equation}
\label{bounce} \varphi_{B}(x) \,=\,\frac{4\pi}{\beta}-
\frac{4}{\beta}\arctan\left[\sqrt{\frac{\lambda}
{4\mu-\lambda}}\,\cosh\left(m_{f}\,x\right)\right]\;,
\end{equation}
where
\begin{equation}
\label{falsecurvature} m^{2}_{f} \,=\, \mu-\frac{\lambda}{4} \;
\end{equation}
is the curvature of the relative minimum. Similarly to the kink
(\ref{doublekink}), it admits an expression in terms of a soliton
and an antisoliton of the unperturbed SG model:
\begin{equation}
\label{bouncesumSG} \varphi_{B}(x)\,=\, \varphi_{\text{SG}}(x+R) +
\varphi_{\text{SG}}(-(x-R))\;,
\end{equation}
where now $\varphi_{\text{SG}}(x) = \frac{4}{\beta}\arctan
\left[e^{m_{f}\,x}\right]$ are the Sine-Gordon solitons with the
deformed mass parameter (\ref{falsecurvature}) whereas their
distance $2R$ is now given by
\begin{equation}
\label{fixedR} R \,=\,
\frac{1}{m_{f}}\,\text{arcsinh}\sqrt{\frac{4\mu}
{\lambda}-1}\,\,\,.
\end{equation}

In the small $\lambda$ limit, it is clear that this background
describes the confined soliton and antisoliton of the SG model,
which become free in the $\lambda=0$ point, i.e. where
$R\to\infty$.

The classical background (\ref{bounce}) is not related to any
stable particle in the quantum theory. This can be directly seen
from equation (\ref{stability}); in fact, Lorentz invariance
always implies the presence of the eigenvalue $\omega_{0}^{2}=0$,
with corresponding eigenfunction $\eta_{0}(x) =
\frac{d}{dx}\varphi_{cl}(x)$. However, in the case of the solution
(\ref{bounce}) the eigenfunction $\eta_{0}$ clearly displays a
node, which indicates that the corresponding eigenvalue is not the
smallest in the spectrum. Hence, there must be a lower eigenvalue
$\omega_{-1}^{2}<0$, with a corresponding imaginary part of the
mass relative to this particle state. Furthermore, the instability
of (\ref{bounce}) can be related to the theory of false vacuum
decay \cite{coleman,langer}: due to the deep physical interest of
this topic, we will discuss it separately in Section
\ref{secfalse}.

\psfrag{phiopi}{$\frac{\phi}{\pi}$}

\begin{figure}[ht]
\begin{tabular}{p{8cm}p{8cm}}
\psfrag{V}{$V(\mu=1,\,\lambda=1)$}
\psfig{figure=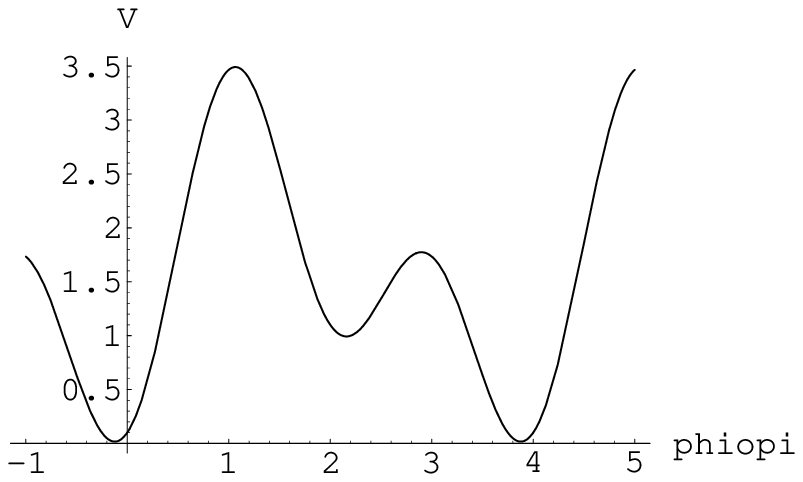,height=5cm,width=7cm} &
\psfrag{V}{$V(\mu=1,\,\lambda=2.3)$}
\psfig{figure=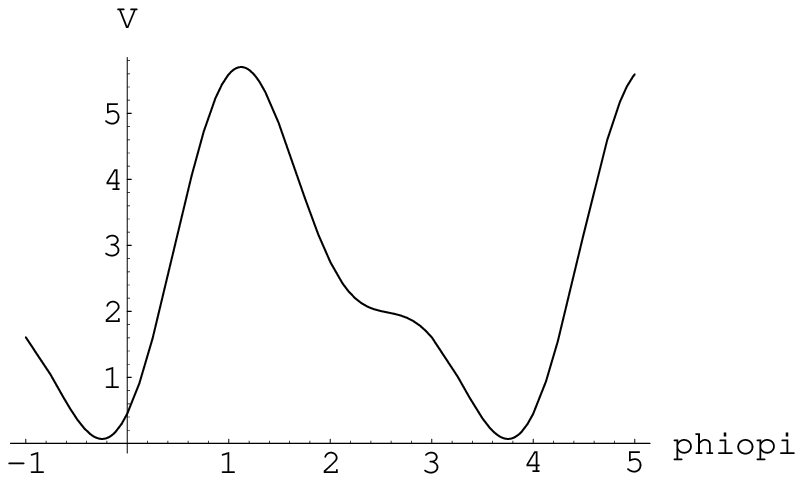,height=5cm,width=7cm}
\end{tabular}
\caption{DSG potential in the case
$\delta=\frac{\pi}{3}$.}\label{figdeltapi3}
\end{figure}

\subsubsection{Comments on generic $\delta$ case}

We have already anticipated that the qualitative features of the
theory relative to $\delta=0$ case are common to all other
theories associated to the values of $\delta$ in the range
$0<\delta<\frac{\pi}{2}$. This can be clearly understood by
looking at the shape of the potential, which is shown in
Fig.\,\ref{figdeltapi3} for the case $\delta=\frac{\pi}{3}$.

In contrast to the $\delta=0$ case, parity invariance is now lost
in these models, and the minima move to values depending on the
couplings. Furthermore, in addition to the change in the nature of
the original vacuum at $\frac{2\pi}{\beta}$, which becomes a
relative minimum by switching on $\lambda$, there is also a
lowering of one of the two maxima. These features make much more
complicated the explicit derivation of the classical solutions, as
we have mentioned at the beginning of the Section.

However, it is clear from Fig.\,\ref{figdeltapi3} that the
excitations of these theories share the same nature of the ones in
the $\delta=0$ case. In fact, the original SG solitons undergo a
confinement, while a new stable topological kink appears,
interpolating between the new degenerate minima. Hence, the
analysis performed for $\delta = 0$ still holds in its general
aspects, i.e. also in these cases the spectrum consists of a kink,
antikink, and three different kinds of neutral particles.

\subsubsection{$\delta=\frac{\pi}{2}$ case}

The value $\delta = \frac{\pi}{2}$ describes the peculiar case in
which no confinement phenomenon takes place, since the two
different vacua of the original two--folded SG remain degenerate
also in the perturbed theory. As a consequence, the original SG
solitons are also asymptotic states in the perturbed theory. By
means of the semiclassical method we can then compute their bound
states, which represent the deformations of the two sets of
breathers in the original two--folded SG. Hence, in this specific
case FFPT and semiclassical method describe the same objects, and
their results can be compared in a regime where both $\beta$ and
$\lambda$ are small.

\footnotesize

\psfrag{phiopi}{$\frac{\phi}{\pi}$}

\begin{figure}[ht]
\begin{tabular}{p{8cm}p{8cm}}
\psfrag{V}{$V(\mu=1,\,\lambda=1)$}\psfig{figure=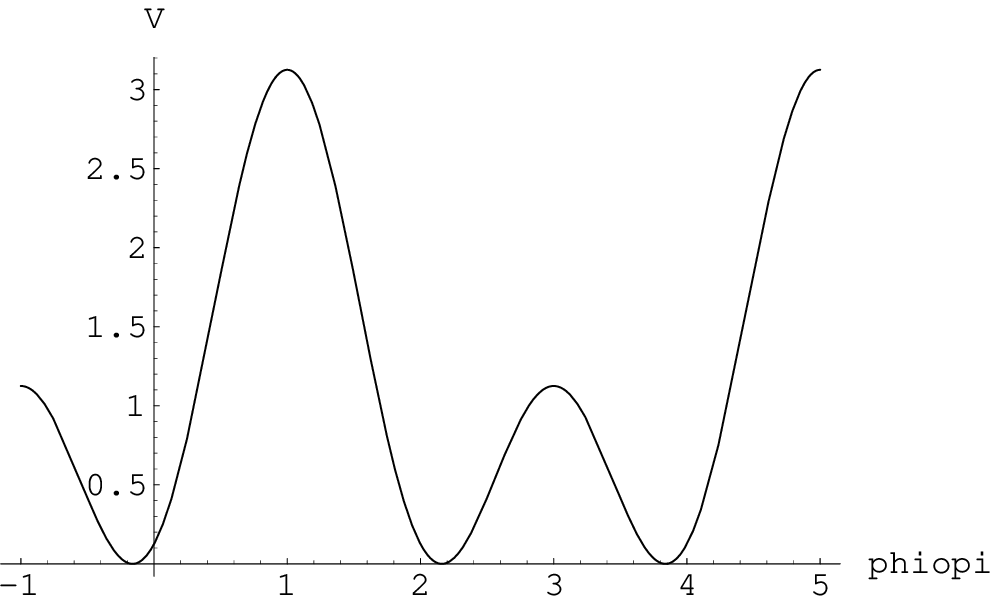,
height=5cm,width=7cm} \vspace{0.2cm}&
\psfrag{V}{$V(\mu=1,\,\lambda=4)$}
\psfig{figure=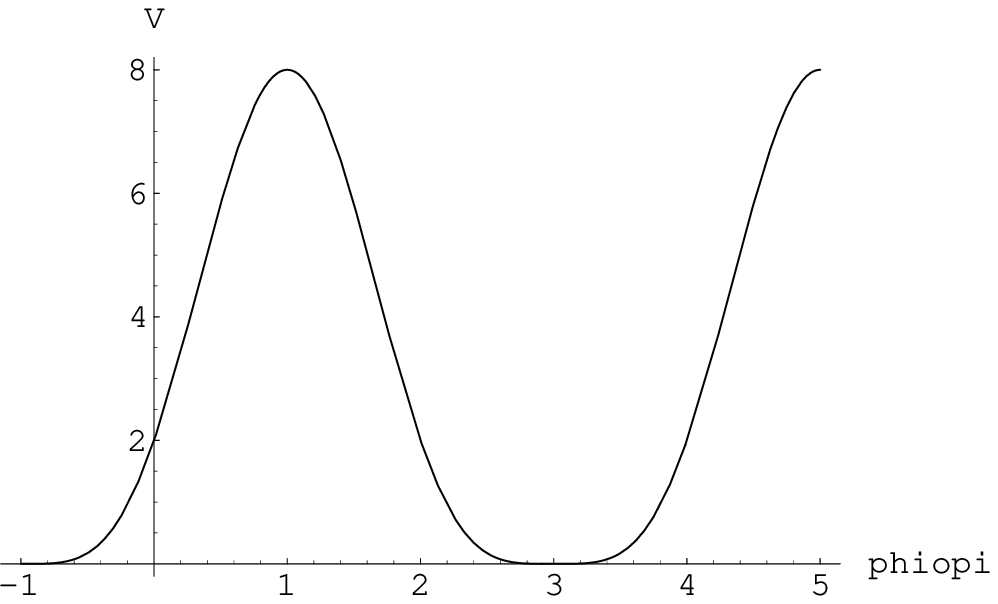,height=5cm,width=7cm}
\end{tabular}
\caption{DSG potential in the $\delta=\frac{\pi}{2}$
case.}\label{figdeltapi2}
\end{figure}

\normalsize

Fig.\,\ref{figdeltapi2} shows the behavior of this DSG potential.
There are two regions, qualitatively different, in the space of
parameters, the first given by $0 < \lambda < 4 \mu$ and the
second given by $\lambda > 4 \mu$. They are separated by the value
$\lambda = 4 \mu$ which has been identified in \cite{dm} as a
phase transition point. We will explain how this identification is
confirmed in our formalism.

Let's start our analysis from the coupling constant region where
$\lambda < 4 \mu$. Switching on $\lambda$, the original
inequivalent minima of the two--folded Sine-Gordon, located at
$\phi_{\text{min}} = 0,\,\frac{2\pi}{\beta}\;
(\text{mod}\,\frac{4\pi}{\beta})$, remain degenerate and move to
$\phi_{\text{min}} = -\phi_{0},\,\frac{2\pi}{\beta} +
\phi_{0}\;(\text{mod}\,\frac{4\pi}{\beta})$, with
$\phi_{0}=\frac{2}{\beta}\arcsin\frac{\lambda}{4\mu}$. The common
curvature of these minima is
\begin{equation}\label{curvaturedeltapi2}
m^{2}\,=\, \mu-\frac{1}{16}\,\frac{\lambda^{2}}{\mu} \;.
\end{equation}
Correspondingly there are two different types of kinks, one called
\lq\lq large kink" and interpolating through the higher barrier
between $-\phi_{0}$ and $\frac{2\pi}{\beta} + \phi_{0}$, the other
called \lq\lq small kink" and interpolating through the lower
barrier between $\frac{2\pi}{\beta} + \phi_{0}$ and
$\frac{4\pi}{\beta} - \phi_{0}$. Their classical expressions are
explicitly given by
\begin{eqnarray}\label{largekink}
\varphi_{L}(x)&=& \,\,\frac{\pi}{\beta}\,\,+\,
\frac{4}{\beta}\arctan\left[\sqrt{\frac{
4\mu+\lambda}{4\mu-\lambda}}\,\tanh\left(
\frac{m}{2}\,x\right)\right]\qquad(\text{mod}\; 4\pi)\;,\\
\label{smallkink} \varphi_{S}(x)&=&\frac{3\pi}{\beta} \,+\,
\frac{4}{\beta}\arctan\left[\sqrt{\frac{ 4\mu - \lambda}{4\mu +
\lambda}}\,\tanh\left(\frac{m}{2}
\,x\right)\right]\qquad(\text{mod}\; 4\pi)\;.
\end{eqnarray}
With the notation previously introduced, the vacuum structure of
the corresponding quantum field theory consists of two sets of
inequivalent minima, denoted by $\mid 0 \,\rangle $ and $\mid 1
\,\rangle$, identified modulo $2$, i.e. $\mid a + 2 n \, \rangle
\equiv \mid a \rangle$. The spontaneous breaking of the symmetry
$T:\,\varphi\,\to\, 2\pi-\varphi$ selects one of these minima as
the vacuum. If we choose to quantize the theory around $\mid 0
\,\rangle $, the admitted quantum kink states are $\mid L
\,\rangle = \mid K_{0,1}\,\rangle$ and $\mid \overline S \,\rangle
= \mid K_{0,-1}\, \rangle $, with the corresponding antikink
states $\mid \overline L \,\rangle =\mid K_{1,0} \,\rangle$ and
$\mid S \,\rangle = \mid K_{-1,0}\,\rangle$, and topological
charges
\begin{equation*}
\begin{array}{c}
Q_{L} = - Q_{\overline L} = 1+\frac{\beta\phi_{0}}{\pi}\;,\\
Q_{S} = -Q_{\overline S} = 1-\frac{\beta\phi_{0}}{\pi}\;. \\
\end{array}
\end{equation*}
Multi--kink states of this theory satisfy the selection rule
coming from the continuity of vacuum indices and are generically
given by
$$
\mid K_{\alpha_1 \alpha_2}(\theta_1) \, K_{\alpha_2
\alpha_3}(\theta_2) \, \cdots K_{\alpha_{n-2}
\alpha_{n-1}}(\theta_{n-2}) \, K_{\alpha_{n-1}
\alpha_{n}}(\theta_{n-1}) \,\rangle
$$
The leading contributions
to the masses of the large and small kink are given by their
classical energies, which can be easily computed
\begin{equation}\label{largesmallkinkmass}
M_{L,S}\,=\,\frac{8\,m}{\beta^{2}}
\left\{1\pm\frac{\lambda}{\sqrt{16\,\mu^{2}-\lambda^{2}}}
\left(\frac{\pi}{2}\pm\arcsin\frac{\lambda}{4\mu}\right)\right\}
\,\,\,.
\end{equation}
The expansion of this formula for small $\lambda$ is given by
\begin{equation}\label{largesmallkinkfirst}
M_{L,S}\;{\mathrel{\mathop{\kern0pt\longrightarrow}
\limits_{\lambda\to 0 }}}\;
\frac{8\sqrt{\mu}}{\beta^{2}}\pm\frac{\lambda}{\beta^{2}}\,
\frac{\pi}{\sqrt{\mu}}+O(\lambda^{2})\;,
\end{equation}
and the first order correction in $\lambda$ coincides with the
result of FFPT in the semiclassical limit (see Appendix
\ref{chapnonint}.\ref{kmasscorr}).

Since two different types of kink $|\,L\rangle$ and $|\,S\rangle$
are present in this theory, one must be careful in applying
eq.\,(\ref{ffinf}) to recover the form factors of each kink
separately. In fact, one could expect that both types of kink
contribute to the expansion over intermediate states
(\ref{intermediate}) used in \cite{GJ} to derive the result. For
instance, starting from the vacuum $|\, 0\,\rangle$ located at
$\phi_{\text{min}} = -\phi_{0}$ there might be the intermediate
matrix elements $_0\langle \bar{S}\, |\, {\cal O} \,|\, L
\rangle_0$ and $_0\langle L \,|\, {\cal O} \,|\, \bar{S}
\rangle_0$. However, if ${\cal O}$ is a non--charged local
operator, it easy to see that these off-diagonal elements have to
vanish for the different topological charges of $|\,L\rangle$ and
$|\,S\rangle$. Hence, the expansion over intermediate states
diagonalizes and one recovers again eq.\,(\ref{ffinf}).

Therefore, from the dynamical poles of the form factor of
$\varphi$ on the large and small kink-antikink states, reported in
Appendix \ref{chapnonint}.\ref{secFF}, we can extract the
semiclassical masses of two sets of bound states:
\begin{equation}
\label{largebound} m_{(L)}^{(n)}\,=\, 2 M_{L}
\sin\left(n_{L}\,\frac{m}{2M_{L}}\right) \,\,\,\,\,\,\, ,
\,\,\,\,\,\,\, 0 < n_{L} < \pi\frac{M_{L}}{m}\;,
\end{equation}
\begin{equation}\label{smallbound}
m_{(S)}^{(n)}\, =\, 2 M_{S}
\sin\left(n_{S}\,\frac{m}{2M_{S}}\right) \,\,\,\,\,\,\, ,
\,\,\,\,\,\,\, 0 < n_{S} < \pi\frac{M_{S}}{m}\;.
\end{equation}
Expanding for small $\lambda$, we can see that these states
represent the perturbation of the two sets of breathers in the
original two--folded Sine-Gordon model:
\begin{equation}
\label{linearlambda}
m_{(L,S)}^{(n)}\;{\mathrel{\mathop{\kern0pt\longrightarrow}
\limits_{\lambda\to 0 }}}\;\frac{16\sqrt{\mu}}{\beta^{2}}
\sin\left(n\,\frac{\beta^{2}}{16}\right)\pm
2\pi\frac{\lambda}{\sqrt{\mu}}
\left[\frac{1}{\beta^{2}}\sin\left(n\, \frac{\beta^{2}}{16}\right)
- \frac{n}{16}\cos\left(n\, \frac{\beta^{2}}{16}\right) \right] +
O(\lambda^{2})
\end{equation}
A discussion of these results, in comparison with previous studies
of this model \cite{hungDSG}, is reported in Appendix
\ref{chapnonint}.\ref{breathers}.

Concerning the stability of the above spectrum, for $\lambda < 4
\mu$ the only stable bound states are the ones with
$m_{(L,S)}^{(n)} < 2 m_{(S)}^{(1)}$; for $\lambda$ close enough to
$4\mu$, however, the small kink creates no bound states, hence the
stability condition\footnote{Specifically, this stability
condition holds for $\lambda>\lambda^*$, where $\lambda^*$ is
defined by $$\frac{\lambda^*}{\sqrt{16
\mu^2-\lambda^{*\,2}}}\left(\frac{\pi}{2}-\arcsin\frac{\lambda^*}{4\mu}\right)=1-\frac{\beta^2}{8\pi}\;.$$}
becomes $m_{(L)}^{(n)} < 2 m_{(L)}^{(1)}$.

Resonances in the classical scattering of the small kinks and
antikinks (\ref{smallkink}) have also been numerically observed in
\cite{campbellDSG}. Our analysis is in agreement with these
results, and it adds in this case another possibility. In fact,
these resonances can be related both to the bound states of small
excited kink--antikink with masses $m_{S^{*}}^{(n)}$ in the range
\begin{equation*}
2M_{S} < m_{S^{*}}^{(n)} < 2M_{S} + \tilde{\omega}_{1}\;,
\end{equation*}
and to the large kink-antikink bound states with masses in the
range
\begin{equation*}
2M_{S} < m_{L}^{(n)} < 2M_{L}\;,
\end{equation*}
where $ m_{L}^{(n)}$ are given by (\ref{largebound}).

In the limit $\lambda\to 4\mu$, $\phi_{0}$ tends to
$\frac{\pi}{\beta}$, the two minima at $\frac{2\pi}{\beta} +
\phi_{0}$ and $\frac{4\pi}{\beta} - \phi_{0}$ coincide and the
small kink disappears, becoming a constant solution with zero
classical energy. All the large kink bound states masses collapse
to zero, and in this limit all dynamical poles of the large kink
form factor disappear. This is nothing else but the semiclassical
manifestation of the occurrence of the phase transition present in
the DSG model (see \cite{dm}).

In the second coupling constant region, parameterized by $\lambda
> 4 \mu$, there is only one minimum at fixed position
$-\frac{\pi}{\beta}\;(\text{mod}\, \frac{4\pi}{\beta})$, with
curvature
\begin{equation}\label{curvaturedeltapi2sec}
m^{2}\,=\,\frac{\lambda}{4}-\mu\;.
\end{equation}
There is now only one type of kink, given by
\begin{equation}\label{commonkink}
\varphi_{K}(x)\,=\,\frac{\pi}{\beta}+
\frac{4}{\beta}\arctan\left[\sqrt{\frac{\lambda}{\lambda-4\mu}}
\,\sinh\left(m\,x\right)\right] \,\,\,.
\end{equation}
Its classical mass, expanded for small $\mu$, is again in
agreement with FFPT (see Appendix
\ref{chapnonint}.\ref{kmasscorr}):
\begin{eqnarray}
&&M_{K}\,=\,\frac{16\,m}{\beta^{2}}\,\left\{1 +
\frac{\lambda}{4\sqrt{\mu(\lambda-4\mu)}}\left(\frac{\pi}{2} -
\arcsin\frac{\lambda - 8\mu}{\lambda}\right)\right\}\;\to\nonumber\\
&&\quad\;{\mathrel{\mathop{\kern0pt\longrightarrow}
\limits_{\lambda\to 0 }}}\;\;
\frac{8\sqrt{\lambda/4}}{(\beta/2)^{2}} -
\frac{\mu}{\beta^{2}}\,\frac{32}{3\sqrt{\lambda}}+O(\mu^{2})\;.\label{commonkinkmass}
\end{eqnarray}
The bound states of this kink (see Appendix
\ref{chapnonint}.\ref{secFF} for the explicit form factors) have
masses
\begin{equation}
m_{(K)}^{(n)}\,=\, 2M_{K}\sin\left(n\,\frac{m}{2M_{K}}\right)
\,\,\,\,\,\,\, , \,\,\,\,\,\,\, 0 < n < \pi\frac{M_{K}}{m}\;.
\end{equation}
For small $\mu$, these states are nothing else but the perturbed
breathers of the Sine-Gordon model with coupling $\beta/2$:
$$
m_{(K)}^{(n)}\;{\mathrel{\mathop{\kern0pt\longrightarrow}
\limits_{\mu\to 0
}}}\;\frac{64}{\beta^{2}}\sqrt{\frac{\lambda}{4}}
\,\sin\left(n\,\frac{\beta^{2}}{64}\right)-\frac{2}{3}\,
\frac{\mu}{\sqrt{\lambda}}
\left[\frac{32}{\beta^{2}}\sin\left(n\,\frac{\beta^{2}}{64}\right)
+ n\,\cos\left(n\, \frac{\beta^{2}}{64}\right) \right] +
O(\mu^{2})
$$

In closing the discussion of the $\delta = \pi/2$ case, it is
interesting to mention another model which presents a similar
phase transition phenomenon, although in a reverse order. This is
the Double Sinh--Gordon Model (DShG), discussed in Appendix
\ref{chapnonint}.\ref{secDShG}. The similarity is due to the fact
that also in this case a topological excitation of the theory
becomes massless at the phase transition point, but the phenomenon
is reversed, because in DSG the small kink disappears when
$\lambda$ reaches the critical value, while in DShG a topological
excitation appears at some value of the perturbing coupling.

\subsection{False vacuum decay}\label{secfalse}
\begin{figure}[h]
\footnotesize\psfrag{phi}{$\varphi$}\psfrag{U(phi)}
{$U(\varphi)$}\psfrag{phi+}{$\varphi_{+}$}
\psfrag{phi-}{$\varphi_{-}$} \psfrag{phi1}{$\varphi_{1}$}
\hspace{4cm}\psfig{figure=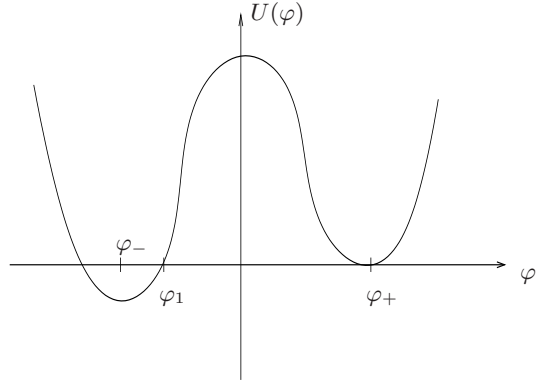,height=5cm,width=7cm}
\normalsize\caption{Generic potential for a theory with a false
vacuum}\label{figfalsevacpot}
\end{figure}

The semiclassical study of false vacuum decay in quantum field
theory has been performed by Callan and Coleman \cite{coleman}, in
close analogy with the work of Langer \cite{langer}. The
phenomenon occurs when the field theoretical potential
$U(\varphi)$ displays a relative minimum at $\varphi_{+}$: this
classical point corresponds to the false vacuum in the quantum
theory, which decays through tunnelling effects into the true
vacuum, associated with the absolute minimum $\varphi_{-}$ (see
Fig. \ref{figfalsevacpot}).

The main result of \cite{coleman} is the following expression for
the decay width per unit time and unit volume:
\begin{equation}
\frac{\Gamma}{V}\,=\,\left(\frac{B}{2\pi\hbar}\right)\,e^{-B/\hbar}
\left|\frac{\det'[-\partial^{2}+U''(\varphi)]}{\det[-\partial^{2}
+ U''(\varphi_{+})]}\right|^{-1/2}\, \left[1+O(\hbar)\right]\;,
\end{equation}
specialized here to the case of two--dimensional space--time.
Omitting any discussion of the determinant, about which we refer
to the original papers \cite{coleman}, we will present here an
explicit analysis of the coefficient $B$.

It has been shown that $B$ coincides with the Euclidean action of
the so--called \lq\lq bounce" background $\varphi_{B}$:
\begin{equation}
\label{Bdecay} B \,=\,
S_{E}\,=\,2\pi\int\limits_{0}^{\infty}d\rho\,
\rho\left[\frac{1}{2}\left(\frac{d\varphi_{B}}{d\rho}\right)^{2} +
U(\varphi_{B})\right]\;.
\end{equation}
This classical solution is the field--theoretical generalization
of the path of least resistance in quantum mechanical tunnelling;
it only depends on the Euclidean radius
$\rho\,=\,\sqrt{\tau^{2}+x^{2}}$ and satisfies the equation
\begin{equation}
\label{bounceeq} \frac{d^{2}\varphi_{B}}{d\rho^{2}} +
\frac{1}{\rho}\,\frac{d\varphi_{B}}{d\rho}\,=\,U'[\varphi_{B}]\;,
\end{equation}
with boundary conditions
\begin{equation*}
\lim\limits_{\rho\to\infty}
\varphi_{B}(\rho)\,=\,\varphi_{+}\;,\qquad
\frac{d\varphi_{B}}{d\rho}(0)=0\;.
\end{equation*}

Although in general one does not know explicitly the bounce
solution, it is possible to set up some approximation to extract a
closed expression for the coefficient $B$. The so-called \lq\lq
thin wall" approximation consists in viewing the potential
$U(\varphi)$ as a perturbation of another potential
$U_{+}(\varphi)$, which displays degenerate vacua at
$\varphi_{\pm}$ and a kink $\varphi_{K}(x)$ interpolating between
them. The small parameter for the approximation is the energy
difference $\varepsilon=U(\varphi_{+})-U(\varphi_{-})$.

In this framework, one can qualitatively guess that the bounce has
a value $\varphi(0)$ very close to $\varphi_{-}$, then it remains
in this position until some vary large $\rho=R$ and finally it
moves quickly towards the final value $\varphi_{+}$. For $\rho$
near $R$, the first--derivative term in eq. (\ref{bounceeq}) can
be neglected; if in addition one also approximates $U$ with
$U_{+}$, then one can express the unknown bounce solution as
\cite{coleman}
\begin{equation}\label{colemanbounce}
\varphi_{B}(\rho)\,=\,
\begin{cases}
\;\varphi_{-}&\qquad \rho\ll R\\
\;\varphi_{K}(\rho-R)&\qquad \rho\approx R\\
\;\varphi_{+}&\qquad \rho\gg R\;.
\end{cases}
\end{equation}
Since the bounce has to represent the path of least resistance,
the parameter $R$, free up to this point, can be fixed by
minimizing the action
\begin{equation*}
S_{E}\,=\,-\pi R^{2}\epsilon+2\pi R \,M_{K}\;,
\end{equation*}
which is given by the sum of a volume term and a surface term.
Hence, the condition $\frac{dS_{E}}{dR}=0$ is realized by the
balance of these two different terms in competition, and it
finally gives
\begin{equation}\label{colemanpred}
R\,=\,\frac{M_{K}}{\varepsilon}\quad\Longrightarrow\quad
B\,=\,\pi\,\frac{M_{K}^{2}}{\varepsilon}\;.
\end{equation}

In the DSG model, however, we know explicitly the bounce
background in the thin wall regime (here we have
$\varepsilon=\frac{2\lambda}{\beta^{2}}$), without any
approximation on the potential. This is given by the solution
(\ref{bounce}) with $x$ replaced by $\rho$, that can be directly
used to estimate the decay width. Unfortunately the integral in
(\ref{Bdecay}) does not admit a simple expression to be expanded
for small $\lambda$, but it is clear from eq.\,(\ref{bouncesumSG})
and Fig. \ref{figbounce} that the leading contribution is given by
\begin{equation}
S_{E}\,\simeq\,2\pi R \int\limits_{R-\Delta r}^{R+\Delta
r}dx\left[\frac{d\varphi_{SG}}{dx}(R-x)\right]^{2}\,\simeq
\,\frac{8\pi}{\beta^{2}}\,\log\left(\frac{16\mu}{\lambda}\right)\;,
\end{equation}
with $R$ given by (\ref{fixedR}). This behavior in $\lambda$ does
not agree with the general prediction (\ref{colemanpred}). The
reason can be traced out in the fact that eq.\,(\ref{bouncesumSG})
explicitly realizes the relation between the bounce and the kink
of the unperturbed theory, but in a more sophisticated way than
(\ref{colemanbounce}). In fact, the mass parameter $m_{f}$ of the
SG kink $\varphi_{SG}$ is dressed to be the one of the DSG theory,
and the parameter $R$ is not free, since (\ref{bounce}) is already
the result of a minimization process, being a solution of the
Euler--Lagrange equations. The thin wall approximation can be
still consistently used because $R$ is very big for small
$\lambda$, while the crucial difference is that the volume term is
now missing from the action, since the value
$\varphi_{B}(0)=\varphi_{1}$ is the so--called classical turning
point (see Fig. \ref{figfalsevacpot}), degenerate with the false
vacuum. It is worth noting that the path of least resistance in
quantum mechanics precisely interpolates between the false vacuum
and the turning point.

Up to the determinant factor, our result for the leading term in
the decay width is then
\begin{equation}
\frac{\Gamma}{V}\,\simeq \,\frac{4}{\beta^{2}}\,
\left(\frac{\lambda}{16\mu}\right)^{8\pi/\beta^{2}}\,
\log\left(\frac{16\mu}{\lambda}\right) \;.
\end{equation}
It will be interesting to investigate whether the above mentioned
difference with the prediction (\ref{colemanpred}) is a particular
feature of the DSG model or it appears for a generic potential if
one improves the approximate description of the bounce along the
lines discussed here.

\section{Summary} \setcounter{equation}{0}

In this Chapter we have shown how the semiclassical method is an
efficient tool for studying non--integrable QFT. In fact, applied
to the broken $\phi^4$ theory, it has provided new simple results
about mass spectrum, three--particle couplings and resonances.

In cases where also the Form Factor Perturbation Theory can be
applied, the semiclassical method may be complementary to this
technique or it may provide results comparable with it. We have
applied both methods for analysing the mass spectrum of the
non--integrable QFT given by the Double Sine--Gordon model, for
few qualitatively different regions of its coupling--constants
space. This model appears to be an ideal theoretical playground
for understanding some of the relevant features of non--integrable
models. By moving its coupling constants, in fact, it shows
different types of kink excitations and confinement phenomena, a
rich spectrum of meson particles, resonance states, false vacuum
decay and the occurrence of a phase transition. In light of the
many applications it finds in condensed matter systems, it would
be interesting to investigate further its properties.

Finally, it is worth mentioning that the semiclassical method can
be easily extended to the analysis of several power--like
potentials, which have interesting physical interpretations along
the lines of the Landau--Ginzburg theory discussed in
Sect.\,\ref{secLG}.

\sezioneapp{Kink mass corrections in the FFPT}\label{kmasscorr}
\setcounter{equation}{0} In this Appendix we compute by means of
the FFPT the corrections to the kink masses in the semiclassical
limit, which is relevant for a comparison with our results.

For small $\lambda$, we have to consider the DSG model as a
perturbation of the two--folded Sine-Gordon \cite{hungDSG}. In the
$\delta = \pi/2$ case, the perturbing operator is $\Psi =
\sin\frac{\beta}{2}\phi$. Its form factors between the vacuum and
the two possible kink-antikink asymptotic states are obtained at
the semiclassical level by performing the Fourier transform of
$\;\sin\left[\frac{\beta}{2}K_{k,k+1}^{cl}(x)\right]$
\cite{finvolff}, with $K_{k,k+1}^{cl}(x)$ given by
eq.\,(\ref{twofoldedkinks}). Hence we obtain
\begin{equation}
F_{K_{k,k+1},\bar{K}_{k,k+1}}^{\Psi}(\theta)\, =\,
\frac{8\pi}{\beta^{2}}\,(-1)^{k}\,\frac{1}
{\cosh\frac{4\pi}{\beta^{2}}(\theta-i\pi)}\;.
\end{equation}
The first order correction in $\lambda$ to the kink masses is then
\begin{equation}
\delta M_{K_{k,k+1}}\,=\,
\frac{\lambda}{\beta^{2}}\,\frac{1}{M_{K}}
\,F_{K_{k,k+1},\bar{K}_{k,k+1}}^{\Psi}(i\pi)
\,=\,(-1)^{k}\frac{\lambda}{\beta^{2}}\,\frac{\pi}{\sqrt{\mu}}\;,
\end{equation}
in agreement with the correction to the classical masses
(\ref{largesmallkinkfirst}), since $K_{0,1}$ is associated with
the large kink, and $K_{1,2}$ with the small one.

In the $\delta = 0$ case, instead, we can explicitly see how the
solitons disappear from the spectrum as soon as $\lambda$ is
switched on. The form factor of the operator $\Psi =
\cos\frac{\beta}{2}\phi$ has, in fact, a divergence at
$\theta=i\pi$
\begin{equation}
F_{K_{k,k+1},\bar{K}_{k,k+1}}^{\Psi}(\theta)\, =\,-i
\frac{8\pi}{\beta^{2}}\,(-1)^{k}\,\frac{1}
{\sinh\frac{4\pi}{\beta^{2}}(i\pi-\theta)}\;.
\end{equation}

\vspace{0.5cm}

The other interesting regime to explore is the small $\mu$ limit.
In the case $\delta=0$, this can be seen as the perturbation of
the SG model at coupling $\tilde{\beta} = \beta/2$ by means of the
operator $\Psi = \cos\,2\tilde{\beta}\varphi$. The semiclassical
form factor is
\begin{equation}
F_{K,\bar{K}}^{\Psi}(\theta)\, =\,
\frac{16}{3}\,\frac{32}{\beta^{2}}\,\frac{i\pi y} {\sin\,i\pi
y}(1-2\,y^{2})\;,
\end{equation}
where we have defined $y =\frac{16}{\beta^{2}}(i\pi-\theta)$. The
corresponding mass correction is given by
\begin{equation}
\delta M_{K}\,=\,\frac{\mu}{\beta^{2}}\,\frac{1}{M_{K}}\,
F_{K,\bar{K}}^{\Psi}(i\pi)\,=\,\frac{\mu}{\beta^{2}}\,
\frac{16}{3}\,\frac{1}{\sqrt{\lambda/4}}\;,
\end{equation}
in agreement with (\ref{doublekinkfirst}).

The case $\delta = \frac{\pi}{2}$ can be described by shifting the
original SG field as $\varphi\, \to\,\varphi + \frac{\pi}{\beta}$.
In this way the perturbing operator becomes $-\Psi$ and we finally
obtain the same mass correction but with opposite sign, as in
(\ref{commonkinkmass}).

\sezioneapp{Semiclassical form factors}\label{secFF}
\setcounter{equation}{0} In this Appendix we explicitly present
the expressions of the two--particle form factors, on the
asymptotic states given by the different kinks appearing in the
DSG theory, of the operators $\varphi(x)$ and $\varepsilon(x)$,
the last one defined by
$$
\varepsilon(x) \,\equiv \, \frac{1}{2}
\left(\frac{d\varphi}{dx}\right)^{2}+V[\varphi(x)] \,\,\,.
$$
These matrix elements are obtained by performing the Fourier
transforms of the corresponding classical backgrounds, as
indicated in (\ref{ffinf}) and (\ref{f2}). We use the notation:
$$
F^{\Psi}_{K\bar{K}}(\theta)\,=\,\langle\,
0\,|\,\Psi(0)\,|\,K(\theta_1)\,\bar{K}(\theta_2)\,\rangle\;,
$$
with $\theta = \theta_1 - \theta_2$. \vspace{0.5cm}

For the kink (\ref{doublekink}) in the $\delta=0$ case we have
\begin{equation}
F^{\varphi}_{K\bar{K}}(\theta)\,=\,\frac{4\pi^{2}}{\beta}\,M_{K}\,
\delta\left[M_{K}(i\pi-\theta)\right] + i\,\frac{4\pi}{\beta}\,
\frac{1}{i\pi-\theta}\;\frac{\cos\left[\alpha
\,\frac{M_{K}}{m}\,(i\pi-\theta)\right]}{\cosh\left[\frac{\pi}{2}
\,\frac{M_{K}}{m}\,(i\pi-\theta)\right]}\;,
\end{equation}
where
$$
\alpha\,=\,\text{arccosh}\sqrt{\frac{\lambda+4\mu}{\lambda}}\;,
$$
while $m$ and $M_{K}$ are given by (\ref{curvaturedeltazero}) and
(\ref{doublekinkmass}), respectively, and
\begin{equation*}
F^{\varepsilon}_{K\bar{K}}(\theta)\,=\,-\,\frac{128\pi}{\beta^{2}}\,
\frac{m^{3}
M_{K}}{\lambda}\;\left\{\frac{1}{\sinh\left[\pi\,\frac{M_{K}}{2m}\,
(i\pi-\theta)\right]}\;\frac{d}{dc}
\left[\frac{\sinh\left[(\text{arccosh}\, c)
\frac{M_{K}}{2m}(i\pi-\theta)\right]}{\sqrt{c^{2}-1}}\right]+\right.
\end{equation*}
\begin{equation}
\left.
-\frac{2\sinh\pi}{\cosh\left[\pi\,\frac{M_{K}}{m}\,(i\pi-\theta)\right]-1}
\;\frac{d}{dc} \left[\frac{c\,\sinh\left[(\text{arccosh} \,c)
\frac{M_{K}}{2m}(i\pi-\theta)\right]}{\sqrt{c^{2}-1}}\right]\right\}\;,
\end{equation}
where $c=1+\frac{8\mu}{\lambda}$.

\vspace{0.5cm}

For the large kink (\ref{largekink}) in the $\delta=\frac{\pi}{2}$
case (with $\lambda<4\mu$) we have
\begin{equation}
F^{\varphi}_{L\bar{L}}(\theta)\,=\,\frac{2\pi^{2}}{\beta}\,M_{L}\,
\delta\left[M_{L}(i\pi-\theta)\right] + i\,\frac{4\pi}{\beta}\,
\frac{1}{i\pi-\theta}\;\frac{\sinh\left[\alpha
\,\frac{M_{L}}{m}\,(i\pi-\theta)\right]}{\sinh\left[\pi\,
\frac{M_{L}}{m}\,(i\pi-\theta)\right]}\;,
\end{equation}
where
$$
\alpha\,=\,2\arctan\sqrt{\frac{4\mu+\lambda}{4\mu-\lambda}}\;,
$$
while $m$ and $M_{L}$ are given by (\ref{curvaturedeltapi2}) and
(\ref{largesmallkinkmass}), respectively, and
\begin{equation}
F^{\varepsilon}_{L\bar{L}}(\theta)\,=\,\frac{8\pi}{\beta^{2}}\,
\frac{m^{3}
M_{L}}{\mu}\;\frac{1}{\sinh\left[\pi\,\frac{M_{L}}{m}\,(i\pi-\theta)\right]}\;
\frac{d}{dc} \left\{\frac{\sinh\left[(\arccos\, c)
\frac{M_{L}}{m}(i\pi-\theta)\right]}{\sqrt{1-c^{2}}}\right\}\;,
\end{equation}
where $c=-\frac{\lambda}{4\mu}$.

\vspace{0.5cm}

For the small kink (\ref{smallkink}) in the $\delta=\frac{\pi}{2}$
case (with $\lambda<4\mu$) we have
\begin{equation}
F^{\varphi}_{S\bar{S}}(\theta)\,=\,\frac{6\pi^{2}}{\beta}\,M_{S}\,
\delta\left[M_{S}(i\pi-\theta)\right] + i\,\frac{4\pi}{\beta}\,
\frac{1}{i\pi-\theta}\;\frac{\sinh\left[\alpha
\,\frac{M_{S}}{m}\,(i\pi-\theta)\right]}{\sinh\left[\pi\,
\frac{M_{S}}{m}\,(i\pi-\theta)\right]}\;,
\end{equation}
where
$$
\alpha\,=\,2\arctan\sqrt{\frac{4\mu-\lambda}{4\mu+\lambda}}\;,
$$
while $m$ and $M_{S}$ are given by (\ref{curvaturedeltapi2}) and
(\ref{largesmallkinkmass}), respectively, and
\begin{equation}
F^{\varepsilon}_{S\bar{S}}(\theta)\,=\,\frac{8\pi}{\beta^{2}}\,
\frac{m^{3}
M_{S}}{\mu}\;\frac{1}{\sinh\left[\pi\,\frac{M_{S}}{m}\,(i\pi-\theta)\right]}\;
\frac{d}{dc} \left\{\frac{\sinh\left[(\arccos \, c)
\frac{M_{S}}{m}(i\pi-\theta)\right]}{\sqrt{1-c^{2}}}\right\}\;,
\end{equation}
where $c=\frac{\lambda}{4\mu}$.

\vspace{0.5cm}

Finally, for the kink (\ref{commonkink}) in the $\delta =
\frac{\pi}{2}$ case (with $\lambda>4\mu$) we have
\begin{equation}
F^{\varphi}_{K\bar{K}}(\theta)\,=\,\frac{2\pi^{2}}{\beta}\,M_{K}\,
\delta\left[M_{K}(i\pi-\theta)\right] + i\,\frac{4\pi}{\beta}\,
\frac{1}{i\pi-\theta}\;\frac{\cos\left[\alpha
\,\frac{M_{K}}{m}\,(i\pi-\theta)\right]}
{\cosh\left[\frac{\pi}{2}\,\frac{M_{K}}{m}\,(i\pi-\theta)\right]}\;,
\end{equation}
where
$$
\alpha\,=\,\text{arccosh}\sqrt{\frac{\lambda-4\mu}{\lambda}}\;,
$$
while $m$ and $M_{L}$ are given by (\ref{curvaturedeltapi2sec})
and (\ref{commonkinkmass}), respectively, and
\begin{equation*}
F^{\varepsilon}_{K\bar{K}}(\theta)\,=\,-\,\frac{128\pi}{\beta^{2}}\,
\frac{m^{3}
M_{K}}{\lambda}\;\left\{\frac{1}{\sinh\left[\pi\,\frac{M_{K}}{2m}\,
(i\pi-\theta)\right]}\;\frac{d}{dc}
\left[\frac{\sinh\left[(\arccos \, c)
\frac{M_{K}}{2m}(i\pi-\theta)\right]}{\sqrt{1-c^{2}}}\right]+\right.
\end{equation*}
\begin{equation}
\left.
-\frac{2\sinh\pi}{\cosh\left[\pi\,\frac{M_{K}}{m}\,(i\pi-\theta)\right]-1}\;
\frac{d}{dc} \left[\frac{c\,\sinh\left[(\arccos\, c)
\frac{M_{K}}{2m}(i\pi-\theta)\right]}{\sqrt{1-c^{2}}}\right]\right\}\;,
\end{equation}
where $c=1-\frac{8\mu}{\lambda}$.

\sezioneapp{Neutral states in the $\delta=\frac{\pi}{2}$
case}\label{breathers} \setcounter{equation}{0} The semiclassical
results reported in the text, i.e. eqs.\,(\ref{largebound}),
(\ref{smallbound}) and (\ref{linearlambda}), pose an interesting
question about the nature of neutral states in the DSG model at
$\delta=\frac{\pi}{2}$. It should be noticed, in fact, that the
first order correction in $\lambda$ obtained by the Semiclassical
Method does not match with the results reported in \cite{hungDSG}
where, by using the FFPT and an extrapolation of numerical data,
the authors concluded that this correction was instead identically
zero\footnote{It is worth stressing that the linear correction
(\ref{linearlambda}) in $\lambda$ is very small even for finite
values of $\beta$ (it is easy to check, indeed, that the first
term of its expansion is $\frac{\pi}{24}
\left(\frac{\beta^2}{16}\right)^2$) and somehow compatible with
the numerical data given in \cite{hungDSG}.}. It is worth
discussing this problem in more detail.

In the standard Sine--Gordon model, the breathers
$|\,b_{n}\rangle$, with $n$ odd (or even), are defined as the
bound states of odd (or even) combinations of
$|\,K\,\bar{K}\rangle$ and $|\,\bar{K}\,K\rangle$, where $K$
represents the soliton and $\bar{K}$ the antisoliton. The
combinations $|\,K\,\bar{K}\pm\bar{K}\,K\rangle$ are eigenstates
of the parity operator $P:\; \phi\to - \phi$, which commutes with
the hamiltonian and acts on the soliton transforming it into the
antisoliton. The above mentioned identification of the bound
states relies on a very peculiar feature of the Sine--Gordon
$S$--matrix in the soliton sector \cite{zams}, whose elements are
defined as
\begin{eqnarray}
K(\theta_{1})\,\bar{K}(\theta_{2})&=&
S_{T}(\theta_{12})\,\bar{K}(\theta_{2})\,K(\theta_{1})+
S_{R}(\theta_{12})\,K(\theta_{2})\,\bar{K}(\theta_{1})\;,\nonumber\\
K(\theta_{1})\,K(\theta_{2})&=&
S(\theta_{12})\,K(\theta_{2})\,K(\theta_{1})\;,\\
\bar{K}(\theta_{1})\,\bar{K}(\theta_{2})&=&
S(\theta_{12})\,\bar{K}(\theta_{2})\,\bar{K}(\theta_{1})\;.\nonumber
\end{eqnarray}
In fact, both the transmission and the reflection amplitudes
$S_{T}(\theta)$ and $S_{R}(\theta)$ display poles at
$\theta_{n}^{*} = i(\pi-n\xi)$, with residua which are equal or
opposite in sign depending whether $n$ is odd or even. Hence, the
diagonal elements
\begin{eqnarray}
S_{-}(\theta)&=&
\frac{1}{2}\left[S_{T}(\theta)-S_{R}(\theta)\right]\;,\\
S_{+}(\theta)&=&
\frac{1}{2}\left[S_{T}(\theta)+S_{R}(\theta)\right]
\end{eqnarray}
have only the poles with odd or even $n$, respectively, and for
each $n$ there is only one bound state with definite parity.

However, this is a special feature of the Sine--Gordon model which
finds no counterpart, for instance, in other problems with a
similar structure. As an explicit example, one can consider the
$(\text{RSOS})_{3}$ scattering theory, which displays a 3-fold
degenerate vacuum and two types of kink and antikink with the same
mass. The central vacuum is surrounded by two other minima, as in
the Sine--Gordon case, and this gives the possibility to define
both a kink-antikink state and an antikink-kink state around it.
The minimal scattering matrix, given in \cite{rsos}, can be
dressed with a CDD factor to generate bound states. It is easy to
check that the common poles in the transmission and reflection
amplitudes have in this case different residua, giving rise to two
distinct bound states, degenerate in mass, over the central
vacuum.

Hence, if we call $|\,b_{n}^{(0)}\rangle$ the bound states of
kink-antikink and $|\,b_{n}^{(1)}\rangle$ the bound states of
antikink-kink, in general we have to consider them as two distinct
excitations, and if they have the same mass we can build two other
states from their linear combinations
\begin{equation}\label{basischange}
|\,b_{n}^{(\pm)}\rangle\,=\,\frac{|\,b_{n}^{(0)}\rangle\,\pm\,
|\,b_{n}^{(1)}\rangle}{\sqrt{2}}\;.
\end{equation}
The peculiarity of the Sine--Gordon model is the removal of this
double multiplicity due to the fact that the states
$|\,b_{2n+1}^{(+)}\rangle$ and $|\,b_{2n}^{(-)}\rangle$ decouple
from the theory. This feature is shared also by the two-folded
version of the model, since the kink scattering amplitudes have
the same analytical form as in SG \cite{foldedSG}.

In the two--folded SG there are two different kink states
$|\,K_{-1,0} \rangle$ and $|\,K_{0,1}\rangle$ (see Sect.
\ref{secDSG} and ref. \cite{foldedSG} for the notation), and the
parity $P$, which is still an exact symmetry of the theory, acts
on them transforming the kink of one type into the antikink of the
other type:
\begin{equation}
P: \;|\, K_{0,1}\rangle\to\,|\,K_{0,-1}\rangle\;,\quad
|\,K_{-1,0}\rangle\to\,|\,K_{1,0}\rangle\;.
\end{equation}
If we quantize the theory around the vacuum $|\,0\rangle$, we can
define $|\,b_{n}^{(0)} \rangle$ as the bound states of
$|\,K_{0,1}\,K_{1,0}\rangle$, and $|\,b_{n}^{(1)}\rangle$ as the
bound states of $|\,K_{0,-1}\,K_{-1,0}\rangle$. These degenerate
states, which transform under $P$ as
\begin{equation}
P: \; |\,b_{n}^{(0)}\rangle\to\,|\,b_{n}^{(1)}\rangle\;,\quad
|\,b_{n}^{(1)}\rangle\to\,|\,b_{n}^{(0)}\rangle\;,
\end{equation}
can be still organized in parity eigenstates
$|\,b_{n}^{(\pm)}\rangle$, and the particular dynamics of the
problem causes the decoupling of half of them from the theory.
Furthermore, it is easy to see that the form factors of an odd
operator between two of these states has to vanish in virtue of
the relation
\begin{eqnarray*}
\langle\,0\,|\,\sin\frac{\beta}{2}\phi\,|\,b_{n}^{(\pm)}b_{n}^{(\pm)}\rangle
& = &
\langle\,0\,|\,P^{-1}P\,\left(\sin\frac{\beta}{2}\phi\right)\,
P^{-1}P|\,b_{n}^{(\pm)}b_{n}^{(\pm)}\rangle \,=\\
& = &-\,
\langle\,0\,|\,\sin\frac{\beta}{2}\phi\,|\,b_{n}^{(\pm)}b_{n}^{(\pm)}\rangle\;,
\end{eqnarray*}
leading to the FFPT result that the breathers receive a zero mass
correction at first order in $\lambda$, as it is claimed in
\cite{hungDSG}.

However, FFPT can be applied by taking into account the nature of
neutral states in the DSG model, where the addition to the
Lagrangian of the term
$\;-\frac{\lambda}{\beta^{2}}\,\sin\frac{\beta}{2}\varphi\;$
spoils the invariance under $P$. The kinks $|\,K_{-1,0} \rangle$
and $|\,K_{0,1} \rangle$ are deformed into the small and large
kinks $|\,S \rangle$ and $|\,L \rangle$, respectively, which are
not anymore degenerate in mass and cannot be superposed in linear
combinations. Hence, the neutral states present in the theory are
$|\,b_{n}^{(L)}\rangle$ and $|\,b_{n}^{(S)}\rangle$, deformations
of  $|\,b_{n}^{(0)}\rangle$ and $|\,b_{n}^{(1)}\rangle$
respectively. In virtue of the general considerations presented
above, one can see that this interpretation does not lead to any
drastic change in the spectrum. In fact, the states
$|\,b_{2n+1}^{(+)}\rangle$ and $|\,b_{2n}^{(-)}\rangle$ have no
reason to decouple in the DSG theory, but they have to carry a
coupling which is a function of $\lambda$ adiabatically going to
zero in the two--folded SG limit.

A proper use of the FFPT on $|\,b_{n}^{(0)}\rangle$ and
$|\,b_{n}^{(1)}\rangle$ reproduces indeed the situation described
by (\ref{linearlambda}), in which the two sets of breathers
receive mass corrections including also odd terms in $\lambda$,
but with opposite signs. This is easily seen by considering the
$P$ transformations in the two--folded SG model:
\begin{eqnarray*}
\langle\,0\,|\,\sin\frac{\beta}{2}\phi\,|\,b_{n}^{(0)}b_{n}^{(0)}\rangle
& = &
\langle\,0\,|\,P^{-1}P\,\left(\sin\frac{\beta}{2}\phi\right)\,
P^{-1}P|\,b_{n}^{(0)}b_{n}^{(0)}\rangle \,=\\
& = &-\,
\langle\,0\,|\,\sin\frac{\beta}{2}\phi\,|\,b_{n}^{(1)}b_{n}^{(1)}\rangle\;,
\end{eqnarray*}
which gives, at first order in $\lambda$,
\begin{equation}
\delta\,m_{(L)}=-\,\delta\,m_{(S)}\;,
\end{equation}
in agreement with our semiclassical result (\ref{linearlambda}).
It is worth noting that also with this interpretation the total
spectrum of the DSG model remains unchanged under the action of
$P$, which corresponds to the transformation $\lambda\to
-\lambda$. In fact, the two types of kinks and breathers are
mapped one into the other. This is consistent with the observation
that $P$, although it is not anymore a symmetry of the perturbed
theory, simply realizes a reflection of the potential, hence the
total spectrum should be invariant under it.

Presently the above symmetry considerations seem to us the correct
criterion to define the neutral states, and find confirmation in
our semiclassical result (\ref{linearlambda}). However, the
available numerical data presented in \cite{hungDSG} pose a
challenge to this interpretation and further studies are needed to
solve this interesting and delicate problem. In fact, although
$\delta\,m_{(L)}$ and $\delta\,m_{(S)}$ are not forced to vanish
by symmetry arguments, there is in principle the possibility that
both of them are identically zero in the complete quantum
computation. This could follow from the use of the exact kink
masses entering eqs.\,(\ref{largebound}) and (\ref{smallbound}),
together with a proper shift of the semiclassical pole in the form
factors, due to higher order contributions. The exact cancellation
of the linear corrections is a very strong requirement, in support
of which we have presently no indication in the theory, but a
careful analysis of this point is neverthless an interesting open
problem.

\sezioneapp{Double Sinh--Gordon model} \label{secDShG}
\setcounter{equation}{0}Among the different qualitative features
taking place in perturbing integrable models, a situation
particularly interesting is the one in which the perturbation is
adiabatic for small values of the parameters but nevertheless a
qualitative change in the spectrum occurs by increasing its
intensity.

This is indeed the situation in the $\delta = \frac{\pi}{2}$ case
of DSG model, where we have two types of kinks for small
$\lambda$, but at $\lambda = 4 \mu$ one of them disappears from
the spectrum. This phenomenon is obviously unaccessible by means
of FFPT, hence the semiclassical method is the best tool to
describe it.

Here we consider another interesting example of this kind,
realized by the Double Sinh-Gordon Model (DShG). In this case the
phenomenon is even more evident, because in the unperturbed
Sinh-Gordon model there are no kinks at all, but just one scalar
particle, while perturbing it, at some critical value of the
coupling a kink and antikink appear, i.e. there is a deconfinement
phase transition of these particles.

\begin{figure}[ht]
\begin{tabular}{p{5cm}p{5cm}p{5cm}}
\psfrag{V}{$V(\mu=1,\,\lambda=0)$}
\psfig{figure=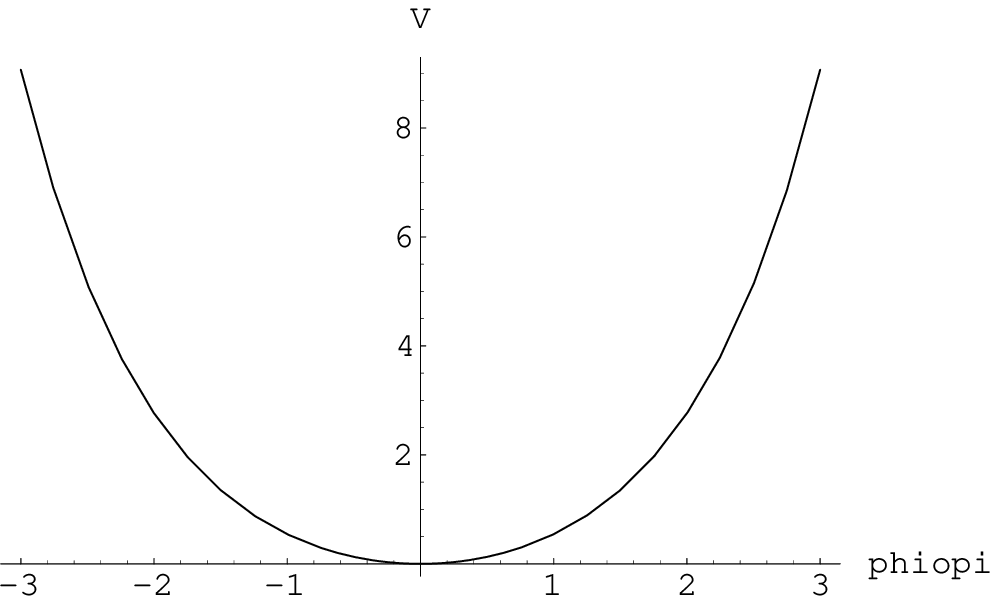,height=4cm,width=4.5cm}&
\psfrag{V}{$V(\mu=1,\,\lambda=4)$}\psfig{figure=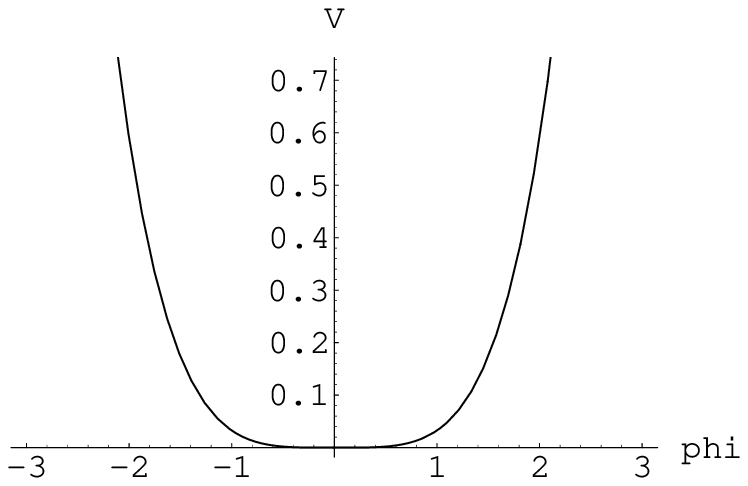,
height=4cm,width=4.5cm}& \psfrag{V}{$V(\mu=1,\,\lambda=5)$}
\psfig{figure=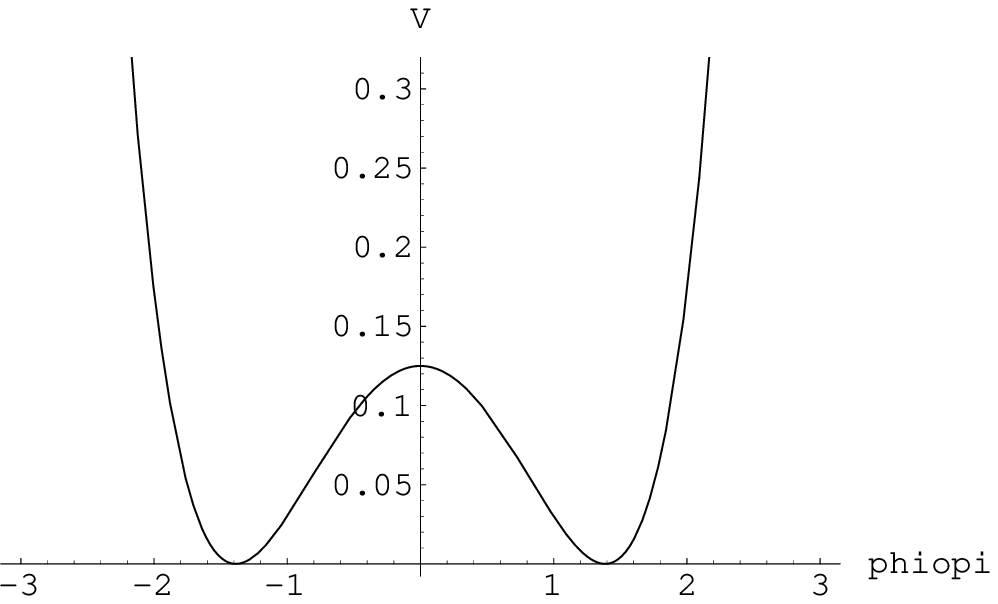,height=4cm,width=4.5cm}
\end{tabular}
\caption{DShG potential}\label{figpotDShG}
\end{figure}

The DShG potential, shown in Fig. \ref{figpotDShG}, is expressed
as
\begin{equation}\label{DShGpot}
V(\varphi) \,= \, \frac{\mu}{\beta^{2}}\, \cosh\beta\,\varphi -
\frac{\lambda}{\beta^{2}}\, \cosh\left(\frac{\beta}{2}\,\varphi
\right)\;.
\end{equation}

\psfrag{phiopi}{$\phi$}

In the regime $\lambda < 4 \mu$ the qualitative features are the
same as in the unperturbed Sinh-Gordon model. At $\lambda = 4
\mu$, however, the single minimum splits in two degenerate minima,
which for $\lambda > 4 \mu$ are located at $\varphi_{\pm} = \pm
\frac{2}{\beta} \text{arccosh}\frac{\lambda}{4\mu}$. A study of
the classical thermodynamical properties of the theory in this
regime has been performed in \cite{DShG} with the transfer
integral method.

The kink interpolating between the two degenerate vacua is
\begin{equation}
\varphi_{K}(x) \, = \, \frac{4}{\beta}\,\text{arctanh}
\left[\sqrt{\frac{\lambda-4\mu}{\lambda+4\mu}}\,
\tanh\left(\frac{m}{2}\,x\right)\right] \; ,
\end{equation}
with
$$
m^{2} \,= \,\frac{\lambda^{2}-16\mu^{2}}{16\mu} \;.
$$
Its classical mass is given by
\begin{equation}
M_{K} \,=\, \frac{8m}{\beta^{2}}\left\{ -1 +
\frac{2\lambda}{\sqrt{\lambda^{2} - 16\mu^{2}}}\,\text{arctanh}
\sqrt{\frac{\lambda - 4\mu}{\lambda + 4\mu}}\right\}\;.
\end{equation}
From the form factor of $\varphi$ on the kink-antikink asymptotic
state, expressed as
\begin{equation}
F_{2}(\theta)\, =\, -i\frac{\pi}{\beta}\,
\frac{1}{i\pi-\theta}\,\frac{\sin\left[\text{arccosh}
\frac{\lambda}{4\mu}\, \frac{M_{K}}{m}\,(i\pi-\theta)\right]}
{\sinh\left[\pi\,\frac{M_{K}}{m}\,(i\pi - \theta)\right]}\;,
\end{equation}
we derive the bound states spectrum
\begin{equation}
m_{(K)}^{(n)}\,=\, 2 M_{K} \sin\left(n\,\frac{m}{2M_{K}}\right)
\,\,\,\,\,\,\, , \,\,\,\,\,\,\, 0 < n < \pi\frac{M_{K}}{m}
\end{equation}
All the kink--antikink bound states disappear from the theory at a
certain value $\lambda^{*}>4\mu$ such that
$\left.\pi\frac{M_{K}}{m}\right|_{\lambda^{*}}=1$, and the kink
becomes a constant solution with zero classical energy when
$\lambda \to 4 \mu$. This is the semiclassical manifestation of a
phase transition, analogous to the one observed in DSG with
$\delta=\frac{\pi}{2}$. As we have already anticipated, here the
phenomenon occurs in a reverse order, since in this case a kink
appears in the theory by increasing the coupling $\lambda$.

\capitolo{Finite--size effects}
\label{chapfinitesize}\setcounter{equation}{0}

Quantum field theory on a finite volume is a subject of both
theoretical and practical interest. It almost invariably enters
the extrapolation procedure of numerical simulations, limited in
general to rather small samples, but it is also intimately related
to quantum field theory at finite temperature. It is therefore
important to increase our ability in treating finite size effects
by developing efficient analytic means. In the last years, a
considerable progress has been registered in particular on the
study of finite size behaviour of two dimensional systems. Also
for these models, however, an exact treatment of their finite size
effects has been obtained only in particular situations, namely
when the systems are at criticality or if they correspond to
integrable field theories. At criticality, in fact, methods of
finite size scaling and Conformal Field Theory permit to determine
many universal amplitudes and to extract as well useful
information on the entire spectrum of the transfer matrix. Away
from criticality, exact results can be obtained only for
integrable theories which, on a finite volume, can be further
analysed by means of Thermodynamical Bethe Ansatz. This technique
provides integral equations for the energy levels, mostly solved
numerically. In all other cases, the control of finite size
effects in two dimensional QFT has been reached up to now either
by perturbative or numerical methods.

The semiclassical method is capable of shedding new light on this
topic \cite{finvolff,SGscaling,SGstrip}. In fact, once the proper
classical solutions are identified to describe a given geometry,
the spectral function in finite volume can be easily estimated by
adapting the Goldstone and Jackiw's result. Furthermore, the
finite--volume kinks can be quantized semiclassically by adapting
the DHN technique, which permits to write in analytic form the
discrete energy levels as functions of the size of the system.
This procedure can be also extended to geometries with boundaries.

The Chapter is organized as follows. A brief overview of the known
non--perturbative techniques to study finite--size effects is
presented in Section\,\ref{finvolintro}. Section\,\ref{sectff}
describes the semiclassical results obtained about form factors
and correlation functions in finite volume, for the Sine--Gordon
and broken $\phi^4$ theories on a cylinder with antiperiodic
boundary conditions. Finally, the semiclassical quantization of
the Sine--Gordon model on a periodic cylinder and on a strip with
Dirichlet boundary conditions are presented in
Sections\,\ref{sectcyl} and \ref{sectstrip}, respectively. The
Chapter also contains few appendices, devoted to some technical
discussions. Appendix \ref{chapfinitesize}.\ref{appreg} presents
the quantization of a free bosonic theory in a finite volume and a
comparison of finite--volume and finite--temperature computations
of the simplest one--point correlation function. Appendix
\ref{chapfinitesize}.\ref{appell} collects relevant mathematical
properties of the elliptic functions used in the text whereas
Appendix \ref{chapfinitesize}.\ref{lame} displays the main
properties of the Lam\'e equation, which plays a central role in
the finite--size quantization.

\section{Known facts}\label{finvolintro}
\setcounter{equation}{0}

\subsection{General results}

The understanding of finite--size effects is an important tool in
the study of QFT. The finite--volume energy spectrum, in fact,
contains a lot of information about the properties of the theory
in infinite volume. This was first pointed out by L\"{u}scher
\cite{luscher}, who showed how the mass of a particle in a large
but finite volume approaches exponentially its asymptotic value in
a way controlled by the scattering data of the infinite--volume
theory. L\"{u}scher's results were obtained in the context of
Monte Carlo simulation of lattice field theories, with the intent
of correctly interpreting the numerical data, unavoidably affected
by finite--size effects. However, these findings proved to be
useful for purposes other than merely controlling a systematic
error source. In fact, they can be applied with the opposite
spirit, for extracting scattering data from the numerical
analyses. It is worth mentioning that this also provides an
independent tool for checking the $S$--matrices of integrable
theories, often obtained on the basis of conjectures about the
conserved quantities and particle content of the considered
theories.

\begin{figure}[ht]\hspace{1.5cm}
\begin{tabular}{p{8cm}p{7cm}}
\psfrag{4}{\hspace{-2.5cm}$(1)$\hspace{2cm}\footnotesize$a$}\psfrag{2}{\footnotesize$b$}\psfrag{3}{\footnotesize$c$}
\psfrag{1}{\footnotesize$a$}\psfrag{s}{\hspace{-0.2cm}$\sim$}
\psfig{figure=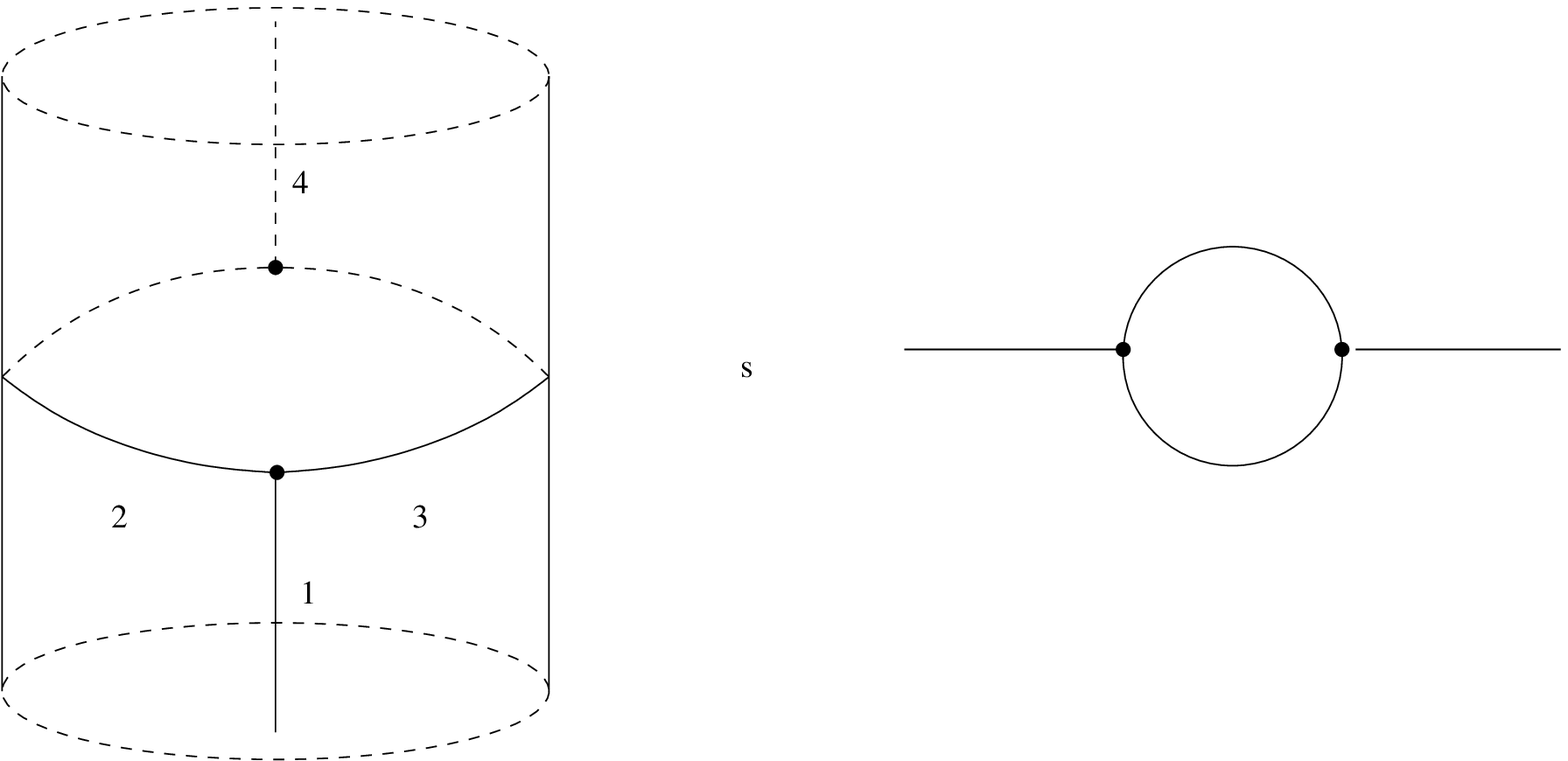,height=3.5cm,width=6.5cm}&
\psfrag{4}{\hspace{-2.5cm}$(2)$\hspace{2cm}\footnotesize$a$}\psfrag{3}{\footnotesize$b$}
\psfrag{1}{\footnotesize$a$}\psfrag{s}{\hspace{-0.2cm}$\sim$}
\psfig{figure=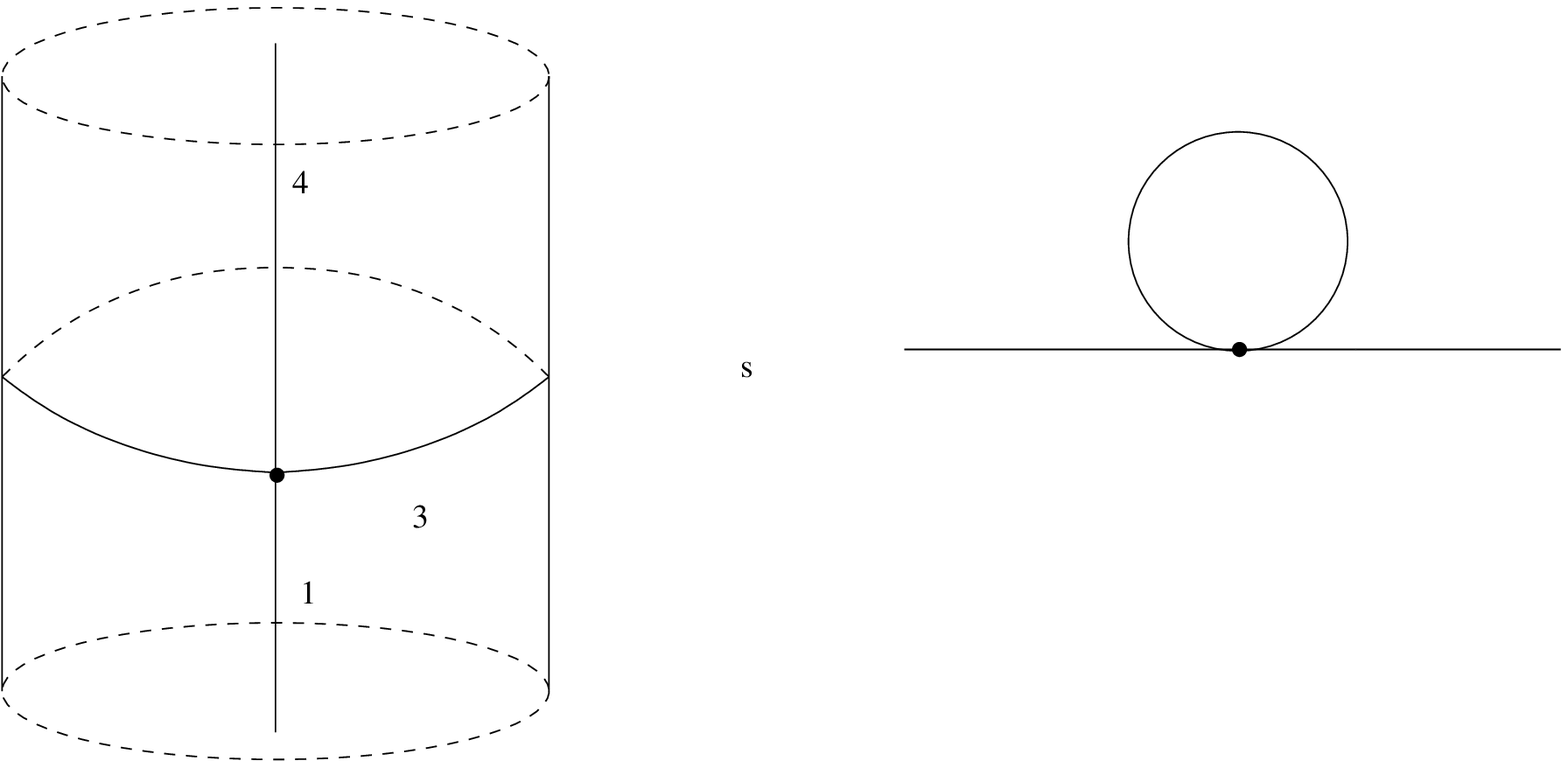,height=3.5cm,width=5.5cm}
\end{tabular}\caption{The two finite--volume contributions to the self--energy,
and their infinite--volume analogs.}\label{figluscher}
\end{figure}

We illustrate here L\"{u}scher's result for the mass corrections
in finite--volume, restricting to the case of $(1+1)$--dimensional
space--time, with the space variable compactified on a periodic
cylinder of circumference $R$. For the general $d+1$--dimensional
result and its proof, see \cite{luscher,kmluscher}. The deviation
of the mass of a particle from its value in infinite volume is
shown to originate from polarization effects. In fact, the
self--energy of the particle receives contributions from processes
in which virtual particles \lq\lq travel around the finite--size
world". The phenomenon can take place at leading order in two
different ways, represented in Fig.\,\ref{figluscher}. The first,
possible only in theories with appropriate cubic couplings and
indicated by $(1)$, is a virtual process in which the particle
splits in two constituents, which travel around the world before
recombining to give back the original particle. The
infinite--volume analog of this process is also depicted in
Fig.\,\ref{figluscher}\,$(1)$. The second process, indicated as
$(2)$ in Fig.\,\ref{figluscher}, is the analog of a tadpole
diagram in infinite volume, and involves the scattering amplitudes
of the given particle with the others in the theory, which travel
around the world before annihilating again.

Therefore, the finite--size mass of a particle $a$ is given by
\begin{equation}
m_a(R)=m_a(\infty)+\Delta m^{(1)}_a(R)+\Delta
m^{(2)}_a(R)+O\left(e^{-\sigma_a R}\right)\;,
\end{equation}
where the constant $\sigma_a$ indicates the order of the errors,
and
\begin{eqnarray}
\Delta m^{(1)}_a(R)&=&-\sum\limits_{b,\,c}
\,\theta\left(m_a^2-|m_b^2-m_c^2|\right)
\mu_{abc}\,R_{abc}\,e^{-\mu_{abc}R}\;,\label{luscherR}\\
\Delta m^{(2)}_a(R)&=&-\sum\limits_b \,\,{\cal
P}\int\limits_{-\infty}^{\infty} d\theta\,e^{-m_b
R\,\cosh\theta}\,m_b\cosh\theta\,\left(S_{ab}^{ab}(\theta+i\pi/2)-1\right)\;\label{luscherS}
\end{eqnarray}
(${\cal P}$ denotes the principal part of the integral).
$S_{ab}^{ab}$ is the two--particle scattering amplitude defined in
(\ref{2partS}) and $\theta$ is the rapidity variable introduced in
(\ref{rapidity}). In correspondence of a bound state pole at
$\theta^*=i u_{ab}^{c}$, the $S$--matrix has the residue $R_{abc}$
introduced in (\ref{resS}). Finally, the quantity $\mu_{abc}$ is
defined as $\mu_{abc}=m_b\,\sin u_{ab}^{c}$, and the step function
$\theta\left(m_a^2-|m_b^2-m_c^2|\right)$ restricts the sum in
(\ref{luscherR}) to those processes in which the virtual particles
$b$ and $c$ can effectively recombine into $a$ in the forward
direction. It is worth noticing that the term $\Delta
m^{(1)}_a(R)$ is of order $O\left(e^{-\mu_{abc} R}\right)$, while
the term $\Delta m^{(2)}_a(R)$ is of order $O\left(e^{-m_{b}
R}\right)$.

\vspace{0.5cm}

As we have already commented, L\"uscher result has been derived in
the context of numerical simulations in lattice QFT. Another
efficient numerical method for studying two--dimensional QFT in
finite--size, which does not require a discretization of space,
was suggested in \cite{TCSA}, and is called Truncated Conformal
Space Approach (TCSA). This method is based on the possibility of
viewing a massive QFT as a perturbation of some CFT, and its
description gives us the possibility of discussing some general
properties of finite--size energy levels.

We have shown in Sect.\,\ref{secCFT} that a conformally invariant
theory can be mapped on a cylindrical geometry, and the
finite--size Hamiltonian can be expressed in terms of the Virasoro
generators and the central charge as
$$
H_{CFT}=\frac{2\pi}{R}\left(L_{0}+\bar{L}_{0}-\frac{c}{12}\right)\;.
$$
Perturbing this CFT  with a relevant operator $\Phi$ with zero
spin and scaling dimension $\Delta=h+\bar{h}$, the Hamiltonian
becomes (see (\ref{Rgflow}))
\begin{equation}\label{CFTpert}
H_{\lambda}\,=\,H_{CFT}+\lambda V\;,\qquad \text{with} \qquad
V=\int\limits_0^R \Phi(x,t)\,dx\;
\end{equation}
(we are considering for simplicity a single perturbing operator,
but the discussion can be straightforwardly generalized to
multiple perturbations). The induced Renormalization Group flow is
parameterized by the dimensionless variable $\lambda
R^{\,2-\Delta}\;$, and the energy eigenvalues $E_i$ can be
expressed in terms of \textit{scaling functions} $f_i$ of this
variable:
$$
E_i(R,\lambda)\,=\,\frac{2\pi}{R}\;f_i\left(\lambda
R^{\,2-\Delta}\right)\;.
$$
These functions interpolate between the massless CFT,
corresponding to the ultraviolet (UV) limit $\lambda
R^{\,2-\Delta}\to 0$, and the massive infrared (IR) theory,
corresponding to $\lambda R^{\,2-\Delta}\to \infty$. In the two
limits the energies are dominated by
\begin{eqnarray}\label{energyUV}
E_i(R,\lambda)&\sim
&\frac{2\pi}{R}\left(h_i+\bar{h}_i-\frac{c}{12}\right)\qquad
\text{for}\qquad \lambda R^{\,2-\Delta}\to 0\;,\\
E_i(R,\lambda)&\sim &\varepsilon_0\,
\lambda^{\frac{2}{2-\Delta}}\,R\,+\,m_i\qquad
\hspace{0.55cm}\text{for}\qquad \lambda R^{\,2-\Delta}\to
\infty\;,\label{energyIR}
\end{eqnarray}
where $h_i,\bar{h}_i$ are the dimensions of the conformal state
related to the $i$th level, $\varepsilon_0$ is the dimensionless
bulk vacuum energy and $m_i$ is the (multi)particle mass term of
the $i$th level.

\begin{figure}[ht]
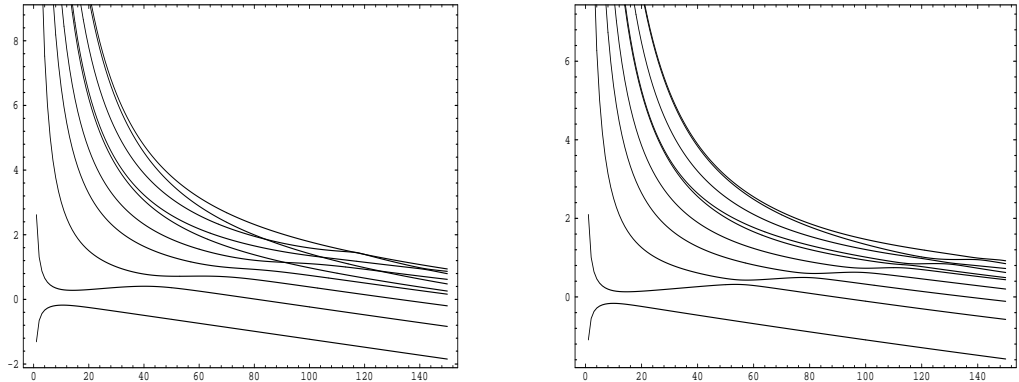
\hspace{1.5cm}
\begin{tabular}{p{7cm}p{7cm}}
\psfig{figure=figure7d.epsi,height=5cm,width=6cm}&
\psfig{figure=figure7e.epsi,height=5cm,width=6cm}
\end{tabular}\caption{TCSA results obtained in \cite{DMS} for the discrete spectrum $E_i(R)$
in the Ising model with magnetic perturbation (left picture), and
in the Ising model with both magnetic and thermal perturbation
(right picture).}\label{figtrunc}
\end{figure}

The TCSA technique takes advantage of the fact that the
eigenstates of $H_{CFT}$ can be chosen as a basis for the Hilbert
space of the perturbed problem. As a consequence, the matrix
elements of $V$ in this basis
$$
\langle\,\phi_i\,|\,V\,|\,\phi_j\,\rangle=\frac{R}{2\pi}\langle\,\phi_i\,|\,\Phi(0,0)\,|\,\phi_j\,\rangle
$$
can be computed in terms of the structure constants which enter
the OPE in CFT (see (\ref{OPE})). The obtained Hamiltonian can be
finally truncated to the desired level, putting an upper bound on
the conformal dimensions of the states in the basis. As a result
of its diagonalization, one typically obtains finite--size spectra
as shown in Fig.\,\ref{figtrunc} (usually, one sets $\lambda=1$
and plots the spectrum as a function of $R$).

From the UV and IR asymptotic behaviours of the obtained energy
spectra it is therefore possible to extract several conformal and
scattering data, according to (\ref{energyUV}) and
(\ref{energyIR}). Furthermore, also the elastic two--particle
$S$--matrix can be measured \cite{luscher2part}, by considering
two--particle states $|\,A_a\,A_b\,\rangle$ with zero total
momentum, relative momentum $k$ and energy
$$
E\,=\,\sqrt{m_a^2+k^2}+\sqrt{m_b^2+k^2}\;.
$$
In fact, the momentum $k$ can be extracted from the above equation
after the numerical determination of $E$, $m_a$ and $m_b$, and the
$S$--matrix can be then deduced by the quantization condition
$$
e^{i k R}\,S_{ab}(k)\,=\,1\;.
$$

TCSA can be applied to both integrable and non--integrable
theories, and this permits to observe an interesting phenomenon.
In fact, in integrable theories it is quite common to observe
crossing of different energy levels at finite values of $R$, while
these intersections are immediately removed as one adds a
non--integrable perturbation (see for instance
Fig.\,\ref{figtrunc}). In non--relativistic quantum mechanics it
is well known that two energy levels can cross only if they belong
to different irreducible representations of the symmetry group of
the Hamiltonian, otherwise, in absence of any symmetry, they have
to repel.
This has a clear counterpart in QFT: the presence of conserved
charges ensures the stability of particles above threshold, whose
lines cross an infinite number of one--particle lines. As soon as
a perturbation is switched on, however, integrability is lost and
inelastic processes become allowed, causing the decay of the
particle. As a consequence, the particles lines above threshold
become broken lines, whose difference in slope with the threshold
is proportional to the decay width of the unstable states
\cite{luscherres}. Therefore, the study of finite--size effects
also gives evidences about the integrability of the theory in
exam.

\subsection{Integrable QFT}

We have seen so far how important are finite--size effects for the
understanding of QFT. It is therefore useful to have the
possibility of controlling them also analytically. Apart from the
case of CFT, where conformal invariance permits a complete
treatment of the problem, interesting results can be obtained for
integrable QFT, passing through a thermodynamic interpretation of
the finite size of the system.

\begin{figure}[ht]
\hspace{2cm}
\begin{tabular}{p{5cm}p{7cm}}
\psfrag{x}{$x$}\psfrag{t}{$t$}\psfrag{L}{\hspace{-0.3cm}$$}\psfrag{R}{$R$}\psfig{figure=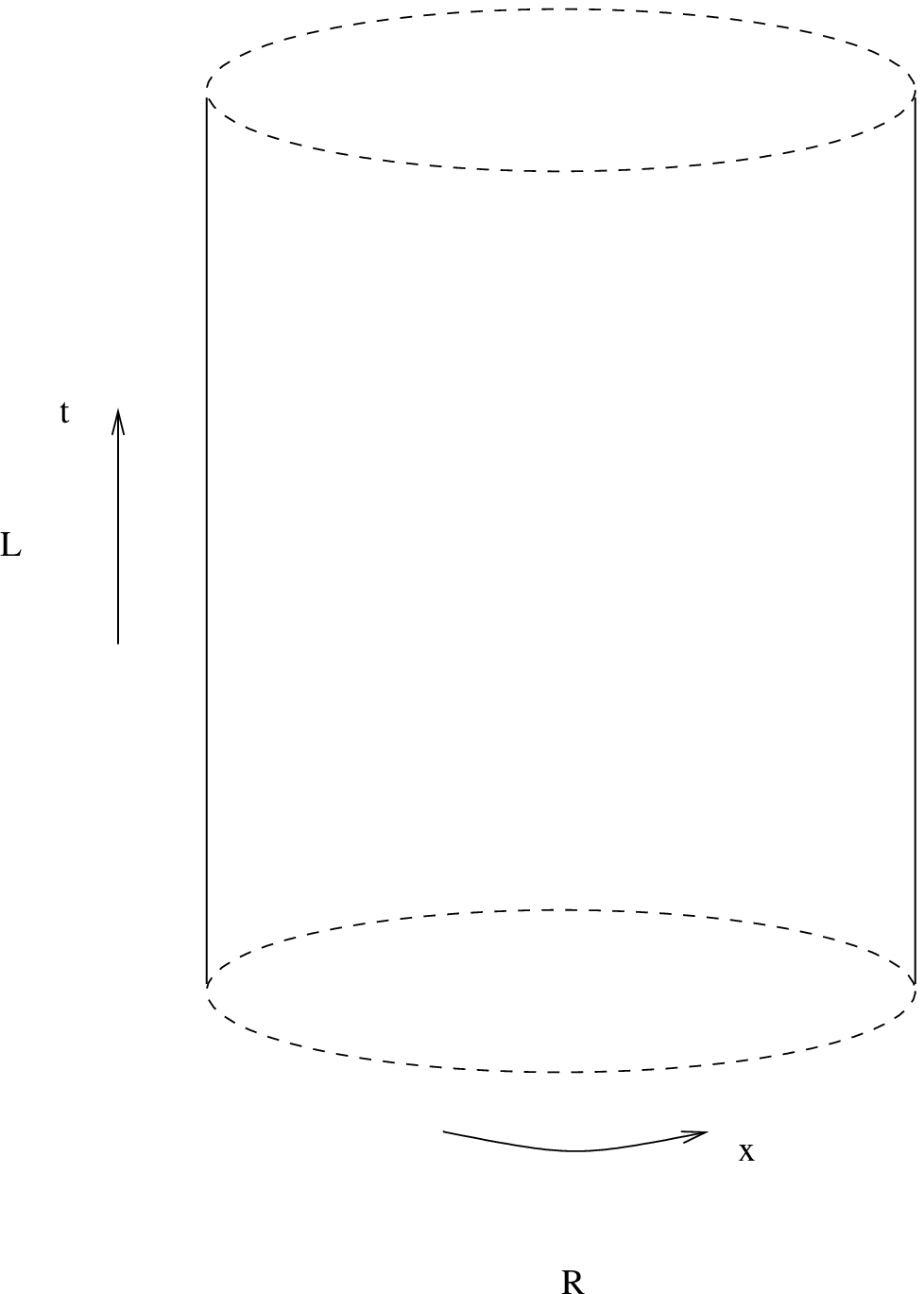,
height=5cm,width=3cm} &
\psfrag{x}{$x$}\psfrag{t}{$\tau$}\psfrag{L}{$L$}\psfrag{R=1/T}{$R\propto
1/T$} \psfig{figure=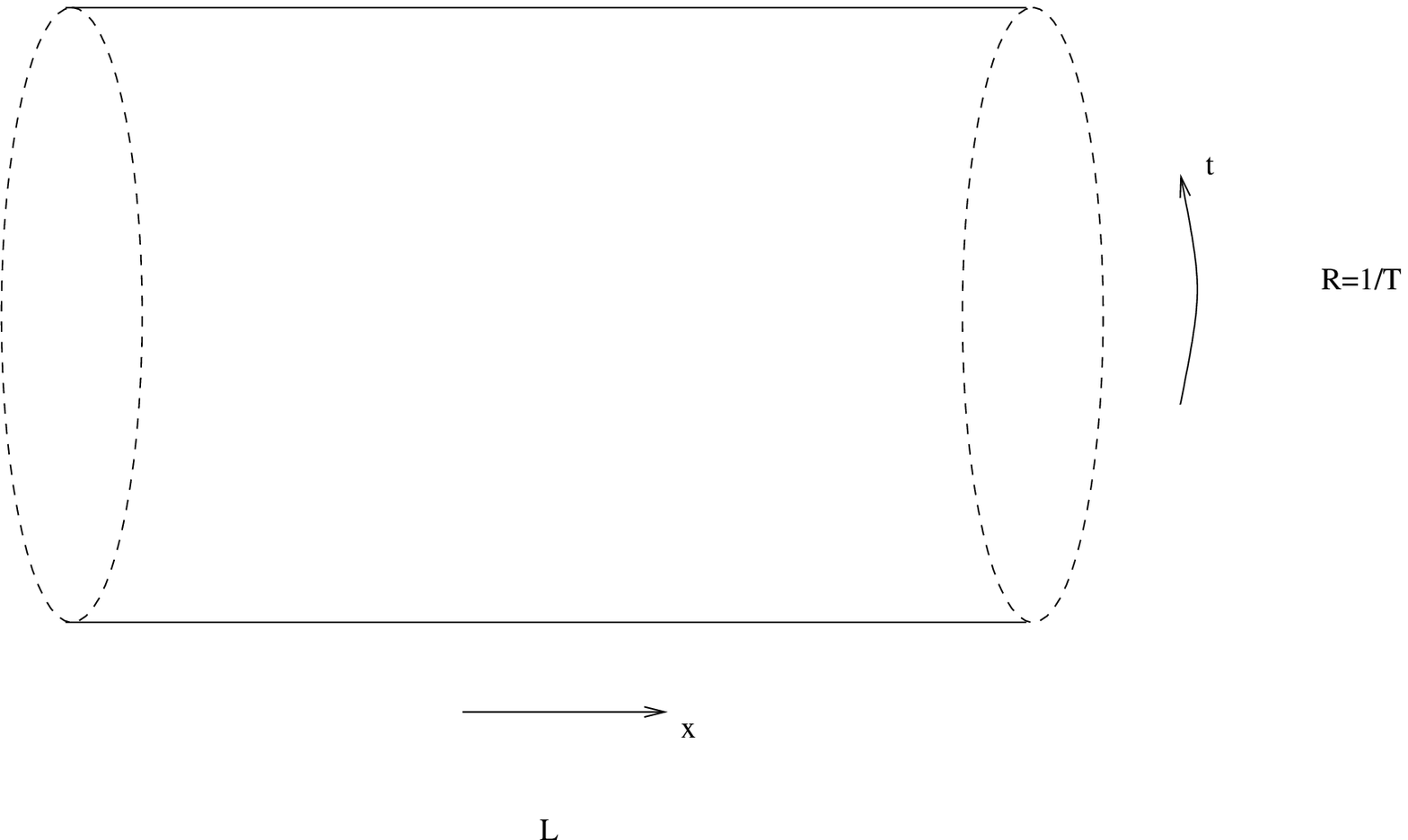,height=3cm,width=6cm}
\end{tabular}
\caption{Finite volume and finite temperature interpretations of
the cylindrical geometry.}\label{figvoltemp}
\end{figure}

The cylindrical geometry can be physically interpreted in two
different ways, illustrated in Fig.\,\ref{figvoltemp}. One is the
finite--volume picture described above, in which the space
variable is compactified on a circle of circumference $R$ while
the time evolution takes place along an unbounded direction. The
other picture is the finite--temperature one, in which the space
variable is unbounded but the (euclidean) time lives on a circle
of circumference $R$. In the Matsubara formalism, this corresponds
to introducing a finite temperature in the system, related to the
compactification size by the inverse proportionality $R\propto
1/T$.

In the case of integrable QFT, a lot of information about the
thermodynamics can be obtained from the knowledge of the
$S$--matrix. In fact, the partition function
$$
Z=\text{Tr}\left(e^{-RH}\right)
$$
can be determined by means of the so--called Thermodynamic Bethe
Ansatz (TBA) technique, proposed in \cite{TBA}. Here we limit
ourselves to illustrate the basic ideas underlying this method,
referring to the original literature for a consistent description.
If we put the theory on a large box $0<x<L$, assuming for
simplicity that the spectrum contains only one type of particle
with mass $m$ and scattering matrix $S$, the quantization
conditions of the momenta are given by the Bethe Ansatz equations
\begin{equation*}
e^{i m L\sinh\theta_i }\prod\limits_{j\neq
i}S(\theta_i-\theta_j)\,=\,1\;,
\end{equation*}
with the rapidity
variable $\theta$ defined in (\ref{rapidity}). The thermodynamic
limit of the equivalent condition
\begin{equation}\label{momentumquant}
m L\sinh\theta_i+\sum\limits_{j\neq
i}\delta(\theta_i-\theta_j)\,=\,2\pi n_i\;,\qquad n_i \in
\mathbb{Z}\;,
\end{equation}
with $\delta(\theta)=-i\log S(\theta)$, is given by
$$
m\cosh\theta+2\pi\,\varphi(\theta) *
\rho_1(\theta)\,=\,2\pi\rho(\theta)\;,
$$
where $\rho(\theta)$ and $\rho_1(\theta)$ represent, respectively,
the density of levels and of occupied states per unit volume, $*$
indicates the convolution operator and
$\varphi(\theta)=\frac{d}{d\theta}\delta(\theta)$. The
minimization of the free energy with respect to the density of
states leads to the integral equation
\begin{equation}\label{eqpsen}
\varepsilon(\theta)\,=\,mR\cosh\theta-\varphi(\theta)
*\log(1+e^{-\varepsilon(\theta)})
\end{equation}
for the so--called pseudo--energy $\varepsilon(\theta)$, defined
as
$$
\frac{\rho_1}{\rho}=\frac{1}{1+e^{\varepsilon}}\;.
$$
The partition function can be then expressed in terms of the
pseudo--energy, which is numerically determined from
(\ref{eqpsen}), as
$$
Z(R,L)\,=\,\exp\left[mL\int \frac{d\theta}{2\pi}\,\cosh\theta\,
\log\left(1+e^{-\varepsilon(\theta)}\right)\right]\;.
$$
If we now come back to the finite--volume picture, performing the
$L\to\infty$ limit we realize that the partition function has to
be dominated by
$$
Z(R,L)\,=\,\text{Tr}\left(e^{-LH_R}\right)\simeq e^{-L E_0(R)}\;,
$$
where $E_0(R)$ is the ground state eigenvalue of the
finite--volume Hamiltonian $H_R$. Parameterizing it conveniently
as $E_0(R)=\frac{2\pi}{R}\, f_0(m R)$, we therefore obtain the
scaling function \cite{TBA,kmTBA}
\begin{equation}\label{TBA}
f_0(m R)\,=\,-\frac{mR}{2\pi}\int
\frac{d\theta}{2\pi}\,\cosh\theta\,
\log\left(1+e^{-\varepsilon(\theta)}\right)\;,
\end{equation}
which in the UV limit has to reduce to $f(m R)\to
h_0+\bar{h}_0-c/12$, where $h_0$ and $c$ are the lowest conformal
dimension and the central charge of the corresponding CFT (see
(\ref{hamcyl})).

In this framework, the determination of excited energy levels is
not possible or extremely involved. To this respect, however,
other techniques exploiting integrability have been developed and
applied for instance in \cite{ddv,BLZ,RavaniniSG}.

\vspace{0.5cm}

Besides energy levels, the understanding of QFT obviously requires
also the knowledge of correlation functions. In the
finite--temperature picture, the matrix elements are computed by
performing a thermodynamic averaging over the usual
infinite--volume particle states:
$$
\langle \,{\cal O}_1\;\ldots\;{\cal
O}_n\,\rangle_R\,\equiv\,\frac{1}{Z}\,\text{Tr}\left(e^{-RH}{\cal
O}_1\;\ldots\;{\cal O}_n\right)\;,
$$
where $H$ is the infinite--volume Hamiltonian and $Z$ is the
partition function. Therefore, in the case of integrable QFT, a
method has been proposed in \cite{lm} for expressing the
finite--temperature correlators in terms of the infinite volume
form factors and some thermodynamical quantities available from
TBA. For instance, the one--point functions are given as
$$
\langle \,{\cal
O}(x,t)\,\rangle_R\,=\,\frac{1}{Z}\,\sum\limits_{n=0}^{\infty}\,\frac{1}{n!}\,\int\,\frac{d\theta_1}{2\pi}\;\ldots\;
\frac{d\theta_n}{2\pi}\,\left(\prod\limits_{i=1}^{n}\,f(\theta_i)\,e^{-\,\varepsilon(\theta_i)}\right)\,\langle\,
\theta_n,\,\ldots\,,\theta_1\,|\,{\cal
O}(0,0)\,|\,\theta_1,\,\ldots\,,\theta_n\,\rangle \;,
$$
where the \lq\lq filling fractions"  are defined as
$f(\theta)=\left(1+e^{-\,\varepsilon(\theta)}\right)^{-1}$, with
the pseudo--energy $\epsilon(\theta)$ given by the solutions of
(\ref{eqpsen}). Higher point correlation functions can be in
principle expressed in a conceptually similar way, although there
is some controversy about their correct determination
\cite{{finTpolemic}}.

Further insight on this topic could come from a comparison with
the finite--volume picture, as it was done for the ground state
energy. In this case, the computation of matrix elements involves
the notion of finite--size ground state $|\,0\,\rangle_R\;$:
$$
\langle \,{\cal O}_1\;\ldots\;{\cal O}_n\,\rangle_R\,\equiv\,
_R\langle\,0\,|\,{\cal O}_1\;\ldots\;{\cal
O}_n\,|\,0\,\rangle_R\;.
$$
Unfortunately, at present the precise structure of this ground
state is not clear, and moreover, the finite--size matrix elements
of operators are still poorly understood. It is precisely from
this problem that we will start, in the next Section, our
discussion of the semiclassical results obtained in finite volume.

\section{Semiclassical form factors in finite volume}\label{sectff}
\setcounter{equation}{0}

A particularly interesting problem in the study of QFT in finite
volume regards the possibility of defining a \lq\lq form factor
representation" for the correlation functions, in analogy to the
infinite volume expression (\ref{spectral}). There are reasons to
expect, in fact, a fast convergent behaviour of these series also
for finite volume correlators, as it happens in infinite volume
(see, for instance \cite{fastconv}). If this would be indeed the
case, accurate estimates of finite volume correlators and other
related physical quantities, could be obtained by just using few
exact terms of their spectral representations, having consequently
a great simplification of the problem. This observation makes
clear that it is worth pursuing the research on finite--volume
form factors, looking in particular for an efficient scheme of
approximation.

In the non--perturbative study of form factors at a finite volume,
there are so far only semiclassical results relative to a
conformal theory \cite{smirnovfinvol} as well as exact
calculations relative to the Ising model \cite{fonszam} (for a
Bethe Ansatz approach see, however, \cite{korepin}). Although
these findings are very interesting, the techniques employed in
the above papers are however strictly related either to the
specific integrable structure of the considered models or to the
free nature of the Majorana fermion field of the Ising model.

In \cite{finvolff}, we have proposed to face this problem with a
semiclassical approach based on Goldstone and Jackiw's result.
This method, contrary to the ones previously mentioned, does not
require integrability and it is then of more general
applicability, of course within its range of approximation. It is
worth to recall an important feature that has come out from the
study of the semiclassical form factors in infinite volume. As we
have seen in Sect.\,\ref{SGinfinite}, their accuracy seems to
extend, somehow, beyond the regime in which they were supposed to
be valid. Together with the known fast convergency properties of
the spectral series, the above result suggests that the
semiclassical method may provide a rather precise estimate of
finite volume correlation functions, an outcome which may be
useful for many applications.

The application of the procedure described in
Sect.\,\ref{secsemFF} to the finite volume case is
straightforward, thanks to the possibility of choosing
$\hat{f}(a)$ as a solution of eq.\,(\ref{firstorder}) with any
constant $A$. As explicitly shown by the examples discussed below,
this is equivalent to define a kink solution configuration on a
finite volume, with the constant $A$ directly related to the size
of the system. We have now to consider the matrix elements of
$\phi(0)$ between two eigenstates $|p_{n_{1}}\rangle$ and
$|p_{n_{2}}\rangle$ of the finite volume hamiltonian $H_R$. These
states can be naturally labelled with the so-called
"quasi-momentum" variable $p_{n}$, which corresponds to the
eigenvalues of the translation operator on the cylinder (multiples
of $\pi/L$), and appears in the space dependent part of
eq.\,(\ref{heis}) in the case of finite volume. The Bethe ansatz
equations (\ref{momentumquant}), valid for large $R$, are exactly
a relation between this variable and the free momentum
$p^{\infty}$ of the infinite volume asymptotic states, through a
phase shift $\delta(p^{\infty})$ which encodes the information
about the interaction:
\begin{equation*}
p_{n}^{\infty}+\frac{\delta(p_{n}^{\infty})}{R}=\frac{2n\pi}{R}\equiv
p_{n}\;.
\end{equation*}
Defining $\theta_{n}$ as the "quasi-rapidity" of the kink states
by
\begin{equation*}
p_{n}=M(R)\sinh\theta_{n}\simeq M(R)\theta_{n}\;,
\end{equation*}
we can now write the form factor at a finite volume by replacing
the Fourier integral transform with a Fourier series expansion:
\begin{equation}\label{ff}
f(\theta_{n})\,=\,\langle
p_{n_{2}}|\,\phi(0)\,|p_{n_{1}}\rangle\,=\,
M(R)\int\limits_{-R/2}^{R/2} da\,e^{i\,M(R)\theta_{n}
a}\phi_{cl}(a)\;,
\end{equation}
\begin{equation}\label{inverseff}
\phi_{cl}(a)\,=\,\frac{1}{R\,M(R)}
\sum\limits_{n=-\infty}^{\infty}\,e^{-i\,M(R)\theta_{n}
a}f(\theta_{n})\;,
\end{equation}
where
\begin{equation*}
M(R)\,\theta_{n}\,\simeq\,
p_{n_{1}}-p_{n_{2}}\,=\,\frac{(2n_{1}-1)\pi}{R}\, -\,
\frac{(2n_{2}-1)\pi}{R}\,\equiv\,\frac{2n\pi}{R}\;.
\end{equation*}
Since the energy eigenvalues of the finite volume hamiltonian
cannot be related to the quasi-rapidity as in
eq.\,(\ref{rapidity}), in principle we are not allowed to express
the crossed channel form factor $F_{2}(\theta)$ via the change of
variable $\theta\rightarrow i\pi-\theta$. However, it is easy to
show that in our regime of approximations the deviations of the
kink energy from (\ref{rapidity}) are of higher order in the
coupling and can be neglected at this stage.

On the cylinder, the spectral function can be expressed as a
series expansion on the form factors:
\begin{equation*}
\rho^{(\phi)}(E_{k},p_{k}) \,=\,
2\pi\sum\limits_{n}\frac{1}{n!}\frac{1}{(2R)^{n}}\sum\limits_{k_{1},...,k_{n}}
\frac{1}{E_{k_{1}}E_{k_{2}}...E_{k_{n}}}\,
\delta(E_{k}-E_{k_{1}}...-E_{k_{n}})\,\delta(p_{k}-p_{k_{1}}...-p_{k_{n}})
\times \end{equation*}
\begin{equation}\label{rhocyl}
\times|\,\langle\,0|\,\phi(0)|\,n\,\rangle\,|\,^{2}\;,
\end{equation}
where $p_{k_{i}}$ are the quasi-momenta of the intermediate
states, and $E_{k_{i}}$ are the finite volume energy eigenvalues,
that will be discussed in the following Sections. For the moment
their knowledge is not necessary, because at leading semiclassical
order the spectral function $\hat\rho^{(\phi)}(E_{k},p_{k})$ is
given by the trivial vacuum term plus the kink--antikink
contribution, and the kink energies can be consistently
approximated with their classical masses $M$, which can be exactly
computed as functions of the volume. We then have
\begin{equation}
\hat\rho^{(\phi)}(E_{k},p_{k}) \,= \,
2\pi\delta(E_{k})\delta(p_{k})|\,\langle\,0|\,\phi(0)|\,0\,\rangle\,|\,^{2}
+ \frac{\pi}{4}\, \frac{\delta\left(\frac{E_{k}}{M}-2\right)}
{M^{2}}\sum\limits_{\theta_{k_{1}}}\left|F_{2}\left(2\theta_{k_{1}}
+ i\pi-\frac{p_{k}}{M}\right)\right|^{2}\,\,\,.
\end{equation}
As in the infinite volume case, the consistency of the
semi--classical approximation selects as the relevant values of
the above series those with $\theta_k \simeq 0$ and therefore it
can can be roughly estimated by simply evaluating $|F_{2}|^{2}$ at
$\theta_{k_{1}}=0$.

We will now explicitly construct the form factors on the kink
states in the sine--Gordon model (\ref{SGpot}) and in the broken
$\phi^4$ theory (\ref{phi4pot}), both defined on a cylindrical
geometry with antiperiodic boundary conditions, respectively given
by
\begin{equation}\label{antiperiodicbc}
\begin{cases}
\phi(x+R)\,=\,\frac{2\pi}{\beta}-\phi(x)\qquad &\text{for sine--Gordon}\;,\\
\phi(x+R)\,=\,-\phi(x)\qquad &\text{for broken $\phi^4$}\;.
\end{cases}
\end{equation}

\subsubsection{Sine--Gordon model}

\begin{figure}[h]
\hspace{3cm} \psfig{figure=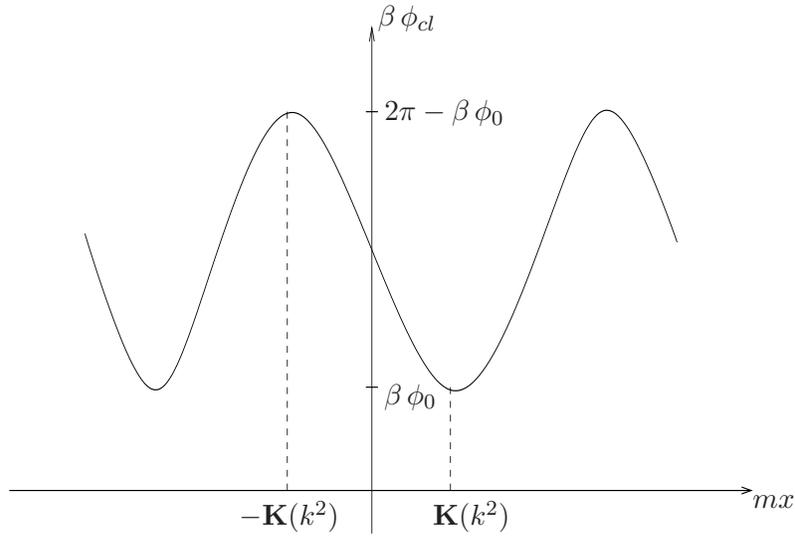,height=7cm,width=10cm}
\vspace{0.1cm} \caption{Solution of eq.\,(\ref{firstSG}) with
$-2<A<0$ and $x_0=0$.} \label{figSGkink}
\end{figure}

In order to identify a kink on the twisted cylinder, we have to
look for a static finite energy solution of the SG model
satisfying the anti--periodic boundary condition
(\ref{antiperiodicbc}). For the first order equation
\begin{equation}
\frac{1}{2}\left(\frac{\partial \phi_{cl}} {\partial x}\right)^{2}
\,=\, \frac{m^{2}}{\beta^{2}} \left(1-\cos\beta\phi_{cl} + A
\right) \,\,\, \,\,\, \label{firstSG}
\end{equation}
a solution with this property can be found for $-2<A<0$, and it is
expressed as
\begin{equation}\label{SGkinkantiper}
\phi_{cl}(x) = \frac{2}{\beta}\,\arccos\left[\,k\;
\textrm{sn}\left( m (x-x_0),\,k^{2}\right)\right]\;,
\end{equation}
where $\textrm{sn}(u,k^{2})$ is the Jacobi elliptic function with
modulus $k^{2}=\frac{A+2}{2}$, and the presence of a free
parameter $x_{0}$, which represents the kink's center of mass
position, is due to the translational invariance of the theory
around the cylinder axis. The plot of (\ref{SGkinkantiper}) as a
function of the real variable $x$ is drawn in Fig.\ref{figSGkink}.
For a given value of $A$, this solution oscillates with a period
$4 \textbf{K}(k^{2})$ between $\phi_{0}$ and
$\frac{2\pi}{\beta}-\phi_{0}$, where $\phi_{0}$ is defined by the
condition $V(\phi_{0})=-\,\frac{m^2}{\beta^2}\,A$, and
$\textbf{K}(k^{2})$ is the complete elliptic integral of the first
kind. For the definitions and properties of elliptic integrals and
Jacobi elliptic functions, see Appendix
\ref{chapfinitesize}.\ref{appell}.

\psfrag{phicl(x)}{$\beta\,\phi_{cl}$}
\psfrag{phi0}{$\beta\,\phi_{0}$} \psfrag{2
pi-phi0}{$2\pi-\beta\,\phi_{0}$}
\psfrag{K(k^2)}{$\textbf{K}(k^{2})$}
\psfrag{-K(k^2)}{$-\textbf{K}(k^{2})$} \psfrag{x}{$m x$}

The solution (\ref{SGkinkantiper}) has been proposed in
\cite{takoka} as a model of a crystal of solitons and antisolitons
in the sine-Gordon theory in infinite volume. In our finite volume
case, the solution (\ref{SGkinkantiper}) has to be interpreted,
instead, as a single (anti)soliton defined on a cylinder of
circumference
\begin{equation}\label{sizeantiper}
R=\frac{1}{m}\,2 \textbf{K}\left(k^2\right)\;.
\end{equation}
Within this interpretation, the periodic oscillations of the
solution represent the soliton circling around the cylinder.
Eq.\,(\ref{sizeantiper}) is the explicit relation between the size
of the system and the integration constant $A$; one can
consistently recover the infinite volume limit for $A\rightarrow
0$: in this limit $R$ goes to infinity and the function
(\ref{SGkinkantiper}) goes to the standard (anti)soliton solution
(\ref{SGkinkinfvol}).

We can now write the finite volume form factor (\ref{ff}) in terms
of the antikink background (\ref{SGkinkantiper}):
\begin{equation}\label{SGffantiper}
f(-\theta_{n})=\frac{2 M }{\beta}\int
\limits_{-R/2}^{R/2}da\,e^{i\,M\theta_{n} a}\arccos\left[\,k \;
\textrm{sn}(m a)\right]=
\end{equation}
\begin{equation*}
=-\frac{2}{\beta\theta_{n}}\left[e^{iM\theta_{n}\frac{R}{2}}
\log\left(k+i k'\right)-
e^{-iM\theta_{n}\frac{R}{2}}\log\left(-k+i k'\right)
\right]-\frac{2\pi i}{\beta}\,\frac{1}{\theta_{n}\,
\cosh\left[\frac{\textbf{K}'}{m}\,M\theta_{n}\right]}\;\;,
\end{equation*}
where $k'=\sqrt{1-k^{2}}$ and
$\textbf{K}'(k^{2})=\textbf{K}(k'^{2})$. The kink mass $M$ is
represented at this order by the classical energy, given by
\begin{equation}\label{SGclassenantiper}
{\cal E}_{cl}(R) = \int\limits_{-R/2}^{R/2}dx\left[\frac{1}{2}
\left(\frac{\partial \phi_{cl} }{\partial x}\right)^{2} +
\frac{m^{2}}{\beta^{2}}(1-\cos\beta\phi_{cl})\right] =
8\frac{m}{\beta^{2}}\left[
\textbf{E}(k^{2})-\frac{1}{2}(1-k^{2})\textbf{K}(k^{2})\right]\;,
\end{equation}
where $\textbf{E}(k^{2})$ is the complete elliptic integral of the
second kind. In order to obtain the result (\ref{SGffantiper}) one
has to use the relation
\begin{equation*}
\arccos\left[k\, \textrm{sn}(m a)\right]=\frac{1}{i}\log\left[k\,
\textrm{sn}(m a)+i\,\textrm{dn}(m a)\right]\;,
\end{equation*}
and, after an integration by parts, finally compare the inverse
Fourier transform (\ref{inverseff}) with the expansion \cite{GRA}
\begin{equation*}
\textrm{cn}(m a) =
\frac{2\pi}{k}\,\frac{1}{mR}\sum\limits_{n=1}^{\infty}
\frac{\cos\left[\frac{(2n-1)\pi}{R}a\right]}
{\cosh\left[\frac{(2n-1)\pi}{R}\,\frac{\textbf{K}'}{m}\right]}\;.
\end{equation*}

The form factor (\ref{SGffantiper}) has the correct IR
limit\footnote{The function $\frac{e^{-ixR/2}}{x}$ can be shown to
tend to $-i\pi\delta(x)$ in the distributional sense for
$R\rightarrow\infty$, and in the same way one can show that
$\frac{\cos(xR/2)}{x}$ tends to zero.}, and leads to the following
expressions for $F_{2}(\theta)$ and for the spectral function:
\begin{equation}\label{SGf2antiper}
F_{2}(\theta_{n})=\frac{4\pi i}{\beta(i\pi-\theta_{n})}\,
\left\{\frac{1}{\cosh\left[\frac{M}{m}\textbf{K}'\,(i\pi-\theta_{n})\right]}+
\left(-1+\frac{2}{\pi}\arctan\frac{k'}{k}\right)\,
\cos\left[M(i\pi-\theta_{n})\frac{R}{2}\right] \right\}\;,
\end{equation}
\begin{equation}\label{SGrhoantiper}
\hat\rho(E_{n},p_{n}) =
4\pi^{3}\,\delta\left(\frac{E_{n}}{M}-2\right)\frac{1}{\beta^{2}(p_{n})^{2}}
\left\{\frac{1}{\cosh\left[\frac{\textbf{K}'}{m}\,p_{n}\right]}+
\left(-1+\frac{2}{\pi}\arctan\frac{k'}{k}\right)\,\cos\left[p_{n}
\frac{R}{2}\right] \right\}^{2}\;.
\end{equation}
Note that the finite volume dependence of both the form factor
(\ref{SGf2antiper}) and the spectral function (\ref{SGrhoantiper})
is not restricted to the second term only. The
$M(R)\,\textbf{K}'(k^{2})$ factor in the first term carries the
main $R$-dependence, although it is not manifest but implicitly
defined by eq.\,(\ref{sizeantiper}).

\subsubsection{Broken $\phi^4$ theory}

The differential equation
\begin{equation*}
\frac{1}{2}\left(\frac{\partial \phi_{cl}} {\partial x}\right)^{2}
\,=\,
\frac{\lambda}{4}\,\left(\phi^{2}-\frac{m^2}{\lambda}\right)^{2}\,+\,A
\,\,\, \,\,\,
\end{equation*}
has a solution given by
\begin{equation}\label{phi4kink}
\bar{\phi}_{cl}(\bar{x}) \,=\,(\pm) \,
\sqrt{2-\bar{\phi}_{0}^{2}}\,\;
\textrm{sn}\left(\frac{\bar{\phi}_{0}}{\sqrt{2}}\,
(\bar{x}-\bar{x}_0),\,k^{2}\right)\;,
\end{equation}
with $k^{2} = \frac{2}{\bar{\phi}_{0}^{2}}-1$, $V(\phi_{0}) = -A$
and $1 <\bar{\phi}_{0} < \sqrt{2}$, where we have rescaled the
variables as
\begin{equation*}
\bar{\phi}\,=\, \frac{\sqrt{\lambda}}{m}\,\phi\,,\qquad \bar{x}\,
=\, m x\;.
\end{equation*}
This function oscillates with period
\begin{equation}\label{sizephi4}
2 R \,=\, \frac{4}{m}\, \frac{\sqrt{2}}{\bar{\phi}_{0}}\,
\textbf{K}(k^{2})
\end{equation}
between the two values $-\sqrt{2-\bar{\phi}_{0}^{2}}\,$ and
$\sqrt{2-\bar{\phi}_{0}^{2}}$, and it satisfies the anti--periodic
boundary condition (\ref{antiperiodicbc}). Moreover, it goes to
the standard (anti)kink solution (\ref{phi4kinkinf}) for
$\bar{\phi}_{0}\rightarrow 1$, i.e. when $R\, \rightarrow \,
\infty$.

From the analytic knowledge of the background (\ref{phi4kink}), we
can immediately extract an important scattering data of the
non--integrable $\phi^4$ theory. In fact, the leading term in the
kink mass is given by the classical energy, expressed for generic
$R$ as
\begin{equation}\label{phi4classen}
{\cal E}_{cl}(R) \, = \, \frac{m^{3}}{\lambda}\,
\frac{\sqrt{2}}{\bar{\phi}_{0}}
\left(-\frac{1}{6}\bar{\phi}_{0}^{4} \, \textbf{K}(k^{2}) +
\frac{1}{3}\bar{\phi}_{0}^{2} \,\left[2 \textbf{E}(k^{2}) -
\textbf{K}(k^{2})\right] + \frac{\textbf{K}(k^{2})}{2}\right)
\,\,\,.
\end{equation}
It is easy to see that for $R\rightarrow\infty$ this quantity
indeed reproduces the infinite--volume energy
(\ref{phi4kinkinfecl}). From its asymptotic expansion for large
$R$, we can obtain the leading order of the kink mass correction
in finite volume, and compare it with L\"{u}scher's result
(\ref{luscherR},\,\ref{luscherS}). Taking into account the
$k\rightarrow 1$ ($k'\rightarrow 0$) expansions of $\textbf{E}$
and $\textbf{K}$ (see Appendix\,\ref{chapfinitesize}.\ref{appell})
and noting from (\ref{sizephi4}) that
$$
e^{-\sqrt{2}m R}=\frac{1}{256}(k')^{4}+\cdots\;,
$$
we derive the following asymptotic expansion of ${\cal E}_{cl}$
for large $R$:
\begin{equation}
{\cal E}_{cl}(R)=M_{\infty}-8\sqrt{2}\,\,\frac{m^{3}}{\lambda}
e^{-\sqrt{2}m R}+O\left(e^{-2 \sqrt{2}m R}\right)\;.
\end{equation}
The counterpart of this behaviour in L\"uscher's theory is given
by the process $(1)$ in Fig.\,\ref{figluscher}, where particle $a$
is the kink, particle $b$ is the elementary meson, and particle
$c$ is another kink. In fact, in the broken $\phi^4$ theory the
elementary boson has semiclassical leading mass $m_b=\sqrt{2}\,m$,
which is much lower than the kink one (\ref{phi4kinkinfecl}),
therefore the dynamical pole relative to this process is located
at $u_{k\,b}^{k}\simeq \frac{\pi}{2}$. From the comparison with
(\ref{luscherR}) we finally extract the leading semiclassical
expression for the residue of this $3$--particle process:
$$
R_{k\,k\,b}\,=\,8\,\frac{m^2}{\lambda}\;.
$$
It is easy to see that this result correctly reproduces, as
required by (\ref{resS}), the square of the value (\ref{gphi4})
previously obtained for the 3-particle coupling
$\Gamma_{k\,\bar{k}\,b}$ by looking at the residue of the
kink--antikink form factor in infinite volume\footnote{Crossing
symmetry implies the equality $R_{k\bar{k}b}=R_{kkb}$.}.

Moving then to the finite--volume form factors, these can be
computed by comparing the inverse Fourier transform
(\ref{inverseff}) with the expansion \cite{GRA}
\begin{equation*}
\textrm{sn}(u) \,=\, \frac{\pi}{k\textbf{K}} \,
\sum\limits_{n=1}^{\infty}
\frac{\sin\left[\frac{(2n-1)\pi}{2\textbf{K}}u\right]}
{\sinh\left[\frac{(2n-1)\pi}{2\textbf{K}}\,\textbf{K}'\right]}\;,
\end{equation*}
obtaining the expression
\begin{eqnarray}
F_{2}(\theta_{n}) & = & M \frac{m}{\sqrt{\lambda}}\,
\sqrt{2-\bar{\phi}_{0}^{2}}\, \int \limits_{-R/2}^{R/2} da \,
e^{i\,M(i\pi-\theta_{n})a}\textrm{sn}
\left(\frac{\bar{\phi}_{0}}{\sqrt{2}}\,m a\right) \, = \, \nonumber \\
& = & i\pi \sqrt{\frac{2}{\lambda}} M \,
\frac{1}{\sinh\left[\frac{\sqrt{2}}{m\bar{\phi}_{0}}
\textbf{K}'M(i\pi-\theta_{n})\right]}\;.
\end{eqnarray}
Therefore, the $1/\lambda$ leading contribution to the spectral
function is given by
\begin{equation}
\hat\rho(E_{n},p_{n}) \, = \, \frac{2\pi}{\lambda}\,
\delta\left(E_{n}/m\right)\delta\left(p_{n}/m\right) +
\frac{\pi^{3}}{2\lambda}\,\delta\left(\frac{E_{n}}{M}-2\right)\frac{1}{
\sinh^{2}\left[
\frac{\sqrt{2}}{m\bar{\phi}_{0}}\textbf{K}'p_{n}\right]} \;.
\end{equation}
Again, as in the Sine-Gordon case, the finite volume dependence of
these quantities comes from the factor $M(R)\textbf{K}'(k^{2})$,
where $M(R)$ is the kink mass given by (\ref{phi4classen}).

\section{Sine--Gordon model on the cylinder}\label{sectcyl} \setcounter{equation}{0}

As we have discussed, the knowledge of form factors permits to
estimate the the spectral density representation of correlation
functions at a finite volume in both integrable and
non--integrable theories. However, correlation functions need
another set of data for their complete determination, precisely
the energies of the intermediate states at a finite volume.

The paper \cite{SGscaling} has been mainly devoted to fill this
gap, that is, to face the problem of a semiclassical computation
of the energies $E_i(R)$ of vacua and excited states as functions
of the circumference $R$ of a cylindrical geometry. The analytic
form of the semiclassical scaling functions for two--dimensional
QFT admitting static kink solutions can be achieved by suitably
adapting the DHN method to the finite geometry. The example
discussed in \cite{SGscaling} is the sine--Gordon model, which is
particularly appealing for its simplified semiclassical results
whereby the significant physical effects are not masked by other
additional complications. Moreover, due to the integrable nature
of this theory, its finite size effects have been previously
studied by means of Thermodynamical Bethe Ansatz
\cite{ddv,RavaniniSG}, and it would be interesting to perform a
quantitative comparison between these results and the
semiclassical ones, in order to directly control their range of
validity. However, as already pointed out, semiclassical methods
apply not only to integrable theories and this opens the way to
describe analytically the finite size effects also in
non--integrable models.

The Sine--Gordon model (\ref{SGpot}) on a cylindrical geometry
admits quasi--periodic boundary conditions (b.c.)
\begin{equation*}
\phi(x + R,t) \,=\,\phi(x,t) + \frac{2n\pi}{\beta} \,\,\,
\end{equation*}
(the arbitrary winding number $n\in \mathbb{Z}$ originates from
the invariance of the potential (\ref{SGpot}) under
$\phi\to\phi+\frac{2n\pi}{\beta}$). In particular, since we are
interested in the one--kink sector, which is defined by $n=1$, we
will impose the b.c.
\begin{equation}
\phi(x + R,t) \,=\,\phi(x,t) + \frac{2\pi}{\beta} \,\,\,.
\label{periodicbc}
\end{equation}
The first step for applying the semiclassical method to this
problem is to find the finite size analog of the kink solution,
satisfying now the b.c.'s (\ref{periodicbc}). However, the success
in constructing the scaling functions depends on whether one is
able to solve the corresponding Schr\"{o}dinger equation
(\ref{stability}) and to derive an analytical expression for its
frequencies $\omega_k$. It turns out that the semiclassical finite
size effects in SG model are intrinsically related to the simplest
($N=1$) Lam\`{e} equation, which admits a complete analytical
study.

The subject is organized as follows: after the introductory
Section \ref{secte0reg}, where we discuss the simplest scaling
function in a finite volume in order to clarify the nature of
divergencies encountered in such computations, in Section
\ref{sectscaling} we present the complete semiclassical analysis
of the energy levels in the one--kink sector. Finally,
Section\,\ref{ffperiodiccyl} is devoted to the finite--size form
factors for the SG model on a periodic cylinder.

\subsection{Ground state energy regularization}\label{secte0reg}

As shown by eq.\,(\ref{e0}), quantum corrections to energy levels
are given by the series on the frequencies $\omega_n$. However,
this series is generally divergent (this is the usual UV
divergence in field theory) and a criterion is needed to
regularize it. It is quite instructive to consider the simplest
example where such divergence occur, i.e. in the calculation of
the ground state energy ${\cal E}_{0}^{\text{vac}}(R)$ of the
vacuum sector of the SG theory on a cylindrical geometry of
circumference $R$. This can be constructed by implementing the DHN
procedure for one of the constant solutions, for instance
$\phi_{\text{cl}}^{\text{vac}} = 0$, imposing periodic boundary
conditions for the corresponding fluctuations
$\eta^{\text{vac}}(x)$. Obviously, what comes out is nothing else
but the Casimir energy of a free bosonic field $\phi(x,t)$ with
mass $m$. In this case the frequency eigenvalues are fixed to be
\begin{equation*}
\omega_{n}\,=\,\sqrt{p_{n}^{2}+m^{2}}\,\,\,,
\end{equation*}
with $p_{n} = 2\pi n/R$ and $n=0,\pm 1,\pm 2,\ldots $.

The ground state energy has to be regularized by subtracting its
infinite--volume continuous limit: this ensures in fact the proper
normalization of this quantity, expressed by
$$
\lim_{R \rightarrow \infty} {\cal E}_0^{\text{vac}}(R)
\,=\,0\,\,\,.
$$
The ground state energy at a finite volume is
therefore defined by
\begin{equation}
{\cal E}_{0}^{\text{vac}}(R) \,=\,
\frac{1}{2}\sum\limits_{n=-\infty}^{\infty}\sqrt{\left(\frac{2 \pi
n}{R} \right)^{2}+m^{2}}\,
-\,\frac{1}{2}\int\limits_{-\infty}^{\infty}dn\,\sqrt{\left(\frac{2
\pi n}{R}\right)^{2}+m^{2}} \,\,\,.
\end{equation}
Isolating the zero mode, it can be conveniently rewritten as
\begin{equation*}
{\cal E}_{0}^{\text{vac}}(R) \,=\, \frac{m}{2} +
\frac{2\pi}{R}\sum
\limits_{n=1}^{\infty}\sqrt{n^{2}+\left(\frac{r}{2\pi}
\right)^{2}}\,
-\,\frac{2\pi}{R}\int\limits_{0}^{\infty}dn\,\sqrt{n^{2} +
\left(\frac{r}{2\pi}\right)^{2}} \;,
\end{equation*}
where $r \equiv m R$. Since the divergence of the series is due to
the large $n$ behaviour of the first two terms in the expansion
$$
\sqrt{n^{2}+\left(\frac{r }{2\pi} \right)^{2}}\,\simeq\,
n+\frac{1}{2}\left(\frac{r}{2\pi}\right)^{2}\frac{1}{n} + {\cal
O}\left(\frac{1}{n^2}\right)\;,
$$
we begin our calculation by
subtracting and adding these divergent terms to it:
\begin{eqnarray}
S(r) \,\equiv
\,\sum\limits_{n=1}^{\infty}\sqrt{n^{2}+\left(\frac{r }{2\pi}
\right)^{2}}&=&\sum\limits_{n=1}^{\infty}\left\{\sqrt{n^{2} +
\left(\frac{r}{2\pi}\right)^{2}}-n-\frac{1}{2}\left(\frac{r}{2\pi}\right)^{2}
\frac{1}{n}\right\} + \nonumber \\
&+&\,\sum\limits_{n=1}^{\infty}n\,
+\,\frac{1}{2}\left(\frac{r}{2\pi}\right)^{2}\sum\limits_{n=1}^{\infty}\frac{1}{n}
\,\,\,. \label{subseriess}
\end{eqnarray}
The first series in the right hand side of the above expression is
now convergent, whereas the last two terms should be coupled to
the analogous ones coming from the integral, whose divergencies
have to be handled in strict correspondence with those coming from
the series. Hence, by subtracting and adding the leading
divergence to the integral
\begin{eqnarray}
I(r) \,& \equiv\,&
\int\limits_{0}^{\infty}dn\,\sqrt{n^{2}+\left(\frac{r }{2\pi}
\right)^{2}} = \nonumber \\
& = & \int\limits_{0}^{\infty}dn\left\{\sqrt{n^{2}+\left(
\frac{r}{2\pi}
\right)^{2}}-n\right\}\,+\,\int\limits_{0}^{\infty}dn\,n\;,
\label{intsub1}
\end{eqnarray}
we can combine the last term in this expression with the one in
(\ref{subseriess}) and implement the well known regularization
\begin{equation}\label{firstsubtraction}
\sum_{n=0}^{\infty} n - \int_0^{\infty} n \,dn \,=\,
\lim_{\alpha\rightarrow 0}\left[ \sum_{n=0}^{\infty} n\,e^{-\alpha
n} - \int_0^{\infty} n \,e^{-\alpha n} \,dn \right] \,=\,
-\frac{1}{12} \,\,\,.
\end{equation}
However, the first term in (\ref{intsub1}) still contains a
subleading logarithmic divergence, as it can be seen by explicitly
computing the integral by using a cut-off $\Lambda$, in the limit
$\Lambda \rightarrow \infty$
\begin{equation}\label{intsub2}
\int\limits_{0}^{\Lambda}dn\left\{\sqrt{n^{2}+\left(\frac{r
}{2\pi} \right)^{2}}-n\right\} =
\frac{1}{2}\left(\frac{r}{2\pi}\right)^{2}\ln
2\Lambda\,+\,\frac{1}{4}\left(\frac{r}{2\pi}\right)^{2}\,-\,
\frac{1}{2}\left(\frac{r}{2\pi}\right)^{2}\ln\frac{r}{2\pi}
\,\,\,.
\end{equation}
This divergence can be cured by subtracting and adding the term
$\frac{1}{2} \left(\frac{r}{2\pi}\right)^2 \ln \Lambda$. By
combining this last term with its analogous in the series we have
\begin{equation}
\lim_{\Lambda\to\infty} \left( \sum_{n=1}^{\Lambda} \frac{1}{n}
\,-\,\ln \Lambda \right) \,=\, \gamma_E \,\,\,,
\label{secondsubtraction}
\end{equation}
where $\gamma_{E}$ is the
Euler-Mascheroni constant, while the remaining part of
(\ref{intsub2}) with the above subtraction is now finite.

Collecting the above results, the finite expression of the ground
state energy on a cylinder is then given by
\begin{equation}
{\cal E}_{0}^{\text{vac}}(R)\,=\,
\frac{1}{R}\left[-\frac{\pi}{6}+\frac{r}{2} +
\frac{r^{2}}{4\pi}\left(\ln\frac{r}{4\pi}+\gamma_{E}-
\frac{1}{2}\right) +\sum_{n=1}^{\infty}\left(\sqrt{(2 \pi n)^{2} +
r^{2}}-2\pi n - \frac{r^{2}}{4\pi n}\right)\right]\;.
\label{Casimiro}
\end{equation}
It is now easy to see that (\ref{Casimiro}) nicely coincide with
the analogous expression obtained in the finite--temperature
picture, given by the TBA result (\ref{TBA}) for the free case, in
which the pseudo--energy is simply given by
$\varepsilon(\theta)=mR\cosh\theta$:
\begin{equation*}
{\cal E}_{0}^{\text{vac}}(R)  \,=\,
-m\,\int_{0}^{\infty}\frac{d\theta}{2\pi}\,
\cosh\theta\,\ln\left(1-e^{-r\cosh\theta}\right)\;.
\end{equation*}
In fact, this integral formula can be expressed in terms of Bessel
functions, which admit a series representation that directly leads
to (\ref{Casimiro}) (see Ref.\,\cite{kmTBA}). Moreover, one can
also check that the above regularization scheme ensures the
agreement between the finite--volume and finite--temperature
calculations of the one--point functions $\langle \phi^{2k}
\rangle$. The interested reader can find the simplest example of
these calculations in Appendix \ref{chapfinitesize}.\ref{appreg}.

Finally, it is worth noting that the result (\ref{Casimiro}) can
also be obtained by using a simpler prescription which
automatically includes the subtraction of the various
divergencies, fastening the calculation. This consists in ignoring
the divergent part of the integral, keeping only its finite part,
and in regularizing the divergent series as
\begin{eqnarray}
\label{z(-1)}\left.\sum_{n=1}^{\infty}
n\;\right|_{\text{reg}}&=&-\frac{1}{12}
\label{riem}\,\,\,, \\
\label{z(1)}\left.\sum_{n=1}^{\infty}\frac{1}{n}\;\right|_{\text{reg}}
& = &\gamma_{E}+\ln\frac{r}{2\pi} \label{magic} \,\,\,.
\end{eqnarray}
Formula (\ref{z(-1)}) is the standard regularization of the
Riemann zeta function $\zeta(-1)$, where $\zeta(s)
\,=\,\sum_{n=1}^{\infty} \frac{1}{n^s}\;$, and usually corresponds
to normal ordering with respect to the infinite volume vacuum
(see, for instance, \cite{birrel}, chapter 4). On the contrary,
the regularization of the second series is a--priori ambiguous due
to its logarithmic divergence, and its finite value (\ref{magic})
was chosen according to the above discussion.

\subsection{Scaling functions}\label{sectscaling}

We will now develop a complete semiclassical scheme to analyse the
energy of the quantum states in the one--kink sector of SG model
on the cylinder. This can be achieved by applying the DHN method
to an appropriate kink background.

\subsubsection{Properties of the periodic kink
solution}

\begin{figure}[ht]
\hspace{3cm} \psfig{figure=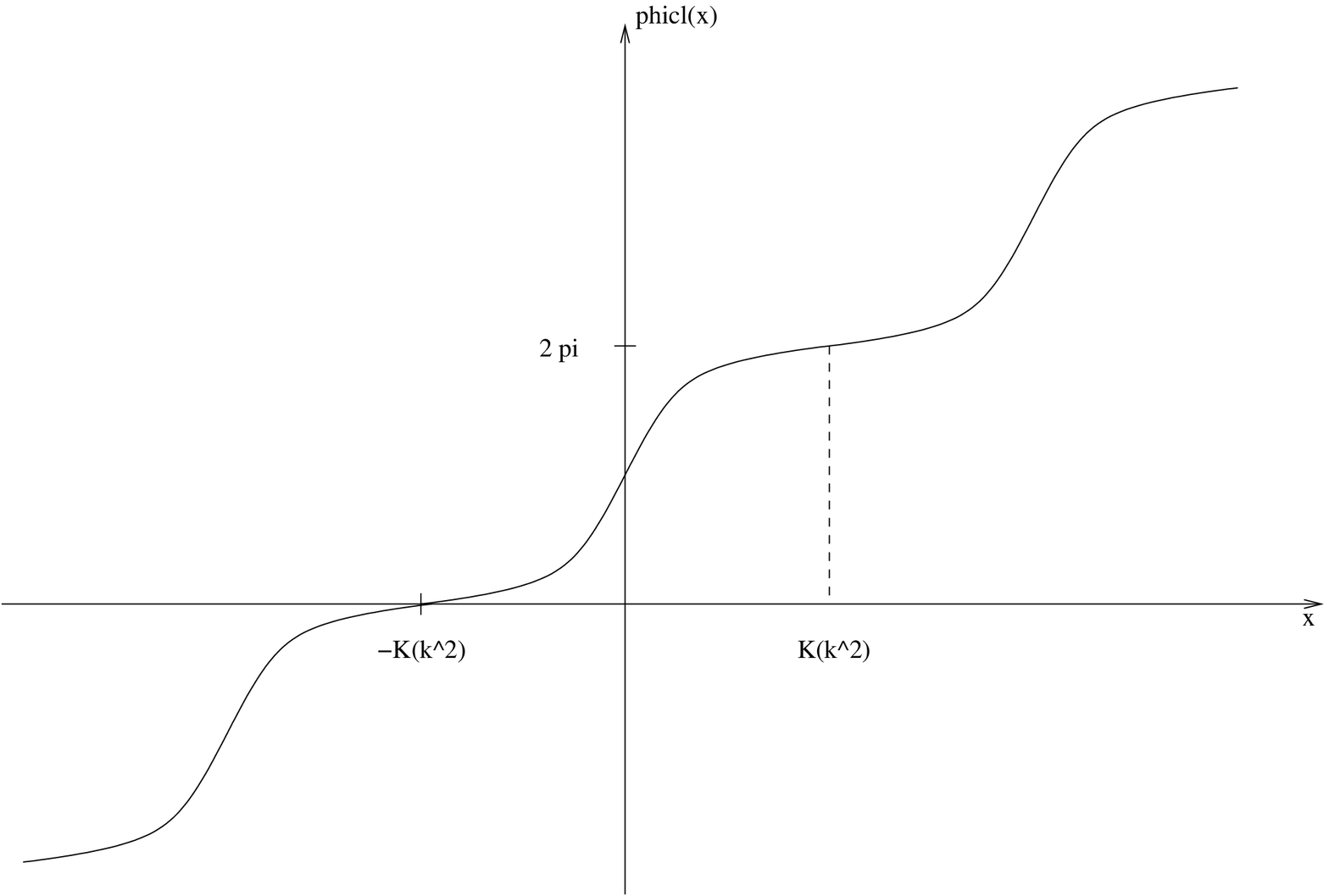,height=7cm,width=10cm}
\vspace{0.1cm} \caption{Solution of eq. (\ref{firstSG}) with $A >
0$ and $x_{0}=0$.} \label{figSGam}
\end{figure}

In order to identify a kink on the cylinder, we have to look for a
static finite energy solution of the SG model satisfying the
quasi--periodic boundary condition (\ref{periodicbc}). For the
first order equation (\ref{firstSG}) a solution which has this
property can be found for $ A > 0$. It can be expressed as
\begin{equation}
\phi_{cl}(x)\,=\,\frac{\pi}{\beta} + \frac{2}{\beta}\,
\textrm{am}\left(\frac{m (x - x_0)}{k},k^{2}\right)\;, \qquad
k^{2} \,=\,\frac{2}{2+A}\,\,\,, \label{SGam}
\end{equation}
provided the circumference $R$ of the cylinder is identified with
\begin{equation}
R \,=\,\frac{2}{m} \,k\,\textbf{K}\left(k^{2}\right)\;\,,
\label{sizeper}
\end{equation}
where $\textbf{K}(k^{2})$ denotes the complete elliptic integral
of the first kind\footnote{The definition and basic properties of
$\textbf{K}(k^2)$ and the Jacobi elliptic
amplitude$\,\text{am}(u,k^{2})\,$ can be found in Appendix
\ref{chapfinitesize}.\ref{appell}.}. The parameter $x_{0}$ in
(\ref{SGam}) represents the kink's center of mass position, and
its arbitrariness is due to the translational invariance of the
theory around the cylinder axis. The behaviour of (\ref{SGam}) as
a function of the real variable $x$ is shown in Figure
\ref{figSGam}.

\psfrag{phicl(x)}{$\beta\phi_{cl}$} \psfrag{2 pi}{$2\pi$}
\psfrag{K(k^2)}{$\textbf{K}(k^{2})$}
\psfrag{-K(k^2)}{$-\textbf{K}(k^{2})$}
\psfrag{x}{$\hspace{0.2cm}\frac{m x}{k}$}

The function (\ref{SGam}) has been first proposed in \cite{takoka}
and interpreted as a crystal of solitons in the sine-Gordon theory
in infinite volume. In our finite volume case, instead,
(\ref{SGam}) has to be seen as a single soliton defined on a
cylinder of circumference $R$ (given by eq.\,(\ref{sizeper})),
while its quasi-periodic oscillations represent winding around the
cylinder. As shown in eq.\,(\ref{sizeper}), there is an explicit
relation between the size of the system and the integration
constant $A$. It is easy to see that the infinite volume solution
(\ref{SGkinkinfvol}) is consistently recovered from (\ref{SGam})
in the limit $A \rightarrow 0$, i.e. when $R$ goes to infinity.

The classical energy of the kink on the cylinder is given by
\begin{equation}
{\cal E}_{cl}(R) \,=\, \int\limits_{-R/2}^{R/2}dx
\left[\frac{1}{2} \left(\frac{\partial \phi_{cl} }{\partial
x}\right)^{2} +
\frac{m^{2}}{\beta^{2}}(1-\cos\beta\phi_{cl})\right] \,=\,
\frac{8m}{\beta^{2}}\left[\frac{\textbf{E}(k^{2})}{k} +
\frac{k}{2}\left(1-\frac{1}{k^{2}}\right)\textbf{K}(k^{2})\right]\;,
\label{classenper}
\end{equation}
where $\textbf{E}(k^{2})$ is the complete elliptic integral of the
second kind. In the $R\rightarrow\infty$ limit (which corresponds
to $k'\rightarrow 0$, with $(k')^{2} \equiv 1-k^{2}$), ${\cal
E}_{cl}(R)$ approaches exponentially the correct value $M_{\infty}
= \frac{8m}{\beta^2}$. This can be seen expanding $\textbf{E}$ and
$\textbf{K}$ for small $k'$ (see Appendix
\ref{chapfinitesize}.\ref{appell}), and expressing the result in
terms of $mR$, which can be itself expanded in $k'$ in virtue of
the relation (\ref{sizeper}):
$$
e^{-m R}\,=\,\frac{1}{16}(k')^{2}+\cdots\;.
$$
Hence the large $R$ expansion of the classical energy is
\begin{equation}
\label{SGclassenexp} {\cal E}_{cl}(R) \,=\, M_{\infty} +
\frac{32}{\beta^{2}}\,m\, e^{-mR} + O\left(e^{-2 m R}\right)\;.
\end{equation}
We will comment more on the interpretation of this result in the
following.

Similarly, one can derive the behaviour of ${\cal E}_{cl}(R)$ for
small $r=mR$, which corresponds to the limit $A \rightarrow
\infty$ (or $k^2 \rightarrow 0$):
\begin{equation}
{\cal E}_{cl}(R) \,=\,\frac{2 \pi}{R}\,\frac{\pi}{\beta^{2}} +
m\frac{r}{\beta^2} - m\left(\frac{r}{2 \pi}\right)^3 \,
\frac{\pi}{2 \beta^2} + \cdots \label{smallLcl}
\end{equation}
This formula will be relevant in the later discussion of the UV
properties of the scaling functions.

Before moving to the quantization of the kink--background
(\ref{SGam}), it is worth recalling that another simple kind of
elliptic function, given in (\ref{SGkinkantiper}), was also
proposed in \cite{takoka} and interpreted as a crystal of solitons
and antisolitons in the infinite volume SG. This background
corresponds as well to a kink on the cylinder geometry but
satisfying the \textit{antiperiodic} boundary conditions
\[
\phi(x + R,t) \,=\,\frac{2\pi}{\beta}-\phi(x,t) \,\,\,.
\]
The associated form factors were obtained in \cite{finvolff} and
have been described in Sect.\,\ref{sectff}. Although the
quantization of this second kink solution is technically similar
to the one of (\ref{SGam}) presented here, it displays however
some different interpretative features that justify its discussion
in a separate future publication \cite{preparation}.

\subsubsection{Semiclassical quantization in finite
volume}

The application of the DHN method to the periodic kink
(\ref{SGam}) requires the solution of eq.\,(\ref{stability}) for
the quantum fluctuations $\eta_{\omega}$, which in this case takes
the form
\begin{equation}
\left\{\frac{d^{2}}{d
\bar{x}^{2}}+k^{2}\left(\bar{\omega}^{2}+1\right)-2
k^{2}\,\textrm{sn}^{2}(\bar{x},k^2)
\right\}\eta_{\bar{\omega}}(\bar{x}) \,=\, 0\;,
\label{SGsemiclper}
\end{equation}
where $\textrm{sn}(\bar{x},k^2)$ is the Jacobi elliptic function
defined in Appendix \ref{chapfinitesize}.\ref{appell}, and we have
introduced the rescaled variables
\begin{equation*}
\bar{x}\,=\,\frac{m x}{k}\,,\qquad
\bar{\omega}\,=\,\frac{\omega}{m}\;.
\end{equation*}
Due to the periodic properties of $\phi_{cl}(x)$ expressed by
eq.\,(\ref{SGam}), the boundary condition (\ref{periodicbc})
translates in the requirement for $\eta_{\bar{\omega}}(\bar{x})$
\begin{equation}
\eta_{\bar{\omega}}\left(\bar{x} + \frac{m R}{k}\right) \,=\,
\eta_{\bar{\omega}}(\bar{x}) \,\,\,. \label{periodiceta}
\end{equation}
Eq.\,(\ref{SGsemiclper}) can be cast in the so--called
Lam\'e form, which admits the two linearly independent solutions
\begin{equation*}
\eta_{\pm a}(\bar{x}) \,=\,\frac{\sigma(\bar{x}+i \textbf{K}' \pm
a)}{\sigma(\bar{x} + i \textbf{K}')}\;e^{\mp\,\bar{x}
\,\zeta(a)}\;,
\end{equation*}
where the auxiliary parameter $a$ is defined as a root of the
equation
\begin{equation*}
{\cal P}(a) \,= \, \frac{2-k^{2}}{3}-k^{2}\bar{\omega}^{2} \,\,\,.
\end{equation*}
The Weierstrass functions ${\cal
P}(u)$, $\zeta(u)$ and $\sigma(u)$ are defined in Appendix
\ref{chapfinitesize}.\ref{lame}, where the Lam\'e equation and its
relation with (\ref{SGsemiclper}) are discussed in detail.

\begin{figure}[ht]
\hspace{3cm} \psfig{figure=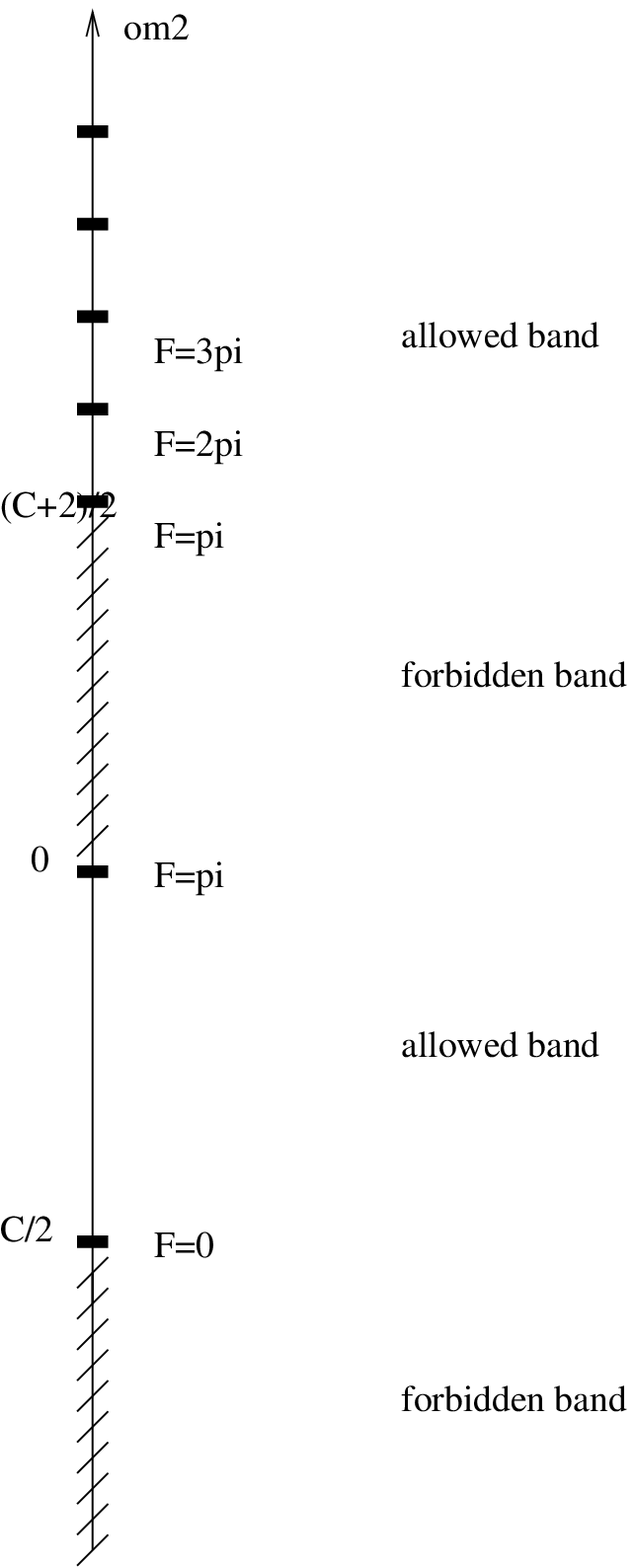,height=7cm,width=10cm}
\vspace{0.1cm} \caption{Spectrum of eq.\,(\ref{SGsemiclper})}
\label{spectrumSGper}
\end{figure}

As it is usually the case for a Schr\"odinger--like equation with
periodic potential, the spectrum of eq.\,(\ref{SGsemiclper}) has a
band structure, determined by the properties of the Floquet
exponent
\begin{equation}\label{floquetper}
F(a) \,=\, 2 i\left[\textbf{K}\, \zeta(a) -
a\,\zeta(\textbf{K})\right] \,\,\,,
\end{equation}
which is defined as the phase acquired by $\eta_{\pm a}$ in
circling once the cylinder
$$
\eta_{\pm a}(\bar{x} + 2 \textbf{K}) \,=\,e^{\pm i F(a)} \,
\eta_{\pm a}(\bar{x}) \,\,\,.
$$
We have two allowed bands for real $F(a)$, i.e.
$$
0 <\bar{\omega}^{2} < \frac{1}{k^{2}}-1 \,\,\,\,\,\,\,\,
\makebox{and} \,\,\,\,\,\,\,\, \bar{\omega}^{2} > \frac{1}{k^{2}}
\,\,\,,
$$
and two forbidden bands for $F(a)$ complex, i.e.
$$
\bar{\omega}^{2} < 0 \,\,\,\,\,\,\,\, \makebox{and}
\,\,\,\,\,\,\,\, \frac{1}{k^{2}} - 1 < \bar{\omega}^{2} <
\frac{1}{k^{2}} \,\,\,.
$$
The band $0 < \bar{\omega}^2 < \frac{1 - k^2}{k^2}$ is described
by $a = \textbf{K} + i y$, where $y$ varies between $0$ and
$\textbf{K}'$ and, correspondingly, $F(a)$ goes from $0$ to $\pi$.
The other allowed band $\bar{\omega}^2 > \frac{1}{k^2}$
corresponds instead to $a = i y$ and, by varying $y$, $F(a)$ goes
from $\pi$ to infinity, as it is shown in Fig.\,
\ref{spectrumSGper}.

\psfrag{om2}{$\bar{\omega}^{2}$}
\psfrag{(C+2)/2}{$\frac{1}{k^{2}}$}
\psfrag{0}{\hspace{-0.8cm}$\frac{1}{k^{2}}-1$} \psfrag{C/2}{$0$}
\psfrag{F=0}{$F=0$}
\psfrag{F=pi}{$F=\pi$}\psfrag{F=2pi}{$F=2\pi$}\psfrag{F=3pi}{$F=3\pi$}
\psfrag{allowed band}{allowed band}\psfrag{forbidden
band}{forbidden band}

By imposing the periodic boundary conditions (\ref{periodiceta})
on the fluctuation $\eta(\bar{x})$, one selects the values of
$\bar{\omega}^{2}$ for which the Floquet exponent is an even
multiple of $\pi$, thus making the spectrum of
eq.\,(\ref{SGsemiclper}) discrete. These eigenvalues are
$\bar{\omega}_{0}^{2}=0$, which is the zero mode associated with
translational invariance and has multiplicity one, and the
infinite series of points
\begin{equation}
\bar{\omega}_{n}^{2} \,\equiv \,\frac{1}{k^{2}}\,
\left[\frac{2-k^{2}}{3} - {\cal P}(i y_{n})\right]\;
\label{omeganper}
\end{equation}
with multiplicity two, placed in the highest band
$\bar{\omega}^{2}
> \frac{1}{k^{2}}$, with $y_{n}$ determined by the equation
\begin{equation}
F(iy_{n}) \,=\, 2\textbf{K}\,i\,\zeta(i y_{n})+2
y_{n}\,\zeta(\textbf{K}) \,=\, 2 n\pi\;, \qquad\quad n=1,2,\ldots
\label{ynper}
\end{equation}
In the IR limit ($A \rightarrow 0 $) the spectrum goes to the one
related to the standard background (\ref{SGkinkinfvol}): the
allowed band $0 < \bar{\omega}^{2} < \frac{1}{k^{2}}-1$, in fact,
shrinks to the eigenvalue $\bar{\omega}_{0}^{2} = 0$, while the
other allowed band $\bar{\omega}^{2} > \frac{1}{k^{2}}$ merges in
the continuous part of the spectrum $\bar{\omega}_{q}^{2} =
1+q^{2}$. Therefore, by recalling the interpretation of the
continuous spectrum in infinite volume as representing the
scattering states of the kink with breathers, we can now give to
the requirement (\ref{ynper}) the physical meaning of a
quantization condition for the momenta of the states constituted
by a kink and a breather on the finite volume, analogous to the
Bethe ansatz equation (\ref{momentumquant}) but valid for any
value of the size of the system. In particular, using the
small--$k'$ expansions of the elliptic integrals and Weierstrass
functions presented in Appendices
\ref{chapfinitesize}.\ref{appell} and
\ref{chapfinitesize}.\ref{lame}, it easy to check that for large
$R$ the condition (\ref{ynper}) indeed reduces to the form
$$
m R \,q\,+\,\delta(q)\,=\,2 n\pi \;,
$$
where $m q$ represents the momentum of the lightest breather, and
$\delta(q)$ is the semiclassical phase shift (\ref{phaseshiftSG}).

\begin{figure}[ht]
\begin{center}
\psfig{figure=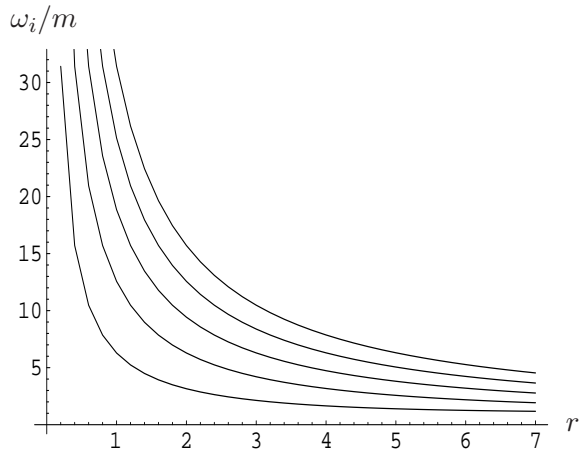,height=6cm,width=8cm} \caption{The first
few levels defined in (\ref{omeganper})}\label{figomegai}
\end{center}
\end{figure}

It is useful to note that, although the $R$ dependence of the
frequencies (\ref{omeganper}) is quite implicit, since it passes
through the inversion of eq.\,(\ref{sizeper}), nevertheless these
are analytic functions of $R$ and it is extremely simple to plot
them. The corresponding curves, shown in Figure \ref{figomegai},
provide an important piece of information, since they are nothing
else but the energies of the excited states with respect to their
ground state ${\cal E}_0(R)$.

\psfrag{omega1}{$\omega_{i}/m$}\psfrag{ell}{$r$}

\begin{figure}[ht]
\begin{center}
\psfrag{omega1}{$\omega_{1}/m$}\psfrag{ell}{$r$}
\psfig{figure=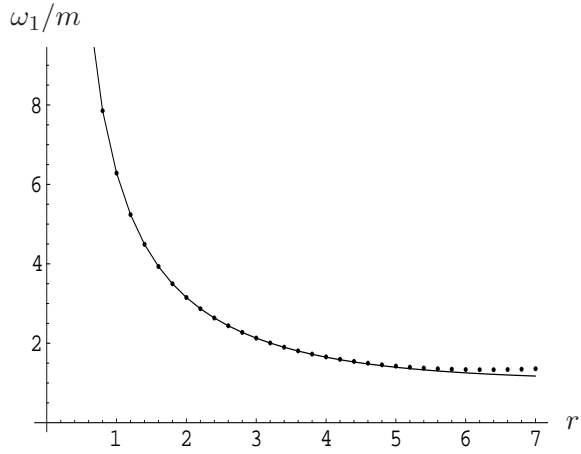,height=6cm,width=8cm}
\end{center}
\caption{Comparison between the exact energy level $\omega_{1}/m$
given by (\ref{omeganper}) (continuous line) and the approximate
expression (\ref{omeganexp}) (dotted line).}\label{figcompar}
\end{figure}

To complete the analysis, it remains then to determine the finite
volume ground state energy ${\cal E}_0(R)$ of the kink sector. In
analogy with the infinite volume case (see
eq.\,(\ref{subinfvol})), this is defined by
\begin{equation}
{\cal E}_0(R) \,=\, {\cal E}_{cl}(R) +  \sum_{n=1}^{\infty}
\omega_n(R) - \frac{\delta
\mu^{2}}{\beta^{2}}\int\limits_{-R/2}^{R/2}dx
\left[1-\cos\beta\phi_{cl}\right] - {\cal E}_0^{\textrm{vac}}(R)
\,\,\,. \label{grstatefinal}
\end{equation}
Before commenting in
detail each of these terms, let's focus first on the main problem
in deriving a closed expression for ${\cal E}_0(R)$, which
consists in the evaluation of the infinite sum on the frequencies
$\omega_n(R)$ or, better, in isolating its finite part. We need
therefore a method for solving the transcendental equation
(\ref{ynper}) for $y_n(k^2)$ in order to make the expression
(\ref{omeganper}) for the frequencies $\omega_n(k^2)$ explicit. As
we have already seen for the classical energy, two kinds of
expansion are possible, one in the elliptic modulus $k$ and the
other in the complementary modulus $k'$, which are efficient
approximation schemes in the small and large $r$ regimes,
respectively. Here for simplicity we only present the small $r$
expansion. By taking into account the series expansion in $k$ for
$\textbf{K}$, $\zeta(u)$ and ${\cal P}(u)$ (see Appendices
\ref{chapfinitesize}.\ref{appell} and
\ref{chapfinitesize}.\ref{lame}), we are led to look for a
solution of eq.\,(\ref{ynper}) in the form
$$
y_n(k^2) \,=\,\sum_{s=0}^{\infty} (k^2)^s \,y_n^{(s)} \,\,\,.
$$
Here we give the result for the first few coefficients
$y_n^{(s)}$, $s=0,1,2$:
\begin{eqnarray*}
y_n^{(0)} & = & \textrm{arctanh}\frac{1}{2 n} \,\,\,,\nonumber \\
y_n^{(1)} & = & \frac{1}{4} \; y_n^{(0)} \,\,\,,\\
y_n^{(2)} & = & \frac{9}{64}\; y_n^{(0)} - \frac{n}{16 (4 n^2
-1)^2}\,\,\,, \nonumber
\end{eqnarray*}
which are those relevant in the later analysis of the UV
properties of the scaling function. As a consequence, we obtain
the following simple expression for the frequencies:
\begin{equation*}
\frac{\omega_n}{m} \,=\, \frac{2 n}{k}\,\left[1 - \frac{k^2}{4} -
\frac{k^{4}}{64}\;\frac{20 n^2 -9}{4 n^2 -1} + O(k^6) \right]
\,\,\,.
\end{equation*}
Comparing it order by order with the small--$k$ expansion of
eq.\,(\ref{sizeper})
\begin{equation*}
r \,=\, mR \,=\, \pi \, k \,
\left[1+\frac{k^{2}}{4}+\frac{9}{64}\,k^{4}+O(k^{6})\right]\;,
\end{equation*}
we finally obtain the explicit $R$--dependence
\begin{equation}\label{omeganexp}
\frac{\omega_{n}(R)}{m}\,=\,\frac{2\pi}{r}
\,n+\left(\frac{r}{2\pi}\right)^{3} \,\frac{n}{4n^{2}-1} + \ldots
\end{equation}
It is worth noting that this series expansion in $r$, which can be
easily extended up to desired accuracy, efficiently approximates
the exact energy levels also for rather large values of the
scaling variable. Fig.\,\ref{figcompar} shows a numerical
comparison between the first energy level given by
(\ref{omeganper}) and its approximate expression
(\ref{omeganexp}), and for the higher levels it is possible to see
that the agreement is even better.

With the above analysis, the sum over frequencies in the ground
state energy (\ref{grstatefinal}) takes the form
\begin{equation}
\sum\limits_{n=1}^{\infty}\,\frac{\omega_{n}(R)}{m}\,=\,\frac{2\pi}{r}
\,\sum\limits_{n=1}^{\infty}\,n+\left(\frac{r}{2\pi}\right)^{3}
\,\sum\limits_{n=1}^{\infty}\,\frac{n}{4n^{2}-1} + \ldots\;
\end{equation}
As we will see below, the subtraction of counterterm and vacuum
energy in (\ref{grstatefinal}) leads to the cancellation of all
the divergencies, producing a finite expression for the ground
state energy in the kink sector.

Moving now to the analysis of the remaining terms in
(\ref{grstatefinal}), a similar series expansion can be easily
performed on each of them. The classical energy ${\cal
E}_{cl}(R)$, given in eq.\,(\ref{classenper}), has already be
treated in this way in eq.\,(\ref{smallLcl}). The finite volume
counterterm (C.T.), where the one--loop mass renormalisation is
given by
$$
\delta \mu^{2}\,=\,\left.-\frac{m^{2}\beta^{2}}{8\pi}\,
\frac{2\pi}{R}\sum_{n=-\infty}^{\infty}
\frac{1}{\sqrt{m^{2}+\frac{(2 n \pi)^{2}}{R^{2}}}}\right.\,\,\,
$$
and $\phi_{cl}$ is given by (\ref{SGam}), takes the explicit form
$$
\text{C.T.}\,=\, m \,
\left[k\textbf{K}\left(k^{2}\right)-\frac{\textbf{K}\left(k^{2}\right)-
\textbf{E}\left(k^{2}\right)}{k}\right]
\left.\sum\limits_{n=-\infty}^{\infty}\frac{1}{\sqrt{(2n\pi)^{2}+r^{2}}}\right.
\;.
$$
The first terms of its expansion in $R$ are then
\begin{equation}
\frac{\text{C.T.}}{m}\,=\,
\frac{1}{4}\,+\,\frac{r}{4\pi}\left.\sum\limits_{n=1}^{\infty}\frac{1}{n}\right.
\,-\,\frac{r^{2}}{32\pi^{2}}\,-\,\frac{1}{4}\left(\frac{r}{2\pi}\right)^{3}
\left.\sum\limits_{n=1}^{\infty}\left(\frac{1}{n}+\frac{1}{n^{3}}\right)\right.
+ \ldots \;,
\end{equation}
Finally, the vacuum energy ${\cal E}_0^{\textrm{vac}}(R)$ is the
one precisely computed in Sec.\,\ref{secte0reg}. Since its role is
to cancel certain divergencies present in the other terms of
${\cal E}_0(R)$, in complete analogy with the infinite volume case
(see eq.\,(\ref{subinfvol})), we will now consider its \lq\lq
naive" formulation, given by
\begin{equation}
\frac{{\cal E}^{\text{vac}}_{0}(R)}{m}\, =
\,\frac{1}{2m}\sum\limits_{n=-\infty}^{\infty}\sqrt{\left(\frac{2n\pi}{R}\right)^{2}+m^{2}}
\,=\,\frac{1}{2}\,+\,\frac{2\pi}{r}
\left.\sum\limits_{n=1}^{\infty}n\right.\,+\,\frac{r}{4\pi}
\left.\sum\limits_{n=1}^{\infty}\frac{1}{n}\right.
\,-\,\frac{1}{8}\left(\frac{r}{2\pi}\right)^{3}
\left.\sum\limits_{n=1}^{\infty}\frac{1}{n^{3}}\right. + \ldots
\end{equation}
Hence, in the final expression for the ground state energy all the
divergent series present in the sum over frequencies, in the
counterterm and in the vacuum energy cancel out, and one obtains
\begin{equation}\label{grstatefinalexp}
\frac{{\cal E}_{0}(R)}{m} \,=\,
\,\frac{2\pi}{r}\,\frac{\pi}{\beta^{2}}\,-\,\frac{1}{4}\,+
\,\frac{1}{\beta^{2}}\,r
\,-\,\frac{1}{8}\,\left(\frac{r}{2\pi}\right)^{2}\,-\,
\left(\frac{r}{2\pi}\right)^{3} \left[\frac{1}{8}\, \zeta(3)
-\frac{1}{4} (2\,\log 2 -1) - \frac{\pi}{2\beta^{2}}\right] +
\ldots\;,
\end{equation}
where we have used \cite{GRA}
\[
\sum_{n=1}^{\infty} \frac{2 n^2 -1}{8 n^3 (4 n^2 -1)} \,=\,
\frac{1}{8}\, \zeta(3) -\frac{1}{4} (2 \,\log 2 -1) \,\,\,
\]
in order to evaluate explicitly the coefficient of the $r^{3}$
term.

Repeating the above calculations, one can also easily write the
finite expressions of the excited energy levels (\ref{tower}),
whose series expansion in $r$ is given by
\begin{eqnarray}\label{eiexp}
\frac{{\cal E}_{\{k_{n}\}}(R)}{m} & = &
\frac{2\pi}{r}\,\left(\frac{\pi}{\beta^{2}} +
\sum\limits_{n}k_{n}\,n\right)
\,-\,\frac{1}{4}\,+\,\frac{1}{\beta^{2}}\,r
\,-\,\frac{1}{8}\,\left(\frac{r}{2\pi}\right)^{2}\,+\, \\
&& \hspace{-3mm} -\, \left(\frac{r}{2\pi}\right)^{3}
\left[\frac{1}{8} \zeta(3) -\frac{1}{4}(2 \,\log 2 -1)
-\frac{\pi}{2\beta^{2}} +
\sum\limits_{n}k_{n}\,\frac{n}{4n^{2}-1}\right] + \ldots \nonumber
\end{eqnarray}
where $\{k_{n}\}$ is a set of integers defining a particular
excited state of the kink.

\subsubsection{UV--IR correspondence}

The semiclassical quantization of the periodic kink (\ref{SGam})
provides us with analytic expressions, albeit implicit, for the
scaling functions in the kink sector for arbitrary values of the
scale $r = m R$. These quantities control analytically the
interpolation between the Hilbert spaces of the ultraviolet (UV)
and infrared (IR) limiting theories. It is worth noting that,
although we obtain them in the framework of a particle--like
description proper of the IR limit, the kink background
(\ref{SGam}) is intrinsically formulated on a finite size, leading
to the possibility of extracting UV data. Hence, it is important
to check whether our scaling functions reproduce both the expected
results for the IR ($r \rightarrow \infty$) and UV ($r \rightarrow
0$) limits.

\vspace{0.5cm}

Concerning the IR behaviour, we have already seen that in the $R
\rightarrow \infty$ limit all the quantities in exam, i.e. the
classical solution, its classical energy and the stability
frequencies, correctly reach their asymptotic values. In addition,
we can now analyse the asymptotic approach of the kink mass to the
infinite--volume limit, in order to compare it with the known
scattering data of the SG theory, according to L\"{u}scher's
theory described in Sect.\,\ref{finvolintro}. Restricting for
simplicity our analysis to the leading term in $\beta$ in the kink
mass (which, in our approach, is simply given by the classical
energy), we have then to compare the expansion presented in
(\ref{SGclassenexp}) with the term that dominates L\"{u}scher's
result for small $\beta$. This is given by formula
(\ref{luscherR}) evaluated for the process in which a kink splits
into the lightest breather and another kink, i.e.
\begin{equation}\label{luschersmallbeta}
\Delta M(R)\,=\,-m_{b}\sin u_{kb}^{k} \,R_{k b k}\;e^{-m_{b} \sin
u_{kb}^{k} R} +\ldots\;,
\end{equation}
Recalling the kink--breather $S$--matrix \cite{zams}
$$
S_{k b}(\theta) \,=\,\frac {\sinh \theta + i
\cos\frac{\gamma}{16}} {\sinh \theta - i \cos\frac{\gamma}{16}}
\,\,\,\,\,\, , \,\,\,\,\,\, \gamma
\,=\,\frac{\beta^2}{1-\beta^2/8\pi}
$$
and selecting its $s$-channel pole $\theta^* = i u_{kb}^k = i
\left(\frac{\pi}{2} + \frac{\gamma}{16}\right)$, we
find\footnote{The negative sign of the residue is due to the odd
parity of the lightest breather, which implies
$\Gamma_{k}^{kb}=-\Gamma_{kb}^k$ in (\ref{resS})}
$$
R_{k b k} \,=\, - 2\,\textrm{cotg} \frac{\gamma}{16} \,\,\,.
$$
Substituting in (\ref{luschersmallbeta}), for small $\beta^2$ we
have
$$
\Delta M(R)\,=\,m\,\frac{32}{\beta^2}\,e^{-mR} + \ldots \;,
$$
which therefore reproduces eq.\,(\ref{SGclassenexp}). It is a
remarkable fact that the classical energy alone, being the leading
term in the mass for $\beta^2 \rightarrow 0$, contains the IR
scattering information which controls the large--distance
behaviour of ${\cal E}_0(R)$.

\vspace{0.5cm}

As we have seen in Sect.\,\ref{finvolintro}, the UV behaviour for
$r \rightarrow 0$ of the ground state energy $E_0(R)$ of a given
off--critical theory is related instead to the Conformal Field
Theory (CFT) data $(\Delta,\bar{\Delta},c)$ of the corresponding
critical theory and to the bulk energy term as
\begin{equation}
E_0(R) \,\simeq\,\frac{2\pi}{R} \left(h + \bar{h} -
\frac{c}{12}\right) + B R + \cdots\; \label{UVCFT}
\end{equation}
where $c$ is the central charge, $h + \bar{h}$ is the lowest
anomalous dimension in a given sector of the theory and $B$ the
bulk coefficient. For the Sine--Gordon model the bulk energy term
is given by \cite{ddv,bulkterm}
\begin{equation}
B\,=\, 16\,\frac{m^{2}}{\gamma^{2}}\,\tan\frac{\gamma}{16}\; ,
\label{bulktermm}
\end{equation}
while its UV limit is described by the CFT given by the gaussian
action (\ref{gaussian}) with the normalization constant fixed to
the value\footnote{Note that the usual normalization adopted in
the CFT literature is instead $g = \frac{1}{4\pi}$.} $g=1$
$$
{\cal A}_{\textrm{G}} \,=\,\frac{1}{2}\, \int d^2 x\;
\partial_{\mu} \phi \,\partial^{\mu} \phi \;,
$$
and the free bosonic field compactified on a circle of radius
${\cal R}=\frac{1}{\beta}$ (see eq.(\ref{compact})). We have
described the main properties of this CFT in Sect.\,\ref{secCFT}.
In particular, the central charge takes the value $c=1$, and the
theory is divided into sectors labelled by two integers $s$ and
$n$. In each sector, the state with lowest anomalous dimension
(\ref{dimvertex}) is created by the vertex primary operator
$V_{s,n}\,$, introduced in (\ref{vertex}). It is worth noticing
that the SG action can be seen as the perturbed gaussian CFT of
the form (\ref{CFTpert}) with $\lambda=\frac{m^2}{\beta^{2}}$ and
$\Phi=V_{1,0}+V_{-1,0}$. These perturbing operators carry
anomalous dimension $\Delta_{\pm 1,0}=h_{\pm 1,0}+\bar{h}_{\pm
1,0}=\frac{\beta^2}{4\pi}$, which is zero in the semiclassical
limit, determining the form of the scaling variable as $r=mR$.

The vacuum sector is described by $s = n = 0$, with $h_{vac} +
\bar{h}_{vac} = 0$, while the kink sector, defined by the boundary
condition (\ref{periodicbc}), naturally corresponds to $s=0,n=1$,
in which the lowest anomalous dimension is
\begin{equation}\label{dimsol}
\Delta_{0,1} =h_{0,1}+ \bar{h}_{0,1} \,=\, \frac{\pi}{\beta^2}
\,\,\,.
\end{equation}
The conformal vertex operator $V_{0,1}$ has been put in exact
correspondence with the soliton--creating operator of SG in
Ref.\,\cite{BerLec}.

The question to be addressed now is whether the small $r$
expansion of ${\cal E}_0^{vac}(R)$ and ${\cal E}_0(R)$ given by
eqs.\,(\ref{Casimiro}) and (\ref{grstatefinalexp}) reproduces, in
semiclassical approximation, the above data controlling the UV
limit of SG model. For the vacuum sector, comparing
(\ref{Casimiro}) with (\ref{UVCFT}), we correctly obtain $c = 1$
and $h_{vac} = \bar{h}_{vac} =0$. We do not expect, however, to
obtain the bulk term $B$ relative to SG model by looking at
(\ref{Casimiro}), simply because the semiclassical expression of
the ground state energy in the vacuum sector applies equally well
to any theory which has a quadratic expansion near the vacuum
state. Namely, apart from the value of the mass $m$,
eq.\,(\ref{Casimiro}) is a universal expression that does not
refer then to SG model. The kink scaling function
(\ref{grstatefinalexp}) has instead a richer structure. The
obtained scaling dimension
$$
h+\bar{h} \,=\,\frac{\pi}{\beta^{2}}\;
$$
is the expected one, given by (\ref{dimsol}), for the
soliton-creating operator in Sine-Gordon, while the central charge
contribution $c=1$ is absent, simply because in
(\ref{grstatefinalexp}) we have subtracted the vacuum ground state
energy from the kink one\footnote{The value $c=1$, coming out from
the regularization of the leading term of the series on the
frequencies (\ref{omeganexp}), is in fact exactly cancelled by the
same term in the vacuum energy.}. Moreover, the bulk coefficient
$B = \frac{m^{2}}{\beta^{2}}$ present in (\ref{grstatefinalexp})
correctly reproduces the semiclassical limit of the exact one,
given in eq.\,(\ref{bulktermm}). In principle, this bulk term
should be present in all the energy levels, included the ground
state energy in the vacuum sector, but its non--perturbative
nature makes impossible to see it in the semiclassical expansion
around the vacuum solution, which is in fact purely perturbative.
Hence it is not surprising that to extract the bulk energy term we
have to look at the kink ground state energy, in virtue of the
non--perturbative nature of the corresponding classical solution.
Finally, the expression (\ref{eiexp}) for the excited energy
levels explicitly show their correspondence with the conformal
descendants of the kink ground state. In fact, their anomalous
dimension is given by
\begin{equation}
h_{\{k_{n}\}}+\bar{h}_{\{k_{n}\}}=\frac{\pi}{\beta^{2}}+
\sum\limits_{n}k_{n}\,n\;.
\end{equation}

\vspace{0.5cm}

The successful check with known UV and IR asymptotic behaviours
confirms the ability of the semiclassical results to describe
analytically the scaling functions of SG model in the one--kink
sector. It would be interesting to further test them at arbitrary
values of $r$ through a numerical comparison with the results of
\cite{ddv, RavaniniSG} in an appropriate range of parameters. This
was not pursued here because the results presently available in
the literature were obtained for values of $\beta$ which are
beyond the semiclassical regime and moreover the energy levels
were plotted as functions of a different scaling variable, i.e.
the one defined in terms of the kink mass. We hope however to come
back to this problem in the future.

\subsection{Form factors and correlation functions}\label{ffperiodiccyl}

The semiclassical scaling functions, derived in
Sect.\,\ref{sectscaling}, provide an important information about
the finite size effects in SG model. As in the infinite volume
case, however, the complete description of the finite volume QFT
requires to find, in addition to the energy eigenvalues
(\ref{eiexp}), the kink form factors and the correlation functions
of local operators. This section is devoted to the analysis of
this problem, i.e. to the determination of the finite volume form
factors and the corresponding spectral functions.

The form factors can be determined with the same procedure
described in Sect.\,\ref{sectff} and applied to the SG model and
the broken $\phi^{4}$ field theory, both defined on a cylindrical
geometry with \textit{antiperiodic} boundary conditions. In what
follows we will apply it instead to the case of SG model with
\textit{periodic} boundary conditions. The corresponding finite
volume form factor (\ref{ff}) can be written in terms of the
soliton background (\ref{SGam}):
\begin{equation}\label{SGff}
f(\theta_{n})=M\int \limits_{-R/2}^{R/2}da\,e^{i\,M\theta_{n}
a}\left[\frac{\pi}{\beta}+\frac{2}{\beta}\,\textrm{am}\left(\frac{m
x}{k},k^{2}\right)\right] =
\end{equation}
\begin{equation*}
=\frac{2\pi}{\beta}\left\{\frac{M}{2}\,R\,
\delta_{M\theta_{n},0}-i\,\frac{1-\delta_{M\theta_{n},0}}{\theta_{n}}
\left[\cos\left(M\theta_{n}\,R/2\right)-
\frac{\sin\left(M\theta_{n}\,R/2\right)}{M\theta_{n}R/2}\right]
+i\,\frac{1}{\,\theta_{n}\,\cosh\left(k\,\textbf{K}'\frac{M}{m}\,
\theta_{n}\right)}\right\}\;.
\end{equation*}
In order to obtain this result one has to compare the inverse
Fourier transform (\ref{inverseff}) with the expansion \cite{GRA}
\begin{equation*}
\textrm{am}(u) \,=\,\frac{\pi \,u}{2\textbf{K}}+
\sum\limits_{n=1}^{\infty} \frac{1}
{\,n\,\cosh\left[n\pi\frac{\textbf{K}'}{\textbf{K}}\right]}\,
\sin\left[n\pi\frac{u}{\textbf{K}}\right]\;.
\end{equation*}

The form factor (\ref{SGff}) has the correct IR limit\footnote{The
functions $\frac{\cos(x R/2)}{x}$ and $\frac{\sin(x R/2)}{x^{2}
R/2}$ can be shown to tend to zero in the distributional sense for
$R\rightarrow\infty$.}, and leads to the following expressions for
$F_{2}(\theta)$ and for the spectral function\footnote{Here we are
considering the matrix elements on the antisymmetric combinations
of kink and antikink.}:
\begin{eqnarray}\nonumber
F_{2}(\theta_{n})&=&\frac{4\pi i}{\beta\, \hat{\theta}_{n}}\,
\left\{\frac{1}{\cosh\left[k\,\textbf{K}'\,\frac{M}{m}\,\hat{\theta}_{n}\right]}\,
\, + \right.\\
&&\left.-\,\left(1-\delta_{\hat{\theta}_{n},0}\right)
\left[\cos\left(M \,\hat{\theta}_{n}\,R/2\right) -
\frac{\sin\left(M \hat{\theta}_{n}\,R/2\right)}
{M \,\hat{\theta}_{n} \,R/2}\right]\right\}\label{SGfinvolf2} \\
\hat\rho(E_{n},p_{n}) & = &
4\pi^{3}\,\delta\left(\frac{E_{n}}{M}-2\right)\frac{1}{\beta^{2}(p_{n})^{2}}
\left\{\frac{1}{\cosh\left[\frac{k\,\textbf{K}'}{m}\,p_{n}\right]}
\,\, + \right.
\nonumber \\
&& \left. - \left(1-\delta_{p_{n},0}\right)
\left[\cos\left(p_{n}\,R/2\right)-\frac{\sin\left(p_{n}\,R/2\right)}
{p_{n}R/2}\right]
 \right\}^{2}\;,
\label{SGfinvolrho}
\end{eqnarray}
where $\hat\theta = i \pi - \theta$. Note that the finite volume
dependence of both the form factor (\ref{SGfinvolf2}) and the
spectral function (\ref{SGfinvolrho}) is not restricted to the
second term only. The $k \,\textbf{K}'(k^{2})\,M(R)$ factor in the
first term carries the main $R$-dependence, although it is not
manifest but implicitly defined by eq.\,(\ref{sizeper}).

Another quantity of interest is the two--point function $\langle\,
0\,|\, \varepsilon(x)\varepsilon(0)\,|\, 0\,\rangle$ of the energy
density operator. One can calculate it by evaluating its spectral
function
\begin{equation*}
\rho^{(\varepsilon)}(p^{2})=\int\limits_{-R/2}^{R/2}dx\,\langle\,
0\,|\, \varepsilon(x)\varepsilon(0)\,|\, 0\,\rangle \, e^{-ip\cdot
x}
\end{equation*}
in terms of the form factors of $\varepsilon(x)$, similarly to
what we have done for the field $\phi$ above. In order to find the
semiclassical form factor
\begin{equation*}
f_{\varepsilon}(\theta_{n})\,=\,\langle
p_{n_{2}}|\,\varepsilon(0)\,|p_{n_{1}}\rangle\;,
\end{equation*}
we need to compute the Fourier transform of
\begin{equation*}
\varepsilon(\phi_{\text{cl}})\,=\,\frac{2
m^{2}}{\beta^{2}k^{2}}(1+k^{2})-\frac{4
m^{2}}{\beta^{2}}\,\text{sn}^{2}\left(\frac{m x}{k}\right)\;.
\end{equation*}
This can be easily obtained from the following expansion
\begin{equation}\label{expffeps}
\text{sn}^{2}u\,=\,\frac{1}{k^{2}\textbf{K}}\left\{\textbf{K}-\textbf{E}-\frac{\pi^{2}}{\textbf{K}}
\sum\limits_{n=1}^{\infty}\frac{n\,\cos\frac{n\pi
u}{\textbf{K}}}{\sinh\frac{n\pi\textbf{K}'}{\textbf{K}}}\right\}\;,
\end{equation}
and we finally have
\begin{equation}
f_{\varepsilon}(\theta_{n})\,=\,M^{2}\left\{\delta_{M\theta_{n},0}\,
+\,\frac{4\pi}{\beta^{2}}\,\frac{\theta_{n}}{\sinh\left(k\,\textbf{K}'\frac{M}{m}\,\theta_{n}\right)}\right\}\;.
\end{equation}
The corresponding semiclassical spectral function is thus given by
\begin{equation}
\hat\rho^{(\varepsilon)}(E_{n},p_{n}) =
\frac{4\pi^{3}}{\beta^{4}}\,\delta\left(\frac{E_{n}}{M}-2\right)\frac{p_{n}^{2}}
{\sinh^{2}\left(\frac{k\,\textbf{K}'}{m}\,p_{n}\right)}\;.
\end{equation}

It is worth mentioning that it is also possible to obtain the
two--point functions of certain vertex operators
$V_{b}^{\pm}(x,t)=e^{\pm i\beta b \phi(x,t)}$ (for $b =
\frac{1}{2},1,\frac{3}{2},2,...$), since the required Fourier
expansion formulas of the type (\ref{expffeps}) are known in these
cases \cite{whit}.

\section{Strip geometry}\label{sectstrip} \setcounter{equation}{0}

In our study of finite--size effects, we have focused so far only
on the cylindrical geometry, since the relevant features we were
interested to underline can be already observed in this simple
case. However, an even richer phenomenology is produced by
including non--trivial boundaries, and the semiclassical
techniques are also suited to describe such geometries. In this
Section, after a brief overview of boundary effects in CFT, we
will describe how the semiclassical method has been applied in
\cite{SGstrip} to study a quantum field theory defined on a strip
of width $R$, with certain boundary conditions at its edges. In
particular, the example discussed is the sine--Gordon model
subjected to Dirichlet boundary conditions at both edges of the
strip.

\subsection{Boundary effects in CFT}

With the purpose of giving an intuitive idea of boundary effects
in QFT, we focus here on the case of conformally invariant
theories, due to the powerful analytical techniques available in
this situation. We limit ourselves to a brief description of the
main results, referring to the original literature for a complete
discussion \cite{finitesize,cardy}. Furthermore, it should be
mentioned that several exact results have been obtained also for
integrable QFT with boundary, in virtue of the factorization
property of the scattering in the bulk and off the boundary
\cite{ghoshzam}.

In a critical system with a boundary, conformal transformations
must map the boundary onto itself and preserve the boundary
conditions. As a consequence, holomorphic and antiholomorphic
fields no longer decouple, and only half of the conformal
generators remain. This can be easily seen in the prototype
geometry for a two-dimensional system with boundary, i.e. the
upper half plane. In fact, infinitesimal local conformal
transformations of the form $z\rightarrow z+\epsilon (z)$ map the
real axis onto itself if and only if $\epsilon
(\bar{z})=\bar{\epsilon} (z)$, i.e. $\epsilon$ is real on the real
axis, and this constraint eliminates half of the conformal
generators.

A powerful tool for describing the half--plane geometry is the
so--called method of images. It consists in regarding the
dependence of the correlators on antiholomorphic coordinates
$\bar{z}_{i}$ on the upper half-plane as a dependence on
holomorphic coordinates $z_{i}^{*}=\bar{z}_{i}$ on the lower
(unphysical) half-plane. Having introduced in this way a mirror
image of the system, the definition of the stress--energy tensor
$T(z)$ can be extended into the lower half-plane as
$T(z^{*})=\overline{T}(z)$. Such an extension is compatible with
the boundary conditions, because
\begin{equation}\label{Tboundary}
T|_{B}=\overline{T}|_{B}
\end{equation}
at the boundary, which in cartesian coordinates means
$T_{xy}|_{B}=0$, i.e. there is no energy or momentum flux across
the surface. In this framework, it is possible to show that a
$n$--point correlation function
$\;\langle\,\phi_1(z_1)\,...\,\phi_n(z_n)\,\rangle_{hp}\;$ on the
half--plane can be identified with the $2n$--point function on the
entire plane
$\;\langle\,\phi_1(z_1)\,\phi_1(z^*_1)\,...\,\phi_n(z_n)\,\phi_n(z^*_n)\,\rangle_{p}\;$,
in which a mirror image is associated to each field. The role of
the boundary is therefore simulated by the interaction between
mirror images of the same holomorphic field.

\begin{figure}[h]
\psfrag{x}{$x$}\psfrag{t}{$t$}\psfrag{R}{$R$}\psfrag{T}{\hspace{-0.6cm}$z\to
w(z)$}\psfrag{H}{$H_{\alpha\beta}$}\psfrag{h}{$H$}\psfrag{A}{$|\,\alpha\,\rangle$}\psfrag{B}{$|\,\beta\,\rangle$}
\psfrag{a}{$\alpha$}\psfrag{b}{$\beta$}
\hspace{0cm}\psfig{figure=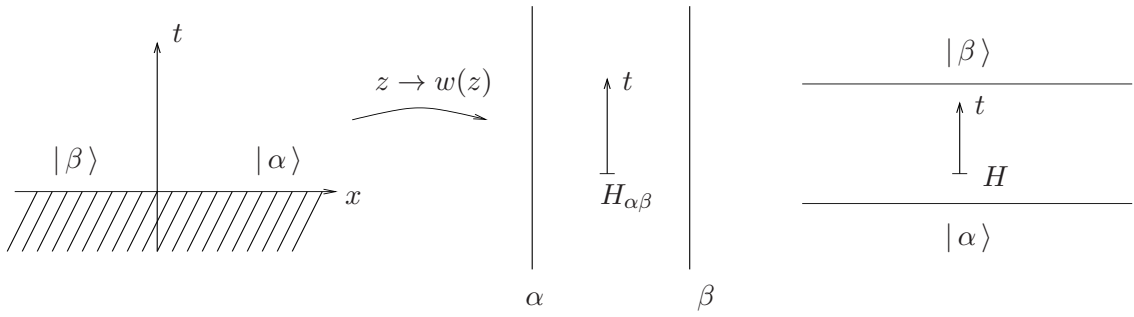,height=4cm,width=15cm}
\caption{Conformal map (\ref{maphpstrip}) from the half plane to
the strip.} \label{fighpstrip}
\end{figure}

The most intriguing feature of boundary CFT is the existence of a
relation between boundary conditions and the bulk operator content
of the theory. In fact, condition (\ref{Tboundary}) is expressed
in terms of the Virasoro generators acting on a boundary state
$|\alpha\rangle$ as
$$
\left(L_{n}-\bar{L}_{-n}\right)|\,\alpha\,\rangle=0\,,
$$
and the solutions are the so-called Ishibashi states
$$
|\,j\,\rangle\rangle\equiv\sum_{N}|j;\,N\rangle\otimes U
\overline{|j;\,N\,}\rangle\,,
$$
where $|\,j;\,N\,\rangle$ and $\overline{|\,j;\,N}\,\rangle$ are
the holomorphic and antiholomorphic states belonging to the
conformal family of the bulk primary operator labelled by $j$, and
$U$ is an antiunitary operator introduced for technical reasons.
The boundary states $|\,\alpha\,\rangle$ are therefore linear
combinations of Ishibashi states associated with different primary
operators
$$
|\,\alpha\,\rangle\,=\,\sum_j
C_{\alpha\,j}\,|\,j\,\rangle\rangle\;,
$$
and the coefficients $C_{\alpha\,j}$ can be determined by
exploiting the modular invariance of the theory. This procedure
consists in first mapping the half plane on a strip through the
conformal transformation
\begin{equation}\label{maphpstrip}
z\,\to\,w(z)=\frac{R}{\pi}\ln z\;,
\end{equation}
and then in interpreting the strip geometry in two different
physical ways, similarly to what we have discussed in
Sect.\,\ref{finvolintro} for the cylinder (see
Fig.\,\ref{fighpstrip}).

\vspace{0.5cm}

In the first physical picture the boundaries are placed at
$x=0,R$, and the partition function is obtained by evolving the
Hamiltonian $H_{\alpha\beta}$ along the time direction, while in
the second scheme space is unbounded, and the infinite--volume
Hamiltonian $H$ evolves from a state $|\,\alpha\,\rangle$ to
another state $|\,\beta\,\rangle$. The equality of the result in
the two different pictures permits to express the coefficients
$C_{\alpha\,j}$ in terms of the so--called modular matrix of the
system, establishing a $\,1:1\;$ correspondence between boundary
states and primary operators of the bulk theory. For instance, in
the Ising model (introduced in Sect.\,\ref{secCFT}) there are
three possible boundary states, symbolically indicated as
$|\,\tilde{j}\,\rangle$ and given by
\begin{eqnarray*}
|\,\tilde{\mathbb{I}}\,\rangle&=&\frac{1}{\sqrt{2}}\,|\,\mathbb{I}\,\rangle\rangle
+\frac{1}{\sqrt{2}}\,|\,\varepsilon\,\rangle\rangle+\frac{1}
{\sqrt[4]{2}}\,|\,\sigma\,\rangle\rangle\\
|\,\tilde{\varepsilon}\,\rangle&=&\frac{1}{\sqrt{2}}\,|\,\mathbb{I}\,\rangle\rangle
+\frac{1}{\sqrt{2}}\,|\,\varepsilon\,\rangle\rangle-
\frac{1}{\sqrt[4]{2}}\,|\,\sigma\,\rangle\rangle\\
|\,\tilde{\sigma}\,\rangle&=&\,|\,\mathbb{I}\,\rangle\rangle-\,|\,\varepsilon\,\rangle\rangle\\
\end{eqnarray*}

Finally, it is useful for the future discussions to illustrate the
features of the Gaussian CFT, described in Sect.\,\ref{secCFT},
defined here on a strip with Neumann or Dirichlet boundary
conditions. The transformation (\ref{maphpstrip}) leads to the
following expression for the Hamiltonian on the strip
\cite{cardy,saleurlect}:
\begin{equation}\label{CFTstrip}
H\,=\,\frac{\pi}{R}\left(L_0-\frac{c}{24}\right)\;.
\end{equation}
In the case of Neumann boundary conditions
$$
\partial_x\,\phi(0,t)\,=\,\partial_x\,\phi(R,t)\,=\,0\;,
$$
there is no winding, and the sectors of the theory are only
labelled by the momentum index $s$. The lowest eigenvalue of $L_0$
in each sector is given by
\begin{equation}
h_{s}  \,=\, \frac{1}{2\pi g} \,\frac{s^2}{{\cal R}^{2}}\;.
\end{equation}
On the contrary, Dirichlet boundary conditions
$$
\phi(0,t)\,=\,\phi_{0}\;,\qquad \phi(R,t)\,=\,\phi_{R}\;
$$
constrain the momentum index to zero, but leave the winding free.
In this case, the lowest conformal dimension in each sector is
expressed as
\begin{equation}\label{CFTdimD}
h_{n}  \,=\, \frac{g}{2\pi} \left[(\phi_R-\phi_0)\,+\,2\pi n {\cal
R}\right]^{2}\;.
\end{equation}

\subsection{Sine--Gordon model on the strip}

The Sine--Gordon model (\ref{SGpot}) can be defined on a strip
$x\in[0,R]$. Among the several possible boundary conditions which
preserve the integrability of the model, a particularly
interesting example is given by the Dirichlet boundary conditions
(D.b.c.)
\begin{equation}\label{Dbc}
\phi(0,t)\,=\,\phi_{0}+\frac{2\pi}{\beta}\,n_{0}\;,\qquad
\phi(R,t)\,=\,\phi_{R}+\frac{2\pi}{\beta}\,n_{R}\;,\qquad \forall
t\;
\end{equation}
with $0\leq \phi_{0,R}< \frac{2\pi}{\beta}$ and $n_{0,R}\in
\mathbb{Z}$. The topological charge of this model is conserved
also in the presence of boundaries and it can be conveniently
defined as
\begin{equation*}
Q\,\equiv\,\frac{\beta}{2\pi}
\left\{\int\limits_{0}^{R}\partial_{x}\phi\,\,dx\,-\,(\phi_{R}-\phi_{0})\right\}\,=\,
n_{R}-n_{0}\;.
\end{equation*}
Hence the space of states is split in topological sectors with
$Q=0,\pm 1,\pm 2...$, and within a given $Q$-sector the states are
characterized by their energies only.

It is worth mentioning that, in recent years, this problem (and
variations thereof) has attracted the attention of several groups:
the case of half--plane geometry, for instance, has been discussed
by bootstrap methods in \cite{ghoshzam,mattdorey,BPTT} and by
semiclassical ones in \cite{SSW,corr,kormospalla} whereas the
thermodynamics of different cases in a strip geometry has been
studied in a series of publications (see
\cite{LMSS,SS,cauxsal,leerim,rim,ahnnep,ravSGstrip}). In
completing the work \cite{SGstrip}, a paper on a (semi)classical
analysis of Sine--Gordon model on a strip \cite{hungstrip} also
appeared, which partially overlaps with it.

The semiclassical quantization presented in \cite{SGstrip} adds
new pieces of information on this subject and it may be seen as
complementary to the aforementioned studies. For the Sine--Gordon
model with periodic boundary conditions, alias in a cylindrical
geometry, this program has been completed in \cite{SGscaling} and
presented in Sect.\,\ref{sectcyl}. Given the similarity of the
outcoming formulas with the ones appearing in \cite{SGscaling}, in
the sequel we will often refer to Sect.\,\ref{sectcyl} for the
main mathematical definitions as well as for the discussion of
some technical details. There is though a conceptual difference
between the periodic example and the one studied here: in the
periodic case, in fact, the vacuum sector is trivial at the
semiclassical level (it simply corresponds to the constant
classical solution) and therefore the semiclassical quantization
provides non-perturbative results just starting from the one-kink
sector. Contrarily, on the strip with Dirichlet b.c., the vacuum
sector itself is represented by a non-trivial classical solution
and its quantization is even slightly more elaborated than the one
of the kink sectors.

\begin{figure}[ht]
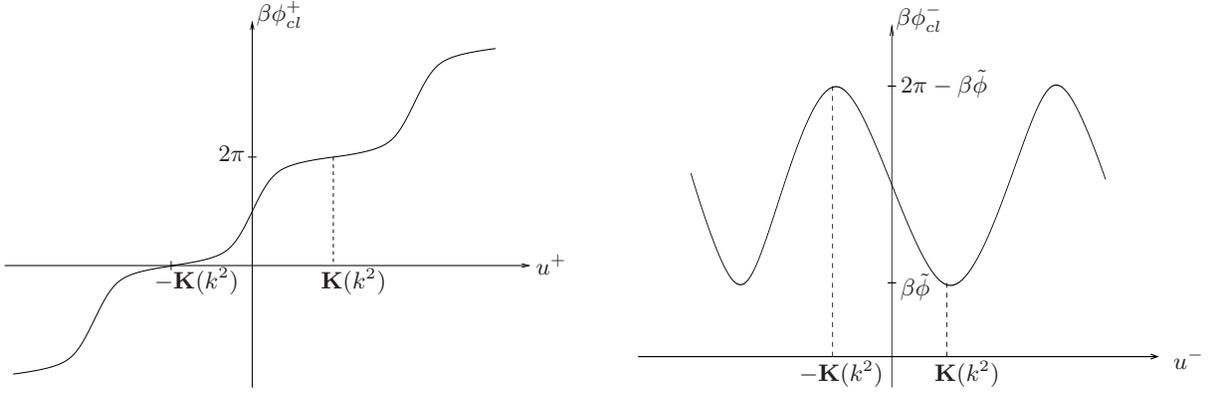

\begin{tabular}{p{8cm}p{8cm}}

\footnotesize

\psfrag{phicl(x)}{$\beta\phi^{+}_{cl}$} \psfrag{2 pi}{$2\pi$}
\psfrag{K(k^2)}{$\textbf{K}(k^{2})$}
\psfrag{-K(k^2)}{$-\textbf{K}(k^{2})$}
\psfrag{x}{$\hspace{0.2cm}u^{+}$}

\psfig{figure=SGkinkper.eps,height=5cm,width=7cm}&

\footnotesize

\psfrag{phicl(x)}{$\beta\phi^{-}_{cl}$}
\psfrag{phi0}{$\beta\tilde{\phi}$} \psfrag{2
pi-phi0}{$2\pi-\beta\tilde{\phi}$}
\psfrag{K(k^2)}{$\textbf{K}(k^{2})$}
\psfrag{-K(k^2)}{$-\textbf{K}(k^{2})$}
\psfrag{x}{$\hspace{0.2cm}u^{-}$}

\psfig{figure=SGkink.eps,height=5cm,width=7cm}
\end{tabular}
\caption{Solutions of eq.\,(\ref{firstSG}), $A > 0$ (left hand
side), $-2 < A < 0$ (right hand side).}
 \label{figSGsol}
\end{figure}

In order to describe the classical solutions of this problem, it
is worth to preliminarily recall some results already obtained. As
we have seen in the previous sections, the SG equation of motions
admit three kinds of static solution, depending on the sign of the
constant $A$ in (\ref{firstSG}). The simplest corresponds to $A=0$
and it describes the standard kink in infinite volume
(\ref{SGkinkinfvol}). In the following, we will be concerned with
the solutions relative to the case $A\neq 0$, which can be
expressed in terms of Jacobi elliptic functions. In particular,
for $ A
> 0$ we have
\begin{equation*}
\phi^{+}_{cl}(x)\,=\,\frac{\pi}{\beta} + \frac{2}{\beta}\,
\textrm{am}\left(\frac{ m (x - x_0)}{k},k\right)\;, \qquad k^{2}
\,=\,\frac{2}{2+A}\,\,\,,
\end{equation*}
which has the monotonic and unbounded behaviour in terms of the
real variable $u^{+}=\frac{ m (x - x_0)}{k}$ shown in
Fig.\,\ref{figSGsol}. For $ -2<A < 0$, the solution is given
instead by
\begin{equation*}
\phi^{-}_{cl}(x)\,=\, \frac{2}{\beta}\,
\arccos\left[k\;\textrm{sn}\left( m (x - x_0),k\right)\right]\;,
\qquad k^{2} \,=\,1+\frac{A}{2}\,\,\,,
\end{equation*}
and it oscillates in the real variable $u^{-}= m (x - x_0)$
between the $k$-dependent values $\tilde{\phi}$ and
$\frac{2\pi}{\beta}-\tilde{\phi}$ (see Fig.\,\ref{figSGsol}).

The SG model with the Dirichlet b.c. (\ref{Dbc}) can be
classically described by using the two building functions
$\phi^{+}_{cl}(x)$ and $\phi^{-}_{cl}(x)$, thanks to their free
parameters $x_0$ and $k$, which can be fixed in terms of $\phi_0$,
$\phi_R$ and $R$. However, in order to simplify the notation, in
writing down our solutions we will rather use $R$ and $x_0$, both
considered as functions of $\phi_0$, $\phi_R$ and $k$ (as a matter
of fact, $k$ can be recovered by inverting the elliptic integrals
which enter the corresponding expression of $R$).

As shown below, both types of solutions $\phi^{+}_{cl}(x)$ and
$\phi^{-}_{cl}(x)$ are needed, in general, to define the classical
background in the vacuum sector whereas only one of them,
$\phi^{+}_{cl}(x)$, is employed for implementing the Dirichlet
b.c. in the kink sector.

\subsubsection{The vacuum sector: $Q=0$}

To discuss the vacuum sector, it is sufficient to restrict the
attention to the case\footnote{All other cases can be described in
a similar way, defining properly $x_{0}$ and $R$, and by using
antikinks when necessary.} $n_0=n_R=0$, $\phi_{0}<\phi_{R}$ and
$\left|\cos\frac{\beta}{2}\phi_{0}\right|>\left|\cos\frac{\beta}{2}\phi_{R}\right|$.
It is also convenient to introduce the compact notation
\begin{equation*}
c\,_{0,R} \,\equiv\, \cos\frac{\beta}{2}\,\phi_{0,R}\;\;.
\end{equation*}

\begin{figure}[ht]
\begin{tabular}{p{4.5cm}p{5.5cm}p{6.5cm}}

\footnotesize

\psfrag{phi}{$\beta\phi^{(1)}_{cl}$} \psfrag{ell}{$x$}

\psfig{figure=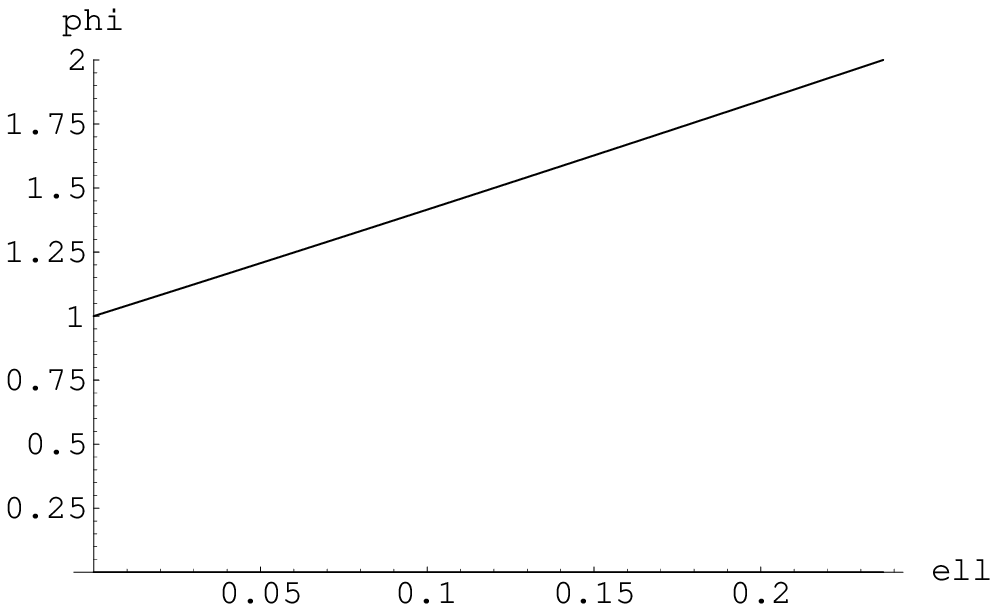,height=4cm,width=4cm}&

\footnotesize

\psfrag{phi}{$\beta\phi^{(2)}_{cl}$} \psfrag{ell}{$x$}

\psfig{figure=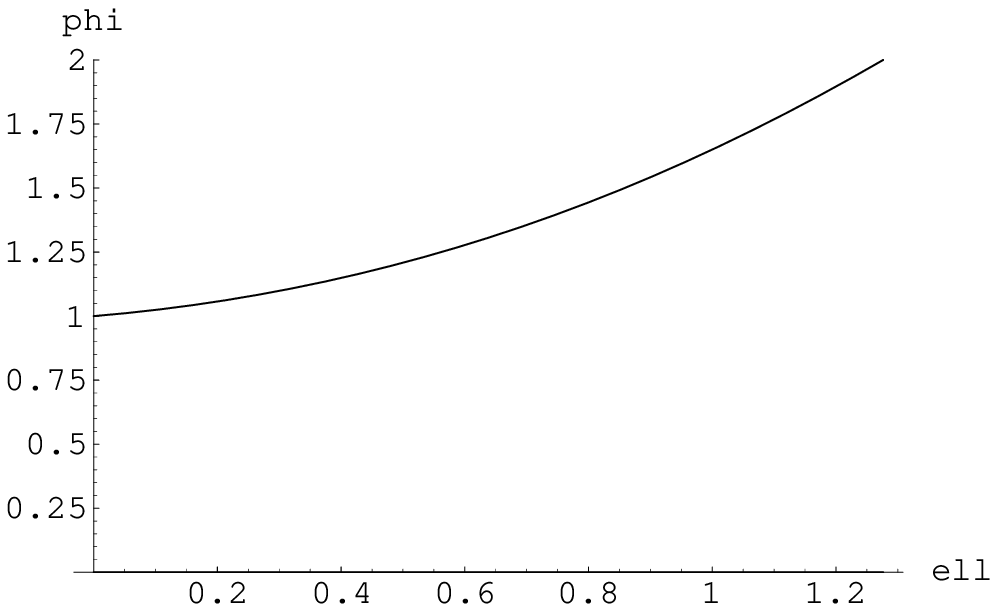,height=4cm,width=5cm}&

\footnotesize

\psfrag{phi}{$\beta\phi^{(3)}_{cl}$} \psfrag{ell}{$x$}

\psfig{figure=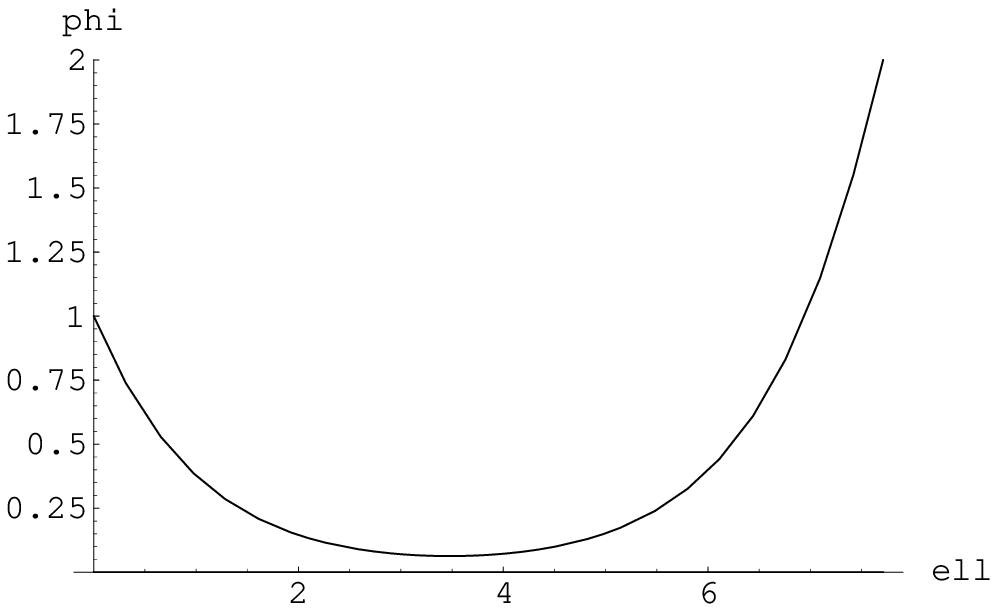,height=4cm,width=6cm}
\end{tabular}
\caption{Classical solution (\ref{vacD}) at some value of $R$, in
the case $\beta\phi_{0}=1$ and $\beta\phi_{R}=2$.} \label{figvac}
\end{figure}

In order to write down explicitly the classical background
corresponding to the vacuum state with Dirichlet b.c., it is
necessary to introduce preliminarily two particular values $R_1$
and $R_2$ of the width $R$ of the strip, which mark a change in
the nature of the solution. They are given by
$$
\begin{cases}
mR_{1}\,=\,\text{arctanh}\left(c\,_{0}\right)-
\text{arctanh}\left(c_{R}\right)\;,\\
mR_{2}\,=\,\textbf{K}(\tilde{k})
-F\left(\arcsin\frac{c_R}{\tilde{k}^{}}
\,,\;\tilde{k}\right)\;,\hspace{1.5cm}
\tilde{k}=\left|\,c\,_{0}\right|\;.
\end{cases}
$$
With these definitions, the classical vacuum solution, as a
function of $x\in[0,R]$, has the following behaviour in the three
regimes of $R$:
\begin{equation}\label{vacD}
\phi^{\text{vac}}_{cl}(x)\,=\,
\begin{cases}
\phi^{(1)}_{cl}(x)\qquad \text{for}\qquad 0<R<R_{1}\\
\phi^{(2)}_{cl}(x)\qquad \text{for}\qquad R_{1}<R<R_{2}\\
\phi^{(3)}_{cl}(x)\qquad \text{for}\qquad R_{2}<R<\infty
\end{cases}
\end{equation}
where
\begin{equation*}
\phi^{(1)}_{cl}(x)\,=\,\phi^{(+)}_{cl}(x)\qquad \text{with}\qquad
\begin{cases}
m x_{0} = -k\,F\left(\frac{\beta}{2}\,\phi_{0}-\frac{\pi}{2}\,,\;k\right)\\
m R =
k\left[F\left(\frac{\beta}{2}\,\phi_{R}-\frac{\pi}{2}\,,\;k\right)-
F\left(\frac{\beta}{2}\,\phi_{0}-\frac{\pi}{2}\,,\;k\right)\right]\\
 0 < k < 1 \end{cases}
\end{equation*}
\begin{equation*}\hspace{-1cm}
\phi^{(2)}_{cl}(x)\,=\,\phi^{(-)}_{cl}(x)\qquad \text{with}\qquad
\begin{cases}
mx_{0} = -2\textbf{K}(k)+F\left(\arcsin\frac{c\,_0}{k}\,,\;k\right)\\
m R = F\left(\arcsin\frac{c\,_0}{k}\,,\;k\right)-
F\left(\arcsin\frac{c_R}{k}\,,\;k\right)\\
\tilde{k} < k < 1\end{cases}
\end{equation*}
\begin{equation*}\hspace{0.5cm}
\phi^{(3)}_{cl}(x)\,=\,\phi^{(-)}_{cl}(x)\qquad \text{with}\qquad
\begin{cases}
mx_{0} = -F\left(\arcsin\frac{c\,_0}{k}\,,\;k\right)\\
m R = 2\textbf{K}(k)-F\left(\arcsin\frac{c\,_0}{k}\,,\;k\right)-
F\left(\arcsin\frac{c_R}{k}\,,\;k\right)\\
\tilde{k} < k < 1\end{cases}
\end{equation*}
In addition to the quantities already defined, we have introduced
here the incomplete elliptic integrals
\begin{equation*}\label{incellint}
F(\varphi,k)\,=\,\int\limits_{0}^{\varphi}\frac{d\alpha}
{\sqrt{1-k^{2}\sin^{2}\alpha}}\;,\qquad
E(\varphi,k)\,=\,\int\limits_{0}^{\varphi}d\alpha
\sqrt{1-k^{2}\sin^{2}\alpha}\;,
\end{equation*}
which reduce to the complete ones at $\varphi=\pi/2$. It is easy
to check that at the particular values $R_1$ and $R_2$, the
different definitions of the background nicely coincide.
Fig.\,\ref{figvac} shows the classical solution at some values of
$R$, one for each of the three regimes\footnote{We have chosen for
the plot the specific values $\beta\phi_0 = 1$ and $\beta\phi_R =
2$, for which $mR_1 = 0.76$ and $mR_2 = 1.49$. The same values
will be considered in all other pictures since their qualitative
features do not sensibly depend on these parameters, except for
few particular values of $\phi_{0,R}$ discussed separately}.

\vspace{5mm}

\begin{figure}[ht]
\footnotesize \psfrag{ec}{$\beta^{2}{\cal E}_{cl}^{\text{vac}}/m$}
\psfrag{ell}{$mR$}
\begin{center}
\psfig{figure=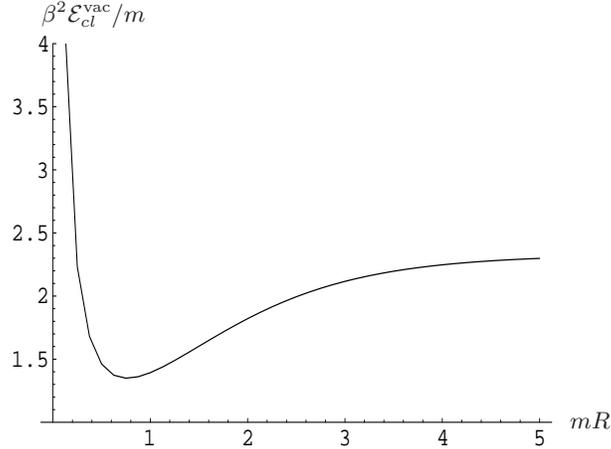,height=6cm,width=8cm}
\end{center}
\caption{Classical energy (\ref{eclD}) for $\beta\phi_{0}=1$ and
$\beta\phi_{R}=2$.} \label{figecl}
\end{figure}

The classical energy of the background (\ref{vacD}) is expressed
as
\begin{equation}\label{eclD}
{\cal E}^{\text{vac}}_{cl}(R)\,=\,
\begin{cases}
{\cal E}^{(1)}_{cl}(R)\qquad \text{for}\qquad 0<R<R_{1}\\
{\cal E}^{(2)}_{cl}(R)\qquad \text{for}\qquad R_{1}<R<R_{2}\\
{\cal E}^{(3)}_{cl}(R)\qquad \text{for}\qquad R_{2}<R<\infty
\end{cases}
\end{equation}
where
\begin{eqnarray}
{\cal
E}^{(1)}_{cl}(R)&=&\frac{2m}{\beta^{2}}\left\{\left(1-\frac{1}{k^{2}}\right)m
R+\frac{2}{k}\left[E\left(\frac{\beta}{2}\,\phi_{R}-\frac{\pi}{2},k\right)-
E\left(\frac{\beta}{2}\,\phi_{0}-\frac{\pi}{2},k\right)\right]\right\}\;,\nonumber\\
{\cal E}^{(2)}_{cl}(R)&=&\frac{2m}{\beta^{2}}\left\{(k^{2}-1)m
R+2\left[E\left(\arcsin\frac{c\,_0}{k}\,,\;k\right)-
E\left(\arcsin\frac{c_R}{k}\,,\;k\right)\right]\right\}\;,\label{eclDexplicit}\nonumber\\
{\cal E}^{(3)}_{cl}(R)&=&\frac{2m}{\beta^{2}}\left\{(k^{2}-1)m
R+2\left[2\textbf{E}(k)-E\left(\arcsin\frac{c\,_0}{k}\,,\;k\right)-
E\left(\arcsin\frac{c_R}{k}\,,\;k\right)\right]\right\}\;,\nonumber
\end{eqnarray}
and it is plotted in Fig.\,\ref{figecl}. As expected, the quantity
(\ref{eclD}) has a smooth behaviour at $R_{1}$ and $R_{2}$, which
correspond to the minimum and the point of zero curvature of this
function, respectively. The non monotonic behaviour of the
classical energy gives an intuitive motivation for the classical
background being differently defined in the three regimes of $R$.

Furthermore, the classical energy can be easily expanded in the
ultraviolet (UV) or infrared (IR) limit, i.e. for small or large
values of $mR$, which correspond to $k\to 0$ in the regime
$0<R<R_1$ or to $k\to 1$ in the regime $R_2<R<\infty$,
respectively.

In fact, expanding the elliptic integrals in (\ref{eclD}) (see
\cite{GRA} and Appendix\,\ref{chapfinitesize}.\ref{appell} for the
relative formulas), and comparing the result order by order with
the small-$k$ expansion of $mR$ defined in the first regime of
(\ref{vacD})
\begin{equation}\label{mRUV}
mR\,=\,k\,\frac{\beta}{2}(\phi_R-\phi_0)\left[1+\frac{k^2}{4}\left(1+
\frac{\sin\beta\phi_R-\sin\beta\phi_0}{\beta(\phi_R-\phi_0)}\right)
+ \cdots \right]\;,
\end{equation}
one obtains the small-$mR$ behaviour
\begin{equation}\label{eclUV}
{\cal E}^{(1)}_{cl}(R)\,=\,\frac{1}{2
R}(\phi_R-\phi_0)^2+R\,\frac{m^2}{\beta^2}
\left[1-\frac{\sin\beta\phi_R-\sin\beta\phi_0}{\beta(\phi_R-\phi_0)}\right]
+ \cdots \;.
\end{equation}
Later we will comment on the meaning of this result in the UV
analysis of the ground state energy. On the other hand, comparing
the expansion for $k\to 1$ of ${\cal E}^{(3)}_{cl}(R)$ in the
third regime with
\begin{equation}\label{mRIR}
mR\,=\,
-\log\left\{\frac{1-k^2}{16}\;\frac{1}{\tan\frac{\beta}{4}\phi_0\,\tan\frac{\beta}{4}\phi_R}\right\}+\cdots
\;,
\end{equation}
one obtains the large-$mR$ behaviour
\begin{equation}\label{eclIR}
{\cal E}^{(3)}_{cl}(R)\,=\, \frac{4m}{\beta^{2}}\,
\left(\,2-\cos\frac{\beta}{2}\phi_R-\cos\frac{\beta}{2}\phi_0\right)-
\frac{32m}{\beta^{2}}\,\tan\frac{\beta}{4}\phi_0\,\tan\frac{\beta}{4}\phi_R
\;e^{-mR}+\cdots \;.
\end{equation}
The first term of this expression is the classical limit of the
boundary energy of the vacuum sector \cite{LMSS}, since it is the
term that needs to be subtracted by choosing to normalise the
energy to zero at $R\rightarrow \infty$.

The classical description of the vacuum sector can be completed by
mentioning the existence of two particular cases in which the
three different regimes of $R$ are not needed. The first is given
by $\phi_{0} = \phi_{R}$, for which the whole range of $R$ is
described by $\phi^{(3)}_{cl}(x)$ in (\ref{vacD}), since $mR_{2} =
0$ in this situation. The second case, defined by $\phi_{0}$
arbitrary and $\phi_{R} = 0$, can be instead described by the
antikink $\bar{\phi}^{(1)}_{cl}(x) = \phi^{(1)}_{cl}(-x)$ alone,
since $m R_{1} = \infty$ for these values of the boundary
parameters (note that $x_0$ and $R$ have to be defined as opposite
to the ones in (\ref{vacD})). As a consequence, these two cases
display a monotonic behaviour of the classical energy, whose UV
and IR asymptotics, respectively, require a separate derivation,
which can be performed by simply adapting the above procedure.

Finally, it is also worth discussing an interesting feature which
emerges in the IR limit of the classical solution (\ref{vacD}). As
it can be seen from Fig.\,\ref{figvac}, by increasing $R$ the
static background is more and more localised closely to the
constant value $\phi(x)\equiv 0$ and this guarantees the
finiteness of the classical energy in the $R\to\infty$ limit,
given by the first term in (\ref{eclIR})\footnote{When
$|c_0|<|c_R|$, the same qualitative phenomenon occurs, but the
constant value is $\phi(x)\equiv \frac{2\pi}{\beta}$ in this
case.}. However, if the IR limit is performed directly on the
classical solution, we obtain one of the static
backgrounds\footnote{Obviously, the same function is obtained as
$\lim\limits_{R\to\infty}\bar{\phi}^{(1)}_{cl}(x)$, in the case
$\phi_R=0$ mentioned above.} studied in \cite{kormospalla}
$$
\phi^{(3)}_{cl}(x)\;{\mathrel{\mathop{\kern0pt\longrightarrow}
\limits_{R\to \infty}}}\;\frac{2}{\beta}\,
\arccos\left[\,\tanh\,m(x-x_{0}^{\infty})\right]\;,\qquad\text{with}\qquad
x_{0}^{\infty} \,=\, -\text{arctanh}(c_0)\;.
$$
The last expression tends to zero as $x\to\infty$ and consequently
has classical energy ${\cal E}_{cl} =
\frac{4m}{\beta^{2}}\,\left(\,1-\cos\frac{\beta}{2}\phi_0\right)$.
This phenomenon can be easily understood by noting that the
minimum of $\phi^{(3)}_{cl}(x)$ (which goes to zero in the IR
limit), is placed at $m\bar{x} = mx_{0} + \textbf{K}(k)$ (see
Fig.\,\ref{figSGsol}) and this point tends itself to infinity as
$k\to 1$. Hence, the information about the specific value of
$\phi_R$ is lost when $R \rightarrow \infty$, i.e. only the states
with $\phi_R = 0$ survive in the IR limit.

\vspace{0.5cm}

We will now perform the semiclassical quantization in the vacuum
sector, around the background (\ref{vacD}). Depending on the value
of $mR$, the stability equation (\ref{stability}) takes the form
\begin{equation}\label{stability1}
\left\{\frac{d^{2}}{d
\bar{x}^{2}}+k^{2}\left(\bar{\omega}^{2}+1\right)-2
k^{2}\,\textrm{sn}^{2}(\bar{x}-\bar{x}_{0},k)
\right\}\eta^{(1)}_{\bar{\omega}}(\bar{x}) \,=\, 0\;,\qquad
\text{with} \quad \bar{x}\,=\,\frac{m x}{k}\,,
\;\;\bar{\omega}\,=\,\frac{\omega}{m}\;,
\end{equation}
when $0<R<R_{1}$, and
\begin{equation}\label{stability2}
\left\{\frac{d^{2}}{d \bar{x}^{2}}+\bar{\omega}^{2}+1-2
k^{2}\,\textrm{sn}^{2}(\bar{x}-\bar{x}_{0},k)
\right\}\eta^{(2,3)}_{\bar{\omega}}(\bar{x}) \,=\, 0\;,\qquad
\text{with} \quad \bar{x}\,=\,m x\,,\;\;
\bar{\omega}\,=\,\frac{\omega}{m}\;,
\end{equation}
when $R_{1} < R < R_{2}$ and $R_{2} < R < \infty$.

Equations (\ref{stability1}) and (\ref{stability2}) can be cast in
the Lam\'e form with $N=1$, which is described in Appendix
\ref{chapfinitesize}.\ref{lame} and has been studied in detail for
the periodic case in Sect.\,\ref{sectcyl}. The only differences
with the periodic case are the presence of a non-trivial center of
mass $x_0$ and the larger number of parameters entering the
expression of the size $R$ of the system: these make more
complicated the so--called \lq\lq quantization condition'' that
determines the discrete eigenvalues, although they do not alter
the general procedure to derive it.

The boundary conditions (\ref{Dbc}), which translate in the
requirement
\begin{equation*} \eta_{\bar{\omega}}(0)
\,=\, \eta_{\bar{\omega}}(R) \,=\,0\,\;,
\end{equation*}
select in this case the following eigenvalues, all with
multiplicity one,
\begin{equation}\label{omegaD}
\omega^{\text{vac}}_{n}(R)\,=\,
\begin{cases}
\omega^{(1)}_{n}(R)\qquad \text{for}\qquad 0<R<R_{1}\\
\omega^{(2)}_{n}(R)\qquad \text{for}\qquad R_{1}<R<R_{2}\\
\omega^{(3)}_{n}(R)\qquad \text{for}\qquad R_{2}<R<\infty
\end{cases}\;,
\end{equation}
where
\begin{eqnarray}\label{omegaDexplicit}
&\omega^{(1)}_{n}(R)&=\;\frac{m}{k}\sqrt{\frac{2-k^{2}}{3}-{\cal P}(i y_{n})} \;\;,\nonumber\\
&\omega^{(2,3)}_{n}(R)&=\;m\sqrt{\frac{2k^{2}-1}{3}-{\cal P}(i
y_{n})}  \;\;,\nonumber
\end{eqnarray}
and the $y_{n}$'s are defined through the \lq\lq quantization
condition''
\begin{equation}\label{quantcond}
2\bar{R}\,i\,\zeta(i y_{n})+i\,\log\left[\frac{\sigma(-\bar{x}_{0}
+ i \textbf{K}'+i y_{n})\,\sigma(\bar{R}-\bar{x}_{0} + i
\textbf{K}'-i y_{n})}{\sigma(-\bar{x}_{0} + i \textbf{K}'-i
y_{n})\,\sigma(\bar{R}-\bar{x}_{0} + i \textbf{K}'+i
y_{n})}\right]\,=\,\,2 n \pi\;,\qquad n=1,2,...
\end{equation}
This equation comes from the consistency condition associated to
the boundary values
$$
\begin{cases}
D_+ \,\eta_{a}(0)+D_-\,\eta_{-a}(0) = 0 \,\,\,\,\;,\\
D_+ \,\eta_{a}(R)+D_-\,\eta_{-a}(R) = 0 \,\,\,,
\end{cases}
$$
where $\eta_{\pm a}$ are the two linearly independent solutions of
the Lam\'e equation which are used to build the general solution
$\eta(x)=D_+\,\eta_{a}(x)+D_-\,\eta_{- a}(x)$ (see Appendix
\ref{chapfinitesize}.\ref{lame} for details). Along the same lines
discussed for (\ref{ynper}) in the periodic case, the requirement
(\ref{quantcond}) can be physically interpreted as a quantization
condition for the momentum of a state containing a neutral
particle between the two Dirichlet boundary states. The agreement
between the large-$R$ limit of this expression and the Bethe
ansatz equations involving the appropriate reflection matrices has
been verified in \cite{hungstrip}.

\begin{figure}[ht]
\footnotesize \psfrag{omega1}{$\omega_{i}^{\text{vac}}/m$}
\psfrag{ell}{$mR$}
\begin{center}
\psfig{figure=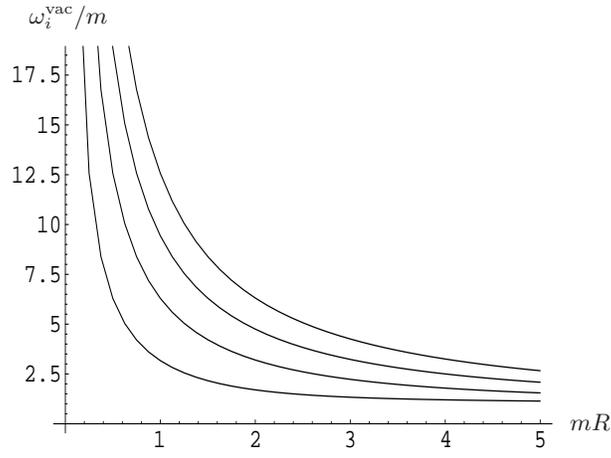,height=6cm,width=8cm}
\end{center}
\caption{The first few energy levels (\ref{omegaD}) for
$\beta\phi_{0}=1$ and $\beta\phi_{R}=2$.} \label{figomegaiD}
\end{figure}

As it can be seen directly from (\ref{tower}), the frequencies
(\ref{omegaD}) are nothing else but the energies of the excited
states with respect to the ground state $E^{\text{vac}}_0(R)$.
They can be easily determined from the above equations and their
behaviour, as functions of $R$, is shown in
Fig.\,\ref{figomegaiD}.

As in the periodic case, a more explicit expression for the energy
levels  (\ref{omegaD}) can be obtained by expanding them for small
or large values of $mR$. The UV expansion, for instance, can be
performed extracting from (\ref{quantcond}) a small-$k$ expansion
for $y_{n}$, inserting it in (\ref{omegaD}), and finally comparing
the result order by order with (\ref{mRUV}). Exploiting the
several properties of Weierstrass functions which follow from
their relation with $\theta$--functions (see for instance
\cite{whit}), one gets
$$
y_n\,=\,\text{arctanh}\,\frac{f}{2n\pi}
+\frac{k^2}{4}\left\{\text{arctanh}\,\frac{f}{2n\pi}+
\,s\;\,\frac{2n\pi(4n^2\pi^2-3f^2)}{(4n^2\pi^2-f^2)^2} \right\} +
\cdots \;,
$$
and
$$
\omega_n^{(1)}\,=\,\frac{m}{k}\,\frac{2n\pi}{f}\left\{1-\frac{k^2}{4}
\left[1+\frac{s}{f}-\frac{2 f s}{4n^2\pi^2-f^2}\right] + \cdots
\right\}\;,
$$
where we have introduced the compact notation
$f\equiv\beta(\phi_R-\phi_0)$,
$s\equiv(\sin\beta\phi_R-\sin\beta\phi_0)$. This leads to the UV
expansion
\begin{equation}\label{omegaDUV}
\omega_n^{(1)}(R)\,=\,\frac{n\pi}{R}+m^2 R \,\;\frac{s}{f}\;
\frac{2n\pi}{4n^2\pi^2-f^2} + \cdots
\end{equation}

In order to complete the above analysis and obtain the reference
value of the energy levels, i.e. the ground state energy
$E^{\text{vac}}_0(R)$ of the vacuum sector, we need the classical
energy (\ref{eclD}) and the sum on the stability frequencies given
in (\ref{omegaD}), i.e.
\begin{equation}\label{grstatevac}
 E^{\text{vac}}_0(R) \,=\, {\cal E}^{\text{vac}}_{cl}(R) +
\frac{1}{2}\,\sum_{n=1}^{\infty} \omega^{\text{vac}}_n(R)\;.
\end{equation}
The above series is divergent and its regularization has to be
performed by subtracting to it a mass counterterm and the
divergent term coming from the infinite volume limit -- a
procedure that is conceptually analogous to the one discussed in
Sect.\,\ref{sectcyl} for the periodic case and therefore it is not
repeated here. Furthermore, as already mentioned, equation
(\ref{grstatevac}) can be made more explicit by expanding it for
small or large values of $mR$. Here, for simplicity, we limit
ourselves to the discussion of the leading $1/R$ term in the UV
expansion since it does not receive contributions from the
counterterm and therefore it can be simply regularised by using
the Riemann $\zeta$--function prescription (see
Sect.\,\ref{secte0reg} for a detailed discussion). The higher
terms, instead, require a technically more complicated
regularization, although equivalent to the one presented in
\cite{SGscaling}.

The UV behaviour of the ground state energy is dominated by
\begin{equation}\label{e0UV}
E^{\text{vac}}_0(R) \,=\,
\frac{\pi}{R}\left[\frac{1}{2\pi}\left(\phi_R-\phi_0\right)^2 -
\frac{1}{24}\right] + \cdots
\end{equation}
where the coefficient $-1/24$ comes from the regularization of the
leading term in the series of frequencies (\ref{omegaDUV}), while
the first term simply comes from the expansion of the classical
energy (\ref{eclUV}). It is easy to see that the above expression
correctly reproduces the expected ground state energy
$(\ref{CFTstrip},\,\ref{CFTdimD})$ for the gaussian Conformal
Field Theory (CFT) on a strip of width $R$ with Dirichlet boundary
conditions.

Finally, it is simple to check that also the excited energy levels
display the correct UV behaviour, being expressed as
\begin{equation}\label{eiUV}
E^{\text{vac}}_{\{k_{n}\}}(R)\, = \,
\frac{\pi}{R}\,\left[\frac{1}{2\pi}\left(\phi_R-\phi_0\right)^2 +
\sum\limits_{n}k_{n}\,n-\frac{1}{24}\right] + \cdots
\end{equation}

\subsubsection{The kink sector: $Q=1$}

In discussing the kink sector we can restrict to
$n_0=0\,,\;n_R=1$, since all other cases, as well as the antikink
sector with $Q=-1$, are described by straightforward
generalizations of the following formulas.

\begin{figure}[ht]
\begin{tabular}{p{8cm}p{8cm}}

\footnotesize

\psfrag{phi}{$\beta\phi_{cl}^{\text{kink}}$} \psfrag{ell}{$x$}

\psfig{figure=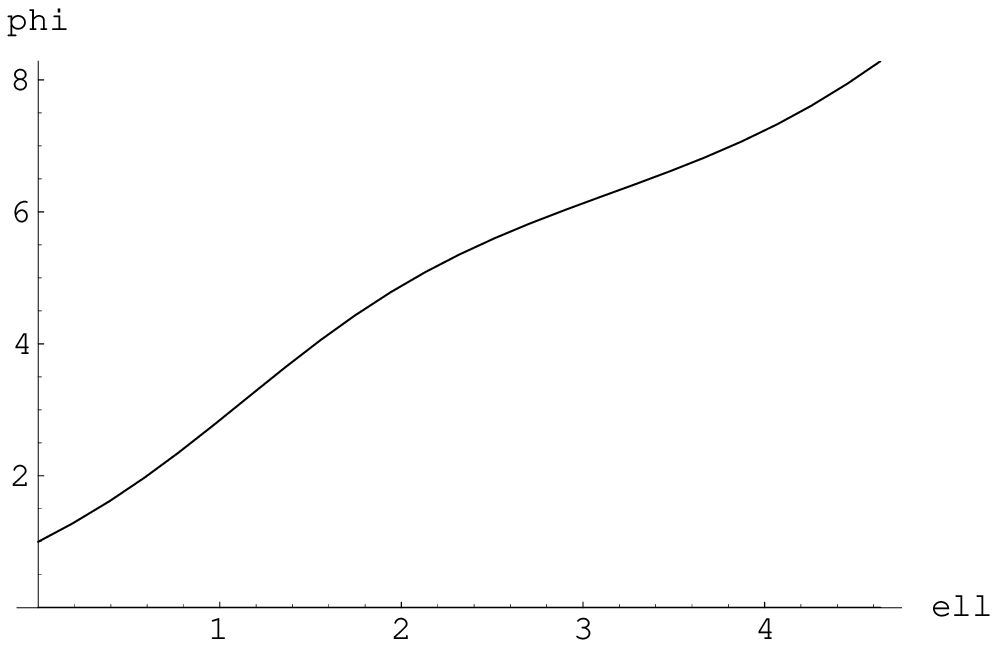,height=5cm,width=7cm}&

\footnotesize

\psfrag{phi}{$\beta\phi_{cl}^{\text{kink}}$} \psfrag{ell}{$x$}

\psfig{figure=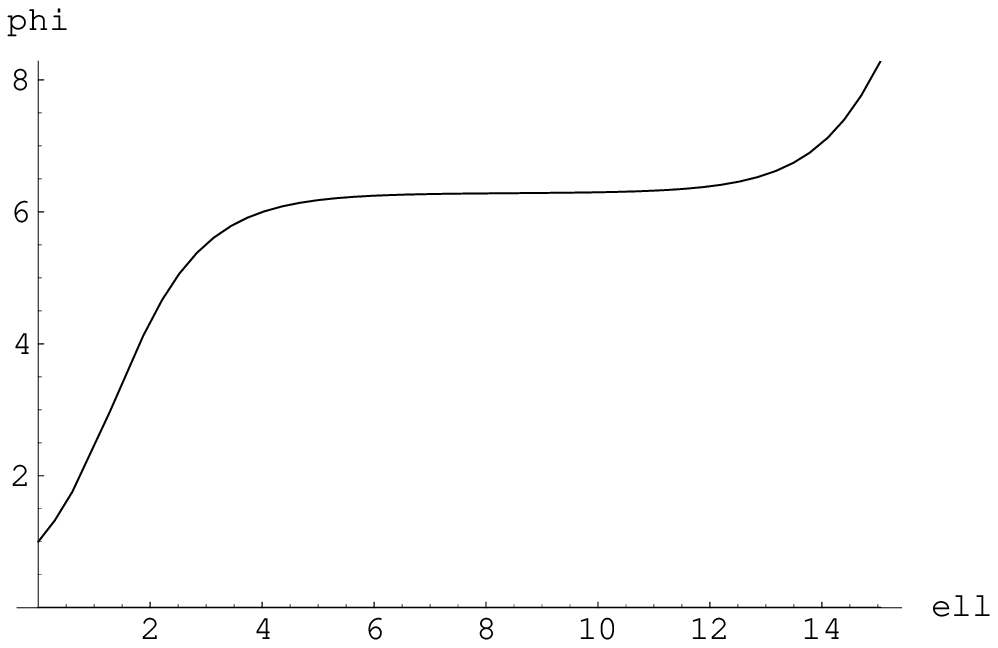,height=5cm,width=7cm}
\end{tabular}
\caption{Classical solution (\ref{kinkD}) at some values of $R$,
in the case $\beta\phi_{0}=1$ and $\beta\phi_{R}=2$.}
\label{figkink}
\end{figure}

The classical solution can be now expressed only in terms of the
function $\phi^{(+)}_{cl}(x)$ as
\begin{equation}\label{kinkD}
\phi_{cl}^{\text{kink}}(x)\,=\,\phi^{(+)}_{cl}(x)\qquad
\text{with}\qquad
\begin{cases}
m x_{0} = -k\,F\left(\frac{\beta}{2}\phi_{0}-\frac{\pi}{2},k\right)\\
m R = k\left[2
\textbf{K}(k)+F\left(\frac{\beta}{2}\phi_{R}-\frac{\pi}{2},k\right)-
F\left(\frac{\beta}{2}\phi_{0}-\frac{\pi}{2},k\right)\right]\\
0 < k < 1
\end{cases}\;,
\end{equation}
since in this case the whole range $0 < m R < \infty$ is spanned
by varying $k$ in $[0,1]$. This can be intuitively understood by
looking at the behaviour of (\ref{kinkD}) in Fig.\,\ref{figkink}.

As a consequence, the classical energy and the stability
frequencies of this sector can be obtained from ${\cal
E}_{cl}^{(1)}$ and $\omega_{n}^{(1)}$ of the vacuum (given
respectively in eq.\,(\ref{eclD}) and (\ref{omegaD})), by simply
replacing $\phi_{R}\to \phi_{R}+\frac{2\pi}{\beta}$. The leading
UV behaviour of the energy levels in this sector, given by
\begin{equation}\label{kinkeiUV}
E^{\text{kink}}_{\{k_{n}\}}(R)\, = \,
\frac{\pi}{R}\,\left[\frac{1}{2\pi}
\left(\left(\phi_R-\phi_0\right)+\frac{2\pi}{\beta}\,Q\right)^2 +
\sum\limits_{n}k_{n}\,n-\frac{1}{24}\right] + \cdots
\end{equation}
with $Q=1$, correctly matches the CFT prediction.

\begin{figure}[ht]
\footnotesize \psfrag{ec}{$\beta^{2}{\cal
E}^{\text{kink}}_{cl}/m$} \psfrag{ell}{$mR$}
\begin{center}
\psfig{figure=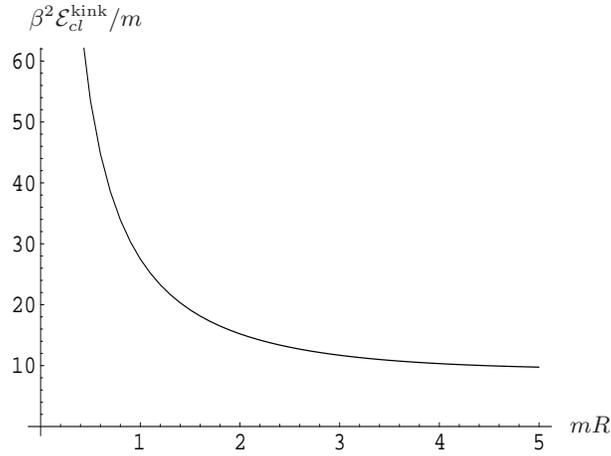,height=6cm,width=8cm}
\end{center}
\caption{Classical energy in the $Q=1$ kink sector for
$\beta\phi_{0}=1$ and $\beta\phi_{R}=2$.} \label{figeclkink}
\end{figure}

The only result which cannot be directly extracted from the vacuum
sector analysis is the IR asymptotic behaviour of the classical
energy, since now the $k\to 1$ limit has to be performed on ${\cal
E}^{(1)}_{cl}$. We have in this case
\begin{equation}\label{mRIRkink}
mR\,=\,
-\log\left\{\frac{1-k^2}{16}\;\frac{\tan\frac{\beta}{4}\phi_0}
{\tan\frac{\beta}{4}\phi_R}\right\}+\cdots \;,
\end{equation}
which leads to
\begin{equation}\label{eclIRkink}
{\cal E}^{(1)}_{cl}(R)\,=\, \frac{4m}{\beta^{2}}\,
\left(\,2-\cos\frac{\beta}{2}\phi_R+\cos\frac{\beta}{2}\phi_0\right)+
\frac{32m}{\beta^{2}}\,\frac{\tan\frac{\beta}{4}\phi_R}
{\tan\frac{\beta}{4}\phi_0}\;e^{-mR}+\cdots \;.
\end{equation}
Analogously to the vacuum sector, the first term of this
expression is related to the classical limit of the boundary
energy in the one--kink sector. Notice that, differently from the
vacuum case, where the asymptotic IR value of the classical energy
was approached from below (see (\ref{eclIR})), the coefficient of
the exponential correction has now positive sign, in agreement
with the monotonic behaviour of the classical energy shown in
Fig.\,\ref{figeclkink}.

When $R\to\infty$, a mechanism analogous to the one discussed for
the vacuum also takes place here: the classical energy is finite
for any value of $\phi_R$, but since $\phi_{cl}^{\text{kink}}(x)$
assumes the value $\frac{2\pi}{\beta}$ at $m\bar{x} = mx_{0} +
k\textbf{K}(k)$ (see Fig.\,\ref{figSGsol}), a point which tends to
infinity as $k\to 1$, only the states with $\phi_R = 0$ survive in
this limit.

\begin{figure}[ht]
\begin{tabular}{p{8cm}p{8cm}}

\footnotesize

\psfrag{phi}{$\beta\phi_{cl}$} \psfrag{ell}{$x$}

\psfig{figure=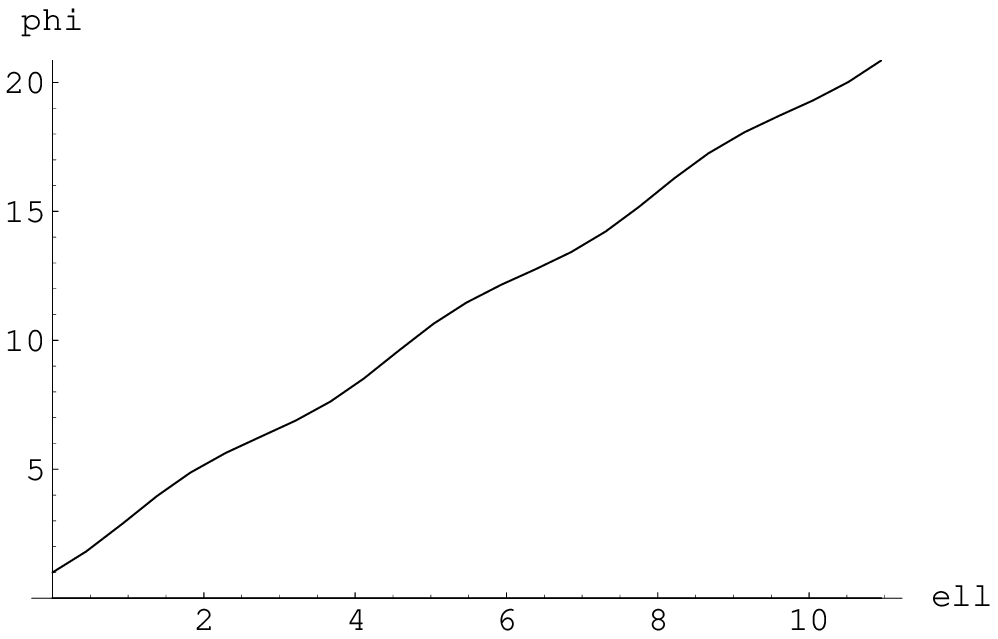,height=5cm,width=7cm}&

\footnotesize

\psfrag{phi}{$\beta\phi_{cl}$} \psfrag{ell}{$x$}

\psfig{figure=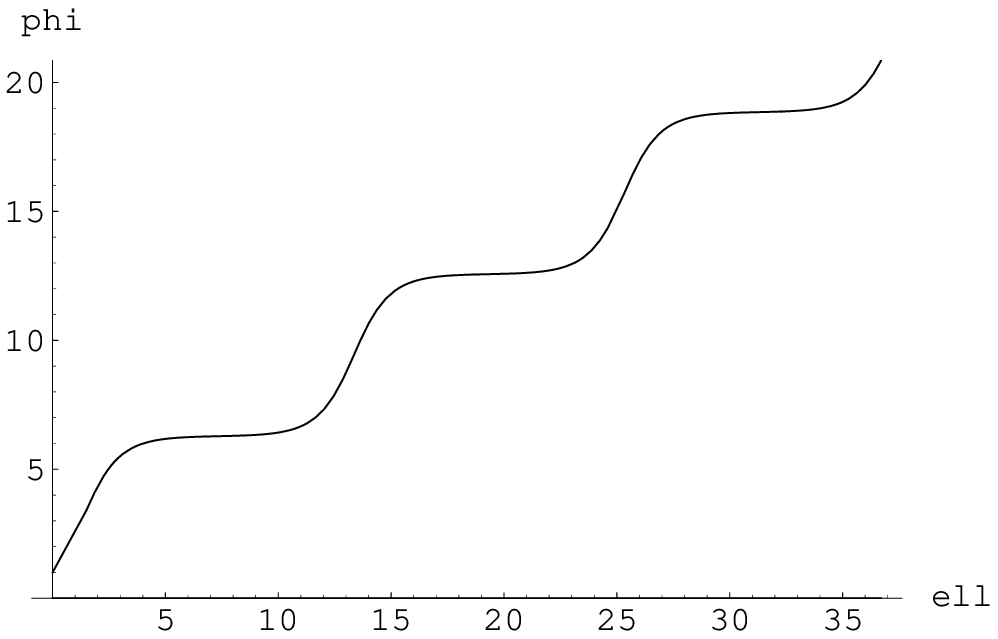,height=5cm,width=7cm}
\end{tabular}
\caption{Classical solution in the $Q=3$ sector ($n_0=0$, $n_R=3$)
at some values of $R$, in the case $\beta\phi_{0}=1$ and
$\beta\phi_{R}=2$.} \label{fig3kink}
\end{figure}

It is worth noticing that $\phi^{(+)}_{cl}(x)$ can be also used to
satisfy, at finite values of $R$, Dirichlet b.c. in sectors with
arbitrary topological charge (see Fig.\,\ref{fig3kink}), giving
rise to the correct UV behaviour (\ref{kinkeiUV}) with $Q = n_R -
n_0$. However, since $\phi^{(+)}_{cl}(x)$ always assumes the value
$\frac{2\pi}{\beta}(n_0+1)$ at $m\bar{x} = mx_{0} +
k\textbf{K}(k)$, which is once again the point going to infinity
when $k\to 1$, in the IR limit it can only correspond to $Q = 1$.
This result seems natural though, since in infinite volume, static
classical solutions can only describe those sectors of the theory
with $Q=0,\pm 1$, while time--dependent ones are needed for higher
values of $Q$. Hence, in the topological sectors with $|Q| > 1$
the space of states will contain, at classical level, the
time--dependent backgrounds, defined for any value of $R$ (which
are not discussed here), plus the static ones of the form
$\phi^{(+)}_{cl}(x)$, which however disappear from the spectrum as
$R\to\infty$.

\section{Summary}

In this Chapter we have shown how the semiclassical methods can
provide an analytic description of finite size effects in
two--dimensional quantum field theories displaying degenerate
vacua. In particular, we have applied the form factors technique
to study the correlation functions in SG and broken $\phi^4$
theories on a cylindrical geometry. Furthermore, implementing on
the cylinder the DHN quantization method, we have derived the
scaling functions of the ground (and excited) states in the
one--kink (with $\,Q=1$) sector of the SG model. The semiclassical
approach provides analytic and non--perturbative expressions for
the energy levels, valid for arbitrary values of the size $R$ of
the system, which permit to link the IR data of the massive theory
with the UV conformal data of CFT. It is particularly interesting
the application to non--integrable models, where the large $R$
behaviour of the obtained scaling functions can be compared with
L\"uscher's theory to extract information about the unknown
scattering properties of the model, as we have shown in the broken
$\phi^4$ case at the level of classical energy.

For the integrable case of sine--Gordon model, the next step in
the semiclassical program is the extension of the DHN method to
describe the multi--kink states ($\,Q=\pm 2,\pm 3,...$) as well as
the non--vacua (\lq\lq breather"--like) part of the $Q=0$ sector.
These states are related to certain time--dependent solutions on
the cylinder, i.e. to the finite volume analog of
soliton--soliton, soliton--antisoliton and breather solutions.
Although more complicated from the technical point of view, the
determination of these classical solutions and the study of their
scaling functions and form factors is a well stated open problem
in the semiclassical framework, which deserves further attention.
Another interesting problem to be studied is related to the the
exactness (or very high accuracy) of the semiclassical results,
which is a peculiar feature of SG model in infinite volume, as we
have observed in Sect.\,\ref{secSGinfvol}. It would be very useful
to understand whether similar phenomena take place for the
semiclassical scaling functions and form factors in finite volume
as well. An indication on this issue could be found by extending
to finite volume the analysis of higher loop quantum corrections
in the semiclassical expansion.

The semiclassical method is suited also for the description of
finite geometries with boundaries, and the example of the SG model
on a strip with Dirichlet b.c. has been considered. The resulting
analytic expressions for the energy levels permit to link the IR
data on the half-line with the UV conformal data of boundary CFT
at $c=1$. In comparison with a cylindrical geometry, an
interesting new feature of the quantum field theory defined on a
strip consists in a non--trivial (and non--perturbative)
semiclassical description of its vacuum sector. Therefore, we have
discussed in detail the classical solutions and energy levels in
the $Q=0$ case, together with the $Q=1$ that can also be described
by static backgrounds. As we have already mentioned, however, the
semiclassical methods are not restricted to static backgrounds
only, and a complete description of the theory in all sectors
requires also the study of time--dependent solutions. Finally, it
is worth noticing that the analysis performed here has natural and
direct extension to other quantum field theories with various
kinds of boundary conditions.

One of the advantages of the semiclassical method is that it works
equally well for both integrable and non--integrable models, if
they admit kink--type solutions. In fact, we have chosen to test
the efficiency of the semiclassical quantization on the example of
SG model, mainly because it leads to the simplest $N=1$ Lam\'e
equation. Static elliptic solutions for other models can be easily
obtained by integrating equation (\ref{firstorder}) with $A\neq 0$
and appropriate boundary conditions. This was done, for instance
in \cite{finvolff}, where we have derived the form factors between
kink states in the broken $\phi^{4}$ model on the cylinder with
twisted boundary conditions. In this case, the quantization of the
finite volume kink involves a Lam\'e equation with $N=2$ and will
be presented in \cite{preparation}. Lam\'e equations with $N>2$
are also expected to enter the quantization of other theories.

\sezioneapp{Free theory quantization on a finite
geometry}\label{appreg}

Let us consider a free bosonic field $\phi(x,t)$ of mass $m$
defined on a cylinder of circumference $R$, i.e. satisfying the
periodic boundary condition \EQ \phi(x + R, t) \,=\,\phi(x,t)
\,\,\,. \EN Imposing the equation of motion and the commutation
relation
$$
[\,\phi(x,t),\Pi(y,t)\,] \,=\,i \delta_P(x-y) \,\,\,,
$$
where
$\Pi(x,t) = \frac{\partial\phi}{\partial t}(x,t)$ is the conjugate
momentum of the field whereas
\[
\delta_P(x) \,=\,\frac{1}{R} \sum_{n=-\infty}^{\infty} e^{\frac{2
\pi i n}{R}\, x} \,\,\,\,\,\,\,\, , \,\,\,\,\,\,\,\, \delta_P(x+R)
\,=\,\delta_P(x)
\]
is the periodic version of the Dirac delta function, we obtain the
mode expansion of the field $\phi(x,t)$. This is given by \EQ
\phi(x,t) \,=\, \sum_{n=-\infty}^{\infty} \frac{1}{2 \omega_n R}
\left[\,A_n \,e^{i(p_n x - \omega_n t)} + A^{\dagger} \, e^{-i
(p_n x - \omega_n t)} \,\right] \,\,\,, \EN where
\[
[A_n,A^{\dagger}_m] \,=\,\delta_{n,m} \,\,\,,
\]
and \EQ \omega_n \,=\,\sqrt{p_n^2 + m^2} \,\,\,\,\,\,\,\, ,
\,\,\,\,\,\,\,\, p_n \,=\,\frac{2 \pi n}{R} \,\,\,\, n =0,\pm
1,\ldots \EN Using the above expansion together with the
commutation relation of $A$ and $A^{\dagger}$, it is easy to
compute the propagator of the field, given by \EQ
\Delta_F(x-x',t-t') \,=\, \langle \phi(x,t) \phi(x',t') \rangle
\,=\, \sum_{n=-\infty}^{\infty} \frac{1}{2\omega_n R}
\,e^{-i[\omega_n (t-t') -p_n (x-x')]} \,\,\,. \EN The vacuum
expectation value of the operator $\phi^2(x,t)$ is then formally
given by \EQ \langle \phi^2(x,t) \rangle \,=\,\Delta_F(0) \,\,\,
\EN and, by translation invariance, is independent from $x$ and
$t$. However this expression is divergent and needs therefore to
be regularized. Analogously to what has been done in the text for
the ground state energy ${\cal E}_0^{\text{vac}}(R)$, we need to
subtract the corresponding expression in the infinite volume, so
that the finite quantity, simply denoted by $\phi^2_0(R)$,
satisfies the usual normalization condition
\[
\lim_{R \rightarrow \infty} \phi^2_0 (R) \,=\,0\,\,\,.
\]
Hence we define \EQ \phi^2_0 (R) \,=\,  \frac{1}{2 R}
\,\sum\limits_{n=-\infty}^{\infty} \frac{1}{\sqrt{\left(\frac{2
\pi n}{R} \right)^{2}+m^{2}}}\, -\,\frac{1}{2 R}
\,\int\limits_{-\infty}^{\infty} dn \frac{1}{\sqrt{\left( \frac{2
\pi n}{R} \right)^{2}+m^{2}}} \;. \EN Isolating its zero mode, the
series needs just one subtraction, i.e.
$$
{\cal S}(r) \,\equiv\,
\sum\limits_{n=1}^{\infty}\frac{1}{\sqrt{n^{2}+\left(\frac{r}{2\pi}
\right)^{2}}} \,=\,
\sum\limits_{n=1}^{\infty}\left\{\frac{1}{\sqrt{n^{2}+\left(\frac{r}{2\pi}
\right)^{2}}}\,-\,\frac{1}{n}
\right\}\,+\,\sum\limits_{n=1}^{\infty}\frac{1}{n}\,\,\,.
$$
($r = m R$). In the above expression, the first series is now
convergent whereas the second series, which is divergent, has to
be combined with a divergence coming from the integral. Indeed we
have
$$
{\cal I}(r) \,\equiv\,
\int\limits_{0}^{\infty}dn\frac{1}{\sqrt{n^{2}+\left(\frac{r}{2\pi}
\right)^{2}}} \,=\, \lim_{\Lambda\to\infty}\left\{\ln
2\Lambda-\ln\frac{r}{2\pi}\right\}
-\lim_{\Lambda\to\infty}\ln\Lambda+\lim_{\Lambda\to\infty}\ln\Lambda\;,
$$
and the last term can be used to compose
(\ref{secondsubtraction}). Collecting the above expressions, it is
now easy to see that $\phi^2 _0(R)$ coincides  with the one
obtained doing the calculation in the other quantization scheme,
i.e. at a finite temperature. In fact, using the results of
Ref.\,\cite{lm}, this quantity can be expressed as \EQ \phi^2
_0(R) \,=\, \int_{-\infty}^{\infty} \frac{d\theta}{2\pi}
\frac{1}{e^{r \cosh\theta} -1} \,\,\,, \EN whose expansion in $r$
is given by \EQ \phi^2 _0(R) \,=\, \frac{1}{2 r} + \frac{1}{2\pi}
\left(\log\frac{r}{2\pi} + \gamma_E -\log 2\right) +
\sum_{n=1}^{\infty} \left( \frac{1}{\sqrt{(2 n \pi)^2 + r^2}} -
\frac{1}{2 n \pi}\right) \,\,\,. \EN

Also this result could have been directly obtained computing only
the finite part of the integral and using the prescription
(\ref{z(1)}).

\sezioneapp{Elliptic integrals and Jacobi's elliptic
functions}\label{appell} \setcounter{equation}{0} In this appendix
we collect the definitions and basic properties of the elliptic
integrals and functions used in the text. Exhaustive details can
be found in \cite{GRA}.

The complete elliptic integrals of the first and second kind,
respectively, are defined as
\begin{equation}\label{ellint}
\textbf{K}(k^{2})\,=\,\int\limits_{0}^{\pi/2}\frac{d\alpha}
{\sqrt{1-k^{2}\sin^{2}\alpha}}\;,\qquad
\textbf{E}(k^{2})\,=\,\int\limits_{0}^{\pi/2}d\alpha
\sqrt{1-k^{2}\sin^{2}\alpha}\;.
\end{equation}
The parameter $k$, called elliptic modulus, has to be bounded by
$k^{2} < 1$. It turns out that the elliptic integrals are nothing
but specific hypergeometric functions, which can be easily
expanded for small $k$:
\begin{eqnarray*}
\textbf{K}(k^{2})&=&\frac{\pi}{2}\;F\left(\frac{1}{2},\frac{1}{2},1;k^{2}\right)=
\frac{\pi}{2}\,\left\{1+\frac{1}{4}\,k^{2}+\frac{9}{64}\,k^{4} +
\ldots + \left[\frac{(2n-1)!!}{2^{n}n!}\right]^{2}k^{2n} + \ldots
\right\}\;,\\
\textbf{E}(k^{2})&=&\frac{\pi}{2}\;F\left(-\frac{1}{2},\frac{1}{2},1;k^{2}\right)=
\frac{\pi}{2}\,\left\{1-\frac{1}{4}\,k^{2}-\frac{3}{64}\,k^{4}+
\ldots -
\left[\frac{(2n-1)!!}{2^{n}n!}\right]^{2}\frac{k^{2n}}{2n-1} +
\ldots \right\}\;.
\end{eqnarray*}
Furthermore, for $k^{2}\to 1$, they admit the following expansion
in the so--called complementary modulus $k' = \sqrt{1-k^{2}}$:
\begin{eqnarray*}
\textbf{K}(k^{2})&=&
\log\frac{4}{k'}+\left(\log\frac{4}{k'}-1\right)\frac{k'^{2}}{4} +
\ldots\;,\\\nonumber \textbf{E}(k^{2})&=&1+\left(\log\frac{4}{k'}
- \frac{1}{2}\right)\frac{k'^{2}}{2} + \ldots \;.
\end{eqnarray*}
Note that the complementary elliptic integral of the first kind is
defined as
$$
\textbf{K}'(k^{2}) \,=\, \textbf{K}(k'^{2})\;.
$$

The function $\text{am}(u,k^{2})$, depending on the parameter $k$,
and called Jacobi's elliptic amplitude, is defined through the
first order differential equation
\begin{equation}\label{ellam}
\left(\frac{d\,\text{am}(u)}{du}\right)^{2}\,=\, 1 - k^{2}
\sin^{2} \left[\text{am}(u)\right]\;,
\end{equation}
and it is doubly quasi--periodic in the variable $u$:
$$
\text{am}\left(u+2n\textbf{K}+2im\textbf{K}'\right) \,= \,
n\pi+\text{am}(u)\;.
$$
The Jacobi's elliptic function $\text{sn}(u,k^{2})$, defined
through the equation
\begin{equation}\label{ellsn}
\left(\frac{d\,\text{sn}u}{du}\right)^{2} \,=\,
\left(1-\text{sn}^{2}u\right)\left(1-k^{2}\text{sn}^{2}u\right)\;,
\end{equation}
is related to the amplitude by
$\text{sn}\,u=\sin\left(\text{am}\,u\right)$, and it is doubly
periodic:
$$
\text{sn}\left(u+4
n\textbf{K}+2im\textbf{K}'\right)\,=\,\text{sn}(u)\;.
$$

\sezioneapp{Lam\'e equation}\label{lame} \setcounter{equation}{0}
The second order differential equation
\begin{equation}\label{lameeq}
\left\{\frac{d^{2}}{d u^{2}}-E-N(N+1){\cal P}(u)\right\}f(u) \,=\,
0\;,
\end{equation}
where $E$ is a real quantity, $N$ is a positive integer and ${\cal
P}(u)$ denotes the Weierstrass function, is known under the name
of $N$-th Lam\'e equation. The function ${\cal P}(u)$ is a doubly
periodic solution of the first order equation (see \cite{GRA})
\begin{equation}\label{defP}
\left(\frac{d{\cal P}}{du}\right)^{2}\,=\,4\left({\cal P} -
e_{1}\right)\,\left({\cal P} - e_{2}\right)\,\left({\cal P} -
e_{3}\right)\;,
\end{equation}
whose characteristic roots $e_{1},e_{2},e_{3}$ uniquely determine
the half--periods $\omega$ and $\omega'$, defined by
$$
{\cal P}\left(u+2n\omega+2 m\omega'\right)\,=\,{\cal P}(u)\;.
$$

The stability equation (\ref{SGsemiclper}) can be identified with
eq. (\ref{lameeq}) for $N=1$, $u=\bar{x}+i\textbf{K}'$ and
$E=\frac{2-k^{2}}{3}-k^{2}\bar{\omega}^{2}$ in virtue of the
relation between ${\cal P}(u)$ and the Jacobi elliptic function
$\text{sn}(u,k)$ (see formulas 8.151 and 8.169 of \cite{GRA}):
\begin{equation}\label{Psn}
k^{2}\text{sn}^{2}(\bar{x},k) \,=\,{\cal P}(\bar{x}+i\textbf{K}')
+ \frac{k^{2}+1}{3}\;.
\end{equation}
Relation (\ref{Psn}) holds if the characteristic roots of ${\cal
P}(u)$ are expressed in terms of $k^{2}$ as
\begin{equation*}
e_{1}\,=\,\frac{2-k^{2}}{3}\;, \qquad
e_{2}\,=\,\frac{2k^{2}-1}{3}\;,\qquad e_{3}
\,=\,-\frac{1+k^{2}}{3}\;,
\end{equation*}
and, as a consequence, the real and imaginary half periods of
${\cal P}(u)$ are given by the elliptic integrals of the first
kind
\begin{equation*}
\omega \,=\,\textbf{K}(k)\;,\qquad \omega' \,=\, i \textbf{K}'(k)
\;.
\end{equation*}
All the properties of Weierstrass functions that we will use in
the following are specified to the case when this identification
holds.

In the case $N=1$ the two linearly independent solutions of
(\ref{lameeq}) are given by (see \cite{whit})
\begin{equation}
f_{\pm a}(u)\,=\,\frac{\sigma(u\pm a)}{\sigma(u)}\;e^{\mp\,u
\,\zeta(a)}\;,
\end{equation}
where $a$ is an auxiliary parameter defined through ${\cal
P}(a)=E$, and $\sigma(u)$ and $\zeta(u)$ are other kinds of
Weierstrass functions:
\begin{equation}
\frac{d\,\zeta(u)}{du} \,=\, - {\cal P}(u)\;,\qquad
\frac{d\,\log\sigma(u)}{du} \,=\, \zeta(u)\;,
\end{equation}
with the properties
\begin{eqnarray}\nonumber
&&\zeta(u+2\textbf{K}) \,=\,\zeta(u) + 2
\zeta(\textbf{K})\;,\\\label{zsprop}
&&\sigma(u+2\textbf{K})\,=\,-\,e^{2(u+\textbf{K})
\zeta(\textbf{K})} \sigma(u)\;.
\end{eqnarray}
As a consequence of eq. (\ref{zsprop}) one obtains the Floquet
exponent of $f_{\pm a}(u)$, defined as
\begin{equation}
f(u+2 \textbf{K}) \,=\, f(u)e^{i F(a)}\;,
\end{equation}
in the form
\begin{equation}
F(\pm a) \,=\, \pm 2 i\left[\textbf{K}\,\zeta(a) -
a\,\zeta(\textbf{K})\right]\;.
\end{equation}
The spectrum in the variable $E$ of eq. (\ref{lameeq}) with $N=1$
is divided in allowed/forbidden bands depending on whether $F(a)$
is real or complex for the corresponding values of $a$. We have
that $E < e_{3}$ and $e_{2} < E < e_{1}$ correspond to allowed
bands, while $e_{3} < E < e_{2}$ and $E > e_{1}$ are forbidden
bands. Note that if we exploit the periodicity of ${\cal P}(a)$
and redefine $a\rightarrow a' = a + 2 n \omega+2 m \omega'$, this
only shifts $F$ to $F' = F+2 m \pi$.

The function $\zeta(u)$ admits a series representation
\cite{hancock} that will be very useful for our purposes in Sect.
\ref{sectscaling}:
\begin{equation}
\zeta(u)\,=\,\frac{\pi}{2\textbf{K}}\,\cot\left(\frac{\pi u}
{2\textbf{K}}\right)+\left(\frac{\textbf{E}}{\textbf{K}}
+\frac{k^{2}-2}{3}\right)\,u + \frac{2\pi}{\textbf{K}}
\sum\limits_{n=1}^{\infty} \frac{h^{2n}}{1-h^{2n}}\,\sin
\left(\frac{n \pi u}{\textbf{K}}\right)\;,
\end{equation}
where $h = e^{-\pi\textbf{K}'/\textbf{K}}$. The small-$k$
expansion of this expression gives
\begin{equation}
\label{expzeta} \hspace{-3.9cm}\zeta(u)\,=\,\left(\cot u +
\frac{u}{3}\right) \,+\,\frac{k^{2}}{12}\left(u - 3 \cot u + 3 u
\cot^{2}u\right)\,+
\end{equation}
\begin{equation*}
+\,\frac{k^{4}}{64}\left(-3 u + ( 4 u^{2} - 5 ) \cot u + u
\cot^{2}u + 4 u^{2} \cot^{3}u + \sin 2u \right) + \ldots
\end{equation*}
(note that $h\approx
\left(\frac{k}{4}\right)^{2}+O\left(k^{4}\right)$). A similar
expression takes place for ${\cal P}(u)$, by noting that ${\cal
P}(u) = - \frac{d\,\zeta(u)}{du}$.

Finally, in the complementary limit $k'=\sqrt{1-k^2}\to 0$, the
Weierstrass functions can be expanded as
\begin{eqnarray*}
\zeta(u)&=&\coth
u-\frac{u}{3}+\frac{(k')^2}{6}\left(u-\frac{3}{2}\coth
u+\frac{3}{2}\,u\;\text{csch}^2 u\right)+...\\
{\cal P}(u)&=&\coth^2
u-\frac{2}{3}+\frac{(k')^2}{6}\left(2-3\coth^2 u+3\,u\;\coth
u\;\text{csch}^2 u\right)+...
\end{eqnarray*}

\chapter*{Conclusions and Outlook} \setcounter{equation}{0}
\addcontentsline{toc}{chapter}{Conclusions and Outlook}

\pagestyle{plain}

In this thesis, we have described some fruitful applications of
the semiclassical methods in the study of non--integrable QFT and
finite--size effects in two dimensions.

The mass spectrum of a generic QFT can be explored
non--perturbatively by exploiting the analytic properties of
semiclassical form factors of the local fields between kink
states, which are available as the Fourier transforms of the
classical kinks. With this technique, we have specifically studied
two non--integrable theories, given by the $\phi^4$ interaction in
the $\mathbb{Z}_2$ broken symmetry phase and by the double
sine--Gordon model, which describe relevant physical phenomena and
find many applications in statistical mechanics and condensed
matter physics. Along similar lines, it would be worthwhile to
extend the semiclassical analysis to other non--integrable
theories which describe the scaling region of interesting
statistical systems, like for instance the $\phi^{6}$ potential
with three degenerate vacua, associated to the tricritical Ising
model with both leading and subleading thermal perturbations.
Moreover, further attention should be dedicated to the false
vacuum decay in presence of unstable classical backgrounds, that
we have studied in the specific case of the double sine--Gordon
model. In fact, the semiclassical method is capable of providing
interesting results about the dynamics of decay processes.

An opportune generalization of the semiclassical quantization
technique has led us to the analytical study of finite--size
effects in QFT. In fact, once the proper classical solutions are
identified on the geometry of interest, one can implement on them
the relativistically refined Goldstone and Jackiw's result and the
DHN quantization method in order to estimate, respectively, the
correlators and the spectrum. In particular, we have studied form
factors and correlation functions for the SG and broken $\phi^4$
theories on a twisted cylinder, and we have derived the scaling
functions of the ground (and excited) states in few sectors of the
SG model defined both on a cylinder and on a strip with Dirichlet
boundary conditions. An interesting open problem is the systematic
investigation of the finite--size spectrum, which consists in
finding time--dependent classical solutions in the remaining
topological sectors of the theory, and in applying to them the
semiclassical quantization procedure. Furthermore, it would be
useful to compare numerically the semiclassical results with the
ones available from different techniques for the SG model in
finite volume. This, in fact, would permit to test quantitatively
the efficiency of the semiclassical approximation, as we have
successfully done in the infinite volume case. A closely related
subject is the study of two-dimensional QFT at finite temperature,
which are known to represent the physical interpretation of the
cylindrical geometry in which the compactified variable is
Euclidean time, instead of space. A more profound understanding of
the relation between finite-volume and finite-temperature pictures
will give the opportunity to gain further insight in these both
quantization schemes.

As a final remark, it is worth to emphasize again that the
semiclassical method is equally suited to describe integrable and
non--integrable models, if they admit kink--type solutions. This
feature, that we have fully exploited in the analysis of
infinite--volume spectra, makes the analytical study of scaling
functions in non--integrable theories a well stated problem, more
complicated only from the technical point of view, but
conceptually analogous to the procedure illustrated for the
sine--Gordon model.

\chapter*{Acknowledgements} \setcounter{equation}{0}
\addcontentsline{toc}{chapter}{Acknowledgements}

\pagestyle{plain}

I'm extremely grateful to Giuseppe Mussardo for introducing and
guiding me into research with enthusiasm and generosity, for
teaching me the importance of method and ideas over notions and
computations, and for making me part of a lively and stimulating
research group.

My deepest gratitude goes also to Galen Sotkov, who shared with me
many months of intense and close work (in spite of the ocean
between us!) and transmitted to me his passion for computations
and intellectual curiosity. I thank Gesualdo Delfino for his
numerous patient explanations, for the illuminating discussions
and for his example in pursuing simplicity and clearness in spite
of any technical complications. Furthermore, I thank Germ\'an
Sierra, who acted as the external referee for this thesis, for the
valuable comments and suggestions. I am grateful to all the
members of the Elementary Particle and Mathematical Physics
sectors in SISSA, for their teaching and for the stimulating
environment they create. Additional special thanks go to Giovanni
Feverati, for his help with LateX subtleties, and to Lorenzo
Brualla, for his uncommon patience and style in listening and
advising me.

It is a pleasure to thank who has been close to me during these
years of study and work, and it is impossible to mention everyone.
In particular, I'm grateful to my father for his guide and
example, and for his deep interest in my activity, often resulted
in endless stimulating discussions. Equally important has been the
support of my mother and my brother Alberto, in sharing the joys
and overcoming the difficulties of these years. I am grateful to
Carlo for giving a special meaning to this experience in Trieste,
but also for his scientific criticism, accompanied by the most
encouraging support. Finally, I thank all the close friends for
having contributed to make so happy and exciting these years, and
all the distant ones for having followed my \lq\lq adventures"
with the warmest participation.

\end{document}